\documentclass[onecolumn, longauth]{aa_modif}

\usepackage[english]{babel}

\usepackage{natbib}
\bibpunct{(}{)}{;}{a}{}{,} 

\usepackage{amsmath}
\usepackage{graphicx}
\usepackage{txfonts}
\usepackage[colorinlistoftodos]{todonotes}
\usepackage[colorlinks=true, allcolors=blue]{hyperref}
\usepackage{enumitem}
\usepackage{verbatim}
\usepackage[hang,small,bf]{caption}
\usepackage{rotating}
\usepackage{url}
\usepackage{times}
\usepackage{color, colortbl}
\usepackage[alsoload=hep]{siunitx}
\usepackage{amsmath}
\usepackage{amsfonts}
\usepackage{amssymb}
\usepackage{bm}
\usepackage{graphicx}
\usepackage{epsfig}
\usepackage{mathtools}
\usepackage{euclid}
\usepackage[normalem]{ulem}
\usepackage{multirow}
\usepackage{tabularx}

\usepackage[nameinlink]{cleveref}

\newcommand{\filepath}[1]{files//#1}

\def\gs{\mathrel{\raise1.16pt\hbox{$>$}\kern-7.0pt %
\lower3.06pt\hbox{{$\scriptstyle \sim$}}}}         %
\def\ls{\mathrel{\raise1.16pt\hbox{$<$}\kern-7.0pt %
\lower3.06pt\hbox{{$\scriptstyle \sim$}}}}

\def\aap{A\&A}

\def\apjl{ApJ}

\def\apjs{ApJS}

\definecolor{amaranth}{rgb}{0.9, 0.17, 0.31}
\definecolor{forestgreen(web)}{rgb}{0.13, 0.55, 0.13}
\definecolor{lavender(web)}{rgb}{0.9, 0.9, 0.98}
\definecolor{cosmiclatte}{rgb}{1.0, 0.97, 0.91}
\definecolor{jonquil}{rgb}{0.98, 0.85, 0.37}
\definecolor{khaki(x11)(lightkhaki)}{rgb}{0.94, 0.9, 0.55}
\definecolor{thistle}{rgb}{0.85, 0.75, 0.85}

\newcommand{\de}{\mathrm{DE}}

\newcommand{\bear}{\begin{eqnarray}}
\newcommand{\eear}{\end{eqnarray}}

\newcommand{\tr}{{\rm tr}}

\definecolor{orange}{rgb}{1,0.5,0}
\definecolor{darkorange}{rgb}{0.69,0.33,0.13}
\definecolor{fidcol}{rgb}{0.7,0,0}
\definecolor{mkcol}{rgb}{0.5,0,0.5}
\definecolor{mmcol}{rgb}{0.7,0.17,0.31}
\definecolor{dscol}{rgb}{0.6,0.1,0.2}
\definecolor{mccol}{rgb}{0.2,0.4,0.6}
\definecolor{darkgreen}{rgb}{0.05,0.5,0.06}
\definecolor{carnelian}{rgb}{0.7, 0.11, 0.11}

\begin{document}

\def\sectionautorefname{Sect.}
\def\subsectionautorefname{Sect.}
\def\figureautorefname{Fig.}
\def\equationautorefname~#1\null{%
  Eq.~(#1)\null
}

\let\subsectionautorefname\sectionautorefname
\let\subsubsectionautorefname\sectionautorefname
\Crefname{table}{Table}{Tables}
\Crefname{figure}{Figure}{Figures}
\Crefname{section}{Sect.}{Sects.}
\Crefname{equation}{Eq.}{Eqs.}
\Crefname{eqnarray}{Eq.}{Eqs.}

\title{\Euclid preparation: VII. Forecast validation for \Euclid cosmological probes}

\author{\Euclid Collaboration\thanks{\email{euclid-istf@mpe.mpg.de}}: A.~Blanchard$^{1}$, S.~Camera$^{2,3}$, C.~Carbone$^{4,5,6}$, V.F.~Cardone$^{7}$, S.~Casas$^{8}$, S.Clesse$^{88,89}$, S.~Ili\'c$^{1,9}$, M.~Kilbinger$^{10,11}$, T.~Kitching$^{12}$, M.~Kunz$^{13}$
% ...
}

\institute{$^{1}$ Institut de Recherche en Astrophysique et Plan\'etologie (IRAP), Universit\'e de Toulouse, CNRS, UPS, CNES, 14 Av. Edouard Belin, F-31400 Toulouse, France\\
$^{2}$ INFN-Sezione di Torino, Via P. Giuria 1, I-10125 Torino, Italy\\
$^{3}$ Dipartimento di Fisica, Universit\'a degli Studi di Torino, Via P. Giuria 1, I-10125 Torino, Italy\\
$^{4}$ Dipartimento di Fisica, Universit\'a degli Studi di Milano "Aldo Pontremoli", Via Celoria 16, I-20133 Milano, Italy\\
$^{5}$ INFN-Sezione di Milano, Via Celoria 16, I-20133 Milano, Italy\\
$^{6}$ INAF-IASF Milano, Via Alfonso Corti 12, I-20133 Milano, Italy\\
$^{7}$ INAF-Osservatorio Astronomico di Roma, Via Frascati 33, I-00078 Monteporzio Catone, Italy\\
$^{8}$ AIM, CEA, CNRS, Universit\'{e} Paris-Saclay, Universit\'{e} Paris Diderot, Sorbonne Paris Cit\'{e}, F-91191 Gif-sur-Yvette, France\\
$^{9}$ CEICO, Institute of Physics of the Czech Academy of Sciences, Na Slovance 2, Praha 8, Czech Republic\\
$^{10}$ CEA Saclay, DFR/IRFU, Service d'Astrophysique, Bat. 709, 91191 Gif-sur-Yvette, France\\
$^{11}$ Institut d'Astrophysique de Paris, 98bis Boulevard Arago, F-75014, Paris, France\\
$^{12}$ Mullard Space Science Laboratory, University College London, Holmbury St Mary, Dorking, Surrey RH5 6NT, UK\\
$^{13}$ Universit\'e de Gen\`eve, D\'epartement de Physique Th\'eorique and Centre for Astroparticle Physics, 24 quai Ernest-Ansermet, CH-1211 Gen\`eve 4, Switzerland\\
$^{14}$ University of California, Berkeley, Berkeley, CA 94720, USA\\
$^{15}$ Institute of Cosmology and Gravitation, University of Portsmouth, Portsmouth PO1 3FX, UK\\
$^{16}$ Institute Lorentz, Leiden University, PO Box 9506, Leiden 2300 RA, The Netherlands\\
$^{17}$ School of Physics and Astronomy, Queen Mary University of London, Mile End Road, London E1 4NS, UK\\
$^{18}$ Universit\'e St Joseph; UR EGFEM, Faculty of Sciences, Beirut, Lebanon\\
$^{19}$ Max Planck Institute for Extraterrestrial Physics, Giessenbachstr. 1, D-85748 Garching, Germany\\
$^{20}$ Departamento de F\'isica, FCFM, Universidad de Chile, Blanco Encalada 2008, Santiago, Chile\\
$^{21}$ Institute of Space Sciences (ICE, CSIC), Campus UAB, Carrer de Can Magrans, s/n, 08193 Barcelona, Spain\\
$^{22}$ Institut d’Estudis Espacials de Catalunya (IEEC), 08034 Barcelona, Spain\\
$^{23}$ Argelander-Institut f\"ur Astronomie, Universit\"at Bonn, Auf dem H\"ugel 71, 53121 Bonn, Germany\\
$^{24}$ INAF-Osservatorio Astronomico di Brera, Via Brera 28, I-20122 Milano, Via E. Bianchi 46, I-23807 Merate, Italy\\
$^{25}$ Departamento de Astrof\'{i}sica, Universidad de La Laguna, E-38206, La Laguna, Tenerife, Spain\\
$^{26}$ Instituto de Astrof\'{i}sica de Canarias. Calle V\'{i}a L\`{a}ctea s/n, 38204, San Crist\'{o}bal de la Laguna, Tenerife, Spain\\
$^{27}$ INAF-Osservatorio di Astrofisica e Scienza dello Spazio di Bologna, Via Piero Gobetti 93/3, I-40129 Bologna, Italy\\
$^{28}$ Dipartimento di Fisica e Astronomia, Universit\'a di Bologna, Via Gobetti 93/2, I-40129 Bologna, Italy\\
$^{29}$ INFN-Sezione di Bologna, Viale Berti Pichat 6/2, I-40127 Bologna, Italy\\
$^{30}$ Universit\"ats-Sternwarte M\"unchen, Fakult\"at f\"ur Physik, Ludwig-Maximilians-Universit\"at M\"unchen, Scheinerstrasse 1, 81679 M\"unchen, Germany\\
$^{31}$ INAF-Osservatorio Astronomico di Trieste, Via G. B. Tiepolo 11, I-34131 Trieste, Italy\\
$^{32}$ INAF-Osservatorio Astrofisico di Torino, Via Osservatorio 20, I-10025 Pino Torinese (TO), Italy\\
$^{33}$ APC, AstroParticule et Cosmologie, Universit\'e Paris Diderot, CNRS/IN2P3, CEA/lrfu, Observatoire de Paris, Sorbonne Paris Cit\'e, 10 rue Alice Domon et L\'eonie Duquet, 75205, Paris Cedex 13, France\\
$^{34}$ Department of Astronomy, University of Geneva, ch. d'\'Ecogia 16, CH-1290 Versoix, Switzerland\\
$^{35}$ INFN-Sezione di Roma Tre, Via della Vasca Navale 84, I-00146, Roma, Italy\\
$^{36}$ Department of Mathematics and Physics, Roma Tre University, Via della Vasca Navale 84, I-00146 Rome, Italy\\
$^{37}$ INAF-Osservatorio Astronomico di Capodimonte, Via Moiariello 16, I-80131 Napoli, Italy\\
$^{38}$ Instituto de Astrof\'isica e Ci\^encias do Espa\c{c}o, Universidade do Porto, CAUP, Rua das Estrelas, PT4150-762 Porto, Portugal\\
$^{39}$ INFN-Bologna, Via Irnerio 46, I-40126 Bologna, Italy\\
$^{40}$ Dipartimento di Fisica e Scienze della Terra, Universit\'a degli Studi di Ferrara, Via Giuseppe Saragat 1, I-44122 Ferrara, Italy\\
$^{41}$ INAF, Istituto di Radioastronomia, Via Piero Gobetti 101, I-40129 Bologna, Italy\\
$^{42}$ Universit\'e C\^{o}te d'Azur, Observatoire de la C\^{o}te d’Azur, CNRS, Laboratoire Lagrange, France\\
$^{43}$ Institut de F\'isica d’Altes Energies IFAE, 08193 Bellaterra, Barcelona, Spain\\
$^{44}$ Instituto de Astrof\'isica e Ci\^encias do Espa\c{c}o, Faculdade de Ci\^encias, Universidade de Lisboa, Tapada da Ajuda, PT-1349-018 Lisboa, Portugal\\
$^{45}$ Department of Physics "E. Pancini", University Federico II, Via Cinthia 6, I-80126, Napoli, Italy\\
$^{46}$ INFN section of Naples, Via Cinthia 6, I-80126, Napoli, Italy\\
$^{47}$ INAF-Osservatorio Astrofisico di Arcetri, Largo E. Fermi 5, I-50125, Firenze, Italy\\
$^{48}$ Centre National d'Etudes Spatiales, Toulouse, France\\
$^{49}$ Institute for Astronomy, University of Edinburgh, Royal Observatory, Blackford Hill, Edinburgh EH9 3HJ, UK\\
$^{50}$ University of Nottingham, University Park, Nottingham NG7 2RD, UK\\
$^{51}$ ESAC/ESA, Camino Bajo del Castillo, s/n., Urb. Villafranca del Castillo, 28692 Villanueva de la Ca\~nada, Madrid, Spain\\
$^{52}$ Universit\'e de Lyon, F-69622, Lyon, France ; Universit\'e de Lyon 1, Villeurbanne; CNRS/IN2P3, Institut de Physique Nucl\'eaire de Lyon, France\\
$^{53}$ Departamento de F\'isica, Faculdade de Ci\^encias, Universidade de Lisboa, Edif\'icio C8, Campo Grande, PT1749-016 Lisboa, Portugal\\
$^{54}$ Instituto de Astrof\'isica e Ci\^encias do Espa\c{c}o, Faculdade de Ci\^encias, Universidade de Lisboa, Campo Grande, PT-1749-016 Lisboa, Portugal\\
$^{55}$ Aix-Marseille Univ, CNRS, CNES, LAM, Marseille, France\\
$^{56}$ Department of Physics, Oxford University, Keble Road, Oxford OX1 3RH, UK\\
$^{57}$ INFN-Padova, Via Marzolo 8, I-35131 Padova, Italy\\
$^{58}$ Department of Physics \& Astronomy, University of Sussex, Brighton BN1 9QH, UK\\
$^{59}$ Centre for Extragalactic Astronomy, Department of Physics, Durham University, South Road, Durham, DH1 3LE, UK\\
$^{60}$ Aix-Marseille Univ, CNRS/IN2P3, CPPM, Marseille, France\\
$^{61}$ Leiden Observatory, Leiden University, Niels Bohrweg 2, 2333 CA Leiden, The Netherlands\\
$^{62}$ von Hoerner \& Sulger GmbH, Schlo{\ss}Platz 8, D-68723 Schwetzingen, Germany\\
$^{63}$ Max-Planck-Institut f\"ur Astronomie, K\"onigstuhl 17, D-69117 Heidelberg, Germany\\
$^{64}$ Department of Physics and Helsinki Institute of Physics, Gustaf H\"allstr\"omin katu 2, 00014 University of Helsinki, Finland\\
$^{65}$ Institut de Physique Nucl\'eaire de Lyon, 4, rue Enrico Fermi, 69622, Villeurbanne cedex, France\\
$^{66}$ Institute of Theoretical Astrophysics, University of Oslo, P.O. Box 1029 Blindern, N-0315 Oslo, Norway\\
$^{67}$ Istituto Nazionale di Astrofisica (INAF) - Osservatorio di Astrofisica e Scienza dello Spazio (OAS), Via Gobetti 93/3, I-40127 Bologna, Italy\\
$^{68}$ Institute for Computational Cosmology, Department of Physics, Durham University, South Road, Durham, DH1 3LE, UK\\
$^{69}$ INAF-Osservatorio Astronomico di Padova, Via dell'Osservatorio 5, I-35122 Padova, Italy\\
$^{70}$ University of Paris Denis Diderot, University of Paris Sorbonne Cit\'e (PSC), 75205 Paris Cedex 13, France\\
$^{71}$ Sorbonne Universit\'e, Observatoire de Paris, Universit\'e PSL, CNRS, LERMA, F-75014, Paris, France\\
$^{72}$ Observatoire de Sauverny, Ecole Polytechnique F\'ed\'erale de Lau- sanne, CH-1290 Versoix, Switzerland\\
$^{73}$ INFN, Sezione di Lecce, Via per Arnesano, CP-193, I-73100, Lecce, Italy\\
$^{74}$ Department of Mathematics and Physics E. De Giorgi, University of Salento, Via per Arnesano, CP-I93, I-73100, Lecce, Italy\\
$^{75}$ Perimeter Institute for Theoretical Physics, Waterloo, Ontario N2L 2Y5, Canada\\
$^{76}$ Department of Physics and Astronomy, University of Waterloo, Waterloo, Ontario N2L 3G1, Canada\\
$^{77}$ Centre for Astrophysics, University of Waterloo, Waterloo, Ontario N2L 3G1, Canada\\
$^{78}$ Space Science Data Center, Italian Space Agency, via del Politecnico snc, 00133 Roma, Italy\\
$^{79}$ European Space Agency/ESTEC, Keplerlaan 1, 2201 AZ Noordwijk, The Netherlands\\
$^{80}$ Jet Propulsion Laboratory, California Institute of Technology, 4800 Oak Grove Drive, Pasadena, CA, 91109, USA\\
$^{81}$ Centro de Investigaciones Energ\'eticas, Medioambientales y Tecnol\'ogicas (CIEMAT), Avenida Complutense 40, 28040 Madrid, Spain\\
$^{82}$ Universidad Polit\'ecnica de Cartagena, Departamento de Electr\'onica y Tecnolog\'ia de Computadoras, 30202 Cartagena, Spain\\
$^{83}$ Kapteyn Astronomical Institute, University of Groningen, PO Box 800, 9700 AV Groningen, The Netherlands\\
$^{84}$ IFPU, Institute for Fundamental Physics of the Universe, via Beirut 2, 34151 Trieste, Italy\\
$^{85}$ SISSA, International School for Advanced Studies, Via Bonomea 265, I-34136 Trieste TS, Italy\\
$^{86}$ INFN, Sezione di Trieste, Via Valerio 2, I-34127 Trieste TS, Italy\\
$^{87}$ Infrared Processing and Analysis Center, California Institute of Technology, Pasadena, CA 91125, USA\\
$^{88}$ CURL \& IRMP , Louvain  University,  2  Chemin  du  Cyclotron,  1348  Louvain-la-Neuve,  Belgium \\
$^{89}$ NaXys, University  of  Namur,  Rempart  de  la  Vierge  8,  5000  Namur,  Belgium\\
}

\date{}
\authorrunning{Euclid Collaboration et al.}

\abstract
{}
{The \Euclid space telescope will measure the shapes and redshifts of galaxies to reconstruct the expansion history of the Universe and the growth of cosmic structures. The estimation of the expected performance of the experiment, in terms of predicted constraints on cosmological parameters, has so far relied on various individual methodologies and numerical implementations, which were developed for different observational probes and for the combination thereof. In this paper we present validated forecasts, which  combine both theoretical and observational ingredients for different cosmological probes. This work is presented to provide the community with reliable numerical codes and methods for \Euclid cosmological forecasts.}
{We describe in detail the methods adopted for Fisher matrix forecasts, which were applied to galaxy clustering, weak lensing, and the combination thereof. We estimated the required accuracy for \Euclid forecasts and outline a methodology for their development. We then compare and improve different numerical implementations, reaching uncertainties on the errors of cosmological parameters that are less than the required precision in all cases. Furthermore, we provide details on the validated implementations, some of which are made publicly available, in different programming languages, together with a reference training-set of input and output matrices for a set of specific models. These can be used by the reader to validate their own implementations if required.} 
{We present new cosmological forecasts for \Euclid. We find that results depend on the specific cosmological model and remaining freedom in each setting, for example flat or non-flat spatial cosmologies, or different cuts at non-linear scales.  The numerical implementations  are now reliable  for these settings. We present the results for an optimistic and a pessimistic choice for these types of settings. We demonstrate that the impact of cross-correlations is particularly relevant for models beyond a cosmological constant and may allow us to increase the dark energy figure of merit by at least a factor of three.}
{}


\maketitle

\newpage


\section{Introduction}
\Euclid\footnote{\url{http://www.euclid-ec.org/}} will explore the expansion history of the Universe and the 
evolution of large-scale cosmic structures by measuring shapes and redshifts of galaxies, 
covering $15\,000\,\deg^2$ of the sky, up to redshifts of about $z=2$. It will be able to 
measure up to 30 million \citep{Pozzetti:2016} spectroscopic redshifts, 
which can be used for galaxy clustering measurements and 2 billion photometric 
galaxy images, which can be used for weak lensing observations (for more details, see \citealt{2018LRR....21....2A} and \citealt{RB}). 
The \Euclid telescope is a 1.2-metre three-mirror anistigmat and it has two instruments that observe in optical and 
near-infrared wavelengths. In the optical, high-resolution images will be observed through a broadband filter with a wavelength range from $500$ to $800\,\nm$ (the VIS band) and a pixel resolution of $0.1$ arcseconds that will result in a galaxy catalogue 
complete down to a magnitude of $24.5$ AB in the VIS band. In the near-infrared, there will be imaging in the \textit{Y}, \textit{J}, and \textit{H} bands with a pixel 
resolution of $0.3$ arcseconds and a grism (slitless) spectrograph with one `blue' grism ($920-1250\,\nm$) and three `red' grisms 
($1250-1850\,\nm$; in three different orientations). This will result in a galaxy catalogue complete to magnitude $24$ in the \textit{Y}, \textit{J}, 
and \textit{H} bands. The optical and near-infrared instruments share a common field-of-view of $0.53$ deg$^2$. 

This paper is motivated by the challenging need to have reliable cosmological forecasts for \Euclid. 
This is required for the verification of \Euclid's performance before launch in order to assess the impact of the 
real design decisions on the final results \citep[for an updated technical description of the mission design, see][]{2016SPIE.9904E..0OR}.  It is also necessary that the different cosmological probes in \Euclid 
use a consistent and well-defined framework for forecasting and that the tools used for such forecasts 
are rigorously validated and verified. These forecasts will then provide the reference for the performance of \Euclid to the scientific community. Forecasting in this context refers to the 
question of how well \Euclid will perform when using the survey data to distinguish the 
standard cosmological model from simple alternative dark energy scenarios, given its current, up-to-date specifications. This paper 
represents the outcome of an intense activity of comparison among different and independent forecasting codes, from various 
science working groups in the \Euclid Collaboration. This work was conducted by a group known as the inter-science taskforce for forecasting (IST:F) and it presents the results of the code comparison for forecasts based on Fisher-matrix analyses. 

\Euclid forecasts were previously made in the Definition Study Report \citep[hereafter `Red Book',][]{RB}, using Fisher matrix predictions, 
similar to those presented in this paper. The description of the Red Book forecasts was not detailed enough 
due to the length constraints of that document. In contrast, in this paper we specify in detail the methodology and the procedures followed 
in order to validate the results and verify their robustness.  
The Red Book did specify the \Euclid instrument, telescope, and survey specifications in detail, and these have been used 
by the community to create a suite of predictions for 
the expected uncertainties on cosmological measurements for a range of models. Many forecasts have been made 
for several scenarios beyond the parameters of the standard $\Lambda$CDM model  (see in particular \citealt{2018LRR....21....2A}): some, for example \citet{2012JCAP...11..052H}, have demonstrated the expected constraints 
on neutrino masses; others have examined the prospect of constraining the properties of different 
dark matter candidates; and most commonly the different parameterisation of dark energy and modified gravity properties have been considered \citep[e.g.][among many 
others]{DiPorto_Amendola_Branchini_2012,Majerotto:2012,2012MNRAS.423.3631W,2013arXiv1308.6070D, 2017PDU....18...73C}. 

In this paper we focus on the primary cosmological probes in \Euclid: weak lensing (WL); photometric galaxy clustering (GC$_{ph}$);  spectroscopic galaxy clustering (GC$_{s}$); their combination; and the addition of the data vector of cross-correlation (XC) between GC$_{ph}$ and WL. For each of the probes we define a recipe, that details the methodology used to make forecasts,
which should serve as a useful documentation to enable the scientific community to be able to implement and reproduce the same methodology.
The comparison presented here has involved several different codes. Therefore, rather than providing
a single open source code, in a specific programming language, we grant a validation stamp for all the codes that
took part in the code comparison and satisfied the requirements. These are listed in  \autoref{sec:allcodes}. 

Since the Red Book the \Euclid baseline survey and instrument specifications have been updated according to the progress in our knowledge of astrophysics, 
and in the evolution of the design of the telescope and instruments. In this paper we also update the Red Book forecasts by including some of these changes, which 
directly impact the statistical precision with which \Euclid can constrain cosmological parameters. Such effects include changes to the survey 
specifications, changes to the instrument that impact the populations of galaxies probed, and changes in the understanding of astrophysical systematic effects. Some of these changes, although not all, have an impact on \Euclid's ability to constrain cosmological parameters \citep{Majerotto:2012,Wang:2013b}.

For the GC observables a significant change since the Red Book has been an increased understanding of
our target galaxy sample \citep{Pozzetti:2016}, which has influenced the instrument design and the consequent changes in the optimisation of the
survey strategy \citep{Markovic:2017}. One of the difficulties in predicting the performance of  \Euclid's spectroscopic survey accuracy has always been the poor knowledge of the number density 
and evolution of the galaxy population that will be detected as H$\alpha$ emitters, and whose redshifts will be measured by the survey. More 
recent observations imply lower number densities at $z>1$ than originally assumed based on information available at the time of the Red Book. 
This has resulted in a re-assessment of performance and the resulting cosmological parameter constraints. The main consequence of a reduced 
number density $n(z)$ of observed galaxies is an increase in the shot noise and a consequent increase of the parameter uncertainties. The science 
reach of an experiment depends heavily on the `effective' volume covered, which in the case of GC studies is a function combining the cosmological 
volume with the product of the mean density of the target population and the amplitude of the clustering power spectrum; this is denoted $nP$. 
A further limitation of the baseline forecasts in the Red Book was that the observing strategy consisted of just two rotations, that is to say  
observing position angles (PA -- also called dithers) for each of the two red and blue channels. 
This had the disadvantage that when a single PA observation is lost, the spectrum corresponding to the affected grism (covering a 
specific blue or red redshift range) could not be reliably cleaned of possible contaminating signals from adjacent spectra. In this paper we 
update the GC forecasts to reflect the changes in the expected PAs and the $n(z)$. 

For WL one of the main difficulties in making cosmological forecasts has been the ability to accurately model the 
intrinsic alignments of galaxies -- a local 
orientation of galaxies that acts to mimic the cosmological lensing signal. In the Red Book a non-parametric model was chosen in redshift and in scale, and prior information 
on the model's nuisance parameters included the expected performance of spectroscopic galaxy-galaxy lensing results at the time of \Euclid's 
launch. Since the time of Red Book there has been an increased attention in this area and several physical models have been proposed \citep{iareview1, iareview2, iareview3} that model intrinsic 
alignments in a more realistic manner. A second area of attention has been in the appreciation of the impact of
the small-scale (high $k$-mode) matter power spectrum on WL two-point statistics \citep[e.g.][]{2018PhRvD..98d3532T,copeland}. 
This has led to improved models of the impact of baryonic feedback, neutrino mass, and non-linear clustering on small scales. 
In this paper we update the WL forecasts to reflect these improved models. 

As well as \Euclid, there are several other large-scale cosmological experiments and survey projects that have produced forecasts since the Red Book, often including \Euclid forecasts for comparison. 
For example the Square Kilometer Array\footnote{\url{http://skatelescope.org/}} (SKA) is expected to deliver a wealth of radio wavelength data. Its Phase 1 Mid array (SKA1-MID) is 
going to be commissioned on similar timescales to \Euclid and is offering a unique opportunity for multi-wavelength synergies. As discussed in \citet{Kitching:2015pta}, a 
multitude of cross-correlation statistics will be provided by comparing \Euclid and SKA clustering and weak lensing data, with the additional advantage of 
the cross-correlations being less affected by systematic effects that are relevant for one type of survey but not the other \citep[for more recent reviews, see also][]{Bull:2018lat,Bacon:2018dui}. Cross-correlation forecasts for an SKA1-MID 21 cm 
intensity mapping survey and \Euclid galaxy clustering have been performed in \citet{Fonseca:2015laa} and \citet{Pourtsidou:2016dzn}. Forecasts for the cross-correlations of shear 
maps between \Euclid and Phase 2 of SKA-MID assuming $15\,000 \,\deg^2$ of sky overlap have been performed in \cite{Harrison:2016stv}. 
Three large optical surveys that will join \Euclid are the Dark Energy Spectroscopic Instrument\footnote{\url{http://desi.lbl.gov}} (DESI), the Large Synoptic Survey Telescope\footnote{\url{https://www.lsst.org}} (LSST), and the Wide Field Infrared 
Survey Telescope\footnote{\url{https://wfirst.gsfc.nasa.gov}} ({WFIRST}). Joint analyses of their data, combined with data coming from cosmic microwave background (CMB) missions can give new insights to a broad 
spectrum of science cases, ranging from galaxy formation and evolution to dark energy and the neutrino mass.  Some of the possibilities of combining \Euclid, LSST, and WFIRST have been described in \citet{Jain:2015cpa} and \citet{2017ApJS..233...21R}; LSST and {WFIRST} have also made their own forecasts e.g. \cite{CCL}. 
Concentrating 
on the merits for large-scale structure measurements, we note that the complementarity of \Euclid, LSST and 
{WFIRST} results in significant improvement in cosmological parameter constraints. Similarly, joint ventures with $\gamma$-ray experiments such as the \textit{Fermi} satellite or the proposed {e-ASTROGAM} mission may lead to an improvement in our understanding of the particle nature of dark matter \citep{2013ApJ...771L...5C,2018JHEAp..19....1D}. In this paper we do not make forecasts for 
other surveys, but the results presented here are an update for the \Euclid forecasts presented in previous inter-survey comparisons. 

This paper is arranged as follows.
In \autoref{Sec:Cosmology} we describe the cosmological models considered in our forecasts and the parameters that characterise
them. In \autoref{Sec:Fisher} we introduce the Fisher-matrix formalism to estimate errors on cosmological parameters and describe 
our methodology to apply it to the different cosmological probes in \Euclid. 
In \autoref{sec:codecomparison} we present the forecasting codes included in our analysis and describe in detail the code-comparison
 procedure that we implemented, as well as the level of agreement found for the different cases and parameter spaces considered. 
 In \autoref{sec:results} we present the final cosmological parameter forecasts for the different probes in \Euclid, considered 
 separately and in combination. Finally, we present our main conclusions in \autoref{sec:conclusions}.
 In this paper we focus on cosmological forecasts that explicitly do no include systematic effects relating to instrument 
 design and performance of data reduction algorithms (although we do include astrophysical systematic effects); the assessment of the impact of these on cosmological parameter estimation is 
 subject to an exercise known as science performance verification (SPV) in the \Euclid Consortium, and will be presented in a 
 series of separate papers.

\section{The cosmological context\label{Sec:Cosmology}}

In the following subsections we describe different cosmological models and their associated parameters. 
We start by introducing and discussing the most commonly used background cosmological quantities.

\subsection{Background quantities} \label{sec:cosmo-bcg}
Applying Einstein's field equations of general relativity
to the metric line element\footnote{In this paper we use a $(-,+,+,+)$ signature for the metric line element.} of a homogeneous and isotropic universe allows us to derive the Friedmann equations, which describe the time evolution of the scale factor $a(t)$ as a function of the curvature parameter $K$ and the energy content of the universe, characterised by the total energy density $\rho$ and pressure $p$. They read
\begin{align}
H^2(t) &\equiv \left[\frac{\dot{a}(t)}{a(t)}\right]^2 =  \frac{8\pi G}{3}\rho(t) - \frac{K c^2}{a^2(t)}\,,\label{eq:GC:Friedmann} \\
\frac{\ddot{a}(t)}{a(t)} &= - \frac{4\pi G}{3}\left[\rho(t) + 3\frac{p(t)}{c^2}\right]\,.\label{eq:GC:acceleration} 
\end{align}
Here the overdot denotes differentiation with respect to the cosmic time $t$, $c$ is the speed of light, $G$ the gravitational constant and $K$ can assume negative, zero, or positive values for open, flat or closed spatial geometry respectively.
A more convenient parameter than the time $t$ to describe the evolution of the scale factor is the redshift, 
$z = a_0/a -1$, where $a_0$ corresponds to the present-day value of the scale factor, normalised as $a_0 = 1$; we 
convert the $t$ variable into $a$ or $z$. The Hubble expansion rate, $H(z)$, can be expressed as
\begin{equation} \label{eq:Hz-general}
H(z) = H_0 E(z)\,,
\end{equation}
with $H_0\equiv H(z=0)$ being the Hubble parameter today, which is commonly written as 
 \begin{equation}
H_0 = 100h\,{\rm km\,s^{-1}\,Mpc^{-1}}\, , \label{eq:GC:H} 
\end{equation} where $h$ is the dimensionless Hubble parameter. The function $E(z)$ will be later specified for each cosmological 
model we use.
For a given value of $H(z)$, there is a value of $\rho$ that results in a spatially flat geometry ($K=0$). That is the critical density 
\begin{equation}
\rho_{\rm crit}(z) = \frac{3H^2(z)}{8\pi G} \, . \label{eq:GC:rhoc} 
\end{equation}
For a generic component labelled $i$, we define the density parameter $\Omega_{i}(z) \equiv \rho_i (z)/\rho_{\rm{crit} }(z)$. Based on \autoref{eq:GC:Friedmann}, we can also introduce 
an effective curvature density parameter 
$\Omega_K(z) = -K c^2/[a^2(z) H^2(z)]$. With these definitions, \autoref{eq:GC:Friedmann} takes the form
\begin{equation}
\sum_{i=1}^N \Omega_i(z) + \Omega_K(z) = 1 \, ,
\label{eq:density_parameters}
\end{equation}
where the sum is over all $N$ species considered in the model. We use present-day values of the density parameters, which, unless specified otherwise, are indicated with a subscript $0$. 

\Cref{eq:GC:Friedmann,eq:GC:acceleration} can be combined into an energy conservation equation
that specifies the relation between $\rho$ (and $p$) and the scale factor. A solution of this equation requires to specify 
the properties of each energy component in the form of an equation of state, $p=p(\rho)$; we specify the latter in
terms of the equation of state parameter $w \equiv p / \rho c^2$, which can be redshift-dependent. 
For the case in which the equation of state parameter is constant in time the energy conservation
equation implies
\begin{equation}
\rho_i(a) \propto a^{-3\left(1+w_i\right)} \, . \label{eq:rho_of_a}
\end{equation}
Once the relations $\rho_i(a)$ are known, these can be used in \autoref{eq:GC:Friedmann} to find a solution 
for $a(t)$. We also note that $-3[1+w_i]$ always gives $\partial \ln \rho_i(a)/ \partial\ln a$. 

The matter energy density at late times is mainly in the form of baryons and cold dark matter (CDM) particles, which
are described by $w_{\rm b} = w_{\text{c}}=0$. The photon radiation density is characterised by $w_{\gamma}=1/3$.
A contribution from massive-neutrinos can be described by a varying equation of state parameter $w_{\nu}$, which matches $w_{\gamma}$
at early times and $w_{\text{c}}$ when they become non-relativistic. 
For the purpose of galaxy clustering and weak lensing measurements, we can consider radiation density to be negligible, 
effectively setting $\Omega_{\gamma,0}=0$, and treat the massive-neutrinos as part of the total matter contribution, 
with $\Omega_{\rm m,0} = \Omega_{\rm c,0}+\Omega_{\rm b,0}+ \Omega_{\nu,0}$, 
since they are non-relativistic at the low redshifts relevant for these probes. 

In the context of general relativity, cosmic acceleration requires a fluid, dubbed `dark energy' (DE), 
with an equation of state $w_{\de}<-1/3$. 
The standard model of cosmology, commonly referred to as the $\Lambda$CDM model,  assumes that this phenomenon is due to 
the presence of a cosmological constant, referred to as $\Lambda$, described by a constant equation of state $w_\Lambda=-1$, which,
 according to  \autoref{eq:rho_of_a}, corresponds to a time-independent energy density $\rho_\Lambda$. 
 The $\Lambda$CDM model currently fits observations very well \citep[see][and references therein]{2018arXiv180706209P} but it suffers from several fundamental problems: the observed value of the
  cosmological constant is many orders of magnitude smaller than the theoretical predictions 
  (the \textit{cosmological constant problem}); the fact that the $\Lambda$ and CDM densities are similar today while they  
  have evolved very differently through time marks our epoch as a special time in the evolution of the Universe 
  (the \textit{coincidence problem}); and in general the model and its parameters cannot be predicted from physical principles.\\
  
  A more general scenario for the component responsible for cosmic acceleration postulates a dynamical DE, with a
redshift-dependent equation of state parameter $w_{\de}(z)$. A commonly used and well-tested parameterisation of the time dependence is
\begin{equation}\label{eq:weos}
w_{\de}(z) = w_0 + w_a \frac{z}{1+z} \, ,
\end{equation}
where $w_0$ is the present ($z=0$) value of the equation of state and $w_a$ is a measure of its time variation. 
In this case, the evolution of the DE density obeys
\begin{equation}
\rho_{\de}(z) = \rho_{\de,0}(1+z)^{3\left(1+w_0+w_a\right)}\,\exp{\left[-3w_a\frac{z}{1+z}\right]}\, .
\label{eq:rho_de}
\end{equation}
Using \Cref{eq:rho_of_a,eq:rho_de} in \autoref{eq:GC:Friedmann}, the function $E(z)$ defined in 
\autoref{eq:Hz-general} becomes
\begin{equation}\label{eq:Ez-w0wa}
E(z)=\sqrt{\Omega_{\text{m},0} \left(1+z\right)^3 + \Omega_{\de,0}\left(1+z\right)^{3\left(1+w_0+w_a\right)}\exp\left[-3w_a\frac{z}{1+z}\right] 
+ \Omega_{K,0} \left(1+z\right)^2}\, ,
\end{equation}
with the current DE density, $\Omega_{\text{DE},0}$, satisfying the relation
$\Omega_{\text{DE},0} = 1-\Omega_{\text{m},0} 
- \Omega_{K,0}$. The $\Lambda$CDM model can be recovered by setting $w_0 = -1$ and $w_a = 0$, in which 
case the function $E(z)$ takes the form
\begin{equation} \label{eq:Ez}
E(z)=\sqrt{\Omega_{\text{m},0} \left(1+z\right)^3 + \Omega_{\Lambda,0}  
+ \Omega_{K,0} \left(1+z\right)^2}\, .
\end{equation}

\subsection{Distance measurements}
\label{sec:distances}

The comoving distance 
to an object at redshift $z$ can be computed as
\begin{equation}
r(z) = \frac{c}{H_0}\int_0^z \frac{{\rm d}z}{E(z)} \, .
\label{eq:comoving_distance}
\end{equation}
Although this quantity is not a direct observable, it is closely related to other distance definitions that are 
directly linked with cosmological observations.
A distance that is relevant for our forecasts is the angular diameter distance, whose definition is based on 
the relation between the apparent angular size of an object and its true physical size  in Euclidean space, 
and is related to the comoving distance by
\begin{equation}
D_\text{A}(z)=
\begin{dcases}
     \left(1+z\right)^{-1}\frac{c}{H_0}\frac{1}{\sqrt{\left|\Omega_{K,0}\right|}}  \sin \left[\sqrt{\left|\Omega_{K,0}\right|} \frac{H_0}{c} r(z) \right] & \text{if}\ \Omega_{K,0} <0 \\
      \left(1+z\right)^{-1} r(z) & \text{if} \ \Omega_{K,0} =0\\
      \left(1+z\right)^{-1} \frac{c}{H_0}\frac{1}{\sqrt{\Omega_{K,0}}} \sinh \left[\sqrt{\Omega_{K,0}} \frac{H_0}{c} r(z) \right] & \text{if}\ \Omega_{K,0} >0.\\
\end{dcases}   \label{eq:GC:Da}
\end{equation}
Also relevant for our forecasts is the comoving volume of a region covering a solid angle $\Omega$ 
between two redshifts  $z_{\rm i}$ and $z_{\rm f}$, which is given by 
\begin{equation}
V(z_{\rm i},z_{\rm f}) = \Omega \int_{z_{\rm i}}^{z_{\rm f}}  \frac{r^2(z)}{\sqrt{1-K r^2(z)}}\frac{c\,{\rm d}z}{H(z)}  \, ;\label{eq:Vcomoving}
\end{equation}
for a spatially flat universe ($K = 0$), this becomes
\begin{equation}
\label{eq:survol}
V(z_{\rm i},z_{\rm f}) = \Omega  \int_{r(z_{\rm i})}^{r(z_{\rm f})} r^2 \,{\rm d}r = \frac{\Omega}{3} \left[ r^3(z_{\rm f}) - r^3(z_{\rm i}) \right] \, .
\end{equation}
These expressions allow us to compute the volume probed by \Euclid within a given redshift interval. 

\subsection{Linear perturbations}
\label{sec:linear_perturbation}
The structure we see today on large scales grew from minute density fluctuations
generated by a random process in the primordial Universe. The evolution of these fluctuations, for
non-relativistic matter on sub-horizon scales, can be described by ideal fluid
equations \citep{Peebles_1980}. Density fluctuations for a given component $i$ are characterised by the
density contrast
\begin{equation}
\delta_i (\vec x, z) \equiv \rho_i(\vec x, z) / \bar \rho_i(z) - 1 \, , 
\label{eq:density_contrast}
\end{equation}
which quantifies the deviations of the density field $\rho_i(\vec x, z)$ around the mean spatial density $\bar \rho_i(z)$ over space, where $\vec x$ is a three-dimensional comoving coordinate at a redshift $z$. To describe these fluctuations statistically, it is convenient to work in Fourier space by decomposing $\delta$ into plane waves,
\begin{equation}
  \delta_i (\vec {x}, z) = \int \frac{{\rm d}^3 k}{(2\pi)^3} \tilde \delta_i(\vec {k}, z)\exp (-{\rm i}\vec {k}\cdot \vec {x}) \, .
  \label{eq:delta_FT}
\end{equation}
The power spectrum, $P_{i}(\vec k, z)$, for the generic component $i$ is defined implicitly as
\begin{equation}
  \left\langle \tilde \delta_i(\vec k, z) \tilde \delta_i(\vec k^\prime, z) \right\rangle
  = (2\pi)^3 \delta_{\rm D}(\vec k + \vec k^\prime) P_{i}(\vec k, z)\,,
  \label{eq:P_i}
\end{equation}
where $\delta_{\rm D}$ is the Dirac delta function. Under the assumptions of statistical homogeneity and isotropy, the power spectrum 
can only depend on $k=\left| \vec k \right|$ and $z$.

The dimensionless primordial power spectrum of the curvature perturbation $\zeta$ generated by inflation is parameterised as a 
power law 
\begin{equation}
 \mathcal{P}_{\rm \zeta}(k) = A_{\rm s} \left(\frac{k}{k_0}\right)^{n_{\rm s}-1} \, ,
  \label{eq:P_large_scales}
\end{equation}
where $A_{\rm s}$ is the amplitude of the primordial scalar perturbation, the scalar spectral index $n_{\rm s}$ 
measures the deviation (tilt) from scale invariance ($n_{\rm s} = 1$), and $k_0$ is a pivot scale. The corresponding power spectrum defined in \autoref{eq:P_i} in terms of density perturbations is related to the primordial one via the transfer function $\mathcal T_i$ through
\begin{equation}
\label{def_Pi}
 P_i(k,z)= {2\pi^2} \mathcal T_i^2(k,z)\mathcal{P}_{\zeta}(k)k \,,
\end{equation}
for a  generic component $i$. In the early Universe during radiation domination, curvature perturbations with comoving scales smaller than the horizon
are suppressed, whereas super-horizon fluctuations remain unaffected, until they enter the horizon. In the matter dominated era, 
curvature perturbations on all scales remain constant. This implies that a characteristic scale corresponding to the epoch of matter-radiation 
equality is imprinted on  the shape of the transfer function, and hence on the matter power spectrum.

The growth of density fluctuations obeys a second-order differential equation. 
At early enough times, when those fluctuations are still small, the fluid equations can be linearised. 
During matter domination, considering matter as a pressureless ideal fluid,
the equation for the evolution of the density contrast becomes
\begin{equation}
  \ddot \delta_{\rm m}\left( \vec {k}, z \right) + 2 H \dot \delta_{\rm m}\left( \vec {k}, z \right) - \frac{3 H_0^2 \Omega_{{\rm m},0}}{2 a^3} \delta_{\rm m}\left( \vec {k}, z \right) = 0 \, .
  \label{eq:delta_ODG}
\end{equation}
In the $\Lambda$CDM scenario with no massive-neutrinos, this equation can be written in terms of the redshift as
\be
\delta_{\rm m}''(\vec {k}, z)+\left[\frac{H'(z)}{H(z)}-\frac{1}{1+z}\right]\delta_{\rm m}'(\vec {k}, z)-\frac{3}{2}\frac{\Omega_\text{m}(z)}{(1+z)^2}\delta_{\rm m}(\vec {k}, z) = 0 \, ,
\label{eq:sec-gc-comparison-ode-delta}
\ee
where we neglect dark energy perturbations, the prime refers to the derivative with respect to $z$, and $\Omega_{\rm m}(z)$ is given by 
\be
\Omega_{\rm m}(z) = \frac{\Omega_{\rm m,0}(1+z)^3}{E^2(z)} \, .
\ee
The solutions $\delta_{\rm m}(\vec {k}, z)$ of \autoref{eq:sec-gc-comparison-ode-delta}, at late times, are scale-independent, which motivates the 
introduction of the growth factor $D(z)$ through
\be
\delta_{\rm m}(\vec {k}, z) = \delta_{\rm m}(\vec {k}, z_i) \frac{D(z)}{D(z_i)} \, ,
\ee
where $z_i$ is an arbitrary reference redshift in the matter-dominated era. A useful quantity is the growth rate parameter, defined as
\be\label{eq:fgrowth}
f(a) = \frac{{\rm d}\ln D(a)}{{\rm d}\ln a} = - \frac{{\rm d}\ln D(z)}{{\rm d} \ln(1+z)}\,.
\ee
Using these definitions,\footnote{We note that throughout, as is common in the community, we use interchangeable arguments of functions where the arguments are also functionally related. For example $D(z)$ or $D(a)$ where $a\equiv 1/(1+z)$. In all cases it should be clear from context of the equation why the function is expressed in the manner that it is.} the growth rate satisfies a first-order differential equation
\be \label{eq:ode-eff}
f'(z)-\frac{f(z)^2}{1+z}-\left[\frac{2}{1+z}-\frac{H'(z)}{H(z)}\right]f(z)+\frac{3}{2}\frac{\Omega_\text{m}(z)}{1+z}=0 \, ,
\ee
with initial condition $f(z=z_{\rm i}) =1$. Using \autoref{eq:fgrowth} the solution for $D(z)$ can be 
expressed in terms of $f(z)$  by the integral
\be
D(z) = D(z=0)\exp\left[-\int_{0}^{z}{\rm d} z'\,\frac{f(z')}{1+z'}\right]\,.
\label{eq:growthfactor-gc-val}
\ee

In $\Lambda$CDM the late-time matter growth is scale-independent, so that the transfer function $\mathcal T_{\rm m}(k,z)$ can be split into a 
scale-dependent part $T_{\rm m}(k)$ (normalised so that $T_{\rm m}\to1$ for $k\to0$) and the scale-independent growth factor $D(z)$ introduced above. 
A convenient way to express the power spectrum defined in \autoref{def_Pi} for matter is 
\begin{equation}
P_{\delta\delta}(k, z) = \left ( \frac{{\sigma_8}}{\sigma_{\rm N}} \right )^2 \left[\frac{D(z)}{D(z=0)}\right]^2 T_{\rm m}^2(k)  k^{n_{\rm s}}\, ,
  \label{eq: pddlinMath}
\end{equation}
with the normalisation constant
\begin{equation}
\sigma_{\rm N}^2 =  \frac{1}{2 \pi^2} \int{{\rm d}k\,T_{\rm m}^2(k) |W_{\rm TH}(k R_8)|^2 k^{n_{\rm s} + 2}}\,,
\label{eq: sigmaN}
\end{equation}
where $W_{\rm TH}(x) = 3(\sin{x} - x \cos{x})/x^3$ is the Fourier transform of the top-hat filter, and $R_8 = 8 h^{-1}$ Mpc.
The pivot scale $k_0$ and the amplitude of the scalar mode $A_{\rm s}$ are absorbed into the normalisation, which is designed to give a desired value of ${\sigma_8}$, the root mean square (rms) of present-day linearly evolved density fluctuations in spheres of 
$8\,h^{-1}$ Mpc, which is given by
\begin{equation}
  \sigma_8^2 = \frac{1}{2 \pi^2} \int{{\rm d}k\,P_{\delta\delta}(k, z = 0) \left|W_{\rm TH}(k R_8)\right|^2 k^2}\,.
  \label{eq: sigma8}
\end{equation}
Hence, the generic power spectrum described in \autoref{def_Pi} relates to the matter power spectrum of \autoref{eq: pddlinMath} when the transfer function of species $i$ is
\begin{equation}
\mathcal T_{\rm m}(k,z)=\frac{\sigma_8k_0^{(n_{\rm s}-1)/2}}{\sigma_{\rm N}\pi\sqrt{2A_{\rm s}}}\frac{D(z)}{D(z=0)}T_{\rm m}(k)\,.
\end{equation}
Since we consider models beyond $\Lambda$CDM, we need to specify the behaviour of the dark energy perturbations. We use a 
minimally-coupled scalar field, dubbed quintessence \citep{Wetterich_1988,Ratra_Peebles_1988}, which is often considered as the standard 
`dynamical dark energy'.  At the level of linear perturbations, this choice corresponds to a fluid with a sound speed equal to the speed of 
light ($c_{\rm s,\de}^2=c^2$), and no anisotropic stress ($\sigma_\de = 0$). This implies that it is smooth deep inside the horizon, and does not 
develop significant fluctuations, in which case the same form of the growth and power spectrum equations given above are still valid. 
In addition, we want to allow for the possibility that $w_\de(z)$ crosses $w=-1$ 
\citep[which is not allowed in quintessence, but could happen in 
multi-field scenarios,][]{2017PhRvD..96f3524P}, and to achieve this we use the Parameterised Post-Friedmann (PPF) prescription \citep{Hu:2007pj, 2008PhRvD..78h7303F, Ade:2015rim}.
In the following, we drop the subscript DE in $w$ and ${ c_{\rm s}^2}$, but keep it, for clarity, in $\sigma_{\rm DE}$.

\subsection{Parameterising the growth of structure: choice of $\gamma$}\label{sec:cosmo-growth}
Outside of the $\Lambda$CDM framework, both the background and the perturbations can be modified \citep[see][for a collection of scenarios and constraints]{2016A&A...594A..14P}. A simple way that was extensively used in the past to model the modified growth of perturbations 
is based on the observation that in $\Lambda$CDM the growth rate of \autoref{eq:fgrowth} is well approximated by
\begin{equation}\label{eq:ggrowth}
f(z) = \left[\Omega_{\rm m}(z)\right]^\gamma \, ,
\end{equation}
with a constant growth index parameter $\gamma \approx 0.55$ \citep{Lahav:1991wc, 2005PhRvD..72d3529L}. A scenario with modified 
growth then corresponds to a different value of $\gamma$. However, for most realistic modified gravity models, a 
constant growth index $\gamma$ is too restrictive if scale dependence exists. 
More importantly, the prescription of \autoref{eq:ggrowth} is incomplete as a general description of the 
evolution of perturbations -- which in general requires, {under the assumption of adiabaticity}, at least two degrees of freedom as a function of time and space \citep{Amendola:2007rr} -- 
and so we prefer a more complete parameterisation, as discussed below. Nonetheless, we include 
$\gamma$ as a parameter in our analysis in order to facilitate comparison with the \Euclid Red Book \citep{RB} where 
this case was explored. We notice that the assumption done in this case of no modifications to the lensing potential, detailed further in the rest of this section, effectively reduces the degrees of freedom to a single one. Generalisations or alternative parameterisations are left for future work.

As mentioned, we need two functions of time and space to describe the evolution of the perturbations in general. 
To ensure that the recipes for different probes consistently implement the same assumptions for the linear evolution of perturbations, 
we need to relate $\gamma$ explicitly to these two functions. In order to understand this relation, it is convenient to make 
these two free functions explicit in the perturbation equations. Since these functions are free, there are different choices 
(equally general) that can be made to define them. Here we express the temporal and spatial metric potentials, $\Psi$ and $\Phi$, in 
terms of the $(\mu_{\rm MG},\Sigma_{\rm MG})$\footnote{The standard notation is $(\mu,\Sigma)$, which variables are, however, already 
used for the cosine of the angle between \vec{k} and the line-of-sight direction and the surface mass density. We therefore label 
with subscript 'MG' the quantities that refer to modifications of gravity.} 
parameterisation \citep[see e.g.][]{2016A&A...594A..14P}. The first free function is 
$\mu_{\rm{MG}}$, which parameterises the growth of structure and is 
implicitly defined as a function of scale factor and scale via
\begin{equation}\label{def_mu}
 k^2\Phi=-\mu_{\rm{MG}}(a,k)4\pi G a^2\sum_i{\left[\rho_i\delta_i+3(\rho_i+p_i/c^2)\sigma_i\right]} \, ,
\end{equation}
where $\rho_i$ is the energy density of the generic component $i$, $p_i$ its pressure and $\delta_i$ its comoving density perturbation, while $\sigma_i$ is the anisotropic stress, which is non-vanishing for relativistic species. 
The second free function $\Sigma_{\rm MG}$ describes the deflection of light and is defined through
\begin{equation}\label{def_sigma}
 k^2\left[\Psi+\Phi\right]=-\Sigma_{\rm MG}(a,k)4\pi G a^2\sum_i{\left[2\rho_i\delta_i-3(\rho_i+p_i/c^2)\sigma_i\right]} \, .
\end{equation}
An alternative option is the pair $(\mu_{\rm MG}, \eta)$, with $\eta$ defined as 
\begin{equation}
 \eta(a,k)=\frac{\Psi}{\Phi}+\frac{3\sum_i (\rho_i+p_i/c^2)\sigma_i}{\sum_i [ \rho_i\delta_i+3(\rho_i+p_i/c^2)\sigma_i]}\,,
\label{eta_def_full}
\end{equation}
which reduces to\footnote{Colloquially known as `slip'.} 
\begin{equation}
\eta(a,k) \approx \Psi/\Phi\,,
\end{equation}
for negligible shear, as happens at the (low) redshifts of interest for \Euclid. Equivalently, combining \Cref{def_mu,eta_def_full}, one has
\begin{equation}\label{eq:neweta}
 k^2\left[\Psi-\eta(a,k)\Phi\right]=\mu_{\rm{MG}}(a,k)12\pi G a^2\sum_i{(\rho_i+p_i/c^2)\sigma_i}\,.
\end{equation}
These definitions differ (at high redshift) from the ones used in \cite{2016A&A...594A..14P}, since they make the anisotropic term explicit in the
equations, separating it from contributions beyond $\Lambda$CDM that may be contained in $\mu_{\rm{MG}}$, $\Sigma_{\rm MG}$, and $\eta$. The case for 
$\mu_{\rm{MG}} = 1$, $\Sigma_{\rm MG} = 1$ (or equivalently $\eta = 1$) corresponds to $\Lambda$CDM, while any deviation -- either due to modified gravity 
or to extra relativistic species -- is encoded in at least one of the functions being different from $1$. 
At low redshifts, the anisotropic stress terms are negligible, and the equations above are simplified, matching the definition used 
in \cite{2016A&A...594A..14P}. 

Linear perturbations for a model are then fixed only once we make a choice for two of these free functions (such as ($\mu_{\rm{MG}}$, $\Sigma_{\rm MG}$) or ($\mu_{\rm{MG}}$, $\eta$)). 
When assuming \autoref{eq:ggrowth}, the function $\mu_{\rm{MG}}$ defined in \autoref{def_mu}, which in general is a function of time and space, can be converted into a function of the scale factor and of $\gamma$. In particular,  (see \citealt{Mueller:2016kpu}, where $\mu_{\rm{MG}}, \Sigma_{\rm MG}$ are called 
$G_{\rm M}, G_{\rm L}$, respectively) $\mu_{\rm{MG}}$ is related to $\gamma$ via the following expression:
\begin{equation}
\mu_{\rm{MG}}(a, \gamma)=\frac{2}{3}\Omega_{\rm m}^{\gamma-1}\left[\Omega_{\rm m}^{\gamma}+2+\frac{H'}{H}+\gamma\frac{\Omega_{\rm m}'}{\Omega_{\rm m}}+\gamma'\ln{\Omega_{\rm m}}\right]\,,\label{eq:mugamma}
\end{equation} 
where the prime denotes differentiation with respect to $\ln a$ and $\Omega_{\rm m}=\Omega_{\rm m}(a)$; for a constant $\gamma$, the last term vanishes and $\mu_{\rm MG}$ only depends on the scale factor $a$ through $\Omega_{\rm m}$ and $H$. Fixing $\gamma$, however, does not fix the second free function defining linear perturbations (such as $\Sigma_{\rm MG}$ or $\eta$ described above), for which there is still a choice to be made to define the model. 

A possible choice for the second condition that fixes linear perturbations is $\Sigma_{\rm MG}=1$, since in this case light deflection (and therefore the lensing potential) 
is the same as in $\Lambda$CDM. This is also a reasonable choice from a theoretical point of view, as it is realised in standard scalar tensor theories, 
such as Brans--Dicke and $f(R)$ \citep[see][]{Amendola:2007rr, Pogosian:2016pwr}. In the limit $\Sigma_{\rm MG}\rightarrow 1$,  the usual 
equations for the lensing power spectrum are valid (see Eq.~\ref{eq: kijiiend}), which means the weak lensing description used for 
$\Lambda$CDM 
can still be applied. This choice is therefore also the one typically implicitly adopted in past analyses of $\gamma$, such as the one described 
in the Red Book. We restrict our analysis to the choice ($\gamma$, $\Sigma_{\rm MG} = $1) here as well to facilitate the comparison with previous analysis.

Finding a value of $\gamma$ different from $\Lambda$CDM (i.e. from a value around $0.55$) is then only related to having $\mu_{\rm{MG}}$ differ from 1 (i.e.\ from the expected value of this function in $\Lambda$CDM), which in turn physically means that the matter spectrum $P_{\delta\delta}(k,z)$ is affected by a different growth rate. Details on how $\gamma$ is specifically treated in 
different observational probes, and in the non-linear regime, is further discussed, separately, 
in \Cref{Sec:model_beyond_gc,Sec:model_beyond_wl,errors_shear}.

\subsection{Impact of neutrinos}
\label{sec:neutrinos}
In the presence of massive neutrinos, 
the definition of the linear growth rate in \autoref{eq:fgrowth} needs to be modified to 
allow for a dependence on both redshift and scale, $f(z,k)$, even at the linear level. 
To handle such dependencies in a semi-analytical way, approximations to the growth rate $f(z,k)$ 
as a function of the neutrino fraction $f_\nu = \Omega_{\nu,0}/\Omega_{{\rm m},0}$ have been developed. 
One approach is the fitting formulae of \cite{Kiakotou:2007pz}, where
 \begin{equation}
 f(z, k; f_\nu, \Omega_{\de,0}, \gamma) \approx \mu_\nu (k,f_{\nu}, \Omega_{\de,0})\Omega_{m}^{\gamma}(z)\,,
 \label{eq:ffull}
 \end{equation}
 with
 \begin{equation}
 \mu_\nu(k,f_{\nu},\Omega_{\de,0}) \equiv 1-A(k)\Omega_{\de,0}f_{\nu}+B(k)f_{\nu}^{2}-C(k)f_{\nu}^{3}\,,
 \label{eq:mu}
 \end{equation}
and the functions $A(k)$ , $B(k)$, and $C(k)$  have been obtained via fits of power spectra 
computed using the Boltzman code \texttt{CAMB} \citep{Lewis:1999bs}. 

\begin{figure*}[h!]
\centering
\includegraphics[width=12.3cm]{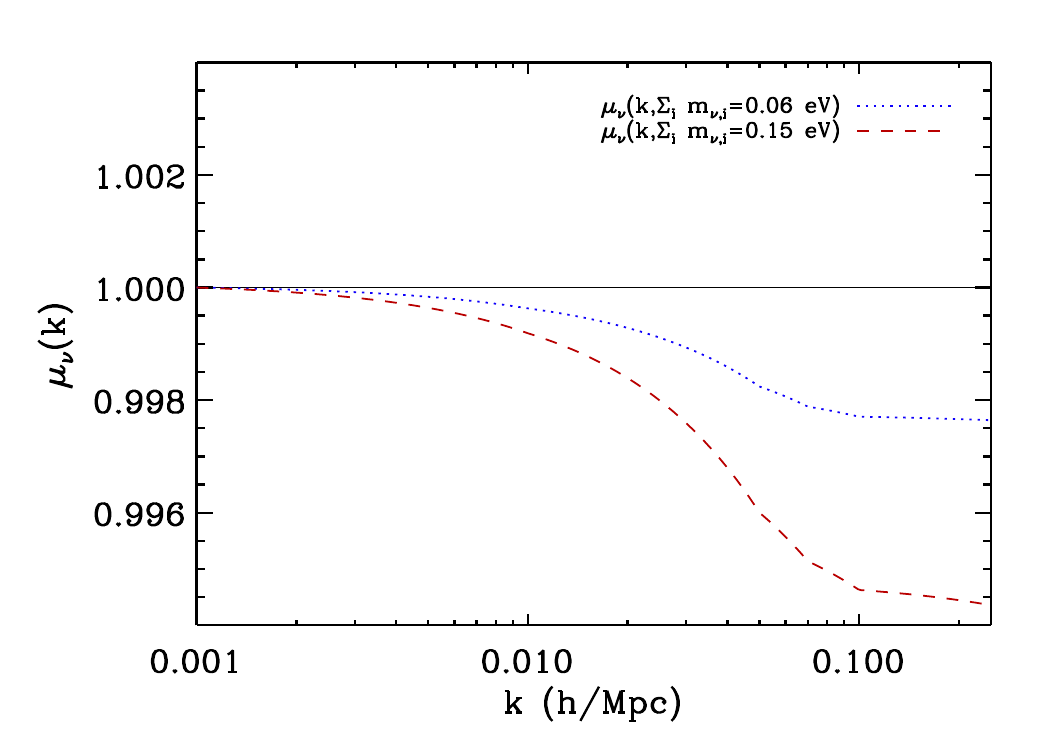}
\caption{Function $\mu_\nu(k)$. It represents the scale dependent correction to $f(z)$ in \Cref{eq:ffull,eq:mu}, evaluated here at $\sum_i m_{\nu, i}=0.06$ and $0.15\,\mathrm{eV}$.}
\label{fig:fkz_nu}
\end{figure*}

\autoref{fig:fkz_nu} shows the effect of massive neutrinos on $f$ in the linear regime, as ratios between
the massive and massless neutrino cases, that it is to say that we show the function $\mu_\nu(k)$, which is independent of redshift. 
As can be seen, in the presence of massive-neutrinos the linear growth rate acquires a scale
dependence, which decreases with decreasing neutrino mass, as the suppression of perturbations due to neutrino free streaming 
decreases, and as the free streaming scale becomes larger. 

Given the current upper limits on the total neutrino mass \citep{2019arXiv191109073P}, the growth suppression is of the order of $0.6\%$ at maximum, and 
mainly affects scales $k>0.1\,h\,{\rm Mpc}^{-1}$. 
For simplicity, we ignore this sub-percent effect in our forecasts, assuming $f$ to be scale independent for values $\sum_i m_{\nu,i}=0.06, 0.15$ eV tested, and computed in the massless limit. For $\sum_i m_{\nu,i}=0.06$ eV solar oscillation experiments constrain neutrinos to be in a normal hierarchy, and for $\sum_i m_{\nu,i}=0.15$ eV either an inverted or a normal hierarchy \citep{2010JCAP...05..035J}; we choose a normal hierarchy in this case.

Concerning the linear matter power spectrum  $P_{\delta\delta}(k,z)$,
in the presence of massive neutrinos, \autoref{eq: pddlinMath} in \autoref{sec:linear_perturbation} can be modified by replacing the 
transfer function $T(k)$ by a redshift-dependent one, $\mathcal{T}(k; z)$, while keeping the scale-independent
linear growth factor $D(z)$ as given in the absence of massive-neutrino free-streaming 
(\citealt{2006PhRvD..73h3520T,2006PhRvD..74d3505T,Eisenstein_Hu_1997}; see equation 25 of \citealt{Eisenstein_Hu_1997}).\footnote{We note that $\mathcal{T}$ is a combination of a transfer function $T$ and a normalisation coefficient, which is why we use a distinct symbol, but care should be taken not to confuse $T$ with $\mathcal{T}$.} 
The fiducial value of $D(z)$ at each redshift can be computed through numerical integration of the differential equations governing 
the growth of linear perturbations in the presence of dark energy \citep{Linder_Jenkins_2003}.
The linear transfer function $\mathcal{T}(k; z)$ depends on matter, baryon, and massive-neutrino densities (neglecting dark energy at early times), 
and is computed via Boltzmann solver codes in each redshift bin. The  forecasts presented in this work will assume a fixed $\Omega_{\rm m, 0}$, 
that is when $\Omega_{\rm \nu, 0}$ is varied, $\Omega_{\rm cb, 0}\equiv\Omega_{\rm c, 0}+\Omega_{\rm b, 0}$ varies as well, in order to keep $\Omega_{\rm m, 0}$ unchanged.

\subsection{Magnification bias}
{We note that we have not included the effects of {magnification
  bias} in our forecasts. Magnification bias refers to the distortion
of the observed galaxy number counts caused by gravitational lensing  (see e.g. \citealt{2016MNRAS.457..764D,2020MNRAS.491.1746T,2020MNRAS.491.4869T}),
and it can be important for Stage IV galaxy surveys. Concerning galaxy
clustering analyses with samples of photometrically detected galaxies,
previous works have shown that neglecting the magnification effects can lead to
significant shifts for several cosmological parameters (see
e.g. \citealt{Cardona:2016qxn,2015JCAP...10..070M}). Very recently it
was shown that magnification bias is also relevant for galaxy
clustering analyses of galaxies detected with spectroscopic
techniques. More specifically neglecting it may lead to biased
parameter estimation, like an incorrect inference of the growth rate
of structures \citep{Jelic-Cizmek:2020pkh}. Furthermore, magnification
can also be important for the cross-correlations between weak lensing
and galaxy clustering of photometrically-detected galaxies \citep{2018JCAP...06..008G}.

In our forecasts we have only focused on the uncertainties on the
different parameters. However, it is important to investigate the effect of
magnification, in both the shift of the best-fit and the
uncertainties on the parameters, in the context of model validation for \Euclid's final data analysis pipeline.}

\subsection{The standard $\Lambda$CDM model and its extensions}
\label{sec:models}

The spatially flat $\Lambda$CDM model is the baseline case considered in this paper, and corresponds to having a cosmological 
constant ($w_0=-1$, $w_a=0$). For a spatially flat cosmology $\Omega_{K,0}=0$ and the value of $\Omega_{\de,0}$ is a derived parameter, since $\Omega_{\de, 0} = 1-\Omega_{{\rm m},0}$. The baseline is then described by a minimal set of 6 parameters: 
\begin{itemize}
\item 
$\Omega_{{\rm b},0}$ and $\Omega_{{\rm m},0}$, the baryon and total matter energy densities at the present time; 
\item 
$h$, the dimensionless Hubble parameter, describing the homogeneous background evolution; 
\item 
$\sigma_8$, describing the amplitude of density fluctuations; 
\item 
$n_{\rm s}$, the spectral index of the primordial density power spectrum; 
\item 
$\sum{m_\nu}$: the sum of neutrino masses. 
\end{itemize} 
In this work, we further study the power of \Euclid primary probes for constraining deviations from the $\Lambda$CDM model by also analyzing extensions of this baseline parameter space. In particular, we consider the following extensions: 
\begin{itemize}
\item
spatially non-flat models, by varying $\{\Omega_{\rm DE,0}\}$ (equivalently one could vary $\Omega_{K,0}$, however we choose the first option in the numerical implementation); 
\item 
dynamical DE models, by varying the background values of $w_0$ and $w_a$; 
\item 
modifications in the growth of structures, by varying the growth index $\gamma$. 
\end{itemize}
The specific fiducial choice of all parameters will be discussed in \autoref{sec:fiducial}.

\section{Fisher matrix formalism for forecasting}
\label{Sec:Fisher}
In this paper we use a Fisher matrix formalism to estimate errors for cosmological parameter 
measurements. In this section we describe the general formalism and define some specific quantities that will 
be used throughout. We also present the detailed recipes used to implement the Fisher matrix formalism 
to compute forecasts of the different cosmological probes in \Euclid.

\subsection{General formalism}  \label{Sec:Fisher:general}

The aim of the analysis presented here is to obtain estimates on the uncertainties on the cosmological parameter measurements, i.e. the posterior 
distribution $P(\vec \theta|\vec x)$ of the vector of (model) parameters 
$\vec \theta$, given the data vector $\vec x$. Using Bayes' theorem, this can be obtained as
\begin{equation}
 P(\vec \theta|\vec x)=\frac{L(\vec x|\vec \theta)P(\vec \theta)}{P(\vec x)}\;,
\end{equation}
where $P(\vec \theta)$ is the prior information on our parameters, $P(\vec x)$ is the Evidence, and $L(\vec x|\vec \theta)$ is the likelihood of the data vector given the parameters.

The Fisher matrix \citep{1995PhDT........19B,1996ApJ...465...34V,TTH97} is defined as the expectation value of the second derivatives of the logarithmic likelihood function (the Hessian) and can be written in the general form 
\begin{equation}
  F_{\alpha\beta} = \left\langle- \left. \frac{\partial^2 \ln L}{\partial\theta_\alpha \partial\theta_\beta}\right|_{\mathbf{\theta}_{\rm ref}} \right\rangle\;, 
  \label{eq:fisher-general}
\end{equation}
where $\alpha$ and $\beta$ label the parameters of interest, $\theta_{\alpha}$ and $\theta_{\beta}$, and the derivatives are evaluated in the point ${\mathbf \theta_{\rm ref}}$ of the parameter space, which coincides with the maximum of the likelihood distribution. The Fisher matrix thus corresponds to the curvature of the logarithmic likelihood, describing how
fast the likelihood falls off around the maximum. For a Gaussian likelihood function this has an analytic expression that depends only on the expected mean and covariance of the data, i.e.
\begin{equation}
    F_{\alpha\beta} =  \frac 1 2 \tr \left[ \frac{\partial \tens{C}}{\partial\theta_\alpha}
    \tens{C}^{-1} \frac{\partial \tens{C}}{\partial \theta_\beta} \tens{C}^{-1} \right]
    + \sum\limits_{pq} \frac{\partial \mu_p}{\partial\theta_\alpha} (\tens{C}^{-1})_{pq} \frac{\partial \mu_q}{\partial\theta_\beta}\;,
  \label{eq:fisher-def}
\end{equation}
where $\vec \mu$ is the mean of the data vector $\vec x$ and $\tens{C} = \langle (\vec x - \vec \mu)(\vec x - \vec \mu)^\intercal \rangle $ is the
expected covariance of the data. 
The trace and sum over $p$ or $q$ here represent summations over the variables in the data vector. 
Often, for Gaussian distributed data $\vec x$ with mean $\langle x \rangle
= \mu$ and covariance $\tens{C}$, either the mean is zero, or the covariance is
parameter-independent. In both cases, one of the two terms in
\autoref{eq:fisher-def} is non-vanishing. In \Cref{sec:recipegc,WLrecipe,sec:recipexc} we specify which of these terms are used, and in the cases that either or both can be used we show both expressions. 

Once the Fisher matrix is constructed, the full expected error covariance matrix of the cosmological 
parameters is the inverse of the Fisher matrix,\footnote{Technically the covariance is the inverse of the Hessian matrix of the likelihood, and the Fisher matrix is the expectation of the Hessian; furthermore the inverse of the Fisher matrix does not give the expectation of the covariance, but yields an upper limit to the errors. Therefore we should use another symbol other than $C$. However for the sake of clarity of expression we use the notation as stated here.}
\begin{equation}
C_{\alpha\beta} = \left(\tens{F}^{-1}\right)_{\alpha\beta} \, . \label{eq:GC:parcov}
\end{equation}
The diagonal elements of the error covariance matrix contain the marginalised errors on the parameters. 
For example, the expected marginalised, $1$-$\sigma$ error on parameter $\theta_\alpha$ (i.e.~having included 
all the degeneracies with respect to other parameters), is
\begin{equation}
\sigma_\alpha = \sqrt{C_{\alpha\alpha}} \, .
\end{equation}
The unmarginalised expected errors, or conditional errors, can be computed 
by $\sigma_{\alpha}=\sqrt{1/F_{\alpha\alpha}}$, i.e. the square root of the reciprocal of the appropriate diagonal 
element of the Fisher matrix.
We can also define the correlation coefficient between the errors on our cosmological parameters as $\rho$, which contributes to the off-diagonal elements in the parameter covariance matrix,
\begin{equation} \label{eq:covariances}
C_{\alpha\beta} = \rho_{\alpha\beta}\sigma_\alpha \sigma_\beta \, .
\end{equation}
It is important to note that 
$\rho_{\alpha\beta} = 0$ if $\alpha$ and $\beta$ are completely independent. 
In order to `marginalise out' a subset of the parameters and obtain a smaller Fisher matrix, 
one should simply remove the rows and columns in the full parameter covariance matrix that 
correspond to the parameters one would like to marginalise over. Re-inverting will then give the 
smaller, marginalised Fisher matrix $\widetilde{F}_{\alpha \beta}$. This is equivalent to taking the {Schur complement} \citep{haynsworth1968, zhang_2005, 2009MNRAS.398.2134K} of the Fisher 
matrix for the smaller subset of parameters.

\subsubsection{Visualising confidence regions} 
The Fisher matrix can be used to plot the marginalised joint posterior 
probability, or projected confidence region, of two parameters $(\theta_\alpha, \theta_\beta)$, 
assuming that the joint posterior probability is well approximated by a Gaussian. 
The log-likelihood function is then locally close to the function of a hyper-dimensional ellipsoid, defined by the Fisher matrix of \autoref{eq:fisher-general}, 
and the projection on any two-dimensional sub-space is simply an ellipse. 
The parametric form of the confidence 
level ellipses is defined by the semi-minor and semi-major axes of the ellipses, $a$ and $b$, and the angle 
of the ellipse $\phi$; $a$ and $b$ are expressed in terms of the larger and smaller eigenvalues of the covariance matrix, and 
$\phi$ is related to the ratio of the $y$-component to the $x$-component of the larger eigenvector (as given by equation 40.72 of the 
Particle Data Group's Review of Particle Properties section on statistics\footnote{\url{http://pdg.lbl.gov/2017/reviews/rpp2017-rev-statistics.pdf}}). They are:
\begin{align}
a =& A \sqrt{ \frac{1}{2}( C_{\alpha\alpha} + C_{\beta\beta} ) + \sqrt{ \frac{1}{4} ( C_{\alpha\alpha} - C_{\beta\beta} )^2 + C_{\alpha\beta}^2 }  } \, , \nonumber \\
b =& A \sqrt{ \frac{1}{2}( C_{\alpha\alpha} + C_{\beta\beta} ) - \sqrt{ \frac{1}{4} ( C_{\alpha\alpha} - C_{\beta\beta} )^2 + C_{\alpha\beta}^2 }  } \, ,  \nonumber \\
\phi =& \frac{1}{2} {\rm atan} \left( \frac{ 2 C_{\alpha\beta} }{ C_{\alpha\alpha} - C_{\beta\beta}} \right)  \, ,
\label{eq:ellipse_params}
\end{align}
where $A$ is a constant factor defined as $A^2 = 2.3,\, 6.18,\, 11.8$ for two-parameter contours at $1$-, $2$-, and $3$-$\sigma$ 
confidence level (C.L.), respectively \citep[see e.g.][]{Press:2007}, where the $C_{\alpha\beta}$ are defined in \autoref{eq:GC:parcov}.

\subsubsection{Figure of merit}\label{sec:fom}
When considering a particular experiment, its performance in constraining specific parameters, e.g. those related to the dark energy 
model, can be quantified
through a figure of merit (FoM). We adopt the FoM as defined in \citet{Albrecht:2006} which is inverse  proportional to the area of the $2$-$\sigma$ contour in the marginalised 
parameter plane for two parameters $\theta_\alpha$ and $\theta_\beta$, 
under the usual Gaussianity assumption inherent in the Fisher forecast formalism. Therefore the FoM can be calculated simply from the 
marginalised Fisher submatrix $\widetilde{\tens{F}}_{\alpha \beta}$ for those two parameters. That is
\begin{equation}
\label{FoM_generic}
\text{FoM}_{\alpha \beta} =  \sqrt{{\rm det}\left(\widetilde{\tens{F}}_{\alpha \beta}\right)} \, .
\end{equation}
The above formula was utilised in the original dark energy FoM definition from \citet{Albrecht:2006}, which considers the $w_0, w_{ a}$ parameterization defined in \autoref{sec:cosmo-bcg}:
\begin{equation}
\label{FoM}
\text{FoM}_{w_0, w_{a}} =  \sqrt{{\rm det}\left(\widetilde{\tens{F}}_{w_0w_{a}}\right)} \, ,
\end{equation}
which is in fact equivalent (within a constant factor) to other definitions used in the literature
 \citep[see, for example,][]{Rassat:2008,Wang:2008, 2012PhRvD..85j3008A, Majerotto:2012,Wang:2013}. 
Different FoMs can also be defined for any 
arbitrary set of parameters by simply taking the determinant of the appropriate Schur complement.
As an example, for a model where the linear growth rate $f(z)$ of density perturbations 
is parameterised as in \autoref{eq:ggrowth}, the FoM 
would read $\smash{\text{FoM}_{\gamma \Omega_{\rm m}} =  [{\rm det}(\widetilde{\tens{F}}_{\gamma \Omega_{\rm m}})]^{1/2}}$, and similarly for any other cosmological parameter pair \citep[e.g.][]{Majerotto:2012}.  
Throughout this paper we refer to the `FoM' as being that defined in \autoref{FoM}. 

\subsubsection{Correlation matrix and the figure of correlation \label{sec:covariance}}
As shown in \citet{2017PDU....18...73C} one can define a correlation matrix $\tens{P}$ for a \emph{d-dimensional }vector
$p$ of random variables as
\begin{equation}
P_{\alpha\beta}=\frac{C_{\alpha\beta}}{\sqrt{C_{\alpha\alpha}C_{\beta\beta}}} \, . \label{eq:correlation_def}
\end{equation}
where $\tens{C}$ is the covariance matrix. 
By definition, $\tens{P}$ is equal to the unit matrix if all parameters are uncorrelated, while it differs from it if some correlation 
is present. We plot this matrix in \autoref{sec:results} to visualise the impact on the attainable cosmological constraints 
of cross-correlations in the photometric survey. 
We further adopt the introduction of a `figure of correlation' (FoC), first defined in \citet{2017PDU....18...73C}, defined 
(here without the logarithm) as: 
\begin{equation}\label{eq:FoC}
\mathrm{FoC} = \sqrt{\det(\tens{P}^{-1})} \, ,
\end{equation}
which is 1 if parameters are fully uncorrelated. Off-diagonal non-zero terms (indicating the presence of correlations among parameters) in  $\tens{P}$ will correspond to $\mathrm{FoC}>1$. The FoC and the FoM are independent quantities, \citep[see][for a geometrical interpretation] {2017PDU....18...73C}.

\subsubsection{Projection into the new parameter space}\label{sec:projection}
The Fisher matrix is defined for a set of parameters $\vec \theta$, but a new Fisher matrix can be constructed 
for an alternative set of parameters $\vec p (\vec \theta)$. 
In this case the new Fisher matrix $\tens{S}$ is related to the original Fisher matrix $\tens{F}$ by a Jacobian transform  
\be
S_{ij} = \sum_{\alpha\beta} \frac{\partial \theta_{\alpha}}{\partial p_{i}}{F}_{\alpha\beta}\frac{\partial \theta_{\beta}}{\partial p_{j}} \, , 
\label{eq:submatrix1}
\ee
where the Jacobian matrices $\partial \theta_{\alpha}/\partial p_{i}$ relate the original to the new parameterization.
Notice that if $\vec p (\vec \theta)$ is not a linear function then the likelihood in the new parameters is in general not 
Gaussian, even if it was in the original parameter space $\vec \theta$. 
Hence the Gaussian approximation inherent in the Fisher formalism may be valid for one choice of parameters but not for another.

\subsubsection{Fiducial parameter values}\label{sec:fiducial}
In this section we detail the choice of fiducial model, about which the derivatives and cosmological quantities used in the Fisher 
matrices considered in this 
paper are computed (for the $\Lambda$CDM case and for its extensions). 
Since we are combining information from the \Euclid galaxy clustering and weak lensing probes, our final Fisher matrix should have 
consistent rows and columns across all the probes. Within the assumption of a minimal cross-covariance between the probes, the 
combination of the Fisher matrices related to the two observables can then be achieved through a direct addition of the matrices. 
The final minimal set of cosmological parameters for both probes is defined as (see \autoref{sec:models})
\be
\theta = \{\Omega_{\rm b,0},\,\Omega_{\rm m,0},\,h,\,n_{\rm s},\,\sigma_8, \, \sum{m_\nu}\}.
\label{eq:9parameters}
\ee
The (minimal) dark energy and modified gravity models discussed in \Cref{sec:cosmo-bcg,sec:cosmo-growth} involve 
the following parameters in addition to the minimal $\Lambda$CDM set:
\begin{align}
\text{non-flat geometry} &:
\{\Omega_{\rm DE,0}\} \label{eq:nonflat} , \\
\text{the evolving equation-of-state parameters} &:
\{w_0, w_{a}\} \label{eq:GC:w0wa} , \\
\text{the (gravitational) growth index parameter} &:
\{\gamma\}. \label{eq:GC:gamma}\ 
\end{align}
These appear in the expansion history and growth of structure respectively (see Eqs.~\ref{eq:weos} and \ref{eq:ggrowth}). The chosen parameter values of our fiducial cosmological model are 
shown in \autoref{tab:cosmofid}.

\begin{table}[h!]
\caption{Parameter values of our fiducial cosmological model, corresponding to those of \citet{2016A&A...594A..14P}, both in the baseline $\Lambda$CDM case and in the extensions considered. All our results have been obtained for two different values of the sum of neutrino masses, 
with higher and lower values of neutrino masses ($\sum{m_\nu}  = 0.15$eV and $0.06$eV) but, unless otherwise specified, we only show results for the $0.06$ eV case. We additionally write the density values {$\omega_\text{b,0}$ and $\omega_\text{m,0}$ }here, since they are varied initially in the calculation for the GC probe alongside several redshift-dependent parameters, whose values are derived from the fiducial model in this table (see \autoref{sec:recipegc} for more details). The final results for all probes will be given in terms of the parameters of \autoref{eq:9parameters} and will therefore refer to $\Omega_\text{b,0}$ and $\Omega_\text{m,0}$. For non-flat cosmologies, $\Omega_{\rm DE,0}$ is also varied. The optical depth $\tau$ is fixed to 0.058 throughout the paper (we do not vary it and use the best fit value from \citealt{2016A&A...594A..14P}).}
\centering
\begin{tabular}{cccccccccccc}
\hline
\multicolumn{6}{c}{Baseline} & & \multicolumn{4}{c}{Extensions}\\
\cline{1-6} \cline{8-11}
 $\Omega_\text{b,0}$ & $\Omega_\text{m,0}$ & $h$ & $n_\text{s}$ & $\sigma_{8}$ & $\sum{m_\nu}$ [eV] & & $\Omega_{\rm DE,0}$ & $w_0$ & $w_{a}$ & $\gamma$ \\ 
({$\omega_\text{b,0}$}) & ({$\omega_\text{m,0}$}) & & & & & & & & & \\
\hline
\hline
 $0.05$ & $0.32$ & $0.67$ & $0.96$ & $0.816$ & $0.06$ & & $0.68$ & $-1$ & $0$ & $0.55$ \\
($0.022445$) & ($0.143648$) & & & & & & & & & \\
\hline
\end{tabular}
\label{tab:cosmofid}
\end{table}
In addition to the fiducial cosmology we also need to define the survey specifications. These are described in more detail in the 
following sections, but we summarise here some of the main characteristics that are common to both GC and WL. 
\begin{table}[h!]
        \centering
\caption{Specifications for the spectroscopic galaxy redshift survey.}
                \label{tab:GC-specifications}
        \begin{tabular}{lll}
                \hline
                & Parameter  & Value \\
                \hline\hline
                Survey area in the sky & $A_{\rm survey}$  & $15\,000\,\deg^2$ \\
                Spectroscopic redshift error & $\sigma_{z}$  & $0.001(1+z)$ \\
                Minimum and maximum redshifts of the sample & $[z_{\rm min},\ z_{\rm max}]$  & $[0.90, 1.80]$ \\
                \hline
        \end{tabular}
\end{table}

\begin{table}[h!]
        \centering
        \caption{Expected number density of observed H$\alpha$ emitters for the \Euclid spectroscopic survey. This number has been updated since the Red Book \citep{RB} to match new observations of number densities and new instrument and survey specifications. The first two columns show the minimum, $z_{\rm min}$, and maximum, $z_{\rm max}$, redshifts of each bin. The third column is the number of galaxies per unit area and redshift intervals, ${\rm d}N(z)/{\rm d}\Omega{\rm d}z$. The fourth column shows the number density, $n(z)$. The fifth column lists the total volume. Finally, in the sixth column we list the galaxy bias evaluated at the central redshift of the bins, $z_{\rm mean} = (1/2)(z_{\rm max} + z_{\rm min})$.}
        \label{tab:gc_nofz}
                \begin{tabular}{cccccc}
                        \hline
                        $z_{\rm min}$ & $z_{\rm max}$ &  ${\rm d}N(z_{\rm mean})/{\rm d}\Omega{\rm d}z\,[\mathrm{deg}^{-2}]$ & $n(z_{\rm mean})\,[h^3\,\mathrm{Mpc}^{-3}]$ & $V_{\rm s}(z_{\rm mean})\,[\mathrm{Gpc}^3\,h^{-3}]$ &$b(z_{\rm mean})$ \\
                        \hline
                        \hline
                        0.90 & 1.10 & 1815.0 & $6.86\times 10^{-4}$ & 7.94   & 1.46 \\
                        1.10 & 1.30 & 1701.5  & $5.58\times 10^{-4}$ & 9.15    & 1.61 \\
                        1.30 & 1.50 & 1410.0  & $4.21\times 10^{-4}$ & 10.05 & 1.75 \\
                        1.50 & 1.80 & 940.97   & $2.61\times 10^{-4}$ & 16.22  & 1.90 \\
                        \hline
                \end{tabular}
\end{table}

\begin{table}[h!]
	\centering
	\caption{Specifications for the \Euclid photometric weak lensing survey.}
	\label{tab:WL-specifications}
	\begin{tabular}{lll}
		\hline 
		& Parameter  & \Euclid \\
		\hline \hline
		Survey area in the sky & $A_{\rm survey}$  & $15\,000\,\deg^2$  \\
		Galaxy number density & $n_{\rm gal}$  & $30\,\mathrm{arcmin}^{-2}$ \\
		Total intrinsic ellipticity dispersion & $\sigma_\epsilon$  & 0.30 \\
		Number of redshift bins & $N_z$ & 10 \\
		\hline 
	\end{tabular}
\end{table}
In \Cref{tab:GC-specifications,tab:gc_nofz,tab:WL-specifications} we describe the specifications used to compute the forecasts 
for GC and WL observations within the \Euclid mission. In this paper we adopt the specifications of the \Euclid
Red Book \citep{RB} as closely as possible, to allow for a direct comparison with those forecasts. We note that during the writing 
of this paper the \Euclid Consortium underwent a process of science performance verification (SPV) -- a review process that also
produced cosmological parameter forecasts that included systematic effects induced by instrumental and data reduction procedures. 
The forecasts in the SPV used the validated IST:F codes, but slightly different areas and depth specifications were used that were motivated from 
the internal \Euclid simulations. The SPV results are not publicly available, and since these simulation-derived quantities may change before 
publication of any SPV results, we choose to remain with the Red Book specifications; however in future work these will also be updated.

\subsubsection{Required accuracy in Fisher matrix forecasts}\label{sec:required-accuracy}
Fisher matrices with a large number of correlated parameters can be difficult to compute and to invert accurately 
due to the numerical complexity involved. 
Therefore one requires accurate estimates of the parameter uncertainties, i.e. ``the error on the errors''. 
One aspect is simply the accuracy of the Fisher matrix approximation itself (i.e. the fact that it assumes a Gaussian likelihood), but others are the precision with which the Fisher matrix is computed. Typically, problems arise in 
the evaluation of the derivatives, the approximation of a sum over redshift planes rather than an integral, and the precision 
of the inversion of the Fisher matrix to obtain the parameter covariance matrix giving the parameter 
uncertainties and their correlations. The assessment of the accuracy with which one needs to compute a Fisher 
matrix is simple in one dimension, with 
\begin{equation}
 \frac{\delta\sigma_{\alpha}}{\sigma_{\alpha}}=\frac{1}{2}\frac{\delta C_{\alpha\alpha}}{C_{\alpha\alpha}}=-\frac{1}{2}\frac{\delta F_{\alpha\alpha}}{F_{\alpha\alpha}} \, ,  
\end{equation}
where $\tens{F}$ is the Fisher matrix and $\tens{C}$ is the correlation matrix, as defined above.
The computed fractional error on the parameter is then of the same order as the numerical fractional error on the Fisher matrix. 

In the following we assume a $10\%$ requirement on the uncertainty in the parameter error arising from numerical inaccuracies. However we note that this is a somewhat arbitrary requirement. We see that for one dimension this does not impose a tight requirement on numerical accuracy, since one would expect Fisher matrix numerical fractional errors $<10^{-4}$. 

For more than one parameter, the covariances between them are important and this will 
be crucial for determining the sensitivity. We can obtain the main effects by considering the amplitude and orientation of the parameter confidence regions and ellipses for two parameters, $\theta_\alpha$ and $\theta_\beta$. In two dimensions, 
\begin{align}
C_{\alpha\alpha}  &= \frac{F_{\alpha\alpha}}{{\rm det}\tens{F}},\\ 
C_{\beta\beta}  &= \frac{F_{\beta\beta}}{{\rm det}\tens{F}},\\  
C_{\alpha\beta}  &= C_{\beta\alpha} = -\frac{F_{\alpha\beta}}{{\rm det}\tens{F}} \,\,, 
\label{eq:fishtocov}
\end{align}
and 
\begin{equation}\label{eq:deltaC}
 \frac{\delta C_{\alpha\beta}}{C_{\alpha\beta}}=\frac{\delta F_{\alpha\beta}}{F_{\alpha\beta}}-\frac{\delta {\rm det}\tens{F}}{{\rm det}\tens{F}} \,\, , 
\end{equation}
where ${\rm det}\tens{F}=F_{\alpha\alpha}F_{\beta\beta}-F_{\alpha\beta}^2$ for a given $\alpha,\beta$ pair where $\alpha\leq 2$ and similarly for $\beta$\footnote{In more than 
two dimensions, the numerator in \autoref{eq:fishtocov} is replaced by the cofactor of element $F_{\alpha\beta}$.}. 
Since
\begin{equation}
 \frac{\delta \sigma_{\alpha}}{\sigma_{\alpha}}=\frac{1}{2}\frac{\delta C_{\alpha\alpha}}{C_{\alpha\alpha}},
\end{equation}
then we see that the first term in \autoref{eq:deltaC} is just what we had in the one parameter case 
and does not impose tight restrictions on the Fisher matrix precision. Therefore, 
it is the determinant term, involving covariances and the potential for degeneracies between parameters, that requires care. 
In two dimensions, for a given $\alpha$ and $\beta$ pair
\begin{equation}
 {\rm det}\tens{F}=\frac{1}{{\rm det}\tens{C}}=\frac{1}{\sigma_\alpha^2\sigma_\beta^2\left(1-r_{\alpha\beta}\right)} \, ,
\end{equation} 
where the correlation coefficient $r_{\alpha\beta} \equiv F_{\alpha\beta}^2/(F_{\alpha\alpha}F_{\beta\beta})$; we also note that $r_{\alpha\beta}=\rho_{\alpha\beta}^2$ from \autoref{eq:covariances}.
From this, one has
\begin{equation}
 \frac{\delta {\rm det}F}{{\rm det}F}=2\frac{\delta F_{\alpha\beta}}{F_{\alpha\beta}}+\frac{F_{\alpha\alpha}F_{\beta\beta}}{F_{\alpha\alpha}F_{\beta\beta}-F_{\alpha\beta}^2}\left[\frac{\delta F_{\alpha\alpha}}{F_{\alpha\alpha}}+\frac{\delta F_{\beta\beta}}{F_{\beta\beta}}-2\frac{\delta F_{\alpha\beta}}{F_{\alpha\beta}}\right].
\end{equation}
Again, we see that the first term indicates that the precision of the Fisher matrix propagates to errors on the parameter uncertainty of the same order, and hence only the second term can cause a change. The final result, in two dimensions, is
\begin{equation}
 \frac{\delta\sigma_{\alpha}}{\sigma_{\alpha}}\approx\frac{\delta_{\rm max}}{1-r_{\alpha\beta}} \, ,
\end{equation}
where $\delta_{\rm max}$ is the maximum fractional imprecision on an element of the Fisher matrix. We note that for two dimensions $r_{\alpha\beta}$ is a single number and so the summation convention does not apply to the above equation (one could write $\sigma_{\alpha\,|\,\beta}$, i.e. the error on $\alpha$ given a fixed $\beta$). 
Assuming $r_{\alpha\beta}\leq 0.999$ for realistic (well chosen) parameters and a requirement that 
$\delta\sigma_{\alpha}/\sigma_{\alpha}\leq 0.1$ we find a required numerical accuracy of 
$\delta_{\rm max}\leq 10^{-4}$, which is a reasonable aim for a good numerical implementation. 
 
We note that the FoM accuracy is simply given by
\begin{equation}
 \frac{\delta{\rm FoM}}{{\rm FoM}}=\frac{1}{2}\frac{\delta{\rm det}F}{{\rm det}F} \, ,
\end{equation}
for any parameter combination given in \autoref{FoM_generic}. Thus if we want the accuracy of FoM to better than $10\%$, this 
becomes the most restrictive constraint.

Another important element is the orientation of the joint parameter contour ellipse. This is especially relevant for combining \Euclid's constraints with other probes. 
The orientation $\phi$ is defined in \autoref{eq:ellipse_params}, and we find its uncertainty is 
\begin{equation}
 \delta\phi=\cos^2(2\phi)\left[\frac{\delta C_{\alpha\beta}}{C_{\alpha\alpha}-C_{\beta\beta}}-\frac{C_{\alpha\beta}\left(\delta C_{\alpha\alpha}-\delta C_{\beta\beta}\right)}{\left(C_{\alpha\alpha}-C_{\beta\beta}\right)^2}\right].
 \label{eq:deltaphi}
\end{equation}
For many parameter pairs, for example $(w_0,w_a)$, one of the parameter uncertainties is much larger than the other. 
Assuming the error on $\theta_\beta$ to be much larger than the one on $\theta_\alpha$ ($\sigma_\beta\gg\sigma_\alpha$) and using \autoref{eq:covariances}, we can rewrite
\begin{equation}
 \delta\phi=\cos^2(2\phi)\left\{\frac{\rho_{\alpha\beta}\sigma_{\alpha}}{\sigma_{\beta}}\left[\frac{\delta C_{\beta\beta}}{C_{\beta\beta}}-\left(\frac{\sigma_{\alpha}}{\sigma_{\beta}}\right)^2\frac{\delta C_{\alpha\alpha}}{C_{\alpha\alpha}}\right]-\left(\frac{\sigma_{\alpha}}{\sigma_{\beta}}\right)^2 \frac{\delta C_{\alpha\beta}}{C_{\alpha\beta}}\right\}.
\end{equation}
The dominant term overall is the first term in the square brackets, and is suppressed with respect to $\delta C_{\beta\beta}/C_{\beta\beta}$ itself by $\sigma_{\alpha}/\sigma_{\beta}$, 
hence it is smaller than the effect on the error on the amplitude of the parameter uncertainty. If we require the parameter uncertainty to be $\delta\sigma_{\alpha}/\sigma_{\alpha}\simeq 0.1$, then the orientation angle will be accurate to $\delta\phi\approx 2\%$. 

The conclusion is that, for $\leq 10\%$ precision on parameter errors and FoM and $\leq 2\%$ precision on the orientation of the contours, the required fractional Fisher matrix precision (both in calculation of elements, and its inversion) 
is approximately $10^{-4}$. In other words, one certainly does not need to push near machine precision. 
We note here that for two-sided or three-point finite differences for evaluation of derivatives with a parameter stepsize $\epsilon$, the errors go as $\epsilon^2$, while five-point differences go as $\epsilon^4$.

\subsection{Recipe for spectroscopic galaxy clustering} \label{sec:recipegc}
This section describes the forecasting procedure recommended for our galaxy clustering observable -- the full, anisotropic, and 
redshift-dependent galaxy power spectrum, $P(k,\mu;z)$. We spend most of the section describing how to model our observable. 
We conclude by specifying how to use our observable in the Fisher matrix calculation, as described generally in \autoref{Sec:Fisher:general}.

The initial goal is to calculate a Fisher matrix for the following full set of cosmological parameters: 
\begin{eqnarray}
\text{power spectrum broadband `shape-parameters'}
&:&
\{{\omega_\text{b,0}}, h, {\omega_\text{m,0}}, n_\text{s}\}\footnotemark \ ,
\nonumber\\
\text{non-linear nuisance parameters}
&:&
\{\sigma_\text{p},\sigma_\text{v}\} \ ,
\label{def:GC:pars}
\\ 
\text{{redshift-dependent} parameters}
&:&
\{\ln D_\text{A}(z_{i}), \ln H(z_{i}), \ln[f\sigma_8(z_{i})], \ln[b\sigma_8(z_{i})], P_\text{s}(z_{i})\} \ . \nonumber
\end{eqnarray}%
The first set of parameters determines the shape of the linear matter power spectrum (as defined in \Cref{sec:cosmo-bcg,sec:linear_perturbation}): the physical densities of baryons, the physical density of total matter, the dimensionless Hubble parameter, and the scalar spectral index. In particular, these parameters control the transfer function and the power law of the spectrum of primordial density perturbations (see \autoref{sec:linear_perturbation} for details).

The second set carries the uncertainty in our theoretical knowledge about late-time non-linearities and redshift-space distortions ({RSDs}), 
which damp the baryon acoustic oscillation (BAO) signal in the galaxy power spectrum, and cause the so-called fingers-of-God effect. 
These are considered nuisance parameters, to be marginalised over. 
Such marginalisation necessarily increases the uncertainty on the other parameters (in the Fisher matrix). 
The non linear recipe adopted in this paper will be discussed in detail later (in \autoref{sec:nonlineargc}).

Like the second set, the {redshift-dependent} set of parameters is also specific to galaxy clustering. 
Anisotropies in the power spectrum with respect to the line-of-sight direction due to the Alcock-Paczynski (AP) effect are parameterised 
in terms of the angular diameter distance $D_\text{A}(z)$ and the Hubble parameter $H(z)$, 
while the pattern of RSDs is described by the combination $f\sigma_8(z) \equiv f(z)\sigma_8(z)$  defined in \autoref{sec:linear_perturbation}; note that here we generalise \autoref{eq: sigma8} to be redshift-dependent by changing the redshift argument to the power spectrum in that integral.  
The additional parameters $b\sigma_8(z)\equiv b(z)\sigma_8(z)$ and $P_\text{s}(z)$, characterise the bias of our target galaxy sample, 
and respectively the shot noise residual, as we describe below. Each parameter in this set is varied freely in each redshift bin of the survey. 
In this way, the constraints on $D_\text{A}(z)$, $H(z)$, and  $f\sigma_8(z)$ obtained at each redshift 
provide us with dark-energy-model-independent constraints on the expansion and structure growth histories of the Universe. 
The key cosmological information contained in these measurements can later be transformed into parameters specific to our chosen 
models. For practical ease, we calculate the natural logarithms of these parameters where possible. 

These three sets of parameters capture all the main contributions of uncertainty in our model, some to be marginalised over, 
and some to be propagated into our final, general parameter set.
After marginalising over our observational and theoretical `nuisance parameters' $\ln[b\sigma_8(z)]$, $P_\text{s}(z)$, 
$\sigma_\text{p}(z_{\rm i})$, and $\sigma_\text{v}(z_{\rm i})$, we transform the constraints on the remaining set of redshift-dependent 
and `shape' parameters to those on the parameters of our chosen specific DE and modified gravity model by using a Jacobian 
matrix (see Eq.~\ref{eq:submatrix1}), in terms of the final set of cosmological parameters 
$\theta = \{\Omega_{\rm b,0},\,h,\,\Omega_{\rm m,0},\,n_{\rm s},\,\sigma_8,\,\sum m_{\nu}\}$ with or without the extensions $\{\Omega_{\rm DE,0},w_0,\,w_{\rm a},\,\gamma\}$, 
as described in \autoref{Sec:Fisher:general}. Finally, we must determine the FoM (see \autoref{sec:fom}) on the DE 
parameters of \Cref{eq:GC:w0wa,eq:GC:gamma}, having marginalised over the rest.

\subsubsection{The observable galaxy power spectrum}\label{sec:GC:pk}

We work in Fourier space to describe the distribution of our target sample -- H$\alpha$-emitting galaxies -- given the statistics of the 
underlying dark matter field. We work in terms of $P(k,\mu;z)$, where $k$ is the modulus of the wave mode in Fourier space, and 
$\mu=\cos{\theta}=\vec{k}\cdot\hat{\mathbf r}/k$ is the cosine of the angle $\theta$ between the wave-vector \vec{k} and the line-of-sight 
direction $\hat{r}$, and $z$ is the redshift of the density field.\footnote{In practice, 
observers measure the power spectrum in redshift bins, so $z$ here becomes the effective redshift of the bin, which for simplicity we 
just approximate here to be the bin's central redshift \citep[see][for a discussion of the effective redshift]{PhysRevD.89.063525}.} 
We give the final, full model 
for the observable galaxy power spectrum in \autoref{eq:GC:pk} and describe all the individual effects in the preceding sections.

In total, five main observational effects must be modelled beyond simply calculating the matter power spectrum, the linear model of which 
is given in \autoref{eq: pddlinMath}. In this brief section we show how we model each effect in the linear, as well as in the non-linear, regime of perturbation theory. 

There are five main effects that we need to model:
\begin{enumerate}
\item \textit{the galaxy bias} (in the case of \Euclid spectroscopy the bias of H$\alpha$-emitting galaxies) with respect to the total matter 
distribution,
\item \textit{anisotropies due to RSD} induced by the peculiar velocity component of the observed redshift causing discrepancies in the 
clustering strength measured at different angles with respect to the line-of-sight.
\item \textit{the residual shot noise} that remains even after the known noise -- due to Poisson sampling by target galaxies of a (theoretically) 
smooth underlying matter density field -- has been subtracted,
\item \textit{the redshift uncertainty} that suppresses the correlation between galaxy positions by smearing out the galaxy field along the line of sight,
\item \textit{distortions due to the AP effect}, which introduces an anisotropic pattern in the power spectrum by rescaling 
wavemodes in the transverse and radial directions by different factors.
\end{enumerate}

The combination of  the above five effects are applied to the power spectrum model. We now describe how we model these effects, 
focusing first on the linear theory predictions. We describe (\autoref{sec:nonlineargc}) how we extend this recipe to take into account the impact 
of non-linearities to derive a final model of the observed galaxy power spectrum.

\paragraph{Effective galaxy bias.} \label{sec:bias-zerr-pshot}

We do not discuss the issue of galaxy bias in detail here, but refer the reader to the comprehensive review of \cite{Desjacques:2016} and use the 
simple linear relation:
\begin{equation}
P_\text{g,lin}(k;z) = b^2(z) P_{\delta\delta}(k;z) \, ,
\label{eq:GC:bpk}
\end{equation}
where $P_{\delta\delta}(k;z)$ (see Eq.~\ref{eq: pddlinMath}) is the linear matter power spectrum, which, when multiplied by the square of the redshift-dependent effective bias of the galaxy sample, $b^2(z)$, yields the galaxy linear power spectrum. In our case, 
the sample in question contains what will be detected by \Euclid as H$\alpha$-line emitter galaxies.

\paragraph{Anisotropies due to RSD.}

The measured galaxy redshifts contain a non-cosmological contribution due to the 
line-of-sight component of the peculiar velocity
\begin{equation}
1+z_{\rm obs} = \left(1+z\right)\left(1+v_{\parallel}/c\right)\, , 
\label{eq: GC_totalz}
\end{equation}
where $v_{\parallel}$ is the galaxy peculiar velocity along the line of sight.
If one assumes that the observed redshifts are entirely cosmological when estimating 
the distances to each galaxy, this distorts the density field in a way that 
imprints a specific pattern of anisotropies onto the observed power spectrum; this is known as the RSD. 
As the peculiar velocity displacements are sourced by the true underlying density field, 
the pattern of the RSDs provides us with an additional source of cosmological information on the relation 
between the density and peculiar velocity fields, which in the linear regime depends on the growth 
rate parameter $f(z)$.

In the linear regime, the relation between the real- and redshift-space galaxy power spectra (here labelled as 'zs' for redshift-space), 
is given by \cite{Kaiser_1987}
\begin{equation}
P_\text{zs,lin}(k,\mu; z) =
\left[b(z)\sigma_8(z)+f(z)\sigma_8(z)\mu^2\right]^2 \frac{P_{\delta\delta}(k;z)}{\sigma_8^2(z)}\, ,
\label{eq:GC:linRSD}
\end{equation}
where $\sigma_8(z)$ is defined in \autoref{eq: sigma8}, but generalised to be redshift-dependent.
Written in this way, \autoref{eq:GC:linRSD} illustrates that the quantities controlling the anisotropies in the 
two dimensional power spectrum are the combinations $b(z)\sigma_8(z)$ and $f(z)\sigma_8(z)$, which we 
treat as free parameters.

\paragraph{Redshift uncertainty and shot noise.} \label{sec:zerr-pshot}

The effect of  redshift uncertainties results in a modification of the power spectrum in the form 
\begin{equation} \label{Pobs_stats_def}
P_\text{zerr,lin}(k,\mu;z) = P_\text{zs,lin}(k,\mu;z)F_z(k,\mu;z) + P_{\rm s}(z) \, ,
\end{equation}
where the factor $F_z(k,\mu;z)$ is given by
\begin{equation}
F_z(k,\mu;z) = \text{e}^{-k^2\mu^2\sigma_r^2(z)},
\end{equation}
and accounts for the smearing of the galaxy density field along the line of sight direction $k_\parallel = k\mu$
due to possible redshift errors.
This error propagates into a comoving distance error \citep[see e.g.][]{Wang:2013b},
\begin{equation}
\sigma_r(z) = \frac{\partial{r}}{\partial{z}} \sigma_z(z) 
= \frac{c}{H(z)} \left(1+z\right) \sigma_{0,z} \, ,
\label{eq:error-z-gc}
\end{equation}
where $\sigma_z(z)=\sigma_{0,z}(1+z)$, and $\sigma_{0,z}$ is a linear 
scaling of the redshift error. 
We note that the damping due to redshift errors does not vary with changes in the expansion history, 
since $k_\parallel \propto H(z)$ and $\sigma_r \propto H^{-1}(z)$ \citep{Wang:2013b}.

As described before, $P_{\rm s}(z)$ is a scale-independent offset due to imperfect removal of shot noise, which is taken 
to be $P_{\rm s}(z)=0$ at all redshifts in our fiducial model, but is allowed to vary with redshift. 
It is standard in galaxy clustering analysis to treat the shot-noise term as a free parameter \citep[see e.g.][]{Gil-Marin2016, 
Beutler2017, Grieb2017}. We therefore include it here to assess the impact that marginalising over this parameter might 
have on the expected parameter constraints from \Euclid.

\paragraph{AP projection effects.} 

The measurement of the galaxy power spectrum requires the assumption of a reference cosmology to transform 
the observed redshifts into distances. Assuming an incorrect cosmology leads to a rescaling of components 
of the wave-vector  $\vec{k}$ in the directions parallel and perpendicular to the line-of-sight, 
$k_{\parallel}$ and $k_{\perp}$, as 
\begin{equation}
k_{\perp} = \frac{k_{\perp,\text{ref}}}{q_\perp}
\qquad \text{and} \qquad
k_{\parallel} = \frac{k_{\parallel,\text{ref}}}{q_\parallel} \, ,
\label{eq:kparperp}
\end{equation}
where the coefficients $q_\perp$ and $q_\parallel$ are given by the ratio of the 
angular-diameter distance and the Hubble parameter in the true and reference cosmologies as
\begin{equation}
q_{\perp}(z) = \frac{D_{\rm A}(z)}{D_{\rm A,\, ref}(z)}
\qquad \text{and} \qquad
q_{\parallel}(z) = \frac{H_\text{ref}(z)}{H(z)} \, .
\label{eq:GC:APqs}
\end{equation}
We label the redshift dependence here, but will suppress this in the following equations for clarity. 
\Cref{eq:kparperp,eq:GC:APqs} can be used to convert from the known reference $k_\text{ref}$ 
and $\mu_\text{ref}$ to the unknown, true $k$ and $\mu$ as
\be
\begin{split}
k(k_\text{ref},\mu_\text{ref}) &= \frac{k_\text{ref}}{q_\perp} \left[ 1 + \mu_\text{ref}^2 \left(\frac{q_\perp^2}{q_\parallel^2} - 1 \right) \right]^{1/2} \ \text{, \,\,\,and}\\
\mu(\mu_\text{ref}) &= \mu_\text{ref} \frac{q_\perp}{q_\parallel} \left[ 1 + \mu_\text{ref}^2 \left(\frac{q_\perp^2}{q_\parallel^2} - 1 \right) \right]^{-1/2} \, . 
\end{split}
\label{eq:kref-muref}
\ee
These relations can be used to model the impact of the particular choice of the reference cosmology on the observed  power spectrum by 
mapping the values of $k_\text{ref}$ and $\mu_\text{ref}$, to the true ones as in \cite{1996MNRAS.282..877B}
\begin{equation}
P_\text{proj,lin}(k_\text{ref},\mu_\text{ref};\,z)=\frac{1}{q_\perp^2 q_\parallel} P_\text{zs,lin}(k(k_\text{ref},\mu_\text{ref}),\mu(\mu_\text{ref});z)\, ,
\label{eq:GC:AP}
\end{equation}
where the overall scaling of the power spectrum by $\left(q_\perp^2 q_\parallel\right)^{-1}$ follows from the dilation of the volume element 
due to the AP effect. 

\paragraph{The full linear model for the observed power spectrum of H$\alpha$ emitters.}

We can now write down the observed, linear anisotropic galaxy power spectrum as
\begin{equation}
P_\text{obs}(k_\text{ref},\mu_\text{ref};z) = 
\frac{1}{q_\perp^2 q_\parallel} 
\left(b\sigma_8(z)+f\sigma_8(z)\mu^2\right)^2 
\frac{P_{\delta\delta}(k;z)}{\sigma_8^2(z)} 
F_z(k,\mu;z) 
+ P_\text{s}(z) \, , 
\label{eq:GC:pklin}
\end{equation}
where all $k = k(k_\text{ref},\mu_\text{ref})$  and $\mu=\mu(\mu_\text{ref})$. 

\subsubsection{Non-linear scales for galaxy clustering} \label{sec:nonlineargc}

At late times and small scales, the matter distribution is affected by the non linear evolution of density fluctuations, which distorts
the shape of the power spectrum beyond the predictions of linear perturbation theory. Most of the \Euclid spectroscopic sample 
will lie in the range $0.9<z<1.8$. We expect that we may gain constraining power by including mildly non-linear scales into the analysis.
However,  this requires modelling non-linear effects that become relevant at the scales of the main feature we are trying to 
measure, namely the BAO signature. We follow a phenomenological model for non-linear effects in galaxy clustering 
similar to other recipes considered in the literature  \citep[e.g.][]{Seo:2007,Wang:2013b}. 
We separately model two main consequences of non-linearities, which are relevant to our 
calculation of the observed galaxy power spectrum. 
Our final model of the observed power spectrum deviates from the linear recipe described in \autoref{sec:GC:pk}  
in two separable ways, which we parameterise using two `non linear parameters', $\sigma_\text{p}(z)$ and $\sigma_\text{v}(z)$. 
Both these parameters have the same physical meaning,   
and represent the linear galaxy velocity dispersion $\left< u_\parallel^2({\bf r},a)\right>$, with 
$u_\parallel({\bf r},a) = v_\parallel({\bf r},a)/(aHf)$, and can thus be 
computed  from the linear matter power spectrum as 
\begin{equation}
\sigma_\text{v}^2(z)=\sigma_\text{p}^2(z)=\frac{1}{6\pi^2}\int P_{\delta\delta}(k,z)\;{\rm d}k\;.
\label{eq:sigmav}
\end{equation}
However, we keep them separate to allow for some theoretical uncertainty in our practical model of non-linearities.

The first effect of non-linearity is that the description of RSD requires an additional factor that we model as a 
Lorentzian, which accounts for the finger-of-God (FoG) effect under the assumption 
of an exponential galaxy velocity distribution function \citep{Hamilton_1998}. 
Secondly, we must account for the damping of the BAO feature, which we calculate with another exponential damping factor, acting 
in both directions: radial and transverse. This is accounted for by using the so-called `de-wiggled' power 
spectrum, $P_\text{dw}(k,\mu;z)$, \cite[see][]{Wang:2013b}. Therefore we model the true non-linear underlying redshift-space 
power spectrum as (cf. equation Eq.~\ref{eq:GC:linRSD} above)
\be
P_\text{zs}(k,\mu;\,z)=
\left\{\frac{1}{1+[f(z)k\mu\ \sigma_\text{p}(z)]^2}\right\}
\left[b(z)\sigma_8(z)+f(z)\sigma_8(z)\mu^2\right]^2\frac{P_\text{dw}(k,\mu;z)}{\sigma_8^2(z)},
\label{eq:pk_zs}
\ee
where the combination $2f(z)\sigma_\text{p}(z)$ would correspond to the pairwise peculiar velocity dispersion along the line of sight, and
$P_\text{dw}(k,\mu;z)$ is given by
\begin{equation}
P_\text{dw}(k,\mu;z) = P_{\delta\delta}(k;z)\,\text{e}^{-g_\mu k^2} + P_\text{nw}(k,\mu;z)\left(1-\text{e}^{-g_\mu k^2}\right) \, .
\label{eq:pk_dw}
\end{equation}
We can imagine this sum as a simple selective suppression of amplitude of the wiggles only, applied first by suppressing the amplitude 
of the whole linear power spectrum in the first term, and the restoration of the power in the broadband power spectrum (excluding the 
BAO wiggles) in the second term.
Therefore, $P_{\delta\delta}(k;z)$ is still the linear power spectrum, whereas $P_\text{nw}(k;z)$ is a `no-wiggle' power spectrum with the 
same broad band shape as $P_{\delta\delta}(k;z)$ but without BAO features, which we compute using the formulae of 
\cite{Eisenstein_Hu_1997b}. The function 
\begin{equation}
g_\mu(k,\mu,z) = \sigma_\text{v}^2(z)\left\{1 - \mu^2 + \mu^2\left[1+f(z)\right]^2\right\}
\label{eq:g_mu}
\end{equation}
is the non-linear damping factor of the BAO signal derived by \cite{EisensteinSeoWhite2007}.
The parameter $\sigma_\text{v}(z)$ controls the strength of the non-linear damping of the BAO signal in all directions in three-dimensions.

We use Eq.~(\ref{eq:sigmav}) to define the fiducial values of $\sigma_\text{p}(z)$ and $\sigma_\text{v}(z)$.
In practice, in \Cref{eq:pk_zs,eq:g_mu} we write the $\sigma_\text{v}(z)$ and $\sigma_\text{p}(z)$ as
\begin{eqnarray}
\sigma_\text{v}(z) &=& \sigma_\text{v}(z_\text{mean})\frac{D(z)}{D(z_\text{mean})}\,, \label{eq:sigv_withGs}\\
\sigma_\text{p}(z) &=& \sigma_\text{p}(z_\text{mean})\frac{D(z)}{D(z_\text{mean})}\,, \label{eq:sigp_withGs}
\end{eqnarray}
where $z_\text{mean}$ is the central redshift of the survey, and $D(z)$ is the growth factor from \autoref{eq:growthfactor-gc-val}, 
treating the values of $ \sigma_\text{v}(z_\text{mean})$ and $\sigma_\text{p}(z_\text{mean})$ as free parameters. We note that we assume in \autoref{eq:sigmav} that $\sigma_\text{v}^2(z)=\sigma_\text{p}^2(z)$, so that the above equation is most likely not the correct scaling in detail; we are making the assumption that we understand the increase with time of the non-linear impact of the parameters as 
being related to the growth of structure. This is in line with the intuition that  $\sigma_\text{v}(z)$ and $\sigma_\text{p}(z)$ are related to 
the velocity dispersion of galaxy populations

An additional effect of the non-linear evolution of density fluctuations is to modify the broadband shape of the power spectrum, increasing the power
on small scales. In the range of scales and redshifts in which we apply our recipe (see \autoref{sec:ccgc}), this effect can be approximately described by 
changing the value of the shot noise term $P_\text{s}(z)$ that we include in our recipe \citep[see e.g.][]{Ho2012}. A more detailed treatment of small scale 
non linearities is out of the scope of our analysis. We note that this is a conservative treatment of non-linear scales; the implementation of a more detailed modelling 
of non linearities in which we could accurately predict these broadband distortions as a 
function of the cosmological parameters would allow us to extract additional information from the shape of the 
observed power spectrum. By marginalizing over $P_\text{s}(z)$ we are conservatively discarding this information from our forecasts. 

\paragraph{The full non-linear model for the observed power spectrum of H$\alpha$ emitters.}

We can now write down the full observed, non-linear anisotropic galaxy power spectrum as
\begin{equation}
P_\text{obs}(k_\text{ref},\mu_\text{ref};z) = 
\frac{1}{q_\perp^2 q_\parallel} 
\left\{\frac{\left[b\sigma_8(z)+f\sigma_8(z)\mu^2\right]^2 }{1 + [f(z)k\mu\sigma_\text{p}(z)]^2}\right\} 
\frac{P_\text{dw}(k,\mu;z)}{\sigma_8^2(z)}F_z(k,\mu;z)+ P_\text{s}(z) \, , 
\label{eq:GC:pk}
\end{equation}
where all $k = k(k_\text{ref},\mu_\text{ref})$ and $\mu=\mu(\mu_\text{ref})$. All the quantities in this expression can be calculated using equations given in this section.
A number of alternatives have been considered in literature, but so far, the phenomenological model of Eq.~(\ref{eq:GC:pk}) has proved to be at least practical, all 
models being in any case only an approximation 
\citep[see e.g.][ for further discussion]{1996ApJS..105...37S, 2012PhRvD..85f3509B,2015PhRvD..92d3514B, 2017PDU....18...73C}.

\subsubsection{Modelling beyond $\Lambda$CDM}
\label{Sec:model_beyond_gc}
In order to calculate the DE FoM (see \autoref{sec:fom}), we must transform our original parameter set (see Eq. \ref{def:GC:pars}) 
into the subset of parameters whose FoM we need. 
We choose to freely vary the {redshift-dependent} parameters in \autoref{def:GC:pars} to avoid limiting ourselves to any particular DE or MG 
model throughout most of our galaxy clustering analysis. We then project our initial Fisher matrix using a Jacobian matrix of derivatives into 
the parameter space specific to our non-$\Lambda$CDM model, as described in \autoref{sec:gc_fisher}.

In particular, the redshift-dependent parameters $D_\textrm{A}(z)$ and $H(z)$ will carry most of the information on the expansion history 
of our universe and therefore contribute most to the constraints on $w_0$ and $w_a$. On the other hand, it is the redshift-dependent 
parameter $f\sigma_8(z)$ that most strongly depends on $\gamma$ via \autoref{eq:ggrowth}. Therefore, this parameter 
(in all its redshift bins) will contribute most to the constraints we achieve on $\gamma$. However, when it comes to $\gamma$, this 
simple picture of dependence is not the whole story. 
Since $\sigma_8(z) = \sigma_{8} D(z)$ depends on the growth factor as well, it directly depends on the $\gamma$ parameter. 
The same is also true for the matter power spectrum itself. For this reason, we always use the ratio $P_\text{dw}(k,\mu;z)/\sigma_8^2(z)$. 
In this way the dependence on $\gamma$ is cancelled as long as the approximation 
$\smash{P_\text{dw}(k,\mu;z) \simeq D^2(z) P_\text{dw}(k,\mu;0)}$ is valid. 
Secondly, the parameter $\sigma_\text{p}$ is calculated from an integral over $P_{\delta\delta}(k;z)$, and hence also depends on $\gamma$. 
This means that when $\sigma_\text{p}$ is considered simply as a nuisance parameter and marginalised over, no constraints on $\gamma$ come from it. On the other hand, if the value of this parameter is computed through \autoref{eq:sigmav}, it
 contributes to the constraint on $\gamma$ as well as the shape parameters\footnote{Massive-neutrinos lead to some scale dependence 
in the growth function. However, as discussed in the next section, the scale dependence of the growth for the minimal mass case 
$\Sigma_i m_{\nu,i} = 0.06$eV is very small and can be neglected here.}.
Finally the function $g_\mu$ from \autoref{eq:g_mu} depends on the cosmological parameters ($\gamma$ and shape) 
through $P_{\delta\delta}$ in $\sigma_{\rm v}$ (calculated in Eq.~\ref{eq:sigmav}) and $f(z)$, which we also include in the modelling.

\subsubsection{Covariance matrix for galaxy clustering}

The statistical constraining power of a galaxy survey depends mainly on the abundance of its target galaxy sample and how the 
survey strategy determines the survey's selection function. Modelled simply, this means that the statistical errors on the power 
spectrum can be estimated by knowing the observed number density of its galaxy targets (in our case H$\alpha$ emitters) and 
the survey volume. 
The expected number of H$\alpha$ emitters per unit area and redshift intervals that will be observed by \Euclid is given 
in \autoref{tab:gc_nofz}. This can be converted to the number density in each redshift bin,
\begin{equation}
n(z) = \frac{{\rm d}N(z)}{{\rm d}\Omega{\rm d}z} \, \frac{A_{\rm survey} }{ V_{\rm s}(z)}  \, {\Delta}z \ ,
\label{eq:gc-ndens}
\end{equation}
where the width of the redshift bin is given in terms of the minimum and maximum redshift of the bin, 
${\Delta}z =z_{\rm max}-z_{\rm min}$. The minimum and maximum redshifts, and the volume of the redshift bin,
$V_\text{s}$,  are given in \autoref{tab:gc_nofz}, 
while the survey area $A_{\rm survey}$ is specified in \autoref{tab:WL-specifications}. 

Assuming that the observed power spectrum  follows a Gaussian distribution, its
covariance matrix can be approximated by
\begin{equation}
C(\vec{k},\vec{k}')\approx\frac{2\left(2\pi\right)^3}{V_{\rm eff}(k,\mu,z)}P_{\rm obs}^2\left(k,\mu;z\right)\delta_{\rm D}\left(\vec{k}-\vec{k}'\right)\, ,
\label{eq: Gaussian_cov_pk}
\end{equation}
where $\delta_{\rm D}$ is the Dirac delta function (reflecting the independence of different Fourier modes) and 
$V_{\rm eff}(\vec{k})$ is the effective volume of the survey given by 
\begin{equation}
V_\text{eff}(k,\mu;z) = V_\text{s} (z)\left[\frac{n(z) P_\text{obs}(k,\mu;z)}{n(z) P_\text{obs}(k,\mu;z) +1}\right]^2\ . \label{eq:gc_effvol}
\end{equation}
Here, $ P_\text{obs}(k,\mu;z)$ is the 
full observed two-dimensional power spectrum, and all quantities are evaluated for the fiducial cosmology.

\subsubsection{Applying the Fisher formalism to galaxy clustering}
\label{sec:gc_fisher}
Following \cite{Tegmark_1997} and \cite{2003ApJ...598..720S} the Fisher matrix for the 
observed galaxy power spectrum in a redshift bin centred on $z_i$ is:
\be
F_{\alpha\beta}^\textrm{bin}(z_{ i}) = \frac{1}{8\pi^2} 
  \int^1_{-1} d\mu
  \int^{k_\text{max}}_{k_\text{min}} k^2{\rm d}k\; 
    \left[\frac{\partial \ln P_\text{obs}(k,\mu ;z_{ i})}{\partial \alpha} 
    \frac{\partial \ln P_\text{obs}(k,\mu ;z_{ i})}{\partial \beta}\right] 
    V_\text{eff}(z_{ i};k,\mu) \ ,
\label{eq:pkfm}
\ee
where $\alpha$ and $\beta$ run over the cosmological parameters we wish to vary, which in turn enter implicitly into 
$P_\text{obs}(z;k,\mu)$ (see Eq.~\ref{eq:GC:pk}).
The values of $k$ and $\mu$ in  \autoref{eq:pkfm} are those obtained in the reference cosmology, $k_\text{ref}$ and $\mu_\text{ref}$, but 
we drop the  subscript ``ref'' from now on for clarity. 
The total Fisher matrix is then calculated by summing over the redshift bins:
\begin{equation}
F_{\alpha\beta} = \sum_{i=1}^{N_{z_\text{bin}}} F_{\alpha\beta}^\textrm{bin}(z_{ i}) \, .
\label{eq:fishersum}
\end{equation}
We note here that for the redshift-dependent parameters, we assume that the redshift bins are independent, 
therefore the sum in \autoref{eq:fishersum} reduces to a single term.
For example, if we take $\alpha~=~H(z_{j})~=~H_{j}$, we have 
\be
F_{H_{j}\beta}^\textrm{bin}(z_{i}) = 0\ \;\; \forall\, { i} \neq {j} \, ,
\ee
since
\be
\frac{\partial \ln P_\text{obs}(z_{ i};k,\mu)}{\partial H(z_{ j})}~=~0\ \;\; \forall\ z_{ i}~\neq~z_{ j} \, .
\ee
The sum in \autoref{eq:fishersum} then reduces to a single term with $i=j$:
\be
F_{H_j\beta} = F_{H_{ j}\beta}^\textrm{bin}(z_{ j}) \, .
\ee
This includes both the off-diagonal terms containing the covariance between different redshift bins of the same {redshift-dependent} parameter, and the covariance between different $z$-bins of different parameters, leaving only the non-zero covariances between {redshift-dependent} parameters within the same redshift bin. In other words,
\be
F_{\alpha(z_{ i})\beta(z_{ j})}^\textrm{bin} = 0\ \;\; \forall\ { i}\neq { j}
\ee
if $\alpha(z_{ i})$ and $\beta(z_{ j})$ are redshift-dependent parameters, and in the following we abbreviate these subscripts to $ij$.

Explicitly, the \textit{model-independent} Fisher matrix is given by \autoref{eq:pkfm} for one single redshift bin, where the subscripts $\alpha$ and $\beta$ run over the shape- and redshift- dependent parameters. 
In this way we have a matrix for one single redshift with $6+(5\times1)$ rows and columns in the full non linear case as reported in \autoref{def:GC:pars} (in the linear case, the number of $z$-independent parameters falls to 4). 
Schematically, let us assume that the Fisher matrix for one redshift bin is 
\be
F_{ij}(z_{k})=
\left[
\begin{array}{cc}
(6\times 6)_{k} & (6\times 5)_{k} \\
(5\times 6)_{k} & (5\times 5)_{k}\\
\end{array}
\right] \, ,
\label{eq:comparison-gc-fm-block}
\ee
where each term corresponds to a block matrix, with each element calculated from derivatives with respect to the parameters of the row and column.  The first $6\times6$ block is composed of derivatives with respect to the shape parameters only, the $5\times5$ block by those with respect to the redshift dependent parameters only, and the $5\times6$ block (which is the transpose of the $6\times5$) contains the mixed derivatives. The total Fisher matrix, including all redshift bins, is then a matrix of dimension $6+5\times n_\text{bin}$. Making use of the formalism in \autoref{eq:comparison-gc-fm-block} we then have
\be
F_{ij}=
\left[
\begin{array}{cccccc}
\sum_{i=1}^{n_\text{bin}}(6\times6)_{i} & (6\times5)_{1}& (6\times5)_{2} & (6\times5)_{3}  & \dots &   (6\times5)_{n_{\rm bin}} \\
(5\times6)_{1} & (5\times5)_{1} & 0 & 0 & \dots & 0\\
(5\times6)_{2} & 0& (5\times5)_{2} & 0 & \dots & 0\\
(5\times6)_{3} & 0& 0&(5\times5)_{3}&\dots&0\\
\vdots & \vdots & \vdots &\vdots & \ddots&\vdots\\
(5\times6)_{n_{\rm bin}} & 0 & 0 & 0& \dots & (5\times5)_{n_{\rm bin}}\\
\end{array}
\right] \ ,
\label{eq:comparison-gc-fm-block-tot}
\ee
where we must sum over the contributions to the shape parameter elements from each redshift, but treat the redshift bins as creating separate {redshift-dependent} parameters.

\paragraph{Derivatives.}\label{sec:gcderivs}
In order to convert the data covariance into the desired parameter covariance in the Fisher matrix calculation in \autoref{eq:pkfm}, we
 need the derivatives of the observable power spectrum $P_\text{obs}$ with respect to the parameters of the chosen parameter space. 
 From \autoref{eq:GC:pk} one may calculate all derivatives numerically using finite differencing methods. In practice, a derivative step 
 should be chosen such that the convergence level towards the true value (generally the analytical one) is reached. However given that 
 the analytical values are inaccessible for most of the derivatives computed here, the best possible test consists in checking the step 
 range for which the derivatives are stable. Choosing large steps raises the truncation error due to the discretization of the parameter 
 values, while using too small steps increases roundoff errors. We note that for the residual shot noise, the following analytical 
 equation may be used instead:
\begin{equation}
\frac{\partial\ln P_\text{obs}\left(z_{ j},k,\mu\right)}{\partial P_{\rm s}(z_{ j})} =
\frac{1}{P_\text{obs}\left(z_{ j},k,\mu\right)} \ 
\label{eq:deriv-shot-noise-gc}
\end{equation}
(See \autoref{sec:ccgc} for practical guidelines on how to calculate the derivatives).

\paragraph{Projection to a new parameter space.}

As discussed in \autoref{sec:recipegc}, we first consider as independent parameters the values of 
$D_A(z_{ i})$, $H(z_{ i})$, $f\sigma_8(z_{ i})$, $b\sigma_8(z_{ i})$ and $P_{\rm s}(z_{ i})$, in each of the $N_{z}$ redshift bins. 
We use two additional nuisance parameters, $\sigma_\text{v}(z_\text{mean})$ and $\sigma_\text{p}(z_\text{mean})$, to account for 
the  uncertainty in our modelling of non-linear effects. 
The full list of parameters, $\{\theta_{\rm i}\}$, of the initial Fisher matrix then includes redshift-independent parameters, 
$\{\theta_{\rm zi}\}$, and redshift-dependent ones $\{\theta_{\rm zd} \}$: $\{\theta_{\rm i} \}= \{\theta_{\rm zd} \}+ \{\theta_{\rm zi}\}$. 
These parameters are listed in \autoref{def:GC:pars} and recalled below for convenience:
\begin{itemize}
\item $5\times N_z$ redshift-dependent parameters,  \\
$\{\theta_{\rm z\text{d}} \}= \{ \ln D_A(z_{ i}),\, \ln H(z_{ i}),\,\ln[f\sigma_8(z_{ i})], \, \ln[b\sigma_8(z_{ i})],\, P_{\rm s}(z_{ i}) \}$
\item 4 cosmological redshift-independent parameters and 2 non-linear model parameters, \\
$\{\theta_{z{\rm i}} \}=\{{\omega_\text{b,0}}, h, {\omega_\text{m,0}}, n_\text{s}, \sigma_\text{v}(z_\text{mean}), \sigma_\text{p}(z_\text{mean})\}$.
\end{itemize}
We must first decide which of these parameters to treat as nuisance parameters, and which we assume contribute to 
our knowledge of our final parameter set. We marginalise out the following nuisance parameters:
\begin{itemize}
\item 4 nuisance parameters to be marginalised over,\\
$\{\theta_\text{nuisance}\} = \{\ln[b\sigma_8(z_{ i})],\, P_{\rm s}(z_{ i}), \sigma_\text{v}(z_\text{mean}), \sigma_\text{p}(z_\text{mean})\}$.
\end{itemize}
We finally wish to forecast the errors on the following final cosmological parameter set in order to be able to combine the
GC-only results with the constraints expected from the weak lensing measurements:
\begin{itemize}
\item 9 cosmological parameters as chosen in \autoref{sec:fiducial},\\
$\{\theta_\text{final}\} = \{\Omega_{\rm b,0},\,h,\,\Omega_{\rm m,0},\,n_{\rm s},\,\Omega_{\rm DE,0}\,w_0,\,w_a,\,\sigma_8,\,\gamma\}$.
\end{itemize}
As described in \autoref{sec:projection}, in order to transform our Fisher matrix from one parameter set to another, we must 
understand the mutual dependence of the remaining initial and final parameters, which is encapsulated in the Jacobian matrix, 
$\tens{J}$ of derivatives of the initial parameters, $\alpha$, with respect to the desired final parameters, $\kappa$, as
\begin{equation}
J_{\alpha\kappa} = \frac{\partial\alpha}{\partial\kappa} \, ,
\end{equation}
and so
\begin{equation}
\widetilde{F}_{\kappa\lambda} = \sum_{\alpha,\beta} J_{\alpha\kappa} F_{\alpha\beta} J_{\beta\lambda} \, ,
\label{gc:eq:projection}
\end{equation}
which defines the final Fisher matrix that we use. 

\paragraph{FoM.}

The specific Fisher submatrices needed to compute the DE FoM (see \autoref{sec:fom}) are calculated through a combination of marginalising 
out the DE-independent parameters and projecting others into the DE parameter space of choice. The exact choices we make, in terms of 
scales included and cosmology considered, can impact the final result. We have tested the GC recipe for a variety of different choices and in \autoref{sec:results} we identify a pessimistic and optimistic case. Different marginalization approaches are described in the literature, \citep[e.g.][]{Wang:2010}. 
We choose to follow what could be referred to as the ''full $P(k)$ method, with growth information included'', where, after marginalizing 
out the observational nuisance parameters,  $b(z_i)\sigma_8(z_i)$ and $P_s(z_i)$, the remaining Fisher matrix is projected into the final 
parameter space of $\{\theta_\text{final}\}$, and the FoM is obtained after again marginalizing out everything except $w_0$ and $w_{ a}$.
We therefore first marginalise out the observational nuisance parameters, $b\sigma_8(z_{ i})$ and $P_{\rm s}(z_{ i})$ by removing their 
corresponding rows and columns from the parameter covariance matrix, $F^{-1}_{ ij}$ 
\citep[as usual, see \autoref{Sec:Fisher:general} and][]{Coe:2009}. 
We are then left 
with $3\times N_z$ redshift-dependent parameters and 4 shape parameters.
We project these $3\times N_z + 4$ parameters into the required final parameter space using \autoref{gc:eq:projection} above.

\subsection{Recipe for weak lensing} \label{WLrecipe}

This section describes the forecasting procedure recommended for weak lensing, using the observable tomographic cosmic shear power spectrum. We first describe the calculation of this observable, and then how it is applied in the Fisher matrix context; that is described in general in \autoref{Sec:Fisher:general}. We note that we refer to {\textit{weak lensing}} as the physical phenomenon, and {\textit{cosmic shear}} as the observable and summary statistic that uses weak lensing to extract cosmological parameters. 

The large-scale cosmic structure deflects the path of photons from distant galaxies, which induces distortions in the images of these galaxies. Locally such distortions can be decomposed at the linear level into convergence and a (complex) shear distortion, which are respectively related to the magnification and shape distortion of the image. More specifically, these are the trace and trace-free parts of the (inverse) amplification matrix that describes the linearised mapping from lensed (image) coordinates to unlensed (source) coordinates \citep[see e.g.][for a general review]{2015RPPh...78h6901K}. Convergence is a change in the observed size of a galaxy, and shear is a change in the observed third flattening or third eccentricity (known as `ellipticity', or polarisation). In this paper we only consider the cosmological signal in the shear field \citep[see e.g.][ for a discussion of the use of the convergence in cosmological parameter estimation]{2015MNRAS.452.1202A}. The shear field caused by large-scale structure has a zero mean because we assume isotropy and homogeneity of the Universe, but its two-point correlation function and its power spectrum (as well as higher-order correlations) contain cosmological information that probes both the background evolution of the Universe and the growth of cosmic structure; these two-point statistics are known as 'cosmic shear'.  

There are several ways in which the cosmic shear can be exploited for cosmological parameter inference. The real/configuration space measurement of the two-point statistic, as a function of angular seperation on the celestial sphere, is known as the cosmic shear correlation function. On the other hand, the angular spherical-harmonic measurement of the two-point statistic is known as the cosmic shear power spectrum. Both of these statistics can be computed in a series of redshift slices/bins that capture the geometry of the three-dimensional shear field. This redshift binning is known as 'tomography' and is required in order to achieve high-precision dark energy measurements \citep{1999ApJ...522L..21H, 2005PhRvD..72b3516C, 2017PDU....18...73C, 2018MNRAS.480.3725S}. Depending on which observable one is interested in, different approximations may be used to make the equations more tractable, but a thorough description of all methods is beyond the scope of this paper. In the following, we shall focus on the cosmic shear (tomographic) power spectrum.

In this section we describe this observable in detail. In addition we quantify the major astrophysical contamination of cosmic shear, which is the intrinsic (local) alignment (IA) of galaxies. The IA signal can be modelled as a projected power spectrum, similar to the cosmic shear observable itself. 

As described in \autoref{Sec:Fisher}, $\theta_\textrm{final} = \{\Omega_{\rm b,0},\,h,\,\Omega_{\rm m,0},\,n_{\rm s},\,\Omega_{\rm \Lambda,\,0},\,w_0,\,w_a,\,\sigma_8,\,\gamma\}$ is the full set of cosmological parameters of interest. In contrast to the GC Fisher matrix, derivatives of the weak lensing observable (cosmic shear) can be computed directly in this parameter space, and therefore no pre-computation, or projection from a different parameter set is required. 

\subsubsection{The observable tomographic cosmic shear power spectrum}\label{sec:wlobs}
In this section we define the primary weak lensing observable. While the cosmological signal with which we are concerned is the additional ellipticity caused by the lensing of the large-scale structure, that we summarise in the cosmic shear power spectrum, we also wish to model the primary astrophysical systematics. We consider five main quantities that must be modelled in order to recover the observable cosmic shear power spectrum:
Firstly, {the (theoretical) cosmic shear power spectrum}, namely the primary cosmological power spectrum;
secondly, {the intrinsic alignment power spectrum}, modelling the local alignment of galaxies, representing the main astrophysical systematics;
as a third quantity, we consider {the small-scale part of the matter power spectrum}, including a halo model describing the clustering of dark matter on small scales beyond linear theory ($k<7h\,\mathrm{Mpc}^{-1}$) to which the cosmic shear power spectrum is particularly sensitive \citep{2018PhRvD..98d3532T}, which includes the impact of baryonic feedback \citep[e.g.] []{Semboloni_etal_2011,copeland};
the fourth item includes {photometric redshifts and number density}, which model the inferred uncertainty in galaxy positions due to the broad-band estimates of the redshifts, where the redshift distribution affects the signal part of the cosmic shear power spectrum but the total number of galaxies does not;
as fifth, and last point, {the shot noise} due to Poisson sampling by galaxy positions of the shear field, which is affected by the total number of galaxies observed.
We now describe each of these effects and show how the observed cosmic shear power spectrum is constructed. The discussion of the non-linear modelling of the matter power spectrum is postponed to \autoref{sec:wlnl}. 

\paragraph{The cosmic shear power spectrum.} 
As defined above, shear is the change in the ellipticity of the image of a background galaxy, caused by the lensing effect of large-scale structure along the line-of-sight. For an individual galaxy we express this, to linear order, as 
\be
\label{eq:shear}
\epsilon=\gamma+\epsilon^{\rm I},
\ee
where $\gamma$ is the cosmological shear and $\epsilon^{\rm I}$ is the intrinsic (unlensed) ellipticity.
This ellipticity is a spin-2 quantity with zero mean over a large survey area and a non-zero two-point correlation function or power spectrum, encoding information on both the expansion history of the Universe and the matter power spectrum. The spherical harmonic transform of the two-point correlation function is the angular power spectrum. While the full computation of the cosmic shear power spectrum is relatively laborious \citep[see e.g.][]{2018arXiv180403668T} due to spherical Bessel functions and several nested integrals, the Limber approximation \citep[expected to be suitable for angular scales of $\ell\gtrsim100$; see][]{2016arXiv161104954K,KH17,2017JCAP...05..014L,Kaiser92,Gi12} allows one to simplify its expression to
\be
\label{eq:cosmicshear}
C^{\gamma\gamma}_{ij}(\ell) \simeq \frac{c}{H_0}\int\mathrm d z\,{\frac{W_{i}^{\gamma}(z) W_{j}^{\gamma}(z)}{E(z) r^2(z)} P_{\delta \delta}\!\left[ \frac{\ell + 1/2}{r(z)}, z \right] }, 
\ee
where $i$ and $j$ identify pairs of redshift bins, $E(z)$ is the dimensionless Hubble parameter of \autoref{eq:Ez}, $r(z)$ is the comoving distance, and $P_{\delta\delta}(k, z)$ is the matter power spectrum evaluated at $k =k_\ell(z)\equiv (\ell + 1/2)/r(z)$ due to the Limber approximation; we define the weight function $W^{\gamma}(z)$ as
\begin{align}
W^{\gamma}_{i}(z) &= 
\frac{3}{2} \left(\frac{H_0}{c}\right)^2 \Omega_{{\rm m},0}(1 + z) r(z)
\int_{z}^{z_{\rm max}}\mathrm d z^{\prime}\,n_i(z^{\prime})\left[ 1 - \frac{\tilde{r}(z)}{\tilde{r}(z^{\prime})} \right]\\
&=\frac{3}{2} \left(\frac{H_0}{c}\right)^2 \Omega_{{\rm m},0}(1 + z) r(z)\widetilde{W}_i(z),
\label{eq: lensweight} 
\end{align}
with $z_{\rm max}$ the maximum redshift of the source redshift distribution. With respect to the standard formalism, we introduce the reduced window function $\smash{\widetilde{W}_i(z)}$ (also colloquially known as the lensing efficiency), which will be important later. In the above we assume that the changes in the formalism due to non-flat spatial geometries can be captured via changes in the cosmological parameters, power spectrum and reduced window function rather than changes to the Limber approximation caused by modifying the spherical Bessel functions in the non-Limber approximated case to hyper-spherical Bessel functions; this assumption is well justified as show in \cite{taylornf}. 

\paragraph{The intrinsic alignment power spectrum.} \label{par:IA} Tidal interactions during the formation of galaxies, and other processes (e.g. spin correlations) may induce a preferred intrinsically correlated orientation of galaxy shapes, affecting the two-point shear statistics \citep{iareview1,iareview3,iareview2}. As a consequence, the observed shear power spectrum will include these additional terms, which cannot be neglected. This effect, referred to as intrinsic alignment (IA), can be considered to be an astrophysical systematic error, contributing to the observed signal and that cannot be removed efficiently by observational strategies. 
The contribution of IA can be seen by considering the two-point correlation function of \autoref{eq:shear},
which results in four terms 
\be 
C^{\epsilon\epsilon}_{ij}(\ell)=C^{\gamma\gamma}_{ij}(\ell)+C^{\rm I\gamma}_{ij}(\ell)+C^{\gamma\rm I}_{ij}(\ell)+C^{\rm II}_{ij}(\ell) \, ,
\ee
where $i<j$, $C^{\rm I\gamma}_{ij}(\ell)$ represents the correlation between background shear and foreground intrinsic alignment, while $C^{\gamma\rm I}_{ij}(\ell)$ represents correlations between foreground shear and background ellipticity. The latter is zero, because a foreground shear should not be correlated with a background ellipticity except if the galaxy redshifts are misassigned. The former, together with the intrinsic-intrinsic (II) alignment autocorrelation power spectrum, can be written in a very similar way to the cosmic shear expression \autoref{eq:cosmicshear}, using a model known as the {linear-alignment} model: 
\begin{align}
C^{\rm I\gamma}_{ij}(\ell) & = \frac{c}{H_0}\int\mathrm d z\,{\frac{W_{i}^{\gamma}(z) W_{j}^{\rm IA}(z) + W_{i}^{\rm IA}(z) W_{j}^{\gamma}(z)}{E(z) r^2(z)} P_{\delta{\rm I}}\!\left[ \frac{\ell + 1/2}{r(z)}, z \right]}, \nonumber\\
C^{\rm II}_{ij}(\ell) & =\frac{c}{H_0}\int\mathrm d z\,{\frac{W_{i}^{\rm IA}(z) W_{j}^{\rm IA}(z)}{E(z) r^2(z)} P_\mathrm{II}\!\left[ \frac{\ell + 1/2}{r(z)}, z \right]} .
\end{align}
The weight function for IA can be conveniently written as
\begin{equation}
W_i^{\rm IA}(z) =  \frac{n_{i}(z)}{c/H(z)} = \left( \frac{H_0}{c} \right) n_i(z) E(z)
\label {eq: iaweight}
\end{equation}
where the term $c/H(z)=\mathrm dr/\mathrm dz$ reflects the choice of integrating with respect to $z$ instead of $r$.

Intrinsic alignment power spectra can be expressed as a function of the matter power spectrum. There are many models available in the literature for describing the effect of IA \citep[e.g.][]{iareview1}; here we follow a simplified yet observationally motivated approach with 
\begin{align}
P_{\delta {\rm I}}(k, z) &= -{\cal{A}}_{\rm IA} {\cal{C}}_{\rm IA} \Omega_{{\rm m},0}\frac{{\cal{F}}_{\rm IA}(z)}{D(z)}P_{\delta \delta}(k,z)\label{eq: pdienla},\\
P_{\rm I I}(k, z) &=  \left[-{\cal{A}}_{\rm IA} {\cal{C}}_{\rm IA} \Omega_{{\rm m},0}\frac{{\cal{F}}_{\rm IA}(z)}{D(z)}\right]^2P_{\delta \delta}(k,z), \label{eq: piienla}
\end{align}
where the function ${\cal{F}}_{\rm IA}(z)$ reads
\begin{equation}
{\cal{F}}_{\rm IA}(z) = (1 + z)^{\eta_{\rm IA}}[\langle L \rangle(z)/L_{\star} (z)]^{\beta_{\rm IA}}  \ . 
\label{eq: effeia}
\end{equation}
We refer to this as the {\it extended} non-linear alignment (eNLA) model. For ${\cal{F}}_{\rm IA}(z) = 1$, the model reduces to the so-called non-linear alignment (NLA) model widely used in the literature \citep{2007NJPh....9..444B}. In \autoref{eq: effeia} $\langle L \rangle(z)$ and $L_{\star}(z)$ are the redshift-dependent mean and the characteristic luminosity of source galaxies, as computed from the luminosity function. With respect to NLA, the extended model therefore includes luminosity dependence of the IA, as hinted at by low redshift studies and hydrodynamical simulations \citep{IAhydro1,IAhydro2,IAhydro3}.  The parameters $\eta_{\rm IA}$, $\beta_{\rm IA}$
and ${\cal{A}}_{\rm IA}$ are free parameters of the model and should be determined by fitting the data and/or via carefully designed simulations. Here, we fix them to the following fiducial values: 
$\{\mathcal A_{\rm IA},\eta_{\rm IA},\beta_{\rm IA}\}=\{1.72,-0.41,2.17\}$, while $\mathcal C_{\rm IA} = 0.0134 $ remains fixed in the Fisher analysis, as it is degenerated with $ \mathcal A_{\rm IA}$.

\paragraph{Photometric redshifts and number density.}
One of the key ingredients in \Cref{eq: lensweight,eq: iaweight} is the number density distribution $n_i(z)$ of the observed galaxies in the $i{\rm th}$ bin. For photometric redshift estimates, this can be written as
\begin{equation}
n_i(z) = \frac{\int_{z_i^-}^{z_i^+}{\rm d} z_{\rm p}\,n(z) p_{\rm ph}(z_{\rm p}|z) }
{\int_{z_{\rm min}}^{z_{\rm max}} {\rm d} z\,\int_{z_i^-}^{z_i^+}{\rm d} z_{\rm p}\,n(z) p_{\rm ph}(z_{\rm p}|z)},
\label{eq: defni}
\end{equation}
where $(z_i^-, z_i^+)$ are the edges of the $i{\rm th}$ redshift bin. The underlying true distribution $n(z)$ appearing in this expression is chosen to be in agreement with the \Euclid Red Book\footnote{Although other more realistic choices are possible, we do not explore them here. We are, however, confident that the choice of $n(z)$ does not impact the results of the comparison significantly \citep{RT}, that is to say, if two codes agree for a given $n(z)$, they still agree if a different $n(z)$ is adopted.} \citep{RB, 1994MNRAS.270..245S}:
\begin{equation}
n(z) \propto  \left( \frac{z}{z_0} \right)^2 \exp{\left[ - \left( \frac{z}{z_0} \right)^{3/2} \right]},
\label{eq: nzredbook}
\end{equation}
where $z_0 = z_{\rm m}/\sqrt{2}$ with $z_{\rm m}$ being the median redshift. According to the \Euclid Red Book, $z_{\rm m} = 0.9$ and the surface density of galaxies is $\bar n_{\rm g} = 30 \ {\rm arcmin}^{-2}$. With this choice, we can set the edges of the $10$ equi-populated bins, which are
\begin{equation}
z_{i} = \{0.0010, 0.42, 0.56, 0.68, 0.79, 0.90, 1.02, 1.15, 1.32, 1.58, 2.50\},
\end{equation}
with $z_i^- = z_i$ and $z_i^+ = z_{i + 1}$. We note that, since we use $10$ equi-populated bins, the number density in each bin is simply $\bar n_{\rm g}/10$. We note that the minimum redshift range is lower than in \cite{RB} where a minimum of $0.2$ was used as a conservative limit to avoid potential catastrophic redshift outliers. We use a more optimistic value of $0.001$, but note that since the lensing kernel peaks around $z=0.3$ the change in inferred parameter errors is small.

The number density $n(z)$ is convolved in \autoref{eq: defni} with the probability distribution function $p_{\rm ph}(z_{\rm p}|z)$, describing the probability that a galaxy with redshift $z$ has a measured redshift $z_{\rm p}$ \citep{2009MNRAS.399.2107K}. A convenient parameterisation for this quantity is given by
\begin{eqnarray}
p_{\rm ph}(z_{\rm p}|z) = \frac{1 - f_{\rm out}}{\sqrt{2 \pi} \sigma_{\rm b} (1 + z)} \exp{\left\{ - \frac{1}{2} \left[ \frac{z - c_{\rm b} z_{\rm p} - z_{\rm b}}{\sigma_{\rm b} (1 + z)} \right]^2 \right\}}
+ \frac{f_{\rm out}}{\sqrt{2 \pi} \sigma_{\rm o} (1 + z)} \exp{\left\{ -  \frac{1}{2}  \left[ \frac{z - c_{\rm o} z_{\rm p} - z_{\rm o}}{\sigma_{\rm o} (1 + z)} \right]^2 \right\}},\label{eq:wlpph},
\end{eqnarray}
which allows us to include both a multiplicative and additive bias in the redshift determination of a fraction $(1 - f_{\rm out})$ of sources with reasonably well measured redshifts, and a fraction $f_{\rm out}$ of catastrophic outliers (i.e., systems with severely incorrect estimate of the redshift). By modifying the parameters of this function, one can mimic different cases of interest. Our choice is summarised in \autoref{tab:photz}.
\begin{table}[h]
\centering
\caption{Parameters adopted to describe the photometric redshift distribution of sources of \autoref{eq:wlpph}.}
\label{tab:photz}
\begin{tabular}{ccccccc}
\hline
$c_{\rm b}$ & $z_{\rm b}$ & $\sigma_{\rm b}$ & $c_{\rm o}$ & $z_{\rm o}$ & $\sigma_{\rm o}$ & $f_{\rm out}$\\
\hline\hline
$1.0$ & $0.0$ & $0.05$ & $1.0$ & $0.1$ & $0.05$ & $0.1$\\
\hline
\end{tabular}
\end{table}
We note that we have fixed these quantities in the Fisher matrix estimate and do not explore their impact on the forecast; in a cosmological parameter inference on data, such parameters should be varied and self-calibrated. {These are nevertheless not entirely flexible and can be inferred independently, with some errors, for instance with clustering-based methods \citep{Gatti:2017hmb}.  The subsample of galaxies for which Euclid will provide spectroscopic redshift measurements will also help to calibrate the photometric redshift error distribution.  The maximal degradation of the FoMs, when redshift errors are considered as free parameters and are self-calibrated from data, has been recently investigated in \citet{Tutusaus:2020xmc}.}

\paragraph{The shot noise.} The uncorrelated part of the intrinsic (unlensed) ellipticity field acts as a shot noise term in the observed power spectrum. This is non-zero for auto-correlation (intra-bin) power spectra, but is zero for cross-correlation (inter-bin) power spectra, because 
ellipticities of galaxies at different redshifts should be uncorrelated. This term can be written as 
\be 
N^\epsilon_{ij}(\ell)=\frac{\sigma^2_\epsilon}{\bar n_i}\delta^{\rm K}_{ij} \,,
\ee
where $\bar n_{i}$ is the galaxy surface density per bin, and has to
be consistently expressed in inverse steradians; $\smash{\delta^{\rm
    K}_{ij}}$ is the Kronecker delta symbol; and $\sigma^2_\epsilon$
is the variance of the observed ellipticities whose square root value is given in \autoref{tab:WL-specifications} which is taken from \cite{Massey04}. We further discuss sources of errors in \autoref{errors_shear}. 

\paragraph{The observed cosmic shear power spectrum.} 
Finally in the flat-sky and Limber approximations, we can write the total observed tomographic shear angular power spectrum as
\be 
C^{\epsilon\epsilon}_{ij}(\ell)=C^{\gamma\gamma}_{ij}(\ell)+C^{\rm II}_{ij}(\ell)+C^{\rm I\gamma}_{ij}(\ell)+N^\epsilon_{ij}(\ell),
\ee
which, in an expanded form reads 
\begin{eqnarray}
C^{\epsilon\epsilon}_{ij}(\ell) & =& \frac{c}{H_0}\int\mathrm d z\,{\frac{W_{i}^{\gamma}(z) W_{j}^{\gamma}(z)}{E(z) r^2(z)} P_{\delta \delta}\!\left[ \frac{\ell + 1/2}{r(z)}, z \right] } 
 + \frac{c}{H_0}\int\mathrm d z\,{\frac{W_{i}^{\gamma}(z) W_{j}^{\rm IA}(z) + W_{i}^{\rm IA}(z) W_{j}^{\gamma}(z)}{E(z) r^2(z)}
P_{\delta{\rm I}}\!\left[ \frac{\ell + 1/2}{r(z)}, z \right]}\nonumber\\
& +& \frac{c}{H_0}\int\mathrm d z\,{\frac{W_{i}^{\rm IA}(z) W_{j}^{\rm IA}(z)}{E(z) r^2(z)} P_\mathrm{II}\!\left[ \frac{\ell + 1/2}{r(z)}, z \right]}
 +N^\epsilon_{ij}(\ell),
\label{eq: defcij}
\end{eqnarray}
where the first three terms give the contributions from the cosmological (theoretical) shear, cosmological shear-IA, and IA-IA correlations, respectively. The $ij$ label the tomographic redshift bin combinations into which the redshift distribution of sources have been divided. Each term in \autoref{eq: defcij} has the same general structure: an integral over redshift of the product of weight functions (depending only on the cosmic evolution) and a power spectrum (probing the growth of structures). The integration formally extends up to the redshift of the horizon, but in practice there is no contribution for $z\ge z_{\rm max}$ with $z_{\rm max}$ the maximum redshift of the source redshift distribution. Moreover, the lower limit must be different from zero, to avoid divergence caused by assumming the Limber approximation in \autoref{eq:cosmicshear}. Therefore, the integration in \autoref{eq: defcij} is in the range $z_{\rm min} \le z \le z_{\rm max}$ with $(z_{\rm min}, z_{\rm max}) = (0.001, 2.5)$ as a reasonable choice.

\paragraph{Compact Notation.} 
For convenience we rewrite the constituents of \autoref{eq: defcij} as
\begin{eqnarray}
C^{\epsilon\epsilon}_{ij}(\ell) = \int_{z_{\rm min}}^{z_{\rm max}}{\rm d}z\,\left\{K_{ij}^{\gamma \gamma}(z) P_{\delta \delta}\!\left[ \frac{\ell + 1/2}{r(z)}, z \right] 
+ K_{ij}^{{\rm I}\gamma}(z) P_{\delta {\rm I}}\!\left[ \frac{\ell + 1/2}{r(z)}, z \right]
+ K_{ij}^{\rm I I}(z) P_{\rm I I}\!\left[ \frac{\ell + 1/2}{r(z)}, z \right]\right\}
+N^\epsilon_{ij}(\ell),
\label{eq: defcijshort}
\end{eqnarray}
having defined the following kernel functions
\begin{eqnarray}
K_{ij}^{\gamma \gamma}(z) &=& \left[ \frac{3}{2} \Omega_{{\rm m},0} (1 + z) \right]^2 \left( \frac{H_0}{c} \right)^3\frac{\widetilde{W}_{i}^{\gamma}(z) \widetilde{W}_{j}^{\gamma}(z)}{E(z)},\label{eq: kijggend}\\
K_{ij}^{{\rm I}\gamma}(z) &=& \left[ \frac{3}{2} \Omega_{{\rm m},0}(1 + z) \right] 
\left( \frac{H_0}{c} \right)^3 
\frac{n_i(z) \widetilde{W}_{j}^{\gamma}(z) + n_j(z) \widetilde{W}_{i}^{\gamma}(z)}{\tilde{r}(z)},\label{eq: kijgiend}\\
K_{ij}^{\rm II}(z) &=& \left( \frac{H_0}{c} \right)^3 \frac{n_i(z) n_{j}(z) E(z)}{\tilde{r}^2(z)}\label{eq: kijiiend}.
\end{eqnarray}
It is worth noticing that the IA kernel function $K^{II}_{ij}(z)$ depends on the product $n_i(z) n_j(z)$ so that 
$K_{ij}^{\rm II}(z)$ is expected to be non-zero only over the redshift range where $n_i(z)$ and $n_j(z)$ are both non-vanishing. 

\subsubsection{Non-linear scales for cosmic shear}\label{sec:wlnl}
The standard cosmic shear power spectrum is sensitive to $k$-modes in the range $0<k \leq 7h \,{\rm Mpc}^{-1}$ \citep{2018PhRvD..98d3532T}, which therefore necessitates a modelling of high-$k$ behaviour of the matter power spectrum. Whilst it is possible to exclude some high-$k$ information by performing a cut in $\ell$-mode this procedure is not a `clean' way to remove modes and is inefficient, particularly at low redshift. As shown in \citet{2018arXiv180903515T}, it is possible to employ a weight function to exactly remove scales above a particular maximum $k$-mode, a method known as '$k$-cut cosmic shear'; however, in this work we wish to present a standard formalism so that results can be more readily compared to the literature, and \citet{RB}. On a related matter, \citet{copeland} have recently found that baryonic effects on the non-linear power spectrum can decrease the FoM by up to 40\%, depending on the choices of model and prior. We therefore need to model the high-$k$ behaviour and we discuss this modelling here. 

The linear power spectrum of \autoref{eq: pddlinMath} is able to match the one measured from numerical simulations only over a limited range in $k$. For cosmic shear, one needs to model the matter power spectrum deep into the non-linear, high-$k$, regime that corresponds to projected angular modes in $\smash{C^{\epsilon\epsilon}_{ij}(\ell)}$ up to multipoles $\smash{\ell \sim 1000}$, and in fact already important at $\ell \sim 100$ \citep{2018PhRvD..98d3532T}. 
A common approach to model the non-linear matter power spectrum is to define it as a general function of the linear power spectrum, $\smash{P_{{\rm lin},\delta\delta}(k, z)}$, that translates it into the full, non-linear power spectrum $\smash{P_{\delta \delta}(k, z)}$. Various recipes are available in the literature, each of which has a different performance on different scales. Here, we implement the two currently most popular recipes, which we shall denote as 'HF' \citep[for \texttt{halofit}, see][]{Smith_etal_2003,2012ApJ...761..152T} and 'HM' \citep[for halo model, see][]{2002PhR...372....1C,Mead16}. These are 
both analogous formalisms, and both define the so-called 1- and 2-halo terms. 
The 1-halo term accounts for correlations between dark matter particles within the same dark matter halo, and dominates on small scales (smaller than the size of the halo); the 2-halo term describes correlations between distinct dark matter haloes, dominates on scales larger than the ones where the 1-halo term dominates and is proportional to the linear matter power spectrum multiplied by the effective bias of the species under consideration (e.g. dark matter haloes, galaxies etc.). The two HF and HM methods differ in how the 1- and 2-halo terms are computed. Additionally, both methods depend on parameters that are fitted to reproduce the matter power spectrum measured from state-of-the-art (at the time they were proposed) simulations. 

The HF method dates back to \citet{Smith_etal_2003}, but here we adopt the revised version from \citet{2012ApJ...761..152T}, that extends the
original recipe (designed for a CDM-dominated universe) to models with
constant equation of state dark energy. The 1- and 2-halo terms are
related to the linear matter power spectrum by a set of empirically
motivated fitting functions, whose parameters have been set in order
to fit the measured power spectrum from 16 different high-resolution
$N$-body simulations. The refined HF method provides an accurate
prediction of the non-linear matter power spectrum over a wide range of
wavenumbers $k \le 30 h \ {\rm Mpc}^{-1}$ at redshifts $0 \le z \le
10$: the precision is $5\%$ for $k  \le  1 h \,{\rm Mpc}^{-1}$ at $0
\le z \le 10$ and $10\%$ for $1 h\,{\rm Mpc}^{-1}\le k  \le
10h\,{\rm Mpc}^{-1}$ at $0 \le z \le 3$. 
Since massive neutrinos are
included in the analysis, we consider the standard HF method with the 
extension for massive-neutrinos described in \citet{Bird2012}.

The HM method has recently been proposed by \citet{Mead16} for CDM with a cosmological constant and was later extended to some non-standard scenarios including massive-neutrinos and some modified gravity theories. It takes into account the halo mass function and the halo density profile for the 1-halo term. The total power spectrum is then modified through the use of empirically motivated functions, whose parameters are determined by fits against the outcome of $N$-body simulations. The resulting non-linear matter power spectrum is accurate to $\sim 5\%$ up to $k \le 10 h \, {\rm Mpc}^{-1}$ for $0 \le z \le 2$. This is a smaller range compared to the HF technique, but it is worth stressing that the simulations used have higher resolution so that the overall performance is actually comparable if not better than for HF. A further advantage of the HM recipe is that it automatically includes nuisance parameters that can be used to correct for the impact of baryonic effects on the small scale power spectrum as verified by a comparison to hydrodynamical simulations. 

We summarise here the non-linear recipes available at the time of this work, which also include massive-neutrino effects.
\begin{itemize}
\item
{\textit{Bird}}. \citet{Bird2012} modified the non-linear HF version from \citet{Smith_etal_2003} to account for massive neutrino effects in the non-linear evolution of the total matter power spectrum.
This is a fitting to N-body simulations that include a massive neutrino component as collisionless particles. However, this version is not used in the present work, since it has been updated by the fitting below.
\item
{\textit{Mead}}. \citet{Mead16} provided a different fitting from the previous ones, to account for massive-neutrino effects at the non-linear level. Also in this case, N-body simulations, which include a massive-neutrino component as a collisionless species, are used, but the parameterisaton is based on a different approach than \citet{Bird2012}. The HM method accounting for massive-neutrinos is not adopted in this work.
\item 
{\textit{TakaBird}}. As reported in the documentation of the \texttt{CAMB} webpage,\footnote{\url{https://camb.info/readme.html}} on March 2014 modified massive-neutrino parameters were implemented in the non-linear fitting of the total matter power spectrum, to improve the accuracy with the updated version of \texttt{halofit} from \cite{2012ApJ...761..152T}. This is the current version adopted in this work when massive-neutrinos are included. The fitting parameters that account for non-linear corrections in the presence of massive-neutrinos are different from the ones implemented by \citet{Bird2012} in the original HF version from \citet{Smith_etal_2003}, and described in the point above.
\end{itemize}
Of the three approaches listed above, we decided to use only the {\textit{TakaBird}} implementation since we deem this to be the most reliable one. The \citet{Bird2012} correction to HF explicitly refers to a $\Lambda$CDM scenario thus preventing us from exploring deviations from it when constraining the DE $(w_0, w_a)$ parameters. On the other hand, the \citet{Mead16} recipe, although based on more recent results, has not been tested against simulations which include neutrinos with a mass as small as the one we are choosing as a fiducial value. Hereafter, therefore, only the Takahashi-Bird, {\textit{TakaBird}}, power spectrum is used to correct for non-linearities in the matter power spectrum used as input for cosmic shear tomography.

\subsubsection{Modelling beyond $\Lambda$CDM}
\label{Sec:model_beyond_wl}
Following the discussion outlined in \autoref{sec:cosmo-growth}, we can recast the growth factor as a function of the growth parameter $\gamma$ according to
\begin{equation}
D(z) \simeq \exp{\left \{ - \int_{0}^{z}{\frac{[\Omega_{\rm m}(x)]^{\gamma}}{1 + x} {\rm d}x} \right \}} \, .\label{eq: dzgamma}
\end{equation}
In the presence of a non-standard growth of structures, as is the case of modified gravity, we effectively deal with a growth factor $D_{\rm MG}(z;\gamma)$ that is different from the $\Lambda$CDM one, and depends on $\gamma$, free parameter of the theory \citep[see e.g ][]{2016JCAP...01..045C, 2017PDU....18...73C}). However we find that small (percent) deviations from $\gamma_{\rm fid}\simeq 6/11$, lead to changes of less than $0.1\%$ in the growth factor over the redshift range of interest, and so one can safely set $D_{\rm MG}(z;\gamma) = D(z)$. Nonetheless, we rescale it as
\begin{equation}
P_{A}^{\rm MG}(k, z) = P_{A}(k, z) \left[\frac{D_{\rm MG}(z;\gamma)}{D(z)}\right]^2,
\label{eq:powmg}
\end{equation}
where $A$ here refers to one of the observables in the set $\{\delta \delta, \delta {\rm I}, {\rm II}\}$. The shear power spectra can now be computed following the cosmic shear recipe detailed above, provided one replaces $\smash{P_{A}(k, z)}$ with its modified version $\smash{P_{A}^{\rm MG}(k, z)}$.

In doing this, we are making two assumptions. First, 
\autoref{eq:powmg} must be considered as a first order approximation. Indeed, while it is exact in the linear regime, its validity in the non linear one depends on the particular MG model being considered. Indeed, by using \autoref{eq:powmg}, we are implicitly stating that MG only rescales the growth of structures, while scale dependent modifications are actually expected. As a second issue, one should also note that, given the way that the IA spectra $P_{\delta {\rm I}}$ and $P_{\rm II}$ are related to the matter one $P_{\delta \delta}$, we are here assuming that MG does not change this relation, which is an assumption that has not been tested. 

\subsubsection{Covariance matrix for cosmic shear}\label{errors_shear}
The error on the observed cosmic shear angular power spectrum can be expressed as 
\begin{equation}
\Delta C^{\epsilon\epsilon}_{ij}(\ell) = 
\sqrt{\frac{2}{(2 \ell + 1) \Delta \ell f_{\rm sky}}}C^{\epsilon\epsilon}_{ij}(\ell) \, ,\label{wlcov}
\end{equation}
where $f_{\rm sky}$ is the fraction of surveyed sky, and $\Delta \ell$ is the multipole bandwith. 
The last term in $C^{\epsilon\epsilon}_{ij}(\ell)$ of \autoref{eq: defcij} is a Poisson noise term, whereas the first three, that is all terms in \autoref{eq: defcijshort}, represent the contribution from cosmic variance.  Lastly, the expression under the square root in \autoref{wlcov} accounts for the limited number of available independent $\ell$ modes, with $f_{\rm sky}$ being the surveyed fraction of the sky. 
In the case that one measures the power spectrum directly from the data, a four point function should be defined to encapsulate the error on the measurement. In this case \autoref{wlcov} the four point function obtained assuming Gaussianity and applying Wick's theorem can be recast as 
\begin{equation}
            {\rm Cov}\left[C^{\epsilon\epsilon}_{ij}(\ell),C^{\epsilon\epsilon}_{kl}(\ell^\prime)\right] =
\frac{C^{\epsilon\epsilon}_{ik}(\ell)C^{\epsilon\epsilon}_{jl}(\ell^\prime)+C^{\epsilon\epsilon}_{il}(\ell)C^{\epsilon\epsilon}_{jk}(\ell^\prime)}{(2 \ell + 1) f_{\rm sky} \Delta \ell}\delta^{\rm K}_{\ell\ell^\prime}.
\end{equation}
Both of these are used in the construction of the Fisher matrix.

A critical ingredient in the estimate of the Fisher matrix is the covariance matrix accounting for the errors in the measured observable. For the shear power spectrum, it can be split into Gaussian and non-Gaussian contributions \citep{2013PhRvD..87l3504T,wlcov1,wlcov2,wlcov3,wlcov4}. The non-Gaussian terms involve the convergence trispectrum, and there are large theoretical uncertainties on how to model this quantity, both in the quasi-linear and in the non-linear regime. An approximate analytical way to carry out this calculation is to rely on the halo model formalism \citep[see e.g. ][]{2001ApJ...554...56C}. Following this approach with the shear specifications given in \autoref{sec:wlobs}, we find that the information content of the shear power spectrum, defined as the total signal-to-noise ratio \citep{2005MNRAS.360L..82R,2009ApJ...701..945S,2009MNRAS.395.2065T,2013MNRAS.429..344K,0004-637X-480-1-22}, decreases by 30\% at $\ell_{\rm max}=5000$ when we add the non-Gaussian contribution. This loss of information content corresponds to an effective cut at $\ell_{\rm max} = 1420$ in a forecast that only uses the Gaussian covariance given by \autoref{wlcov}. Motivated by this result and in order to avoid uncertainties due to non-Gaussian modelling, we decide not to implement full non-Gaussian terms in our analysis and instead investigate the impact of a cut at $\ell_{\rm cut} = 1500$ (defined as our {pessimistic} choice) in our Gaussian forecasts. The impact of a different choice in this scale cut (for example with respect to the alternative value of $\ell_{\rm cut} = 5000$, corresponding to our {optimistic} choice) is discussed in our numerical results in \autoref{sec:results}.

We note that our approach is practical, but has limited applicability. For example an equivalent $\ell$-cut that preserves signal-to-noise is not directly linked to constraints on parameters beyond a global amplitude; parameter sensitivities are generally scale-dependent \citep{copeland} and therefore respond differently to small-$\ell$ cuts and the correlation introduced by non-Gaussian covariances. However, the amplitude of the impact of non-Gaussian modelling used here is small compared to other assumptions made. 

\subsubsection{Applying the Fisher matrix formalism to cosmic shear}\label{sec: fmwl}
For cosmic shear the mean of the shear field is zero, therefore in \autoref{eq:fisher-def} one can either use the first term on the right-hand side where the signal is in the covariance, or one can redefine the mean as being the power spectrum itself. In the former case only the covariance (power spectrum) appears in the expression, while in the latter case the covariance of the covariance (i.e. the four point function) appears. If one were measuring the cosmic shear signal from data the former would represent a measurement of the spherical harmonic coefficients, and a likelihood construction where the signal was in the covariance; the latter would then represent the case that the power spectrum was measured directly. We define both here, since two of the codes tested cover both these choices. 

In the case that one assumes the signal is the covariance (i.e. a measurement of the spherical harmonic coefficients) the Fisher matrix reads
\begin{equation}
F_{\alpha \beta} = \sum_{\ell = \ell_{\rm min}}^{\ell_{\rm max}}\sum_{ij,mn}\frac{\partial C^{\epsilon\epsilon}_{ij}(\ell)}{\partial \theta_\alpha}\left\{\left[\Delta C^{\epsilon\epsilon}(\ell)\right]^{-1}\right\}_{jm}\frac{\partial C^{\epsilon\epsilon}_{mn}(\ell)}{\partial \theta_\beta}\left\{\left[\Delta C^{\epsilon\epsilon}(\ell)\right]^{-1}\right\}_{ni},\label{eq: fmwl}
\end{equation}
where ${C}^{\epsilon\epsilon}(\ell)$ is defined in \autoref{eq: defcij}. No cross-correlation between modes with different $\ell$s is included, which is true under the Gaussian covariance assumption. We shall discuss the quantities $\ell_{\rm min}$, $\ell_{\rm max}$, and $\Delta \ell$ in \autoref{sec:ccwl}. Here $\smash{\{[\Delta C^{\epsilon\epsilon}(\ell)]^{-1}\}_{ij}}$ is the $ij$ element of the inverse of the matrix defined in \autoref{wlcov}.

In the case that one assumes the signal is the mean power spectrum (i.e. a measurement of the power spectrum coming directly from data) the Fisher matrix reads 
\begin{equation}\label{eq:wlfisher_second}
F_{\alpha\beta} = \sum_{\ell=\ell_{\rm min}}^{\ell_{\rm max}}\sum_{ij,mn}
             \frac{\partial C^{\epsilon\epsilon}_{ij}(\ell)}{\partial \theta_{\alpha}} {\rm Cov}^{-1}\left[C^{\epsilon\epsilon}_{ij}(\ell),C^{\epsilon\epsilon}_{mn}(\ell)\right]
\frac{\partial C^{\epsilon\epsilon}_{mn}(\ell)}{\partial \theta_{\beta}}. 
\end{equation}
Both of these definitions are used in the code comparison, and we find agreement between the two approaches; although this may be expected to hold analytically in any case \citep{2013A&A...551A..88C}.

\paragraph{Derivatives}
What is needed to compute the Fisher matrix elements is then not only the tomographic matrix of the ellipticity signal, $C^{\epsilon\epsilon}(\ell)$, but also its derivatives with respect to the cosmological (and nuisance) parameters. Using the compact notation introduced in the previous section \autoref{eq: defcijshort}, we can now compute the shear power spectrum derivatives: 
\begin{equation}
\frac{\partial C^{\epsilon\epsilon}_{ij}(\ell)}{\partial \theta_{\mu}} = \sum_{\{A,a\}}\left\{
\int_{z_{\rm min}}^{z_{\rm max}}{\rm d}z\,\frac{\partial K^{a}_{ij}(z)}{\partial \theta_{\mu}}P_{A}[k_{\ell}(z), z]+\int_{z_{\rm min}}^{z_{\rm max}}{\rm d}z\,K^{a}_{ij}(z) \frac{\partial P_{A}[k_{\ell}(z), z]}{\partial \theta_\mu}\right\},
\label{eq: dcijdpmu}
\end{equation}
where $\{A,a\} \in \{\{\delta \delta,\gamma\gamma\}, \{\delta {\rm I},{\rm I}\gamma\}, \{{\rm II},{\rm II}\}\}$. We note that the first term is only present when differentiating with respect to the background parameters $\{\Omega_{\rm m}, w_0, w_a, h\}$, whilst in the second term the derivatives of the power spectrum may be further decomposed into
\begin{equation}
\frac{\partial P_{A}[k_{\ell}(z), z]}{\partial \theta_{\mu}} = 
\left.\frac{\partial P_{A}(k,z)}{\partial \theta_{\mu}} \right|_{k=k_{\ell}(z)}+\frac{\partial k_\ell(z)}{\partial \theta_{\mu}}\left.\frac{\partial P_{A}(k,z)}{\partial k} \right|_{k=k_{\ell}(z)},
\label{eq: dpdpmu}
\end{equation}
where we remind the reader that $k_{\ell}(z)=(\ell+1/2)/r(z)$.  Again, we note that the second contribution is only present for derivatives with respect to $\{\Omega_{{\rm m},0}, w_0, w_a, h\}$. Finally, we stress that, different from the galaxy clustering case, the lensing Fisher matrix is directly computed with respect to the cosmological parameters of interest, so that there is no need for any projection. 

\subsection{Recipe for the combination of galaxy clustering and weak lensing}\label{sec:recipexc}
While the use of complementary probes is well known to
be essential to provide tight limits on cosmological parameters, the
joint analysis of correlated probes has been relatively little explored until
now, and thus represents a rather new field in cosmology. In this section
we describe our approach for combining galaxy clustering (from the spectroscopic redshift survey and photometric survey) with weak lensing for \Euclid.

For the specific combination of weak lensing and galaxy clustering 
coming from the photometric redshift survey (GC$_{\text{ph}}$), we can use
the angular power spectrum formalism described in \autoref{sec:wlobs}: this allows us to show the impact 
of including or not cross-correlation terms (for every redshift bin) between WL and the GC$_{\text{ph}}$.

\subsubsection{Observables for the photometric survey and probe combination}\label{sec:combobs}

The maximum set of observables that may be used for probe combinations are all individual
\Euclid probes, the cross-correlation (XC) between those probes (which can be used
either as observable or as covariance), and correlations between external data
and \Euclid probes. In the latter case the external data can also be used as individual probes 
in the combined set.

For this work however, we focus on the combination of only cosmic shear and 
galaxy clustering, the latter using both the photometric and spectroscopic
galaxy surveys. In addition to simply combining these probes, we also 
include in our analysis the cross-correlation of photometric galaxy clustering
and cosmic shear. Both latter probes are quantified as 2D projected
observables, which allows us to compute the cross-observables and
cross-covariance matrix in a simple, unified scheme. 
The cross-correlation of cosmic shear with spectroscopic
galaxy clustering is expected to be small given that
the spectroscopic \Euclid survey selects galaxies at
high redshift ($z > 0.9$, see \autoref{sec:fiducial}) and thus has a small
overlap with the lensing kernel. However, when we combine the
photometric and spectroscopic galaxy clustering probes, the two galaxy
distributions overlap and, therefore, we consider the following
two different cases: 
\begin{itemize}
\item 
In the {{pessimistic}} case we only consider galaxies below $z=0.9$ in
the photometric sample, and neglect the scatter from high redshifts
into the photometric sample due to photometric redshift errors: these probes are
uncorrelated, at the price of discarding some parts of the observations. 
\item 
In the {{optimistic}} case we do not perform any cut in
redshift space for the photometric sample and we combine the
spectroscopic and photometric galaxy clustering probes neglecting any
cross-correlation that might appear between them.
\end{itemize}
For the photometric survey we use the specifications of \autoref{tab:WL-specifications}
and the galaxy distribution function $n(z)$ of \autoref{eq: nzredbook}, convolved with the probability
distribution function $p_{\rm ph}(z)$ of \autoref{eq:wlpph}. For the spectroscopic redshift survey
we use instead the specifications of \autoref{tab:GC-specifications} and we follow the approach 
described in \autoref{sec:recipegc} to model this observable.

We first denote the angular density contrast of galaxies in redshift
bin $i$ by $\smash{\delta^i_{\rm g}}$, the shear field of source tomography
bin $j$ as $\gamma^j$, and the normalised galaxy number density in each tomographic bin $i$ of the photometric survey
as $\smash{n_i(z)}$, given by \autoref{eq: defni}. Notice that we assume here the same galaxy distribution for the 
photometric shear and galaxy clustering surveys; this is not guaranteed to be the case for the real photometric survey, 
since the sampled galaxies for GC$_{\text{ph}}$ might be a subset of the full galaxy sample. Therefore, more realistic
galaxy distributions are still being investigated.

We can now define the radial weight function for galaxy clustering as
\begin{equation}\label{eq:photoGCwin}
W^{\rm G}_i (k,z)=b_i(k,z) n_i (z)\frac{H(z)}{c} \, ,
\end{equation}
where $H(z)$ is the Hubble parameter defined in \autoref{eq:Hz-general}, and
$b_i(k,z)$ is the galaxy bias in tomography bin $i$. For the latter, we assume a constant value in each redshift
bin and no dependence on the scale $k$; the values of $b_i$ in each bin are considered as nuisance parameters over which we marginalise when producing 
our results. Their fiducial values are given by
\begin{equation}\label{eq:biasfid}
 b_i(z) = \sqrt{1+\bar{z}} \, , 
\end{equation}
with $\bar{z}$ the mean redshift of each bin. The main assumptions made on the galaxy bias are therefore that it 
appears linearly in \autoref{eq:photoGCwin} and that it does not depend on the scale $k$. Removing these 
assumptions would require the addition of extra nuisance parameters to control the uncertainty on the galaxy bias
and would lead to less constraining power from the photometric galaxy clustering survey. 
Given these assumptions, in the following we drop the $k$ dependence from the expression of $W^{G}_i$.

We use the same radial weight function for cosmic shear $W^\gamma_i(z)$ given by \autoref{eq: lensweight},
and we include the effect of intrinsic alignment described in Section \autoref{sec:wlobs}; combining \autoref{eq: defcij}
with \Cref{eq: pdienla,eq: piienla}, we can define a cosmic shear power spectrum as 
\begin{equation}
 C^{\rm LL}_{ij}(\ell)=c\int_{z_{\rm min}}^{z_{\rm max}}\mathrm dz\,\frac{
 W^{\rm L}_i(z)W^{\rm L}_j(z)}{H(z)r^2(z)} P_{\delta\delta}\left(\frac{\ell+1/2}{r(z)},z\right),
\end{equation}
where we have defined the overall weight function $W^{\rm L}_i(z)$, which includes intrinsic alignment contributions
\begin{equation}
W^{\rm L}_i=W_i^\gamma(z)-\frac{\mathcal{A}_{\rm IA}\mathcal{C}_{\rm IA}\Omega_{\rm m,0}\mathcal{F}_{\rm IA}(z)}{D(z)}W_i^{\rm IA}(z),
\end{equation}
with $W_i^{\rm IA}(z)$ defined in \autoref{eq: iaweight}. This is exactly equivalent to \autoref{eq: defcijshort}, except that we do not include the noise term for clarity. We change notation here to make the relationship between the galaxy clustering power spectra and the cosmic shear power spectrum clearer. 
Finally, using the Limber approximation, the other observables are given by
\begin{align}\label{ClsXC}
C^{\rm GL}_{ij}(\ell) &=
                              c\int\mathrm dz\,\frac{
 W^{\rm G}_i(z)W^{\rm L}_j(z)}{H(z)r^2(z)}P_{\delta\delta}\left(\frac{\ell+1/2}{r(z)},z\right)\,,\\
C^{\rm GG}_{ij}(\ell) &= c\int
                                            \mathrm dz\,\frac{
 W^{\rm G}_i(z)W^{\rm G}_j(z)}{H(z)r^2(z)}P_{\delta\delta}\left(\frac{\ell+1/2}{r(z)},z\right)\,,
\end{align}
and represent, respectively, the cosmic shear power spectrum, the shear-galaxy power spectrum and the galaxy-galaxy power spectrum (using photometric galaxy clustering).

\subsubsection{Non-linear scales for probe combination} \label{subsec:XCnonlinear}

For the modelling of non-linear scales in the cosmic shear, photometric
galaxy clustering, and cross-correlations observables, we proceed as discussed in \autoref{sec:wlnl}. We apply it for
cosmic shear up to multipole $\ell_{\rm max}=5000$ for the optimistic
case and $\ell_{\rm max}=1500$ for the pessimistic case (see
\autoref{sec:xccases} for the detailed description of the optimistic
and pessimistic cases presented in the results). For the photometric galaxy clustering and
cross-correlations, we consider multipoles up to $\ell_{\rm max}=3000$
for the optimistic case and $\ell_{\rm max}=750$ for the pessimistic case.

\subsubsection{Modelling beyond $\Lambda$CDM}
In order to calculate the DE FoM we extend our cosmological model beyond $\Lambda$CDM by adding a dark energy fluid parameterised by two equation of state parameters, $w_0$ and $w_a$, as was done in \Cref{Sec:model_beyond_gc,Sec:model_beyond_wl}. However, when we take cross correlations into account we do not consider cosmological models with deviations from GR, that is to say we do not consider the $\gamma$ parameterisation of the growth of structure that is considered for single probes. There are two main reasons for this. The first one is that the codes used in this work to include cross correlations into the analyses are coupled to Boltzmann codes (contrary to the codes for single probes, which use external input spectra) and they obtain the growth of structure directly from them. Therefore, it is not possible for these codes to include the $\gamma$ parameterisation by only modifying the growth of structure and rescaling the standard matter power spectrum, as is done for single probes. The second main reason is that the public version of the Boltzmann code allowing for the $\gamma$ parameterisation, {\tt MGCAMB}, does not allow for this parameterisation together with an evolving dark energy equation of state, which is the case considered here. As a consequence, we do not consider models beyond GR for cross-correlations in this work.

\subsubsection{Covariance matrix for probe combination} \label{sub:covariancexc}

The contribution of spectroscopic galaxy clustering is added as
a separate Fisher matrix (both in the optimistic and the pessimistic cases) and hence additional covariance terms are not included. 

In the case of 
cosmic shear, photometric galaxy clustering and their cross-correlation, a joint covariance matrix is required, that can be defined in two different ways (see also \autoref{errors_shear}); we present both and their outcome is compared in \autoref{sec:ccxc}. One can define a covariance as 
       \begin{align}
             \Delta{C}^{AB}_{ij}(\ell) &= \frac{1}{\sqrt{f_{\rm sky} \Delta \ell}}\left[C^{AB}_{ij}(\ell) + N^{AB}_{ij}(\ell)\right],
             \label{eq: covsecond}
       \end{align}
where $A,B$ run over the observables L and G, $\Delta \ell$ is the width of the multipoles bins used when computing the angular power spectra, and $i,j$ run over all tomographic bins. Otherwise, one can define the fourth-order covariance as
       \begin{equation}
            {\rm Cov}\left[C^{AB}_{ij}(\ell),C^{A^\prime B^\prime}_{kl}(\ell^\prime)\right] =\frac{\left[C^{AA^\prime}_{ik}(\ell) + N^{AA^\prime}_{ik}(\ell)\right]\left[C^{BB^\prime}_{jl}(\ell^\prime) + N^{BB^\prime}_{jl}(\ell^\prime)\right]+\left[C^{AB^\prime}_{il}(\ell) + N^{AB^\prime}_{il}(\ell)\right]\left[C^{BA^\prime}_{jk}(\ell^\prime) + N^{BA^\prime}_{jk}(\ell^\prime)\right]}{(2 \ell + 1) f_{\rm sky} \Delta \ell}\delta^{\rm K}_{\ell\ell^\prime}. \label{eq: covfourth}
       \end{equation}
As above, $A,B,A^\prime,B^\prime={\rm L},{\rm G}$ and $i,j,k,l$ run over all tomographic bins. In both approaches, the noise terms $N^{AB}_{ij}(\ell)$ take the form
\begin{align}\label{eq:xcnoise}
N^{\rm LL}_{ij}(\ell) &= \frac{\sigma_\epsilon^2}{\bar{n}_i}\delta^{\rm K}_{ij} ,\\
N^{\rm GG}_{ij}(\ell) &= \frac{1}{\bar{n}_i}\delta^{\rm K}_{ij} , \\
N^{\rm GL}_{ij}(\ell) &= N^{\rm LG}_{ij}(\ell) =0,
\end{align}
with $\sigma_\epsilon^2$ and $\bar{n}_i$ introduced in \autoref{sec:wlobs}.
Notice that we assume here that the Poisson errors on cosmic shear and galaxy clustering from the photometric surveys are 
uncorrelated, which yields $\smash{N^{\rm GL}_{ij}(\ell)=0}$. 

As for the cosmic shear case (see \autoref{errors_shear}), the covariance matrix of photometric galaxy clustering can be split into the sum of Gaussian and non Gaussian contributions. The non Gaussian terms involve the galaxy trispectrum, whose modelling beyond the linear regime suffers from significant theoretical uncertainties. As for cosmic shear, we rely on the halo model formalism \citep{Lacasa2016,Lacasa2017} to estimate the impact of the non Gaussian contribution to the photometric galaxy clustering covariance matrix. It is important to note that the photometric galaxy clustering is more non Gaussian than shear due to sampling high density regions instead of total matter, being integrated over a smaller volume (bin width versus whole light cone), and because shot noise is subdominant for clustering while it has a high impact on the shear Gaussian errors. Following this approach with the galaxy specifications given in \autoref{sec:combobs}, we find that the signal-to-noise ratio saturates for $\ell\gtrsim500$ when non Gaussian contributions are included. We mimic this effect using a conservative cut on the multipoles used for GC$_{\text{ph}}$, removing the contributions of multipoles $\ell>750$ from the analysis (pessimistic setting). We still show the impact of varying this cut by testing an ideal (unrealistic) case in which non-Gaussian contributions are entirely neglected up to $\ell_{\rm max} = 3000$ (optimistic setting).

\subsubsection{Applying the Fisher matrix formalism to probe combination}
The final expression of the combined Fisher matrix for the angular power spectra, including the contributions of photometric galaxy clustering, cosmic shear, and their cross-correlation, is given in the case of the fourth order covariance by 
\begin{equation}
F^{\rm XC}_{\alpha\beta} = \sum_{\ell=\ell_{\rm min}}^{\ell_{\rm max}}\sum_{ABCD}\sum_{ij,mn}
             \frac{\partial C^{AB}_{ij}(\ell)}{\partial \theta_{\alpha}} {\rm Cov}^{-1}\left[C^{AB}_{ij}(\ell),C^{CD}_{mn}(\ell)\right]
\frac{\partial
  C^{CD}_{mn}(\ell)}{\partial \theta_{\beta}}\,,\label{eq:xcfisher_fourth}
\end{equation}
and in the case of the second-order covariance as 
\begin{equation}
F^{\rm XC}_{\alpha \beta} =  \frac{1}{2} \sum^{\ell_{max}}_{\ell=\ell_{min}} (2\ell+1)\sum_{ABCD}\sum_{ij,mn}
\frac{\partial C^{AB}_{ij}(\ell)}{\partial \theta_\alpha}\left[\Delta C^{-1}(\ell)\right]^{BC}_{jm}\frac{\partial C^{CD}_{mn}(\ell)}{\partial \theta_\beta}\left[\Delta C^{-1}(\ell)\right]^{DA}_{ni}.\label{eq:xcfisher_second}
\end{equation}
The block descriptors $A,B,C,D$ run
over the combined probes L and G, thus including the three
observables described in \autoref{ClsXC}, that is to say cosmic shear auto-correlation, cross-correlation
between galaxy clustering and cosmic shear galaxy clustering auto-correlation. The indices in the sum $ij$ and $mn$ run over all unique pairs of
tomographic bins $(i\leq j, m\leq n)$ for the cosmic shear
auto-correlation and the galaxy clustering auto-correlation, while all
pairs of tomographic bins are considered to take into account all the
cross-correlations between galaxy clustering and cosmic shear.

Finally, in addition to this combination, we include also the contribution of the spectroscopic galaxy
clustering. Since we are neglecting its correlation with the photometric survey 
observable (both in the pessimistic and optimistic cases), we simply add the spectroscopic Fisher matrix $F^{\rm spec}_{\alpha\beta}$ 
computed as described in \autoref{sec:recipegc}. Therefore, our final matrix is
\begin{equation}
F_{\alpha\beta}=F^{\rm spec}_{\alpha\beta}+F^{\rm XC}_{\alpha\beta}.
\end{equation}

\section{Code comparison}\label{sec:codecomparison}

In this section we describe the code comparison procedure and results for \Euclid's main 
cosmological probes, namely spectroscopic galaxy clustering and weak gravitational lensing, as well as the cross-correlation of photometric galaxy clustering with weak lensing. We have compared and validated six different Fisher matrix codes for galaxy clustering, {seven} for weak lensing, and {five} of the codes also implement cross-correlations.  All the codes implement the recommended recipes for GC$_{\text{s}}$, WL, GC$_{\text{ph}}$ and XC discussed in \autoref{Sec:Fisher}. For clarity and in order to isolate and compare different cases of interest (for example linear modelling versus non-linear corrections, Fisher matrix projection to different parameters, including dark energy and modified gravity parameterisations) we have validated the codes across various intermediate cases of increasing complexity. 

This section aims at describing the implementation and code comparison among the different codes and represents our main deliverable: we include detailed instructions in \autoref{sec:appendix} for the reader who may want to validate another Fisher Matrix forecast code; we make publicly available the recipe used, the reference output Fisher matrices, and plotting routines to reproduce your own comparison plot, in the form of \autoref{fig:resu-GC-proj3}, to check that your own code is validated. 
In \autoref{sec:allcodes} we describe the codes used in our comparison, in the WL, GC and XC cases. As the comparison was done among codes which had been already partially developed before this work, and was motivated by several codes providing different results in both amplitude and orientation, code ownership belongs to the authors who developed them. We have however encouraged all code owners to make these codes publicly available: \href{https://bitbucket.org/joezuntz/cosmosis/wiki/Home}{\texttt{CosmoSIS}} and \href{https://github.com/santiagocasas/cosmomathica}{\texttt{CosmoMathica}} (used to produce the input files for this work) are already publicly available, the authors of four codes (\texttt{FisherMathica}, \texttt{SOAPFish}, \texttt{CCCPy}, \texttt{TotallySAF}) have expressed intention to make them publicly available, five codes (\texttt{BFF}, \texttt{fishMath}, \texttt{CosmicFish}, \texttt{CarFisher} and \texttt{STAFF}) can be available upon request. Requests can be made directly to the authors, as listed below in \autoref{sec:allcodes}.

In \Cref{sec:ccgc,sec:ccwl} we describe specific implementation steps regarding the galaxy clustering and weak lensing code comparisons, in particular the construction of the background cosmological quantities, the implementation of the survey geometry and galaxy sample properties, and the perturbed quantities, in order to construct the observed power spectrum. We give specific and detailed information on the parameters we vary, and the calculation of the derivatives entering the Fisher matrix. We also describe how we implement non-linear corrections and how we project our Fisher matrices to derive constraints on commonly used dark energy and gravity parameters, such as $w_0,w_a$ and $\gamma$. We define the different cases we use to validate our codes and finally show the results of the code comparison, that is to say how well the codes match. In \autoref{sec:ccxc} we repeat this procedure for the probe combination case. Overall, we find very good (percent level) agreement between the different codes for all cases. In \autoref{sec:lessons} we comment on critical lessons we learnt during the code comparison validation, and give some advices for writing a Fisher matrix code for cosmology from scratch.

\subsection{The codes used in the comparison}\label{sec:allcodes}

Here we describe each of the codes used in the comparison, for the cosmic shear, galaxy clustering and cross-correlation observables. The comparison started from codes that were already available at the time of this paper, summarised in \autoref{tab:forecastcodes}. There are a variety of languages and approaches used, some are privately written codes and others use publicly available packages. We further provide in this paper instructions and reference input/output files to proceed with the validation of any other external code the reader may have. 
We emphasise that these codes were validated in their current versions, as of February 2019, and with regard to their specific Fisher matrix implementation and output. 

\paragraph{\texttt{BFF}} 

(\textsc{Bonn Fisher Forecast}) is a code developed by Victoria Yankelevich and is written in \textsc{Fortran 90}. It was designed to comply with the IST specifications by following the guidelines.
However, it is based upon the code developed in the higher-order statistics work package to make forecasts for the galaxy bispectrum (and its combination with the power spectrum; see \cite{2019MNRAS.483.2078Y}.
Due to technical requirements, these two codes have different structures but give consistent results. The main advantage of BFF is computational time, which is about a couple of minutes on a one core laptop.  

\paragraph{\texttt{CarFisher}}
is a Fisher matrix code written by Carmelita Carbone in \texttt{IDL}, which computes cosmological parameter forecasts for a 3D galaxy power spectrum. It has been used, in the linear regime version, to provide results in \citet{RB} and \citet{arXiv:1107.1211,arXiv:1012.2868,arXiv:1112.4810}. Now it also includes the non-linear modelling described in this work. It interfaces with \texttt{CAMB} to produce matter power spectra files, and with a F90 routine for the computation of scale independent growth factor and growth rate. The projection onto the final parameter space is performed by a separate module, that enables one to choose different settings (e.g. $k$-, $z$- or $\mu$-intervals) without the need to rerun the Boltzmann code for power spectra production.

\paragraph{\texttt{FisherMathica}}
is a Fisher matrix code written by Santiago Casas in \texttt{Mathematica} in a modular way, including packages for different applications. 
Based on the code used in \cite{2012PhRvD..85j3008A}, it has been used for forecasts of modified gravity and dark energy \citep{2016JCAP...01..045C, 2017PDU....18...73C, 2018PhRvD..97d3520C}. 
 
The particular advantage of this code is flexibility in terms of cosmological parameters. It accepts any basis and can convert between them, such as $A_{\rm s}$ versus $\sigma_8$ or $\omega_{\rm m,0}$ versus $\Omega_{\rm m,0}$.  It can accept also new parameters, which can be useful for extended models of modified gravity. The \texttt{CosmologyTools} package takes care of computing Hubble functions, distances, coordinate and unit transformations, analytical growth rate equations and other cosmological observables.
The {publicly available \href{https://github.com/santiagocasas/cosmomathica}{\texttt{CosmoMathica}} package\footnote{\url{https://github.com/santiagocasas/cosmomathica}} deals with the interaction with common codes in the cosmological community, such as the Boltzmann codes \texttt{CLASS} and \texttt{CAMB} and its modified gravity version \texttt{MGCAMB}, \citet{Eisenstein_Hu_1997b} transfer functions, the \texttt{Coyote Cosmic Emulator} \citep{Heitmann:2013bra} and other fitting formulas for non-linear power spectra. \texttt{CosmoMathica} has been used to produce the input files for this work.}
The \texttt{FisherGC} package takes care of the galaxy clustering Fisher matrix, by computing the derivatives of the power spectrum with respect to the shape parameters and with respect to the redshift dependent parameters. It reads the GC specifications for different surveys from given text files and computes all needed quantities for the Fisher matrix. 
The integration settings are very flexible and are the ones provided by the \texttt{Mathematica NIntegrate} function.
The \texttt{FisherWL} package computes the kernels, the window functions and the $C(\ell)$ functions for weak lensing. 
It also includes several models of intrinsic alignment. The Fisher matrix is calculated in a parallelised way for each of the parameters and redshift bins.
The \texttt{UsefulTools} package takes care of input/output, creation of tables, Fisher ellipses plots and post-processing. 
It contains many useful matrix and vector operations that are needed to analyse and interpret Fisher matrices results.
Finally, the code can be called interactively using \texttt{Mathematica} notebooks or from the command terminal, using \texttt{.m } file scripts.

\paragraph{\texttt{fishMath}}
is a code that computes galaxy clustering forecasts, written in \texttt{Mathematica} by Elisabetta Majerotto. A previous version of it, which did not include forecasts on shape parameters, 
was used in various papers to produce forecasts \citep{RB,Majerotto:2012, Bianchi:2012za}, and more recent versions of it have been used to make various cross-checks 
for other papers \citep{2013PhRvD..88d3503S, Sapone:2014nna,2016MNRAS.456..109M}. The code allows one to switch from linear modelling to the two different parameterisations of non-linear modelling. It consists of different modules, so that it is possible, for example to start by producing the Fisher matrix or to read the Fisher matrix from a file and marginalise it or project it over the preferred parameters. It also computes errors on parameters and the FoM for dark energy.

\paragraph{\texttt{SOAPFish}}
is a code implemented using a combination of \texttt{Python\,3} and \texttt{Mathematica 10.0.0.0}, by Domenico Sapone. It is a modified and improved version of the code used in \citet{Sapone:2009mb,Sapone:2010uy}, \citet{2013PhRvD..88d3503S}, \citet{Sapone:2014nna}, \citet{2016MNRAS.456..109M}, \citet{Hollenstein:2009ph}, \citet{Belloso:2011ms} and \citet{Sapone:2012nh}, where the authors forecasted the sensitivity with which the next generation of surveys will measure the cosmological parameters with different cosmological models. The code is divided into cells and three main parts. The first one specifies the values of the cosmological parameters and the survey properties, such as survey area, number of bins and error on redshift. Once these input quantities have been called, the code evaluates the cosmological functions, such as the Hubble parameter and the angular diameter distance, and the survey specifications, such as the bin volumes and the galaxy number density in each bin. The third and last part consists of evaluating the derivatives of the observed galaxy power spectrum in a fully and optimised numerical manner from which the Fisher matrices are evaluated.

{\paragraph{\texttt{TotallySAF}} (\texttt{Totally Super Accurate Forecast} )
is a code that has been implemented using \texttt{Python} (compatible with \texttt{Python 2} and \texttt{Python 3}) and \texttt{C++}, developed by Safir Yahia-Cherif and Isaac Tutusaus. The code consists on 2 different packages: \texttt{SpecSAF} and \texttt{XSAF}. The former is a modified and improved version of the code used in \cite{Isaac:2016}, where forecasts for the spectroscopic galaxy clustering probe of \Euclid have been computed for theoretical models with varying dark matter and dark energy equation of state parameters. In order to compute the integrals this code implements a quick and precise method that consists of making sums over four cells in the $k$ and $\mu$ space 
(the one under consideration, the right nearest-neighbour, the down nearest-neighbour and the right-down nearest-neighbour); this whole part of the code is parallelised. Concerning \texttt{XSAF}, it allows for the computation of weak lensing and photometric galaxy clustering forecasts, taking into account their cross-correlations. 
Another feature of this code is that the user can choose higher point numerical derivatives than the 3 and 5 points stencil. The 7, 9, 11, 13 and 15 point stencils 
are available for users who require high precision calculation. For each parameter used it is possible to choose the number of points and the step of the derivatives. The code can also provide the input files and produce the visualization plots.}

\paragraph{\texttt{CCCP}} (\texttt{Camera's Code for Cosmological Perturbations}) 
is a code developed by Stefano Camera and has been used as a standalone code, or in order to compute either angular power spectra only or Fisher matrices only, in various analyses \citep[e.g.][]{Camera:2013xfa,Camera:2014rja,Camera:2014sba,Camera:2014bwa,Bonaldi:2016lbd}. At the time of the present code validation, it is written in \texttt{Mathematica 11.2.0.0} and it is kept updated from version to version. It consists of two main routines: $(i)$ one calculating cosmological background and perturbation quantities for a given set of cosmological parameters; and $(ii)$ another computing the Fisher matrix for the parameters of interest. The former routine employs, whenever is possible, analytical expressions to avoid interpolations and the propagation of numerical errors. Nonetheless, \texttt{CCCP} also allows one to load the growth, the transfer function, the power spectrum or all functions from an external file, and it can be easily interfaced with a Boltzmann solver to obtain these quantities on the fly. Derivatives of angular power spectra with respect to cosmological parameters are computed according to the method discussed here in \autoref{sec:fisher-caveat-WL}, and first proposed by \cite{Camera:2016owj}. It is found that it is in good agreement with MCMC results. Regarding the latter routine, that responsible for Fisher matrix computations, it can be used either with input data provided by the former routine, or with external spectra in form of tables of multipoles and $\smash{C^{ij}_\ell}$ values as inputs. The stability of the Fisher matrix is kept under control via several tests, and it is enforced, when necessary, for example \ by considering the logarithm of the parameter instead of the parameter itself: this reduces the dynamic range between the maximum and minimum eigenvalue, thus stabilising matrix inversion \citep[see][]{Fonseca:2015laa}. Despite the reduction of the dynamic range, matrix inversions may still propagate numerical instabilities into the resulting Fisher matrix. Therefore, \texttt{CCCP} checks the stability of the inverse matrix $\mathbf F^{-1}$ (computed by default via one-step row reduction) by comparing it to the Moore-Penrose pseudo-inverse and to the inverse obtained via eigen-decomposition \citep{Albrecht:2006,Camera:2012ez,Camera:2018jys}.

{\paragraph{\texttt{CCCPy}} was born as a Python version of \texttt{CCCP}. In its state at the time of the publication of this paper, it is based on the Core Cosmology Library (\texttt{CCL}, \citealt{Chisari:2018vrw}), a standardised library of routines to calculate basic observables used in cosmology. Specifically, \texttt{CCCPy} employs \texttt{CCL} for the computation of the various angular power spectra (WL, GC$_{\rm ph}$, and XC), adjusted so to implement the same recipes as discussed in the previous sections, like binning, galaxy bias, and IA. Moreover, it has utilities implementing the SteM derivatives, with and without accuracy checks. It can be run as a single jupyter notebook, from the computation of the observables, to the derivatives, to the final Fisher matrices and the corresponding parameter constraints.}

\paragraph{\texttt{CosmicFish}} 
is a Fisher matrix code written in the \texttt{Fortran90} and \texttt{Python} languages by Marco Raveri and Matteo Martinelli. It is interfaced with \texttt{CAMB\_sources} and its 
generalizations are based on common \texttt{CAMB} modifications, such as \texttt{MGCAMB} \citep{Zhao:2008bn,Hojjati:2011ix} and \texttt{EFTCAMB} \citep{Hu:2013twa,Raveri:2014cka}. 
\texttt{CosmicFish} uses as primary parameters for $\Lambda$CDM cosmology the set $\{\Omega_{{\rm b},0}h^2,\,\Omega_{{\rm c},0}h^2,\,h,\,n_{\rm s},\, \log{10^{10}A_{\rm s}},\, (w_0,\, w_a,\, \gamma)\}$
with the parameters in parentheses added when considering extensions to $\Lambda$CDM. Numerical derivatives are computed with 
with respect to these parameters, with the options of either 3 or 5 points stencil. Finally it projects the results on the common parameter basis combining the original Fisher matrix with a Jacobian matrix. 
\texttt{CosmicFish} has been used in previous works to investigate the information gain given by past and future cosmological surveys \citep{Raveri:2016xof}, as well as to
obtain forecasts for proposed CMB missions \citep{Abazajian:2016yjj}. Also, a preliminary validation of the code was done reproducing results available in the literature at the time of the code release \citep{Raveri:2016qns}. The version validated in this paper contains key features and calculations not available in the current public code, e.g. the effect of intrinsic alignments
and the possibility of working in the common parameter basis. These new features will be included in the next update of the public version of this code.

\paragraph{\texttt{CosmoSIS} }
is a publicly available cosmological parameter estimation code\footnote{\url{https://bitbucket.org/joezuntz/cosmosis/wiki/Home}} \citep{Zuntz:2014csq}. It has a modular structure and this enables the use of existing codes for observable predictions, and different experimental likelihoods, written in different languages, like \texttt{Fortran90}, \texttt{Python} and \texttt{C/C++}. The public version of \href{https://bitbucket.org/joezuntz/cosmosis/wiki/Home}{\texttt{CosmoSIS}} provides a standard
library with different modules for the Boltzmann solver, like \texttt{CAMB} \citep{Lewis:1999bs,Howlett:2012mh}, \texttt{CLASS} \citep{2011arXiv1104.2932L,2011JCAP...07..034B,2011arXiv1104.2934L,2011JCAP...09..032L}, or \texttt{MGCAMB} \citep{Zhao:2008bn,Hojjati:2011ix}, the background quantities, or nuisance parameters (like the galaxy bias, and intrinsic alignments). More importantly for us, it also contains a sampler that computes Fisher matrices.
\texttt{CosmoSIS} runs as a pipeline. Some modifications have been made to the standard implementation by Isaac Tutusaus and Martin Kilbinger for the purposes of this paper, 
and the pipeline used works as follows: the Boltzmann solver used (\texttt{CAMB}) uses $A_{\rm s}$ instead of $\sigma_8$, fixing $A_{\rm s}$ to a dummy value and adding a third module 
into the pipeline to rescale the matter power spectrum. Afterwards the fourth module is added to take into account the non-linear correction
to the matter power spectrum, and the fifth to consider the intrinsic alignments. Two more modules are needed to load the
distribution of galaxies, $n(z)$, and apply the galaxy bias model to
the power spectrum. The projection to the $C_{ij}(\ell)$ values is performed for each of the considered observables, adding the
intrinsic alignment power spectra and finally the full covariance matrix for the observables is computed. 
Once this is done, the Fisher matrix sampler calls the pipeline iteratively for all the variations of the parameters required to
compute the numerical derivatives, and finally the Fisher matrix. Two
derivative methods are available (3-point and 5-point stencils) with a
user-defined step size per parameter. \texttt{CosmoSIS} has been widely used in the literature to analyse
real data \citep{Abbott:2017wau,Abbott:2017smn,Baxter:2018kap}, and to compute forecasts \citep{Harrison:2016stv,Park:2015sxj,Olivari:2017bfv}. 

\paragraph{\texttt{STAFF}} (\texttt{Shear Tomography Accuracy from Fisher Forecast}) 
is a \texttt{Mathematica} code developed by Vincenzo F.\ Cardone for the code comparison challenge. It comes as a 
single notebook (\textit{staff}) performing all the necessary steps to output the final Fisher matrix, or as three separate notebooks for intermediate computations, 
namely \textit{pscalc} for the estimate of the matter power spectrum and its derivatives, \textit{cijdercalc} to obtain the shear, photometric galaxy clustering, and cross-correlation power spectra and derivatives, 
and \textit{fmcalc} for the Fisher matrix evaluation. The main virtues of \texttt{STAFF} are its simplicity and versatility which allows the user to quickly modify the code for any non standard cosmological model of interest. Unfortunately, being written in \texttt{Mathematica}, the code is slower than some others, and hence it is not ideal from this point of view. Nevertheless, the most time consuming step is when computing the matter power spectrum and its derivatives and this operation needs only to be performed once. This is because since the semi-analytical approach implemented for the derivatives of the different power spectra of interest make it possible to express all of them as operations on the matter power spectra. \\

\begin{table}
\centering
\caption{Summary of the codes involved in the comparison. The table reports, for each code involved in the comparison, the contact person(s), the available probes, the capability of the code to consider spatially non-flat models, the interfaced Boltzmann solver(s), the 
possibility to work with external input, the code's language, and whether or not {the validated version of the code is publicly available at the time of the release of this paper. Four codes (\texttt{FisherMathica}, \texttt{SOAPFish}, \texttt{CCCPy}, \texttt{TotallySAF}) are planned to be made publicly available, while five codes (\texttt{BFF}, \texttt{fishMath}, \texttt{CosmicFish}, \texttt{CarFisher} and \texttt{STAFF}) can be available upon request. Requests can be made directly to the authors, as listed in this Table}. For the comparison performed here, we rely on common input cosmological quantities for all the forecast codes except for \texttt{CosmicFish} and \texttt{CosmoSIS}; for these two codes we do not use the common input, but rather exploit their interface with a version of \texttt{CAMB} compatible with the one used to generate it. The orange crosses stand for features of the codes that are implemented but have not been validated in this work.}
\label{tab:forecastcodes}
\scalebox{0.7}{%
\begin{tabular}{lcccccccccc}
\hline
Name       & Contact    &GC$_{\text{s}} $             & GC$_{\text{ph}}$            & WL         & XC$^{\rm (GC_{\text{ph}},WL)}$     & Curvature   & Boltzmann code     & External Input    & Language   & public\\
\hline
\hline
\textsc{BFF} &Yankelevich & \cellcolor{ forestgreen(web)}  \textcolor{white}  \checkmark                   &   \text{\sffamily X}      &  \text{\sffamily X}     & \text{\sffamily X}       &\cellcolor{ forestgreen(web)}  \textcolor{white} \checkmark     & \text{\sffamily X} & \cellcolor{ forestgreen(web)}  \textcolor{white} \checkmark        & Fortran90  & \text{\sffamily X} \\
\hline
\textsc{CarFisher}  & Carbone    & \cellcolor{ forestgreen(web)}  \textcolor{white} \checkmark  &   \text{\sffamily X}     & \text{\sffamily X}    &  \text{\sffamily X}    & \cellcolor{ forestgreen(web)}  \textcolor{white}\checkmark & \texttt{CAMB}               &  \cellcolor{ forestgreen(web)}  \textcolor{white}\checkmark        & IDL        &  \text{\sffamily X}  \\
\hline
\textsc{FisherMathica}& Casas     &  \cellcolor{ forestgreen(web)}  \textcolor{white} \checkmark& \text{\sffamily \textcolor{orange}{X}}         & \cellcolor{ forestgreen(web)}  \textcolor{white} \checkmark   & \text{\sffamily \textcolor{orange}{X}}   &\cellcolor{ forestgreen(web)}  \textcolor{white} \checkmark   & \texttt{CAMB}              &\cellcolor{ forestgreen(web)}  \textcolor{white}  \checkmark        &Mathematica &  \text{\sffamily \textcolor{orange}{X}} \\
            &           &                    &                    &                    &    & &  \texttt{CLASS}              &                   &            & \\
                        &           &                    &                    &       &    & &  \texttt{MGCAMB}              &                   &            & \\
\hline
\textsc{fishMath}        & Majerotto  & \cellcolor{ forestgreen(web)}  \textcolor{white} \checkmark &    \text{\sffamily X}      & \text{\sffamily X} & \text{\sffamily X} &\cellcolor{ forestgreen(web)}  \textcolor{white} \checkmark & \text{\sffamily X} &\cellcolor{ forestgreen(web)}  \textcolor{white}  \checkmark        & Mathematica&  \text{\sffamily X}\\            
\hline
\textsc{SOAPFish}   & Sapone     & \cellcolor{ forestgreen(web)}  \textcolor{white} \checkmark &    \text{\sffamily X}      & \text{\sffamily X}  & \text{\sffamily X} &\cellcolor{ forestgreen(web)}  \textcolor{white} \checkmark & \text{\sffamily X} & \cellcolor{ forestgreen(web)}  \textcolor{white} \checkmark        & Mathematica& \text{\sffamily \textcolor{orange}{X}} \\
           &            &                    &                    &                    &                    &                   &  & &  Python3  &  \\
\hline
\textsc{TotallySAF}        & Tutusaus   & \cellcolor{ forestgreen(web)}  \textcolor{white} \checkmark &   \cellcolor{ forestgreen(web)}  \textcolor{white} \checkmark       & \cellcolor{ forestgreen(web)}  \textcolor{white} \checkmark   & \cellcolor{ forestgreen(web)}  \textcolor{white} \checkmark  &\cellcolor{ forestgreen(web)}  \textcolor{white} \checkmark & \texttt{CAMB}               & \cellcolor{ forestgreen(web)}  \textcolor{white} \checkmark        & Python2/3  & \text{\sffamily \textcolor{orange}{X}} \\
           &Yahia-Cherif&                    &                    &                    &   & &  \texttt{CLASS}               &                   &    C++        &  \\
\hline
\textsc{CCCP}       & Camera     &     \text{\sffamily X}    & \cellcolor{ forestgreen(web)}  \textcolor{white} \checkmark  & \cellcolor{ forestgreen(web)}  \textcolor{white} \checkmark      & \text{\sffamily \textcolor{orange}{X}} &\cellcolor{ forestgreen(web)}  \textcolor{white} \checkmark   & \text{\sffamily X} & \cellcolor{ forestgreen(web)}  \textcolor{white} \checkmark        &Mathematica & \text{\sffamily X}\\
\hline
\textsc{CCCPy}       & Camera     &     \text{\sffamily X} & \text{\sffamily \textcolor{orange}{X}}  &\cellcolor{ forestgreen(web)}  \textcolor{white}  \checkmark     & \text{\sffamily \textcolor{orange}{X}} & \text{\sffamily \textcolor{orange}{X}} & \texttt{CLASS}  & \cellcolor{ forestgreen(web)}  \textcolor{white} \checkmark        & Python &  \text{\sffamily \textcolor{orange}{X}} \\
           &            &                    &                    &                    &  &  &  \texttt{CAMB}             &                   &            &            \\
\hline
\textsc{CosmicFish} & Martinelli &   \text{\sffamily X}      & \cellcolor{ forestgreen(web)}  \textcolor{white}  \checkmark  & \cellcolor{ forestgreen(web)}  \textcolor{white} \checkmark  & \cellcolor{ forestgreen(web)}  \textcolor{white} \checkmark   & \text{\sffamily X}    & \texttt{CAMB}               &\text{\sffamily X} & Fortran 90 & \text{\sffamily X}  \\
           &  Raveri    &                    &                    &                    &   & & \texttt{EFTCAMB}            &                   & Python2.7  &            \\
           &            &                    &                    &                    &  &  &  \texttt{MGCAMB}             &                   &            &            \\
\hline
\textsc{CosmoSIS}   & Kilbinger  &    \text{\sffamily X}      &   \cellcolor{ forestgreen(web)}  \textcolor{white}  \checkmark               &\cellcolor{ forestgreen(web)}  \textcolor{white}  \checkmark  & \cellcolor{ forestgreen(web)}  \textcolor{white} \checkmark    & \text{\sffamily \textcolor{orange}{X}}    & \texttt{CAMB}               & \text{\sffamily \textcolor{orange}{X}}                 & Fortran 90 & \cellcolor{ forestgreen(web)}  \textcolor{white}  \checkmark \\
           & Tutusaus   &                    &                    &            &        & &  \texttt{CLASS}              &                   & C / C++        & \\
           &            &                    &                    &             &       &    &     \texttt{MGCAMB}            &                   & Python  & \\           
\hline
\textsc{STAFF}      & Cardone    &  \text{\sffamily X}  & \cellcolor{ forestgreen(web)}  \textcolor{white} \checkmark &\cellcolor{ forestgreen(web)}  \textcolor{white}  \checkmark     &\cellcolor{ forestgreen(web)}  \textcolor{white}  \checkmark      &\cellcolor{ forestgreen(web)}  \textcolor{white} \checkmark & \text{\sffamily X} &\cellcolor{ forestgreen(web)}  \textcolor{white}  \checkmark        &Mathematica & \text{\sffamily X }\\           
\hline
\end{tabular}
}
\end{table}

\autoref{tab:forecastcodes} presents a summary the forecasting codes involved in the comparison, including:
\begin{itemize}
\item name: label or name of the code;
 \item contact: contact person(s) for the corresponding code;   
 \item GC$_{\text{s}}$: whether the code has been validated to forecast galaxy clustering from the spectroscopic survey;  
 \item GC$_{\text{ph}}$: whether the code has been validated to forecast galaxy clustering from the photometric survey; 
 \item WL: whether the code has been validated to forecast cosmic shear weak lensing from the photometric survey;
 \item XC$^{\rm (GC_{\text{ph}},WL)}$: whether the code has been validated to forecast cross-correlations between galaxy clustering and weak lensing, both from the photometric survey; 
 \item curvature: whether the code has been validated to forecast non-flat models; 
 \item Boltzmann code: which Boltzmann code(s) the code can interface with (if wished); 
 \item external input: whether the code can interface with externally provided fiducial observables and derivatives;  
 \item language: programming language(s) used in the code;  
 \item public: whether the code is public at the time of the release of this paper.
\end{itemize}
Since the objective of this paper is to present validated forecasts that are independent of the code used, we do not express a preference for 
any of the codes described here.  Users can decide which code to use based on their specific needs, considering aspects such as 
availability, cosmological probes considered, programming language, or any other specific feature(s).

\subsection{Galaxy clustering code comparison}\label{sec:ccgc}

In this section we describe the procedure to compare the cosmological forecasts for GC measurements from the 
\Euclid spectroscopic galaxy sample that are predicted by the different codes described in \autoref{sec:allcodes}. 
All codes employ the recipe described in \autoref{sec:recipegc}. The  Fisher matrix for GC is given by \autoref{eq:pkfm} and it depends 
on the model for the observed galaxy power spectrum given by \autoref{eq:GC:pk}, and on the characteristics of the \Euclid survey 
encoded in the effective volume, given by \autoref{eq:gc_effvol}. Here, we describe 
additional issues to take care of when implementing this recipe. If the reader wishes to compare and validate their own 
GC Fisher matrix code for \Euclid GC, they should follow these instructions in order to do so.
As part of the code-comparison process, all codes were adapted to use common inputs. If the reader wishes to follow suit, we make 
these inputs, alongside our outputs, available along with this paper, describing them and providing instructions for their use in 
\autoref{sec:appendix}. 

We first comment on specific issues within the implementation, focussing particularly on the derivatives needed for the Fisher matrix. 
We then prepare a list of comparison cases with increasing complexity.

\subsubsection{Implementation of the galaxy clustering recipe}\label{sec:inputs-baselinegc}

The first step in the implementation of the GC recipe described in \autoref{sec:recipegc} is to construct background 
unperturbed quantities that enter into the Fisher matrix calculation. The Hubble parameter, $H(z)$, the angular diameter 
distance, $D_\text{A}(z)$, and the comoving volume, $V(z_{\rm i},z_{\rm f})$, are given by \Cref{eq:Hz-general,eq:GC:Da,eq:Vcomoving}, respectively.
These functions should be general enough to allow for deviations from the standard flat $\Lambda$CDM cosmology, including
 non-zero curvature, dynamical DE models parameterised as in \autoref{sec:cosmo-bcg}, and deviations from GR as 
 described in \autoref{sec:cosmo-growth}. 

Once these background functions are defined, we can proceed with the estimation of quantities describing 
the \Euclid survey that are required to compute the effective volume $V_\text{eff}(k_\text{ref},\mu_\text{ref};z)$, 
of each redshift bin, defined in \autoref{eq:gc_effvol}. To this end, we first evaluate their comoving volumes, $V_{\rm s}(z)$, 
using \autoref{eq:Vcomoving}. This requires the knowledge of the area of the survey, which is given in \autoref{tab:GC-specifications}. 
As a second step we evaluate the galaxy number density $n(z)$, computed using \autoref{eq:gc-ndens} and 
the fractional number densities listed in \autoref{tab:gc_nofz}.

We must then calculate the quantities that define our observable, namely the two-dimensional 
galaxy power spectrum, 
$P_\text{obs}(k_\text{ref},\mu_\text{ref};z)$, in \autoref{eq:GC:pk}. 
This is achieved using the full non-linear model discussed in \autoref{sec:nonlineargc}. 
In particular, the following quantities are required to compute the observed power spectrum 
(see \autoref{sec:recipegc} for more details): 
\begin{itemize}
\item $f\sigma_8(z)$ and $b\sigma_8(z)$ enter through the modelling of RSDs. The former is 
evaluated using the growth rate, $f(z)$, defined in \autoref{eq:ggrowth}, and the values of $\sigma_8(z)$ from our input 
files (see \autoref{ISTF:input} for details).
The latter can be computed using the 
galaxy bias factors listed in \autoref{tab:gc_nofz} and the same values of $\sigma_8(z)$.
\item The pure matter power spectrum normalised by $\sigma^{2}_8(z)$,
$P_{\delta\delta}(k,z)/\sigma^{2}_8(z)$. Although these redshift-independent ratios are again given as input files (see 
\autoref{ISTF:input}), they can be estimated directly from the output of any Boltzmann code 
(e.g. \citealt{Lewis:1999bs} or \citealt{2011arXiv1104.2932L}).
\item The ``no-wiggles'' power spectrum, which we obtain using the formulae of \cite{Eisenstein_Hu_1997b}. The necessary ratios of 
$P_\text{nw}(k,z)/\sigma^2_8(z)$ are treated as inputs to the codes (see \autoref{ISTF:input}).
\item The absolute error in the distances expressed in terms of the redshift errors  $\sigma_0 = 0.001$, given 
by \autoref{eq:error-z-gc}.
\item The residual shot noise term is represented by $P_\text{shot}(k, z)$, which is set to zero at the reference cosmology. That is, 
our fiducial case assumes a perfect Poisson shot noise subtraction. 
\end{itemize} 
Our initial set of cosmological parameters is given in \autoref{def:GC:pars}. At the linear level, they are divided into four 
redshift-independent shape parameters and five redshift-dependent parameters. The corresponding Fisher matrix has a 
dimension $n_\text{tot} = 4+5\times N_z$, where $N_z $ is the number of redshift bins considered in the survey. 
When considering the full non-linear model,  we add two parameters to the set and hence two rows and columns to the Fisher matrix. 
We then project these initial `model-independent' matrices into the final, model-specific parameter spaces 
following \autoref{sec:gc_fisher}.

\paragraph{A note about derivatives.}

Derivatives of $\log_{10} P_{\rm obs}(k,\mu, z)$ with respect to cosmological parameters translate the covariance matrix of 
the power spectrum into that of the model parameters. Uncertainties in the derivatives can have a significant impact on the final 
results (as they enter the Fisher matrix that has to inverted). This is particularly true for derivatives that depend on numerical estimates of the slope of the power spectrum 
around the BAO wiggles. It is therefore imperative to pay close attention to the implementation of the numerical derivatives 
and to ensure that the parameter steps used lie well within the convergence region. 
More specifically, the parameter steps used should be small enough to be able 
to accurately represent the dependence of the underlying function (the power spectrum) on the various parameters, but large 
enough to avoid machine precision errors. We have checked this by
ensuring we got stable results when the step size is varied in the range $0.0001-0.01$ and  with the different codes. 

As already discussed, most of the derivatives of the observed galaxy power spectrum are calculated numerically. 
We start by considering the derivatives of the matter power spectrum with respect to the shape parameters. 
To this end, we use input files corresponding to matter power spectra computed for different parameter values. 
We evaluate the matter power spectrum at the fiducial cosmology 
and at individually incremented parameter values, keeping all other parameters fixed. The relative increments with respect to 
the reference value of a given shape parameter, $\theta_\text{ref}$, are computed as $\theta_\text{ref}(1+\epsilon)$ and 
$\theta_\text{ref}(1-\epsilon)$.
These matter power spectra are computed for all redshifts bins. In total, we then have $N_z$ matter power spectra for 
the reference cosmology and $2\times N_z$ matter power spectra corresponding to the relative increments. 

We evaluate derivatives with respect to the shape parameters using three-point rules:
\bear
\left.\frac{\partial \ln P_{\text{obs}}(k,\mu, z)}{\partial \theta_\text{shape}}\right |_{\theta_\text{ref}} &=& \left.\frac{\partial \ln \left(P_{\rm dw}(k, z)\sigma^{-2}_8(z)\right)}{\partial\theta_\text{shape}}\right |_{\theta_\text{ref}}\nonumber \\ 
&=&\frac{\ln P\sigma^{-2}_8\left[k\,,z\,;\theta_\text{shape,\,ref}(1+\epsilon)\right]-\ln P\sigma^{-2}_8\left[k\,,z\,;\theta_\text{shape,\,ref}(1-\epsilon)\right]}{2\epsilon\,\theta_\text{shape,\,ref}}\,,
\label{eq:GC:shapederiv}
\eear
where $\theta_\text{shape}$ is any of our four shape parameters, and for simplicity we wrote  $P\sigma^{-2}_8 = P_{\rm dw}(k,z)/\sigma^{2}_8(z)$. 
We remind the reader that the only quantity depending  on the shape parameters is the 
power spectrum with the wiggles removed, which means that the $\mu$ dependence cancels out.

Similarly for the two non-linear parameters, we also use 3-point derivatives to calculate the following:
\bear
\left.\frac{\partial \ln P_\text{obs}}{\partial \sigma_\text{v}}\right |_{\sigma_\text{v,fid}} &=& 
\left.\frac{\partial \ln (P_\text{dw}(k,\mu;z)\sigma^{-2}_8(z))}{\partial \sigma_\text{v}}\right |_{\sigma_\text{v,fid}} \, \text{; }
\nonumber \\
\left.\frac{\partial \ln P_\text{obs}}{\partial \sigma_\text{p}}\right |_{\sigma_\text{p,fid}} &=& 
\left.\frac{\partial}{\partial \sigma_\text{p}} \ln\left[\frac{1}{1 + (f(z)k\mu\sigma_\text{p}(z))^2}\right]\right |_{\sigma_\text{p,fid}} \ .
\label{eq:GC:nlderiv}
\eear
We use different approaches for the derivatives with respect to redshift-dependent parameters. The derivative of the observed 
galaxy power spectrum with respect to the residual shot noise can be computed analytically as the inverse of the 
observed galaxy power spectrum, see \autoref{eq:deriv-shot-noise-gc}. For the rest we use numerical rules at either 3 or 5 points 
in the parameter space.

For derivatives with respect to $\ln f\sigma_8(z)$ and $\ln b\sigma_8(z)$ we use the 3-point rule:
\begin{align}
\frac{\partial \ln P_\text{obs}(k,\mu,z)}{\partial \ln f\sigma_8} &= \frac{\ln P_\text{obs}\left[k\,,z\,;f\sigma_\text{8,\,ref}^{(1+\epsilon)}\right]-\ln P_\text{obs}\left[k\,,z\,;f\sigma_\text{8,\,ref}^{(1-\epsilon)}\right]}{2\epsilon\ln f\sigma_\text{8,\,ref}} \text{; } \label{eq:GC:dlnPdlnfs8}\\
\frac{\partial \ln P_\text{obs}(k,\mu,z)}{\partial \ln b\sigma_8} &= \frac{\ln P_\text{obs}\left[k\,,z\,;b\sigma_\text{8,\,ref}^{(1+\epsilon)}\right]-\ln P_\text{obs}\left[k\,,z\,;b\sigma_\text{8,\,ref}^{(1-\epsilon)}\right]}{2\epsilon\ln b\sigma_\text{8,\,ref}}\,.\label{eq:GC:dlnPdlnbs8}
\end{align}
For $\ln H$ and $\ln D_\text{A}$ we choose 5-point derivative rules in order to better capture the oscillatory behaviour of the galaxy 
power spectrum near the BAO, which is modified by the AP distortions via these two parameters.\footnote{We have also 
tested the results by using the standard 3-point derivative rule and we have found deviations of the order of $0.005 \%$ as long as the 
step used for evaluating the derivatives is sufficiently small.}
The derivatives of the galaxy power spectrum with respect to the Hubble parameter and the angular diameter distance are:
\begin{align}
\frac{\partial \ln P_\text{obs}(k,\mu,z)}{\partial \ln H} &= 8\frac{\ln P_\text{obs}\left[H_\text{ref}^{(1+\epsilon)}\right]-\ln P_\text{obs}\left[H_\text{ref}^{(1-\epsilon)}\right]}{12\epsilon\ln H_\text{ref}}
-\frac{\ln P_\text{obs}\left[H_\text{ref}^{(1+2\epsilon)}\right]-\ln P_\text{obs}\left[H_\text{ref}^{(1-2\epsilon)}\right]}{12\epsilon\ln H_\text{ref}}\,;\\
\frac{\partial \ln P_\text{obs}(k,\mu,z)}{\partial \ln D_A} &= 8\frac{\ln P_\text{obs}\left[D_\text{A,ref}^{(1+\epsilon)}\right]-\ln P_\text{obs}\left[D_\text{A,ref}^{(1-\epsilon)}\right]}{12\epsilon\ln D_\text{A,ref}} - \frac{\ln P_\text{obs}\left[D_\text{A,ref}^{(1+2\epsilon)}\right]-\ln P_\text{obs}\left[D_\text{A,ref}^{(1-2\epsilon)}\right]}{12\epsilon\ln D_\text{A,ref}}\,.
\label{eq:GC:dlnPdlnDa}
\end{align}
Here we omitted the direct dependence on $k$ and $\mu$ for simplicity. It is important to remember that also the wavenumber $k$ 
and the direction $\mu$ are modified through the cosmology (due to projection effects). This modification is taken into account 
using \autoref{eq:kref-muref}, where the subscript `ref' refers to the reference cosmology. It is clear from \autoref{eq:kref-muref} that
$k$ and $\mu$  will differ from the corresponding reference  values when $H_\text{ref}(z)/H(z)$ or 
$D_{\rm A}(z)/D_\text{A,ref}$ deviate from unity, that it is when we take derivatives of the observed galaxy power spectrum with 
respect to $\ln H(z)$ and $\ln D_{\rm A}(z)$. 
We would also like to notice that the independent variables we use are in fact logarithms of $H(z)$ and $D_\text{A}(z)$, therefore the input 
power spectra correspond to parameter increments given by, for example
\be
\ln H_\text{ref}\rightarrow \ln H_\text{ref} +\epsilon \ln H_\text{ref} = (1+\epsilon)\ln\left[H_\text{ref}\right] = \ln\left[ H_\text{ref}^{(1+\epsilon)}\right]\, .
\ee
Therefore the relative increment on the variable $H(z)$ appearing explicitly in the observed galaxy power spectrum is 
$H_\text{ref}^{(1+\epsilon)}$; the same discussion applies to $D_\text{A}(z)$.

It is clear from the above expressions that the derivatives of the observed galaxy power spectrum need to be evaluated at 
each redshift bin. 

\subsubsection{Settings definition}\label{sec:definition-cases-gc}

In order to compare the outputs of the different forecasting codes in a controlled way, we consider a series of settings 
with increasing complexity, which we describe in this section.
We compare our codes at each step and provide the resulting Fisher matrix files in \autoref{sec:appendix} for the reader 
who wishes to perform the comparison with their own code. We recommend taking these steps as they have been optimised for easy 
de-bugging. 
 We now describe each of these settings in detail. A summary of all settings is presented in  \autoref{tab:GC:cases}.

\paragraph{Linear setting.}

In this case we only consider the observed power spectrum for scales  $k<k_\text{max}=0.25\,h\,{\rm Mpc}^{-1}$ for all redshift bins
and use the linear-theory prediction for $P_{\rm obs}(k,\mu)$. This is achieved by fixing the non-linear parameters 
$\sigma_\text{p} = \sigma_\text{v} = 0$ in the non-linear recipe of \autoref{eq:GC:pk}, which is then reduced to
its linear version of \autoref{eq:GC:pklin}. We must compute derivatives of this model with respect to the remaining cosmological parameters.
We note that, when we estimate derivatives with respect to {$\omega_\text{m,0}$} by varying the matter density, we must also vary the 
energy density of the remaining component, $\Omega_{\Lambda,0}$ in this case, in order to keep the cosmology flat. This does not 
mean that we cannot later project into a new parameter set allowing for curvature, since non-flat models should still be allowed, 
which we ensure by keeping $D_\text{A}(z)$ and $H(z)$ free. We use as input files the ratio $P_{\delta\delta}(k;z)/\sigma^2_8(z)$, (calculated  using \texttt{CAMB}),
as well as for the growth rate $f(z)$. This is described in detail in \autoref{sec:inputs-baselinegc}.

The Fisher matrix that we obtain has dimensions $4 + 5\times N_z$, composed of the derivatives of the four shape parameters given by 
\autoref{eq:GC:shapederiv}, and the derivatives of the redshift-dependent parameters given in \Crefrange{eq:GC:dlnPdlnfs8}{eq:GC:dlnPdlnDa}, as well as the analytical derivative with respect to the shot noise,  given in \autoref{eq:deriv-shot-noise-gc}. 

\paragraph{Pessimistic setting.}

As a second case we implement the full non-linear model for the observed power spectrum of \autoref{eq:GC:pk} and consider scales  
$k<k_\text{max}=0.25\,h\,{\rm Mpc}^{-1}$. As this is the same range of scales of above linear setting, the comparison of the results obtained in these
two cases illustrates the degradation of the constraining power of $P_{\rm obs}(k,\mu)$ due to non-linearities.\footnote{Here we use $k<k_\text{max}=0.25\,h\,{\rm Mpc}^{-1}$  as ``pessimistic'', but one may argue that it is not pessimistic or even optimistic, depending on ones confidence in modelling of the non-linear power spectrum. For example the 1-halo term is a 30\% correction to linear theory at this scale. If the model beyond linear scales is insufficient then this is an issue of potential bias in the results; however, this should still enable us to capture the expected uncertainties in fitted parameters correctly.} This configuration requires $P_\text{nw}(k;\,z)$, which is also given as a common input to all codes  (see \autoref{sec:inputs-baselinegc} for 
details). 

In this step we also need to vary the non-linear parameters, $\sigma_{\rm p}(z_{\rm mean})$ and  $\sigma_{\rm v}(z_{\rm mean})$, where 
$z_\text{mean}$ is the mean redshift of the survey.
In this case, we now allow for variations of the growth rate in the FoG term. To this end we rewrite the term containing the growth rate 
in \autoref{eq:pk_zs} as 
\begin{equation}
k\mu f(z)\sigma_\text{p}(z) = k\mu f\sigma_8(z)\frac{\sigma_{\rm p}(z)}{\sigma_8(z)} = k\mu f\sigma_8(z) \frac{\sigma_\text{p}(z_\text{mean})}{\sigma_8(z)}\frac{D(z)}{D(z_\text{mean})} \, .
\label{eq:k_mu_f_sigma}
\end{equation}
Here, we have assumed that $\sigma_{\rm p}$ can be written as $ \sigma_{\rm p}(z_\text{mean}) D(z)/D(z_\text{mean})$. 
Furthermore $\sigma_8(z)$ can also be written in terms of a $\sigma_8(z_\text{mean})$, 
as it was the case for $\sigma_\text{p}(z)$ and $\sigma_\text{v}(z)$, and \autoref{eq:k_mu_f_sigma} then reads:
\begin{equation}
k\mu f(z)\sigma_\text{p}(z)  =  k\mu f\sigma_8(z) \frac{\sigma_\text{p}(z_\text{mean})}{\sigma_8(z_\text{mean})}=k\mu f\sigma_8(z) \tilde{\sigma}_\text{p}(z_\text{mean}) \,,  
\label{eq:k_mu_f_sigma_final}
\end{equation}
where we have now defined a new parameter 
$\tilde{\sigma}_\text{p}(z_\text{mean})=\sigma_\text{p}(z_\text{mean})\sigma^{-1}_8(z_\text{mean})$.  Hence the two independent 
parameters become $\sigma_\text{v}(z_\text{mean})$ and $\tilde{\sigma}_\text{p}(z_\text{mean})$. The FoG term in 
\autoref{eq:pk_zs} is written as 
\be
\frac{1}{1+\left(k\mu f(z)\sigma_\text{p}(z)\right)^2} = \frac{1}{1+\left(k\mu f\sigma_8(z) \tilde{\sigma}_\text{p}(z_\text{mean})\right)^2}\, ;
\label{eq:new_fog_factor}
\ee
whereas the damping term remains as
\be
g_\mu(z) = \sigma_\text{v}^2(z_\text{mean})\frac{D^2(z)}{D^2(z_\text{mean})}\left[1-\mu^2+\mu^2\left(1+f(z)\right)^2\right]\,.
\label{eq:new_damping_factor}
\ee
We calculate the derivatives with respect to the two new non-linear parameters using the same three-point numerical method 
as those used for the shape parameters (see Eq.~\ref{eq:GC:nlderiv}). The Fisher matrix now has its largest dimensions:  
$4 + 2 + 5\times N_z$ rows and columns.

\paragraph{Intermediate setting.}

This case corresponds to the pessimistic setting but the maximum wavenumber considered in the analysis is extended up to $k_\text{max}=0.30\,h\,{\rm Mpc}^{-1}$. Therefore, the dimension of the Fisher matrix is still $4 + 2 + 5\times N_z$.

\paragraph{Optimistic setting.}

This case corresponds to intermediate setting where the maximum wavenumber considered in the analysis is  $k_\text{max}=0.30\,h\,{\rm Mpc}^{-1}$; 
however, we keep $\sigma_{\rm p} = \sigma_{\rm v}$ fixed to their reference non-zero values, computed using \autoref{eq:sigmav}. 
This means that non-linear corrections are included into our analysis, but we assume that we have perfect knowledge of them. 
The Fisher matrix in this step has dimensions $4 + 5\times N_z$. In practice, the Fisher matrix in this  case is the same than in the intermediate setting where we throw the rows and columns corresponding to the parameters $\sigma_{\rm p}$ and  $\sigma_{\rm v}$.


\paragraph{Projecting to the final parameter space.} 

Once the full Fisher matrices have been computed, they can be projected into the same final cosmological parameter spaces as the 
other probes described in this paper (see \autoref{sec:gc_fisher}).

The choice of the parameters in $\theta_\text{final}$ defines the cosmological models to be considered. 
We explore extensions of the $\Lambda$CDM parameter space by allowing for curvature, dynamical dark energy, and deviations 
from general relativity. 
\begin{description}
\item[PR1] As a first step, we restrict the analysis to the $\Lambda$CDM cosmological model, extended to allow also for 
non-flat models. In this case, the Jacobian only contains the 
rows and columns corresponding to the derivatives of the original parameter set with respect to the following~:
\be
\theta_\text{final, PR1} = \{\Omega_{\rm b,0},\,h,\,\Omega_{\rm m,0},\,n_{\rm s},\,\Omega_{\rm DE,\,0},\,\sigma_{8}\}\, ,
\label{eq:proj-gc-case1}
\ee
corresponding to a $(4+3\times N_z)\times 6$ matrix. \smallskip
\item[PR2] In this step we extend the parameter space considered in PR1 by allowing also for dynamical dark energy models,
in which case
\be
\theta_\text{final, PR2} = \{\Omega_{\rm b,0},\,h,\,\Omega_{\rm m,0},\,n_{\rm s},\,\Omega_{\rm DE,\,0}, \,w_0,\,w_a,\,\sigma_{8}\}\,.
\label{eq:proj-gc-case2}
\ee
Now the Jacobian is given by a $(4+3\times N_z)\times 8$ matrix.

The function $\sigma_8(z)$  can be expressed as 
\be
\sigma_8(z)  = \sigma_{8}D(z)\,,
\ee
and consequently $f\sigma_8(z)$ is given by
\be
f\sigma_8(z) = f(z)\sigma_{8}D(z) \, ,
\label{eq:effsigma8-gc}
\ee
where $f(z)$ is the numerical solution of the first-order differential equation in \autoref{eq:ode-eff} and the growth factor 
$D(z)$ is given in terms of $f(z)$ in  \autoref{eq:growthfactor-gc-val}.\smallskip

\item[PR3] 
Finally, we consider our most general cosmological models by extending the  parameter space of PR2 by including
 the growth rate exponent parameter, $\gamma$: 
\be 
 \theta_\text{final, PR3} = \{\Omega_{\rm b,0},h,\,\Omega_{\rm m,0},\,n_{\rm s},\,\Omega_{\rm DE,\,0}, \,w_0,\,w_a,\,\sigma_{8},\,\gamma\}\, .
\label{eq:proj-gc-case3}
\ee
In practice we substitute the expression of the growth rate given by \autoref{eq:fgrowth} into \autoref{eq:effsigma8-gc}. 
This case requires one to take into account the derivatives  of the observed power spectrum with respect to $\gamma$ through 
\autoref{eq:fgrowth}, which means that  the dimensions of the Jacobian matrix are  $(4+3\times N_z)\times 9 $.\smallskip

\item[PRn-flat] 
When  we consider flat cosmologies, which correspond to the same parameter spaces as in cases PR1, PR2, and PR3, we eliminate $\Omega_{\rm DE,\,0}$ by assuming into the equations that $\Omega_{\rm DE,\,0} = 1 - \Omega_{\rm m,\,0}$.
\end{description}
As mentioned in \autoref{sec:gc_fisher}, before projecting the Fisher matrix into a new parameter space, we need to marginalise over some parameters, in particular these are $\sigma_{\rm p}$, $\sigma_{\rm v}$, $\ln b\sigma_8(z)$, and $P_{\rm shot}$. Hence, the dependence of the new parameters $\theta_{\rm final}$ enter in the four shape parameters and in the redshift dependent parameters $\ln D_{\rm A}(z)$, $\ln H(z)$, and $\ln f\sigma_{8}(z)$. 

In order to project the marginalised Fisher matrix into a new parameter space, we need to create a Jacobian matrix; here we report the most general transformation (which correspond to the $\Lambda$CDM non-flat $w_0+w_a+\gamma$ cosmology), which is equal for all the four cases listed above.\footnote{We would like to remind the reader that the cases PR1--3  all have the same number of parameters in their marginalised versions.} The Jacobian is explictly
\be
\begin{bmatrix}
  h^2 & 0 &  2h\Omega_\text{b,0} &0 & 0 & 0 & 0 & 0& 0 \\
  0 & 1 & 0 & 0 & 0 & 0 & 0 & 0 & 0\\ 
  0 & 2h\Omega_\text{m,0} & h^2 & 0 & 0 & 0 & 0 & 0& 0\\
  0 & 0 & 0 & 1 & 0 & 0 & 0 & 0& 0\\
  0 & \left.\frac{\partial \ln D_{\rm A}}{\partial h}\right|_{z_1} & \left.\frac{\partial \ln D_{\rm A}}{\partial \Omega_{\rm m,0}}\right|_{z_1} & 0 & \left.\frac{\partial \ln D_{\rm A}}{\partial \Omega_\text{DE,0}}\right|_{z_1} & \left.\frac{\partial \ln D_{\rm A}}{\partial w_0}\right|_{z_1} & \left.\frac{\partial \ln D_{\rm A}}{\partial w_a}\right|_{z_1} & 0& 0\\
  0 & \left.\frac{\partial \ln H}{\partial h}\right|_{z_1} & \left.\frac{\partial \ln H}{\partial \Omega_{\rm m,0}}\right|_{z_1} & 0 & \left.\frac{\partial \ln H}{\partial \Omega_\text{DE,0}}\right|_{z_1} & \left.\frac{\partial \ln H}{\partial w_0}\right|_{z_1} & \left.\frac{\partial \ln H}{\partial w_a}\right|_{z_1} & 0& 0\\
  0 & 0 & \left.\frac{\partial \ln f\sigma_8}{\partial \Omega_{\rm m,0}}\right|_{z_1} & 0 & \left.\frac{\partial \ln f\sigma_8}{\partial \Omega_\text{DE,0}}\right|_{z_1} & \left.\frac{\partial \ln f\sigma_8}{\partial w_0}\right|_{z_1} & \left.\frac{\partial \ln f\sigma_8}{\partial w_a}\right|_{z_1} & \left.\frac{\partial \ln f\sigma_8}{\partial \sigma_8}\right|_{z_1} & \left.\frac{\partial \ln f\sigma_8}{\partial \gamma}\right|_{z_1} \\
  0 & \left.\frac{\partial \ln D_{\rm A}}{\partial h}\right|_{z_2} & \left.\frac{\partial \ln D_{\rm A}}{\partial \Omega_{\rm m,0}}\right|_{z_2} & 0 & \left.\frac{\partial \ln D_{\rm A}}{\partial \Omega_\text{DE,0}}\right|_{z_2} & \left.\frac{\partial \ln D_{\rm A}}{\partial w_0}\right|_{z_2} & \left.\frac{\partial \ln D_{\rm A}}{\partial w_a}\right|_{z_2} & 0& 0\\
  0 & \left.\frac{\partial \ln H}{\partial h}\right|_{z_2} & \left.\frac{\partial \ln H}{\partial \Omega_{\rm m,0}}\right|_{z_2} & 0 & \left.\frac{\partial \ln H}{\partial \Omega_\text{DE,0}}\right|_{z_2} & \left.\frac{\partial \ln H}{\partial w_0}\right|_{z_2} & \left.\frac{\partial \ln H}{\partial w_a}\right|_{z_2} & 0& 0\\
  0 & 0 & \left.\frac{\partial \ln f\sigma_8}{\partial \Omega_{\rm m,0}}\right|_{z_2} & 0 & \left.\frac{\partial \ln f\sigma_8}{\partial \Omega_\text{DE,0}}\right|_{z_2} & \left.\frac{\partial \ln f\sigma_8}{\partial w_0}\right|_{z_2} & \left.\frac{\partial \ln f\sigma_8}{\partial w_a}\right|_{z_2} & \left.\frac{\partial \ln f\sigma_8}{\partial \sigma_8}\right|_{z_2} & \left.\frac{\partial \ln f\sigma_8}{\partial \gamma}\right|_{z_2}\\
 \ldots & \ldots & \ldots &\ldots & \ldots & \ldots & \ldots & \ldots & \ldots \\
  \ldots & \ldots & \ldots &\ldots & \ldots & \ldots & \ldots & \ldots & \ldots \\
   0 & \left.\frac{\partial \ln D_{\rm A}}{\partial h}\right|_{z_n} & \left.\frac{\partial \ln D_{\rm A}}{\partial \Omega_{\rm m,0}}\right|_{z_n} & 0 & \left.\frac{\partial \ln D_{\rm A}}{\partial \Omega_\text{DE,0}}\right|_{z_n} & \left.\frac{\partial \ln D_{\rm A}}{\partial w_0}\right|_{z_n} & \left.\frac{\partial \ln D_{\rm A}}{\partial w_a}\right|_{z_n} & 0& 0\\
  0 & \left.\frac{\partial \ln H}{\partial h}\right|_{z_n} & \left.\frac{\partial \ln H}{\partial \Omega_{\rm m,0}}\right|_{z_n} & 0 & \left.\frac{\partial \ln H}{\partial \Omega_\text{DE,0}}\right|_{z_n} & \left.\frac{\partial \ln H}{\partial w_0}\right|_{z_n} & \left.\frac{\partial \ln H}{\partial w_a}\right|_{z_n} & 0& 0\\
  0 & 0 & \left.\frac{\partial \ln f\sigma_8}{\partial \Omega_{\rm m,0}}\right|_{z_n} & 0 & \left.\frac{\partial \ln f\sigma_8}{\partial \Omega_\text{DE,0}}\right|_{z_n} & \left.\frac{\partial \ln f\sigma_8}{\partial w_0}\right|_{z_n} & \left.\frac{\partial \ln f\sigma_8}{\partial w_a}\right|_{z_n} & \left.\frac{\partial \ln f\sigma_8}{\partial \sigma_8}\right|_{z_n} & \left.\frac{\partial \ln f\sigma_8}{\partial \gamma}\right|_{z_n} \\  
  \end{bmatrix}
  \, .
\ee
The first four rows of the matrix correspond to the derivatives of the shape parameters with respect to the new set of parameters 
$\theta_\text{final}$, whereas all remaining rows correspond to the derivatives or the redshift-dependent parameters with respect to 
 $\theta_\text{final}$, which are evaluated at the mean redshift of each bin. 

We note that, since background functions do not depend on perturbed quantities, the derivatives of 
$\ln D_{\rm A}$ and $\ln H$ with respect to $n_{\rm s}$ and $\gamma$ are zero, as well as those with respect  to $\sigma_{8}$. 
Moreover, since we are using {$\omega_{\rm b,0}$ and $\omega_{\rm m,0}$} as independent parameters, $H(z)$ and 
$D_{\rm A}(z)$ only depend on the total matter density, represented by {$\omega_{\rm m,0}$}. Furthermore, the derivatives of the growth rate $f(z)$ with respect to $n_{\rm s}$ and $h$ are zero.

\begin{table*}
\centering
\caption{Summary of the cases considered in our code comparison. We note that we use $N_z=4$ redshift bins and each model-independent parameter appears in each of the bins, $z_i$.}
\label{tab:GC:cases}
\begin{tabular}{lcl}
\hline
 Id & No. of parameters & \multicolumn{1}{c}{Parameters}\\
\hline\hline
{ Linear} & 24 & ${\omega_\text{b,0}}, h,  {\omega_\text{m,0}}, n_\text{s}, \ln D_\text{A}(z_i), \ln H(z_i), \ln f\sigma_8(z_i), \ln b\sigma_8(z_i), P_\text{s}(z_i)$  \\
{ Pessimistic} & 26 & ${\omega_\text{b,0}}, h,  {\omega_\text{m,0}}, n_\text{s}, \ln D_\text{A}(z_i), \ln H(z_i), \ln f\sigma_8(z_i), \ln b\sigma_8(z_i), P_\text{s}(z_i)$ \\
{ Intermediate} & 26 & ${\omega_\text{b,0}}, h,  {\omega_\text{m,0}}, n_\text{s}, \sigma_\text{p},\sigma_\text{v}, \ln D_\text{A}(z_i), \ln H(z_i), \ln f\sigma_8(z_i), \ln b\sigma_8(z_i), P_\text{s}(z_i)$ \\
{ Optimistic} & 24 & ${\omega_\text{b,0}}, h,  {\omega_\text{m,0}}, n_\text{s}, \ln D_\text{A}(z_i), \ln H(z_i), \ln f\sigma_8(z_i), \ln b\sigma_8(z_i), P_\text{s}(z_i)$ \\
\hline
{ PR1} & 6 & $\Omega_{\rm b,0}, h, \Omega_{\rm m,0}, n_{\rm s}, \Omega_\text{DE,0}, \sigma_{8}$ \\
{ PR2} & 8 & $\Omega_{\rm b,0}, h, \Omega_{\rm m,0}, n_{\rm s}, \Omega_\text{DE,0}, w_0, w_a, \sigma_{8}$ \\
{ PR3} & 9 & $\Omega_{\rm b,0}, h, \Omega_{\rm m,0}, n_{\rm s}, \Omega_\text{DE,0}, w_0, w_a, \sigma_{8}, \gamma$ \\
{ PR1-flat} & 5 & $\Omega_{\rm b,0}, h, \Omega_{\rm m,0}, n_{\rm s}, \sigma_{8}$ \\
{ PR2-flat} & 7 & $\Omega_{\rm b,0}, h, \Omega_{\rm m,0}, n_{\rm s}, w_0, w_a, \sigma_{8}$ \\
{ PR3-flat} & 8 & $\Omega_{\rm b,0}, h, \Omega_{\rm m,0}, n_{\rm s}, w_0, w_a, \sigma_{8}, \gamma$ \\
\hline
\end{tabular}
\end{table*}

\subsubsection{Results of the code comparison}\label{sssec:ccgc}

In this section we compare the output of the six codes capable of estimating cosmological forecasts for GC measurements 
described in \autoref{sec:allcodes}. These are  \texttt{BFF}, \texttt{CarFisher}, \texttt{fishMath}, \texttt{FisherMathica}, 
\texttt{SOAPFish}, and \texttt{TotallySAF}. We have compared the results of these codes for each of the cases 
summarised in \autoref{tab:GC:cases}. 
The main conclusion of this exercise is that all codes are able to provide consistent cosmological forecasts for \Euclid. 

\begin{figure}[h]
\center 
\includegraphics[width=0.48\textwidth]{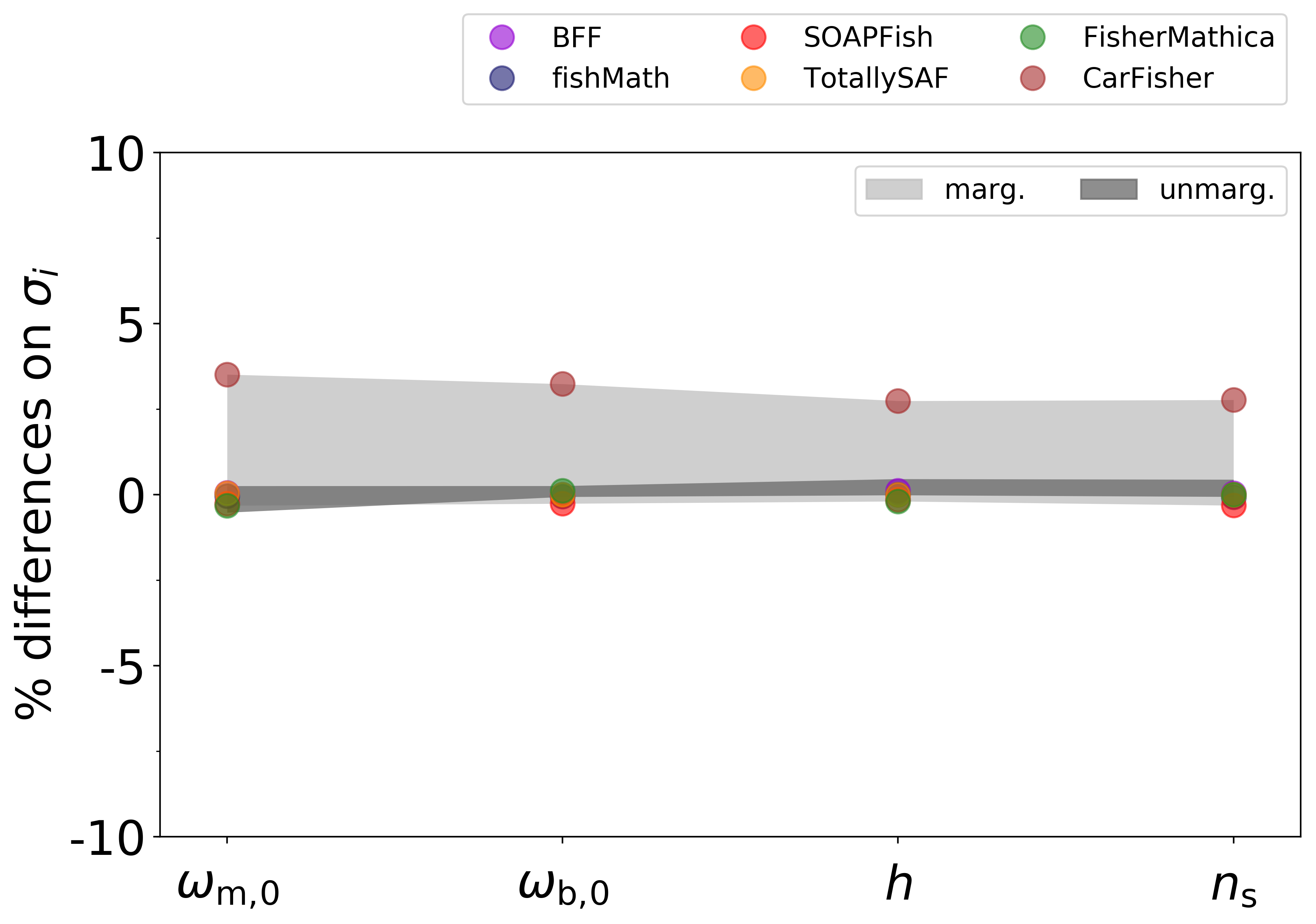}
\includegraphics[width=0.48\textwidth]{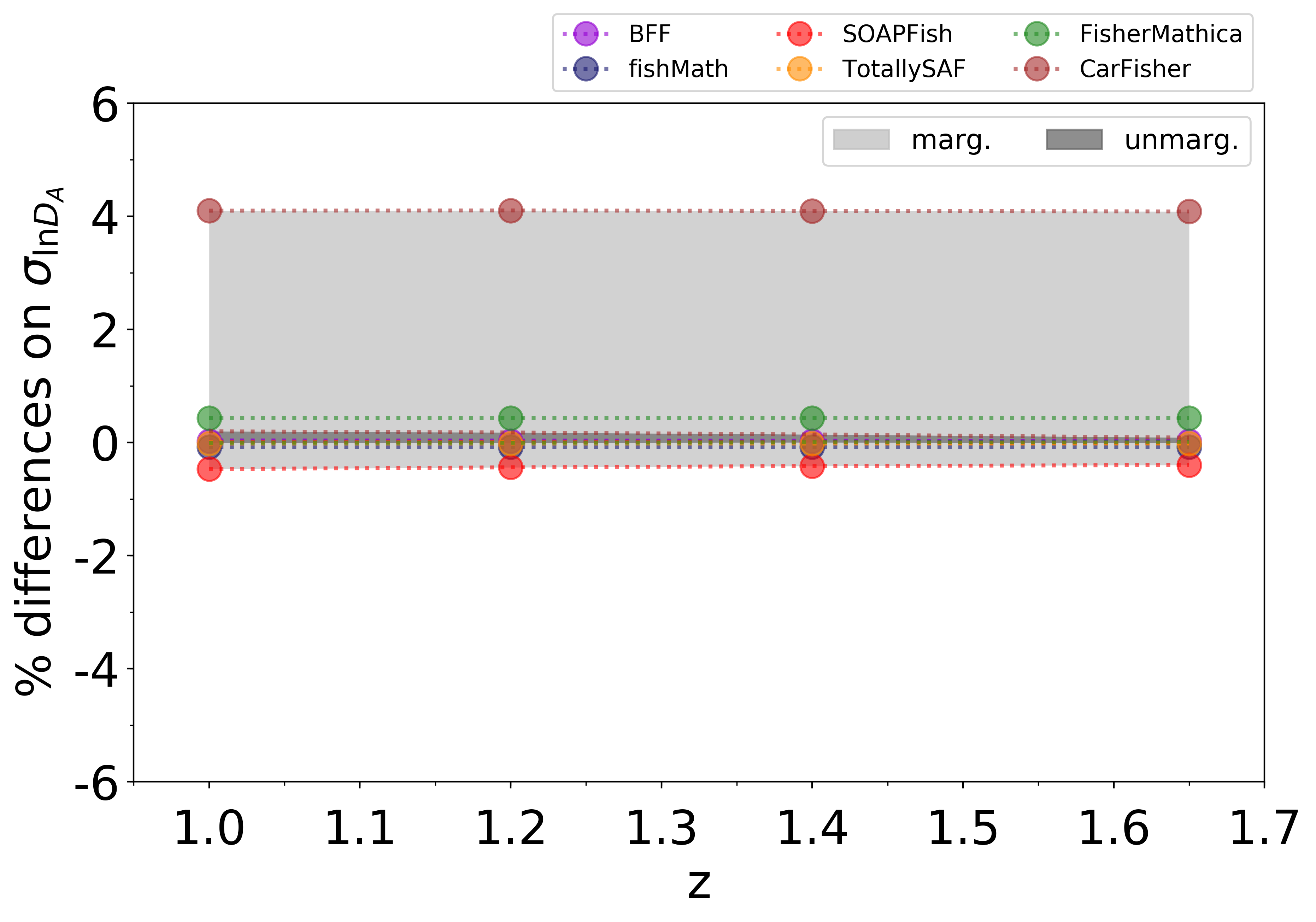}\\
\includegraphics[width=0.48\textwidth]{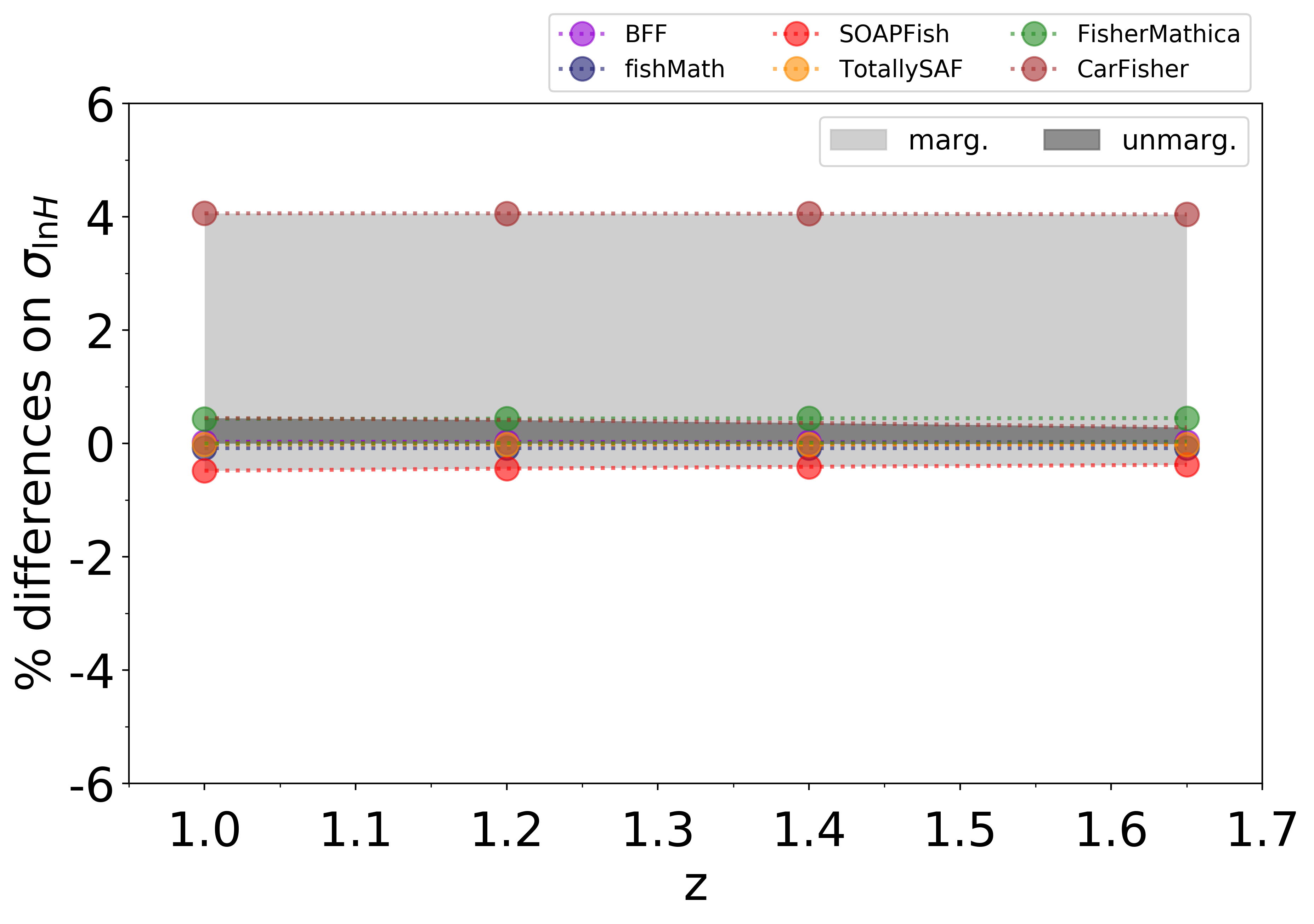} 
\includegraphics[width=0.48\textwidth]{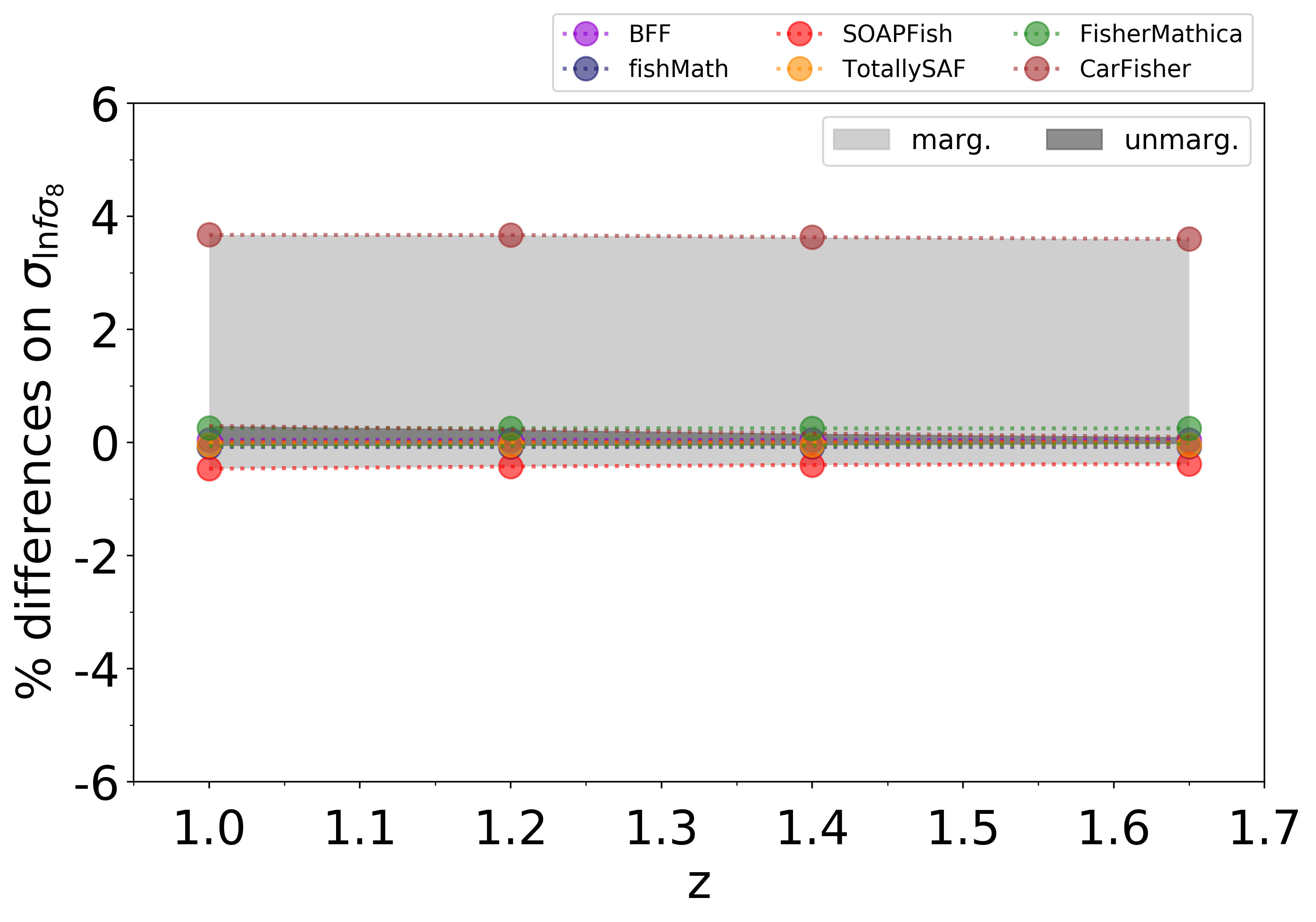}
\caption{Comparison of the errors  for the intermediate case with $k_\text{max}=0.30 \, h\,\textrm{Mpc}^{-1}$. 
The points indicate deviation of each code from the median {of the codes}, the light grey area indicates the maximum deviation from the median of the marginalised errors, while the dark grey area indicates the same but for the unmarginalised errors. 
In the top-left panel the four shape parameters are shown. 
The rest of the panels show redshift-dependent parameters. Specifically, the top-right panel is for the angular diameter distance logarithm, $\ln D_A$, the bottom-left for the Hubble function, $\ln H$, and the bottom-right for the growth of structure, $\ln f\sigma_{8}$.
As explained in the text, the values of these parameters in each redshift bin are treated as separate, independent parameters. However, to indicate that the quantities would be in reality expected to evolve in redshift, we connect the dots with dotted lines in the marginalised case. In the unmarginalised case, we don't plot lines, since the shaded area connects the points already.
The non-linear parameters $\sigma_{\rm p}$  and $\sigma_{\rm v}$, as well as $\ln b\sigma_{8}$  and $P_{\rm s}$, have been marginalised over.}
\label{fig:resu-GC-zdep}
\end{figure}

In \autoref{fig:resu-GC-zdep}, top-left panel, we show the relative difference between the full marginalised errors of the four shape parameters produced by the six codes considered in the current comparison. To produce this plot---and all the similar plots henceforth---we compare, for the constraint on each cosmological parameter, the performance of each code to the median of the constraints obtained by the set of all the available codes. Then, we show the relative difference between a given code and such median in percentage. We note that, by definition of median, in the case an odd number of codes is available, there will always be a code that actually provides the median constraints, meaning its corresponding marker will be placed at $0\%$. The agreement between the outcomes is better than a few percent for the Intermediate case. In \autoref{fig:resu-GC-zdep} we also show the relative difference between the full marginalised errors for the redshift dependent parameters $[\ln D_{\rm A}(z)$, $\ln H(z)$, $\ln f\sigma_{8}(z)]$ produced by the six codes considered in the current comparison. Also for these set of parameters the difference among all the outcomes is of the order of $4\%$. 

\begin{figure}[h]
\center \includegraphics[width=0.48\textwidth]{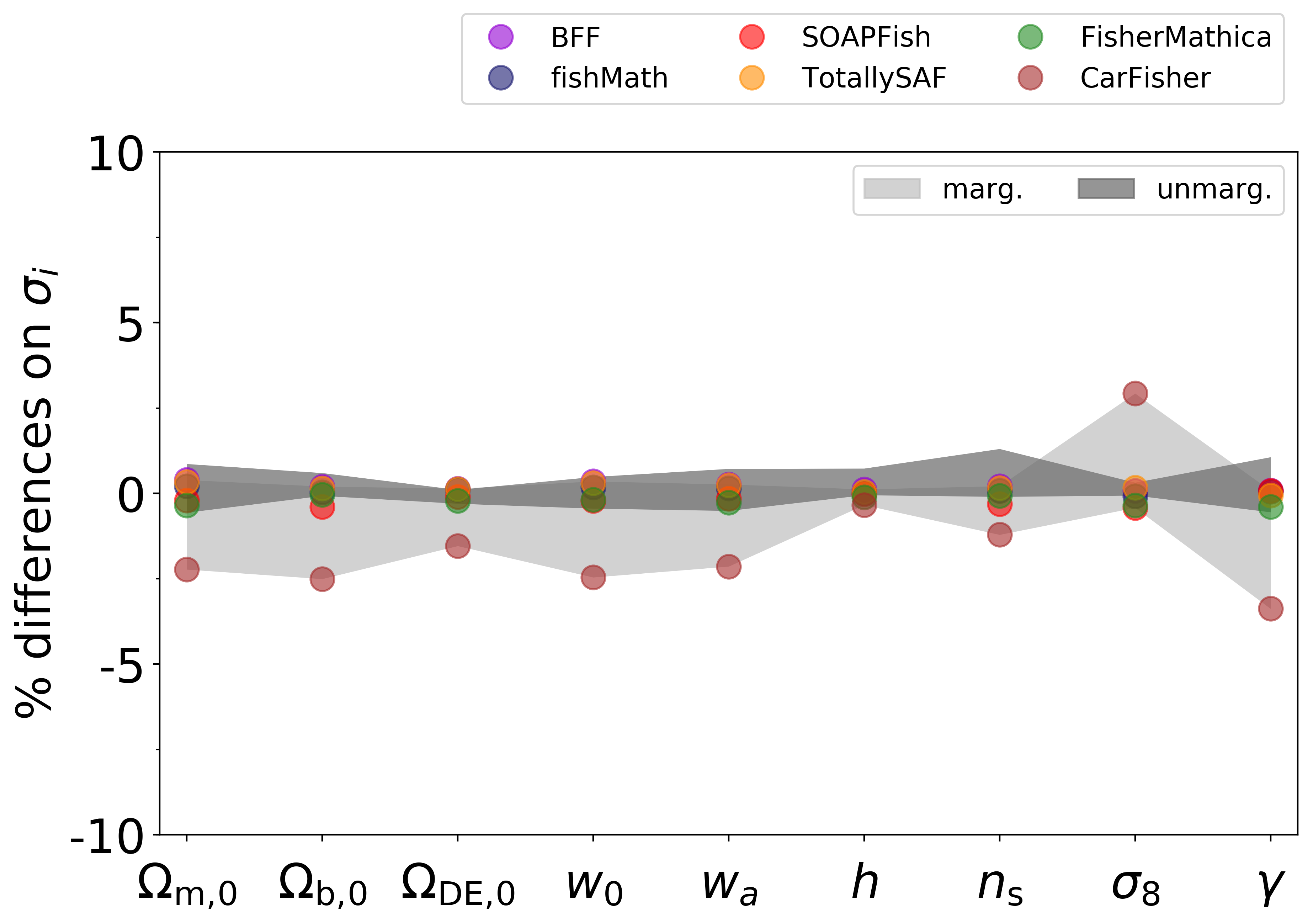}
\caption{Comparison for the projected parameters. Comparison of the errors for the fully-marginalised final parameters among six GC codes, for the non-flat $\Lambda\text{CDM} + w_0 + w_a + \gamma$ case, using the non-linear recipe up to $k_\text{max}=0.30 \, h\,\textrm{Mpc}^{-1}$, corresponding to the optimistic case described in the text.}
\label{fig:resu-GC-proj3}
\end{figure}

Finally in \autoref{fig:resu-GC-proj3} we also show the relative difference of the full marginalised projected parameters among six GC codes, for the non-flat $\Lambda\text{CDM}+ w_0 + w_a + \gamma$ case, using the non-linear recipe up to $k_\text{max}=0.30 \, h\,\textrm{Mpc}^{-1}$, corresponding to the optimistic setting. The agreement between the outcomes are at most of the order of a few percent.

\subsection{Weak-lensing code comparison}\label{sec:ccwl}
We now describe the code comparison and validation of the cosmic-shear weak-lensing Fisher-matrix codes available, from the implementation of the recipe to the final results, showing how well the output from the different codes now match. As for the GC case, we began the comparison process from codes whose results differed in both amplitude and orientation of the expected probability contours, with errors as large as 100$\%$, and reduced them to be at least within $10\%$, as required and discussed in \autoref{sec:required-accuracy}.

\subsubsection{Implementation of the weak-lensing recipe}
We implement the Fisher matrix definition in \autoref{eq: fmwl}, with the \Euclid specifications described in \autoref{tab:WL-specifications}. 

\paragraph{Binning}
The sum over $\ell$ modes in the implementation of \autoref{eq: fmwl} is performed within the range $\ell_{\rm min}=10$ and $\ell_{\rm max} = 1500 (5000)$ for the pessimistic (optimistic) case, as discussed in \autoref{errors_shear}. We use $N_{\ell} = 100$ logarithmically equi-spaced bins. All the quantities entering \autoref{eq: fmwl} are meant to be evaluated at the linear centre of the bin. We compute the cosmic shear power spectrum and its derivatives in each multipole bin $k$ and define $\lambda = \log{\ell}$, such that
\begin{equation}
\log_{10}\ell_k =(\lambda_{k + 1} + \lambda_{k})/2,
\label{eq: lkval}
\end{equation}
with $\lambda_k = \lambda_{\rm min} + (k - 1) \Delta \lambda$, and $\Delta \lambda = (\lambda_{\rm max} - \lambda_{\rm min})/N_\ell$. The width of the bin is  
\begin{equation}
\Delta \ell = 10^{\lambda_{k + 1}} - 10^{\lambda_k},
\label{eq: deltaelle}
\end{equation}
which is therefore different depending on the $\ell$ value.
                                      
\paragraph{Derivatives}
A fundamental aspect of each Fisher forecasting code is the computation of the derivatives, which has been discussed and introduced in \Cref{eq: dcijdpmu,eq: dpdpmu}. A typical procedure to compute derivatives suggests that one should first compute the integral of the cosmic shear power spectrum, and then calculate the derivatives by numerical differentiation using, e.g.\ a finite differencing technique (i.e. a {\it numerical} implementation). However, since the integration limits do not depend on the parameters with respect to which one is deriving, the derivative operator can be taken inside the integral, as done in \autoref{eq: dcijdpmu}; in this case we can skip numerical differentiation to a large extent, since most of the derivatives can be performed analytically (what we call a `{\it semi-analytical}' implementation). Mathematically, the two approaches are equivalent, but in practice this is not always the case because numerical differentiation can introduce a certain degree of noise and some unphysical features. These effects can be reduced to an acceptable level by carefully setting the specifics of any particular implementation. The codes used here adopt different implementation choices, which are then compared. 

We detail below the derivatives needed to implement the semianalytical approach. 
\autoref{eq: dcijdpmu} shows that there are two sets of derivatives one must compute to implement the 
semi-analytical recipe, namely derivatives of the kernel and of the matter and IA power spectra.  
We address them separately in the following discussion.

\paragraph{Semi-analytical approach: derivatives with respect to background quantities}
Derivatives of the kernel and of the matter power spectrum will include derivatives of the Hubble rate and the comoving distance. 

With regard to the Hubble rate, for the non-flat case, the latter can be written in terms of $E(z)$ as in 
 \autoref{eq:Hz-general}, whose expression as a function of the cosmological parameters is given in \autoref{eq:Ez-w0wa}; we then obtain\footnote{We give the derivative of $\ln{E(z)}$ instead of $E(z)$, since these will be more useful in the following.} 
\begin{align}
\frac{\partial \ln{E(z)}}{\partial \Omega_{{\rm m},0}} &= 
\frac{1}{2} \frac{(1 + z)^3 - (1 + z)^2}{E^2(z)},\label{eq: dlnEdOm}\\
\frac{\partial \ln{E(z)}}{\partial \Omega_{\rm{DE},0}} &= 
\frac{1}{2} \frac{(1 + z)^{3(1 + w_0 + w_{ a})} {\rm e}^{-3 w_{a} z/(1 + z)} - (1 + z)^2}{E^2(z)},\label{eq: dlnEdOx}\\
\frac{\partial \ln{E(z)}}{\partial w_0} &= 
\frac{3}{2} \frac{\Omega_{\Lambda,0} (1 + z)^{3(1 + w_0 + w_{ a})} {\rm e}^{-3 w_{ a} z/(1 + z)} \ln{(1+ z)}}{E^2(z)},\label{eq: dlnEdwz}\\
\frac{\partial \ln{E(z)}}{\partial w_{ a}} &= 
\frac{3}{2} \frac{\Omega_{\Lambda,0} (1 + z)^{3(1 + w_0 + w_{ a})} {\rm e}^{-3 w_{a} z/(1 + z)} {\cal{E}}(z)}{E^2(z)},\label{eq: dlnEdwa}
\end{align} 
where ${\cal{E}}(z) \equiv \ln{(1+ z)} - z/(1 + z)$. In the flat case we enforce the constraint $\Omega_{\rm{DE}} = 1 - \Omega_{{\rm m}}$, at all redshifts, so that $\Omega_{{\rm K}} = 0$. Hence, the $(1 + z)^2$ term in \autoref{eq:Ez-w0wa} drops, while the derivative with respect to $\Omega_{{\rm m},0}$ has a further contribution coming from expressing $\Omega_{\rm{DE}, 0} = 1 - \Omega_{{\rm m}, 0}$. As a consequence, the derivatives $\partial{\ln{E(z)}}/\partial{\Omega_{\rm{DE},0}}$ do not appear and the derivative with respect to $\Omega_{{\rm m},0}$ becomes
\begin{equation}
\frac{\partial \ln{E(z)}}{\partial \Omega_{{\rm m},0}} = 
\frac{1}{2} \frac{(1 + z)^3 - (1 + z)^{3(1 + w_0 + w_{ a})} {\rm e}^{-3 w_{ a} z/(1 + z)}}{E^2(z)}.
\label{eq: dlnEdOmFlat}
\end{equation}

With regard to the comoving distance, we give below the logarithmic derivatives of interest. For the general non-flat case, these are 
\begin{align}
\frac{\partial \ln{\tilde{r}(z)}}{\partial \Omega_{{\rm m},0}} &= - \frac{1}{2 \tilde{r}(z)}
\int_{0}^{z}\!\!{\rm d}z^\prime\,{\frac{(1 + z^{\prime})^3 - (1 + z^{\prime})^2}{E^3(z^{\prime})}},\label{eq: dchidOm}\\
\frac{\partial \ln{\tilde{r}(z)}}{\partial \Omega_{\rm{DE},0}} &= - \frac{1}{2 \tilde{r}(z)}
\int_{0}^{z}\!\!{\rm d}z^\prime\,{\frac{(1 + z^{\prime})^{3(1 + w_0 + w_{ a})} {\rm e}^{-3 w_{ a} z^{\prime}/(1 + z^{\prime})} - (1 + z^{\prime})^2}{E^3(z^{\prime})}},\label{eq: dchidOx}\\
\frac{\partial \ln{\tilde{r}(z)}}{\partial w_0} &= - \frac{3}{2 \tilde{r}(z)}
\int_{0}^{z}\!\!{\rm d}z^\prime\,{\frac{\Omega_{\rm{DE},0} (1 + z^{\prime})^{3(1 + w_0 + w_{ a})} {\rm e}^{-3 w_{ a} z^{\prime}/(1 + z^{\prime})} \ln{(1+ z^{\prime})}}{E^3(z^{\prime})}},\label{eq: dchidwz}\\
\frac{\partial \ln{\tilde{r}(z)}}{\partial w_{ a}} &= - \frac{3}{2 \tilde{r}(z)}
\int_{0}^{z}\!\!{\rm d}z^\prime\,{\frac{\Omega_{\rm{DE},0} (1 + z^{\prime})^{3(1 + w_0 + w_{ a})} {\rm e}^{-3 w_{ a} z^{\prime}/(1 + z^{\prime})} {\cal{E}}(z^{\prime})}{E^3(z^{\prime})}},\label{eq: dchidwa}\\
\frac{\partial \ln{\tilde{r}(z)}}{\partial h} &= - \frac{1}{h},\label{eq: dchidhMath}
\end{align}
where $\tilde{r}(z) \equiv r(z)/(c/H_0)$. Again, when the flatness constraint is enforced, the derivative with respect to $\Omega_{{\rm m},0}$ includes an extra term and becomes
\begin{equation}
\frac{\partial \ln{\tilde{r}(z)}}{\partial \Omega_{{\rm m},0}} = - \frac{1}{2 \tilde{r}(z)}
\int_{0}^{z}\!\!{\rm d}z^\prime\,{\frac{(1 + z^{\prime})^3 - (1 + z^{\prime})^{3(1 + w_0 + w_{ a})} {\rm e}^{-3 w_{ a} z^{\prime}/(1 + z^{\prime})}}{E^3(z^{\prime})}},\label{eq: dchidOmFlat} 
\end{equation}
while all the others remain the same, provided $\Omega_{\rm{DE},0} = 1 - \Omega_{{\rm m},0}$. 

\paragraph{Semi-analytical approach: kernel functions}
The derivatives of the kernel functions with respect to the parameters $p_\mu$, needed for \Cref{eq: dcijdpmu,eq: dpdpmu} are conveniently expressed as derivatives of the logarithm of the kernel functions:
\begin{equation}
\frac{\partial K_{ij}^{a}(z)}{\partial p_{\mu}} = K_{ij}^{a}(z) 
\frac{\partial \ln{K_{ij}^{a}(z)}}{\partial p_{\mu}} \,\, ,
\label{eq: dkijggfull}
\end{equation}
where we recall that the index $a \in \{ \gamma \gamma, \rm{I} \gamma, \rm{II} \}$. 
Let us consider first the lensing kernel ($a = \gamma \gamma$). We then obtain
\begin{align}
\frac{\partial \ln{K_{ij}^{\gamma \gamma}(z)}}{\partial \Omega_{{\rm m},0}} &= \frac{2}{\Omega_{{\rm m},0}} - \frac{\partial \ln{E(z)}}{\partial \Omega_{{\rm m},0}} + \frac{\partial \ln{\widetilde{W}_{i}^{\gamma}(z)}}{\partial \Omega_{{\rm m},0}} + \frac{\partial \ln{\widetilde{W}_{j}^{\gamma}(z)}}{\partial \Omega_{{\rm m},0}},\label{eq: dkijggdOm}\\
\frac{\partial \ln{K_{ij}^{\gamma \gamma}(z)}}{\partial \Omega_{\rm{DE},0}} &= - \frac{\partial \ln{E(z)}}{\partial \Omega_{\rm{DE},0}} + \frac{\partial \ln{\widetilde{W}_{i}^{\gamma}(z)}}{\partial \Omega_{\rm{DE},0}} + \frac{\partial \ln{\widetilde{W}_{j}^{\gamma}(z)}}{\partial \Omega_{\rm{DE},0}},\label{eq: dkijggdOx}\\
\frac{\partial \ln{K_{ij}^{\gamma \gamma}(z)}}{\partial w_0} &= - \frac{\partial \ln{E(z)}}{\partial w_0} + \frac{\partial \ln{\widetilde{W}_{i}^{\gamma}(z)}}{\partial w_0} + \frac{\partial \ln{\widetilde{W}_{j}^{\gamma}(z)}}{\partial w_0},\label{eq: dkijggdwz}\\
\frac{\partial \ln{K_{ij}^{\gamma \gamma}(z)}}{\partial w_{ a}} &= - \frac{\partial \ln{E(z)}}{\partial w_{ a}} + \frac{\partial \ln{\widetilde{W}_{i}^{\gamma}(z)}}{\partial w_{ a}} + \frac{\partial \ln{\widetilde{W}_{j}^{\gamma}(z)}}{\partial w_{ a}},\label{eq: dkijggdwa}\\
\frac{\partial \ln{K_{ij}^{\gamma \gamma}(z)}}{\partial h} &= \frac{3}{h},\label{eq: dkijggdh}
\end{align}
with
\begin{align}
\frac{\partial \ln{\widetilde{W}_{i}^{\gamma}(z)}}{\partial p_\mu}=&\left[\int_{z}^{z_{\rm max}}\!\!{\rm d}z^\prime\,{n_i(z^{\prime})\frac{\tilde{r}(z^{\prime})-\tilde{r}(z)}{\tilde{r}(z^{\prime})}} \right]^{-1} 
\int_{z}^{z_{\rm max}}\!\!{\rm d}z^\prime\,{n_i(z^{\prime}) 
\left [ \frac{\partial \ln{\tilde{r}(z^{\prime})}}{\partial p_{\mu}} - \frac{\partial \ln{\tilde{r}(z)}}{\partial p_{\mu}} \right ]
\frac{\tilde{r}(z)}{\tilde{r}(z^{\prime})}} \,\,\, .\label{eq: dlnwdpmu}
\end{align}
All the relevant derivatives of $\ln{E(z)}$ needed for such kernel derivative have been given above.

In order to compute the derivatives of the shear-IA kernel (i.e.\ $a = {\rm I}\gamma$), it is convenient to first split it as
\begin{equation}
K_{ij}^{{\rm I}\gamma}(z) = K_{ij}^{{\rm I}\gamma}(z; i, j) + K_{ij}^{{\rm I}\gamma}(z; j, i)
\end{equation}
with

\begin{align}
K_{ij}^{{\rm I}\gamma}(z; i, j) &=\frac{3}{2} \Omega_{{\rm m},0} (1 + z) \left ( \frac{H_0}{c} \right )^3 
\frac{n_i(z) \widetilde{{W}}_{j}^{\gamma}(z)}{{\tilde{r}(z)}},\\
K_{ij}^{{\rm I}\gamma}(z; j, i) &= \frac{3}{2} \Omega_{{\rm m},0} (1 + z) \left ( \frac{H_0}{c} \right )^3 
\frac{n_j(z) \widetilde{{W}}_{i}^{\gamma}(z)}{{\tilde{r}(z)}}.
\end{align}
The non-null derivatives can then be written as
\begin{align}
\frac{\partial K_{ij}^{{\rm I}\gamma}(z)}{\partial \Omega_{{\rm m},0}} &=  K_{ij}^{{\rm I}\gamma}(z; i, j) \left [ \frac{1}{\Omega_{{\rm m},0}} - \frac{\partial \ln{\tilde{r}}(z)}{\partial \Omega_{{\rm m},0}} + 
  \frac{\partial \ln{\widetilde{{W}}_{j}^{\gamma}(z)}}{\partial \Omega_{{\rm m},0}} \right ] 
+ K_{ij}^{{\rm I}\gamma}(z; j, i) \left [ \frac{1}{\Omega_{{\rm m},0}} - \frac{\partial \ln{\tilde{r}}(z)}{\partial \Omega_{{\rm m},0}} + \frac{\partial \ln{\widetilde{{W}}_{i}^{\gamma}(z)}}{\partial \Omega_{{\rm m},0}} \right ],\label{eq: dkijgidOm}\\
\frac{\partial K_{ij}^{{\rm I}\gamma}(z)}{\partial \Omega_{\rm{DE},0}} &=  K_{ij}^{{\rm I}\gamma}(z; i, j) \left [ -\frac{\partial \ln{\tilde{r}}(z)}{\partial \Omega_{\rm{DE},0}} + \frac{\partial \ln{\widetilde{{W}}_{j}^{\gamma}(z)}}{\partial \Omega_{\rm{DE},0}} \right ] 
+ K_{ij}^{{\rm I}\gamma}(z; j, i) \left [ - \frac{\partial \ln{\tilde{r}}(z)}{\partial \Omega_{\rm{DE},0}} + 
\frac{\partial \ln{\widetilde{{W}}_{i}^{\gamma}(z)}}{\partial \Omega_{\rm{DE},0}} \right ],\label{eq: dkijgidOx}\\
\frac{\partial K_{ij}^{{\rm I}\gamma}(z)}{\partial w_0} &=  K_{ij}^{{\rm I}\gamma}(z; i, j) \left [ - \frac{\partial \ln{\tilde{r}}(z)}{\partial w_0} + \frac{\partial \ln{\widetilde{{W}}_{j}^{\gamma}(z)}}{\partial w_0} \right ] 
+ K_{ij}^{{\rm I}\gamma}(z; j, i) \left [ - \frac{\partial \ln{\tilde{r}}(z)}{\partial w_0} + \frac{\partial \ln{\widetilde{{W}}_{i}^{\gamma}(z)}}{\partial w_0} \right ],\label{eq: dkijgidwz}\\
\frac{\partial K_{ij}^{{\rm I}\gamma}(z)}{\partial w_{ a}} & =  K_{ij}^{{\rm I}\gamma}(z; i, j) \left [ - \frac{\partial \ln{\tilde{r}}(z)}{\partial w_{ a}} + \frac{\partial \ln{\widetilde{{W}}_{j}^{\gamma}(z)}}{\partial w_{ a}} \right ] 
+ K_{ij}^{{\rm I}\gamma}(z; j, i) \left [ - \frac{\partial \ln{\tilde{r}}(z)}{\partial w_{ a}} + \frac{\partial \ln{\widetilde{{W}}_{i}^{\gamma}(z)}}{\partial w_{ a}} \right ],\label{eq: dkijgidwa}\\
\frac{\partial K_{ij}^{{\rm I}\gamma}(z)}{\partial h} & =  3\frac{K_{ij}^{{\rm I}\gamma}(z)}{h}.\label{eq: dkijgidh}
\end{align}
Finally, the derivatives of the IA-IA kernel ($a ={\rm II}$) are 

\begin{equation}
\frac{\partial K_{ij}^{\rm II}(z)}{\partial p_{\mu}} = K_{ij}^{\rm II}(z) \ \times \ \left \{
\begin{array}{lr}
\displaystyle{\partial_\mu \ln E(z)  - 2\partial_\mu\ln{\tilde{r}(z)}},
& p_{\mu} \neq h, \\
 &  \\
\displaystyle{3/h}, & p_{\mu} = h, \\
\end{array}
\right . \ 
\label{eq: dkijiidpmu}
\end{equation}
where the first row holds for derivatives with respect to parameters $\Omega_{{\rm m},0}$, $\Omega_{\rm{DE},0}$, $w_0$, and $w_{ a}$, and $\partial_\mu$ is a short-hand notation for $\partial/\partial p_\mu$.

\paragraph{Semi-analytical approach: matter and IA power spectrum derivatives}
As from \autoref{eq: dpdpmu}, the derivatives of the matter power spectrum $\partial P_{A}(k, z)/\partial p_{\mu}$ (with $A \in \{\delta \delta, \delta {\rm I}, {\rm II}\}$) contain two terms, the latter being present only for background parameters $\{\Omega_{{\rm m},0}, \Omega_{\rm{DE},0}, w_0, w_{ a}, h\}$. The second factor in the second term is computed for fixed values of the model parameters, while the first factor in the second term uses the expressions derived in \Crefrange{eq: dlnEdOm}{eq: dlnEdOmFlat} for the derivative of the comoving distance. On the other hand, the first term can only be evaluated numerically, and to do this we rely on a procedure that proves to be more stable than the usual finite differences. Therefore, we adopt the technique proposed by \cite{Camera:2016owj} (see their Appendix~A). For each $k$-$z$ pair, we evaluate $P_{\delta \delta}(k, z)$ for different values of the parameter of interest $p_{\mu}$. We choose values that span the $\pm10\%$ range around the fiducial value of the parameter. If the neighbourhood is small enough, we can approximate the dependence of $P_{\delta \delta}$ on $p_{\mu}$ as linear (following the definition of derivative). Then we fitted a line through these values and take its slope as our final estimate of the $\partial{P_{\delta \delta}(k, z)}/\partial{p_{\mu}}$ derivative. 

Once the matter power spectrum derivatives have been computed, the IA related ones can be analytically 
related to them. Again, it is convenient to move to logarithmic derivatives so that we obtain 
\begin{align}
\frac{\partial P_{\delta{\rm I}}(k, z)}{\partial p_{\mu}} &= P_{\delta{\rm I}}(k, z) \frac{\partial \ln{P_{\delta{\rm I}}(k, z)}}{\partial p_{\mu}},\\
\frac{\partial P_{\rm II}(k, z)}{\partial p_{\mu}} &= P_{\rm II}(k, z) \frac{\partial \ln{P_{\rm II}(k, z)}}{\partial p_{\mu}},
\end{align}
and, from \Cref{eq: pdienla,eq: piienla}, we find the following expression for the derivative with respect to the cosmological parameters $\{\Omega_{{\rm m},0}, \Omega_{\rm{DE},0}, \Omega_{{\rm b},0}, w_0, w_{ a}, h, n_{\rm s}, \sigma_8\}$:
\begin{align}
\frac{\partial \ln{P_{\delta{\rm I}}(k, z)}}{\partial p_{\mu}} &= \frac{\partial \ln{P_{\delta \delta}(k, z)}}{\partial p_{\mu}} 
- \frac{\partial \ln{D(z)}}{\partial p_{\mu}} + \frac{\delta^{\rm K}_{\mu, \Omega_{{\rm m},0}}}{\Omega_{{\rm m},0}};
\label{eq: dlnpdi}\\
\frac{\partial \ln{P_{\rm II}(k, z)}}{\partial p_{\mu}} &= \frac{\partial \ln{P_{\delta \delta}(k, z)}}{\partial p_{\mu}} 
- 2 \frac{\partial \ln{D(z)}}{\partial p_{\mu}} +2 \frac{\delta^{\rm K}_{\mu, \Omega_{{\rm m},0}}}{\Omega_{{\rm m},0}}.
\label{eq: dlnpii}
\end{align}

We note that the second term in \Cref{eq: dlnpdi,eq: dlnpii} is only present for $\Omega_{{\rm m},0}$, $\Omega_{{\rm DE},0}$, $w_0$, $w_{ a}$, and $\gamma$, since the growth factor $D(z)$ depends on these parameters only. 

The derivatives with respect to the IA parameters read
\begin{equation}
\frac{\partial \ln{P_{\delta{\rm I}}(k, z)}}{\partial p_{\mu}} = \left \{
\begin{array}{ll}
\displaystyle{1/{\cal{A}}_{\rm IA}}, & \displaystyle{p_{\mu} = {\cal{A}}_{\rm IA}}, \\
 & \\
\displaystyle{\ln{(1 + z)}}, & \displaystyle{p_{\mu} = \eta_{\rm IA}}, \\
 & \\
\displaystyle{\ln{[\langle L \rangle(z)/L_{\star}(z)]}}, & \displaystyle{p_{\mu} = \beta_{\rm IA}},
\end{array}
\right.
\label{eq: dlnpdidpmu}
\end{equation} 
for the $\delta{\rm I}$ term, and 
\begin{equation}
\frac{\partial \ln{P_{\rm II}(k, z)}}{\partial p_{\mu}} = \left \{
\begin{array}{ll}
\displaystyle{2/{\cal{A}}_{\rm IA}} & \displaystyle{p_{\mu} = {\cal{A}}_{\rm IA}}, \\
 & \\
\displaystyle{2 \ln{(1 + z)}}, & \displaystyle{p_{\mu} = \eta_{\rm IA}}, \\
 & \\
\displaystyle{2 \ln{[\langle L \rangle(z)/L_{\star}(z)]}}, & \displaystyle{p_{\mu} = \beta_{\rm IA}},
\end{array}
\right. 
\label{eq: dlnpiidpmu}
\end{equation} 
for derivatives of the IA-IA power spectrum.

\begin{table*}[h]
\centering
\caption{Summary of a selection of the cases investigated for the weak-lensing code comparison, ordered according to the number of parameters involved. In non-flat cases, $\Omega_{{\rm DE},0}$ is included as a free cosmological parameter (or equivalently $\Omega_{{\rm K},0}$). The quantity $\ell_{\rm max}$ is the maximum multipole included in the analysis. MG corresponds to the modified gravity $\gamma$ parameter (see Eq.~\ref{eq:ggrowth}), changing the growth of structure. IA refers to intrinsic alignment parameters (discussed in  \autoref{par:IA}).}
\label{tab: cases}
\begin{tabular}{cclccccc}
\hline
Case & No. of parameters & \multicolumn{4}{c}{Parameters} & $\ell_{\rm max}$ \\
\cline{3-6}
& & Flat & Non-flat & MG & IA & \\
\hline\hline
1 & 5 & $\Omega_{{\rm m},0}\,\Omega_{{\rm b},0}\,h\,n_{\rm s}\,\sigma_8$ & --- & --- & --- & 1500 or 5000  \\

2 & 6 & $\Omega_{{\rm m},0}\,\Omega_{{\rm b},0}\,h\,n_{\rm s}\,\sigma_8$ & $\Omega_{{\rm DE},0}$ & - & - & 1500 or 5000  \\

3 & 7 & $\Omega_{{\rm m},0}\,\Omega_{{\rm b},0}\,w_0\,w_{ a}\,h\,n_{\rm s}\,\sigma_8$ & --- & --- & --- & 1500 or 5000 \\

4 & 8 & $\Omega_{{\rm m},0}\,\Omega_{{\rm b},0}\,w_0\,w_{a}\,h\,n_{\rm s}\,\sigma_8$ & $\Omega_{\rm{DE},0}$ & --- & --- & 1500 or 5000  \\

5 & 8 & $\Omega_{{\rm m},0}\,\Omega_{{\rm b},0}\,w_0\,w_{a}\,h\,n_{\rm s}\,\sigma_8$ & --- & $\gamma$ & --- & 1500 or 5000 \\

6 & 9 & $\Omega_{{\rm m},0}\,\Omega_{{\rm b},0}\,w_0\,w_{a}\,h\,n_{\rm s}\,\sigma_8$ & $\Omega_{\rm{DE},0}$ & $\gamma$ & - & 1500 or 5000 \\

7 & 10 & $\Omega_{{\rm m},0}\,\Omega_{{\rm b},0}\,w_0\,w_{ a}\,h\,n_{\rm s}\,\sigma_8$ & --- & --- & $ {\cal{A}}_{\rm IA} \, \eta_{\rm IA} \, \beta_{\rm IA}$ & 1500 or 5000 \\

8 & 11 & $\Omega_{{\rm m},0}\,\Omega_{{\rm b},0}\,w_0\,w_{ a}\,h\,n_{\rm s}\,\sigma_8$ & $\Omega_{\rm{DE},0}$ & --- & $ {\cal{A}}_{\rm IA} \, \eta_{\rm IA} \, \beta_{\rm IA}$ & 1500 or 5000 \\

9 & 11 & $\Omega_{{\rm m},0}\,\Omega_{{\rm b},0}\,w_0\,w_{ a}\,h\,n_{\rm s}\,\sigma_8$ & --- & $\gamma$ & $ {\cal{A}}_{\rm IA} \, \eta_{\rm IA} \, \beta_{\rm IA}$ & 1500 or 5000 \\

10 & 12 & $\Omega_{{\rm m},0}\,\Omega_{{\rm b},0}\,w_0\,w_{ a}\,h\,n_{\rm s}\,\sigma_8$ & $\Omega_{\rm{DE},0}$ & $\gamma$ & $ {\cal{A}}_{\rm IA} \, \eta_{\rm IA} \, \beta_{\rm IA}$ & 1500 or 5000 \\
\hline
\end{tabular}
\end{table*}

\subsubsection{Settings definition}
For the purpose of validating different codes, we have compared their output in many different scenarios. A selection of the cases tested is in \autoref{tab: cases}. We consider $\Lambda$CDM and the extensions described in \autoref{Sec:Cosmology} with neutrino mass $m_\nu=0.06\,\mathrm{eV}$. We implement the `TakaBird' \citep{Bird2012,Takahashi:2012em} linear to non-linear mapping, as discussed in \autoref{sec:wlnl}. 
We proceeded by steps of increasing complexity, starting with the simplest assumptions and then adding features after validation. We first compared spatially flat models by setting $\Omega_{{\rm DE},0} = 1 - \Omega_{{\rm m},0}$ (hence $\Omega_{{\rm K},0} = 0$), neglecting contribution from IA in \autoref{eq: defcijshort} and keeping only standard cosmological parameters (plus neutrino mass). Then we tested extensions to non-flat models, as well as  including modelling of IA systematics (eNLA model described in  \autoref{par:IA}), and implemention of  non-standard parameters such as $w$, $w_a$, $\gamma$ discussed in \autoref{eq:ggrowth}. 

The large number of cases offered the possibility to investigate how theoretical uncertainties impact the cosmic shear forecasts and how they degrade as the number of nuisance parameters increases. 
\begin{figure}[h]
\center 
\includegraphics[width=0.47\textwidth]{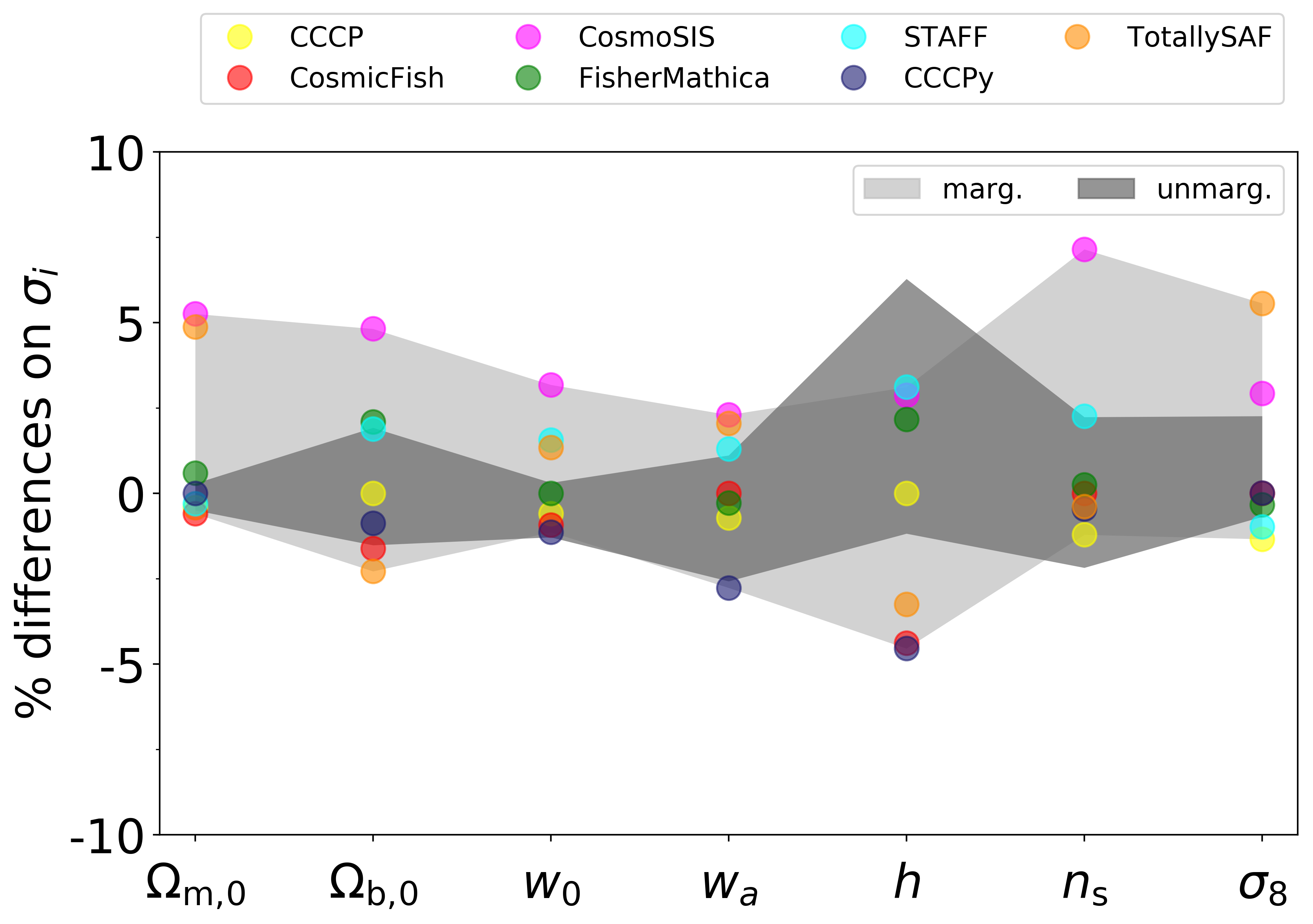}
	\includegraphics[width=0.47\textwidth]{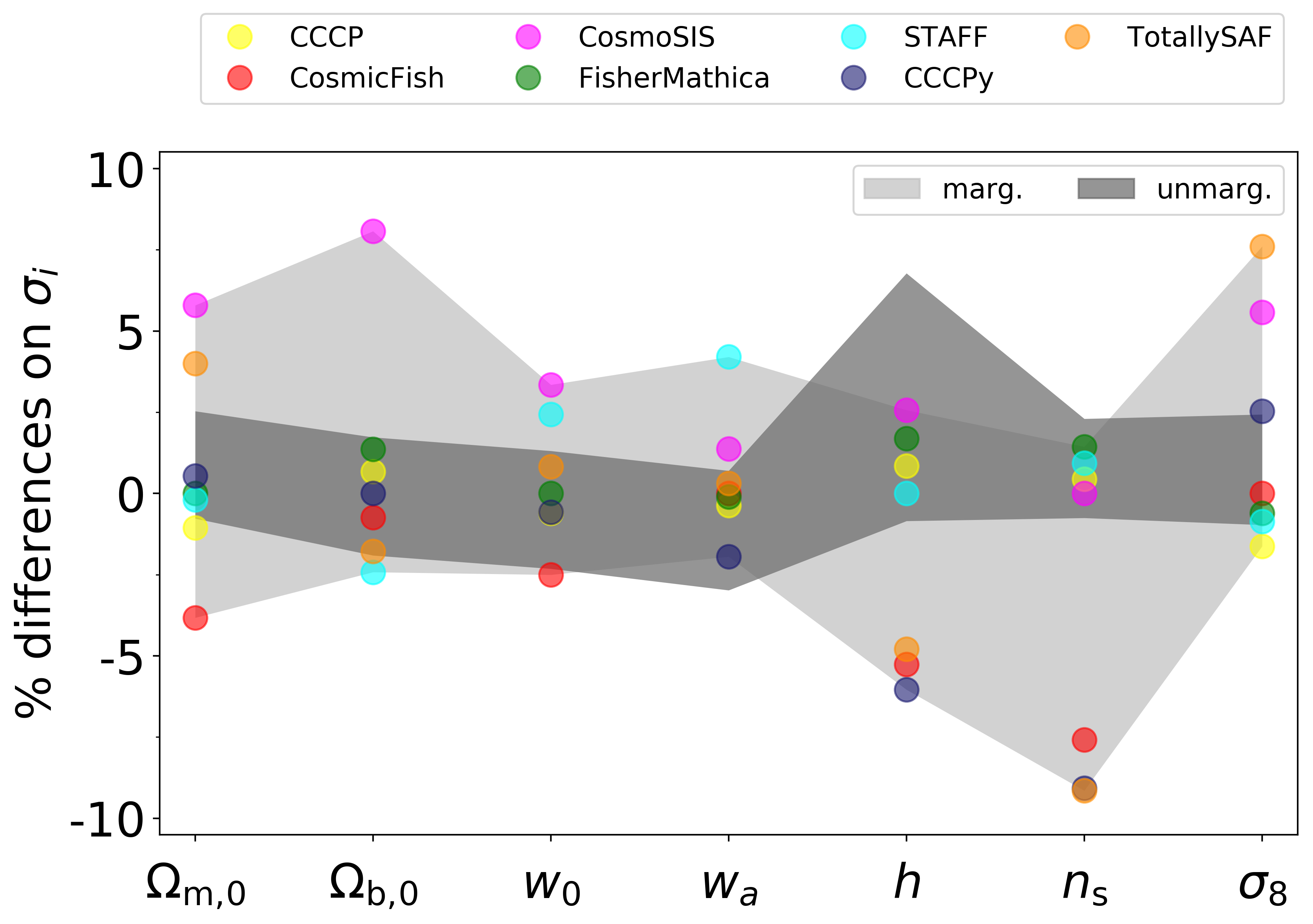}
	\includegraphics[width=0.47\textwidth]{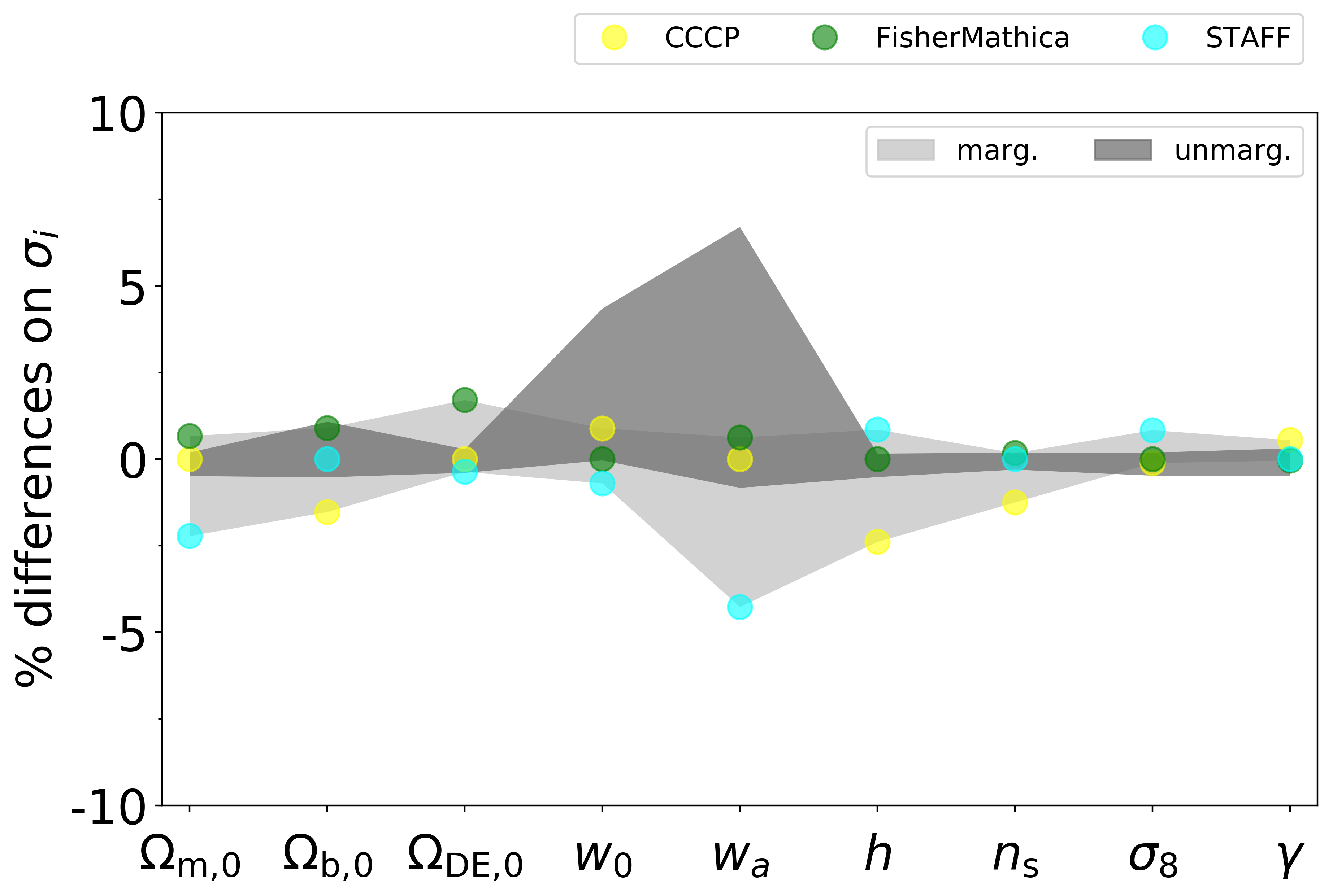}
	\includegraphics[width=0.47\textwidth]{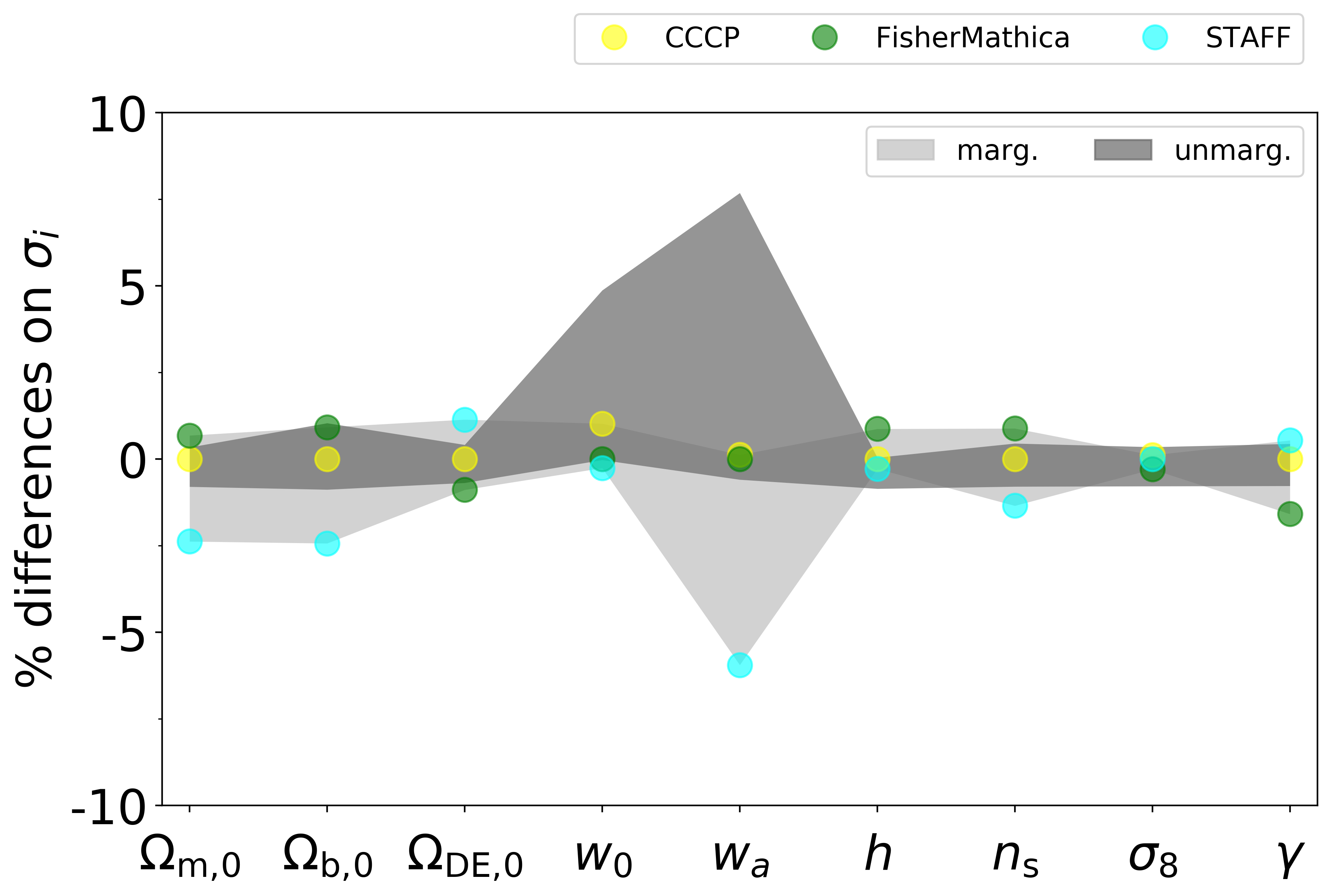}
\caption{Comparison of the errors on the marginalised (light grey) and unmarginalised (dark grey) cosmological parameters. Top panels: we compare seven WL codes for the flat $w_0$, $w_a$ cosmology, using the `TakaBird' recipe for the non-linear power spectrum and a maximum multipole of $\ell_{\rm max}=1500$ (pessimistic cut, top left panel) or $\ell_{\rm max}=5000$ (optimistic cut, top right panel). This is  case 3 in \autoref{tab: cases}. Bottom panels~: we compare three WL codes (the ones available for a non-flat geometry that include $\gamma$), for a $\gamma$, $w_0$, $w_a$ cosmology, using the `TakaBird' non-linear recipe for the power spectrum and a maximum multipole of $\ell_{\rm max}=1500$ (pessimistic cut, bottom left panel) and a maximum multipole of $\ell_{\rm max}=5000$ (optimistic cut, bottom right panel). This is case 6 in \autoref{tab: cases}.}\label{fig:resu-WL-1}
\end{figure}

\subsubsection{Results of the code comparison}

The codes used for the code comparison of the cosmic shear weak lensing are described in \autoref{sec:allcodes}, and include, in particular, {seven codes:} \texttt{CCCP}, {\texttt{CCCPy}}, \texttt{CosmicFish},  \texttt{CosmoSIS}, \texttt{FisherMatica},  \texttt{STAFF}, {\texttt{TotallySAF}}. {One of them is publicly available, three of them are planned to be made available, and the features of all codes are summarised in \autoref{tab:forecastcodes}.}
Computing the Fisher matrix for the different cases in \autoref{tab: cases} is conceptually simple using all the relevant formulae in \Cref{Sec:Cosmology,Sec:Fisher}, but there many  steps to be implemented, and also many potentials errors that can hamper the implementation. Moreover, the critical role played by the numerical derivatives led us to compare the results from codes implementing different recipes. Among the codes described in \autoref{sec:allcodes}, for example, it is worth stressing that \texttt{CCCP}, {\texttt{CCCPy}}, and \texttt{STAFF} adopt the semi-analytical approach, while \texttt{CosmicFish}, \texttt{CosmoSIS}, and {\texttt{TotallySAF}} rely on fully numerical derivatives; \texttt{FisherMatica} follows a mixed approach (the derivatives of the shear power spectrum are computed by first differentiating and then integrating -- as for the semi-analytical case -- but the differentiation of the argument of the integrals is performed numerically). A further important difference among the {seven} codes concerns the way in which derivatives of the matter power spectrum are implemented. While \texttt{CosmicFish} and \texttt{CosmoSIS} compute them internally, the other {five} codes take them as external input (using the same input). As a consequence, we expect better agreement among these {five} codes. The external input for \texttt{CCCP}, {\texttt{CCCPY}}, \texttt{FisherMatica}, \texttt{STAFF}, {and \texttt{TotallySAF}} has been produced using \texttt{CAMB} with the same settings for \texttt{CosmicFish} and \texttt{CosmoSIS}, in order to be consistent and validated as described in \autoref{sec:xccases}. Any discrepancy among them should therefore not originate from the inputs, but rather from how the derivatives are computed.

In \autoref{fig:resu-WL-1} we show how the output of the {seven} codes used for cosmic-shear weak lensing compare among each other, for two representative cases from \autoref{tab: cases}. In particular, we plot the percentage difference on the error on each parameter, as found by each code, when marginalising (light grey band) or not  marginalising (dark grey contours) on the other parameters. Dots correspond to the different codes used in the comparison.   What matters is how close they are to each other, for each parameter, i.e. the hight of the grey bands for each parameter. In all cases (including the ones tested but not shown in the plots for brevity) the percentage difference on errors is within $10 \%$, as required. {We note that we plot the relative difference with respect to the median, as explained in \autoref{sssec:ccgc}.}

\subsection{Probe combination code comparison}\label{sec:ccxc}

In this section we compare the codes used to obtain results for the combination of \Euclid's probes, including galaxy clustering coming from the spectroscopic survey (GC$_{\rm s}$), 
weak lensing (WL) and photometric galaxy clustering (GC$_{\rm ph}$). For the two latter probes, we also include cross-correlation terms (XC$^{(\rm GC_{\rm ph},WL)}$).
When cross terms are neglected the Fisher matrices are added. However in the following we discuss the specific implementation of the recipe for cross-correlation terms discussed in \autoref{sec:recipexc} and the validation of the {four} codes that took part in the comparison. 

\subsubsection{Implementation of the probe combination recipe}                               

The codes used in the comparison, \texttt{CosmicFish}, \texttt{CosmoSIS}, \texttt{STAFF}, {and \texttt{TotallySAF}}, consider different sets of cosmological primary parameters. 
\texttt{CosmoSIS}, \texttt{STAFF},  {and \texttt{TotallySAF}} are able to work directly with the parameter basis of \autoref{eq:9parameters}, while \texttt{CosmicFish}
uses \texttt{CAMB}'s physical parameters $\vartheta=\{\Omega_{\rm b}h^2,\Omega_{\rm c} h^2,h,n_{\rm s},A_{\rm s},w_0,w_a\}$ as the primary set, and then projects on the
chosen set of derived parameters. 
For \texttt{CosmicFish}, \texttt{CosmoSIS}, {and \texttt{TotallySAF}} we use the three-point
stencil method to compute the numerical derivatives; in order to have the same step size for the
different parameters in {all} codes, \texttt{CosmicFish} uses the steps
for the $\vartheta$ parameters that are equivalent to the steps used
by \texttt{CosmoSIS} {and \texttt{TotallySAF}} in the set of parameters of \autoref{eq:9parameters}. A more detailed
explanation of the problems that may arise from the change of the
parameter basis is presented in \autoref{sec:lessons}. 
The \texttt{STAFF} code instead uses the semi-analytical approach described in \autoref{sec:ccwl},
where the numerical part of the derivatives is obtained following the method proposed by \cite{Camera:2016owj} (see their Appendix~A).

It is also important to note that the {four} codes are completely independent of each other and use very different approaches. \texttt{CosmicFish}, \texttt{STAFF}, {and \texttt{TotallySAF}}
implement a second order covariance given in \autoref{eq: covsecond} and build the Fisher matrix as in \autoref{eq:xcfisher_second}, while \texttt{CosmoSIS} considers a fourth order covariance given in \autoref{eq: covfourth} and builds the Fisher matrix as in \autoref{eq:xcfisher_fourth}.

Moreover, the codes differ also in how they obtain the cosmological quantities needed to compute the observables used to construct the Fisher matrix. \texttt{STAFF} {and \texttt{TotallySAF}} obtain these from external input, produced as described
in \autoref{sec:ccwl}. \texttt{CosmicFish} and \texttt{CosmoSIS} are instead directly connected to Boltzmann codes, \texttt{CAMB\_sources} for the former and \texttt{CAMB} for the latter; while \texttt{CosmoSIS} fetches from \texttt{CAMB} the cosmological quantities and computes the spectra as described in \autoref{sec:fisher-caveat-XC}, \texttt{CosmicFish} obtains the $C(\ell)$ directly from the internal computation of \texttt{CAMB\_sources} and can therefore in principle avoid using the Limber approximation, which is instead implicit in the recipe of \autoref{sec:fisher-caveat-XC}. The fiducial $C(\ell)$ obtained with these methods have been compared, finding an agreement to within approximately $1\%$.

\subsubsection{Settings definition}\label{sec:xccases}
For the purpose of the code validation, as for GC and WL single probes, we have investigated several different choices of parameters and settings. We discuss here some of them, as presented below. We use for the nonlinear power spectrum the `TakaBird' modified \texttt{HaloFit} (HF) prescription of \cite{Takahashi:2012em}, with additional
corrections for massive-neutrinos following \cite{Bird2012}, as discussed in \Cref{subsec:XCnonlinear,sec:wlnl}. We use the fiducial discussed in \autoref{sec:fiducial} and, if not otherwise specified, a baseline neutrino mass $\sum{m_\nu}=0.06\ {\rm eV}$ (with one massive-neutrino and two massless neutrinos) and we include intrinsic alignment in the weak-lensing calculations.

As discussed in \Cref{errors_shear,sub:covariancexc}, and as will be further clarified in \autoref{sec:results}, in the following we identify two configurations, corresponding to a stronger or weaker cut of the maximum 2D Fourier mode, $\ell_{\rm max}$, for the $C_\ell$ of weak lensing, photometric galaxy clustering, and the cross-correlation of the 
two probes; we further consider optimistic and pessimistic cuts when adding spectroscopic galaxy clustering as follows.
\begin{align}
 {\textit{\textbf{Pessimistic settings:}}} & \,\,\, k_{\rm max} ({\rm GC_{\rm s}}) = 0.25\ h\ {\rm Mpc}^{-1} \,\,\, ; \nonumber \\
 &\,\,\, \ell_{\rm max}({\rm WL})=1500 \,\,\,,  \nonumber \\ 
 &\,\,\, \ell_{\rm max}({\rm GC_{\rm ph}})=\ell_{\rm max}({\rm XC^{(\rm GC_{\rm ph},WL)}})=750 \,\,\, ;  \nonumber \\
& \,\,\, {\rm GC_{\rm ph}\,\,\,} \text{for  } z < 0.9  \,\,\, {\rm{when}} \,\, {\rm{combined }} \,\, {\rm{with}} \,\, {{\rm GC_{\rm s}}} \,\,. \\
 {\textit{\textbf{Optimistic settings:}}}  &\,\,\,  k_{\rm max} ({\rm GC_{\rm s}}) = 0.3\ h\ {\rm Mpc}^{-1}, {\rm{with}} \,\, {\rm{fixed }} \,\, \sigma_p \,\, {\rm{and }} \,\, \sigma_v \,\,\, ; \nonumber \\
 &\,\,\, \ell_{\rm max}({\rm WL})=5000 \,\,\, ; \nonumber \\ 
 &\,\,\, \ell_{\rm max}({\rm GC_{\rm ph}})=\ell_{\rm max}({\rm XC^{(\rm GC_{\rm ph},WL)}})=3000 \,\,\, .
\end{align}
We note that when combining with the GC$_{\rm s}$ in the pessimistic case we cut GC$_{\rm ph}$ at redshift $z<0.9$ in order to avoid overlap of the two surveys. This is done under the conservative assumption that we are not able to model the correlation of the two. In the optimistic case instead, we perform no cut in redshift and use the full GC$_{\rm ph}$ survey, assuming that the correlation of the two clustering surveys can be neglected given the relatively small common redshift range. Finally, in the optimistic case the GC$_{\rm s}$ Fisher matrix is obtained fixing the non-linear parameters $\sigma_{\rm p}(z_{\rm mean})$ and $\sigma_{\rm v}(z_{\rm mean})$ (see \autoref{sec:ccgc} for details), rather than marginalising over them as is done in the pessimistic case.

\subsubsection{Results of the code comparison}
\begin{figure}[h!]
\begin{tabular}{cc}
\includegraphics[width=0.48\textwidth]{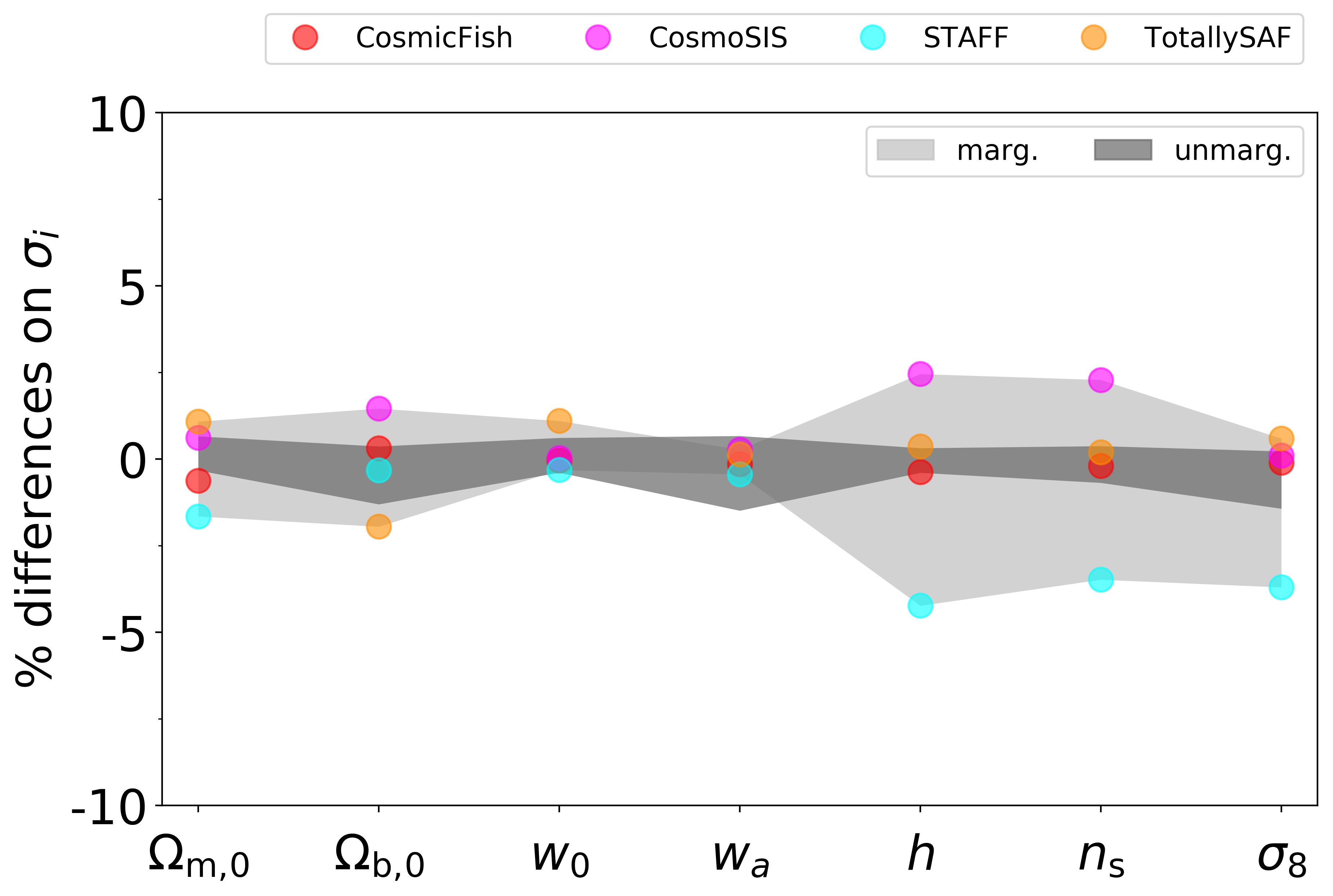} &
\includegraphics[width=0.48\textwidth]{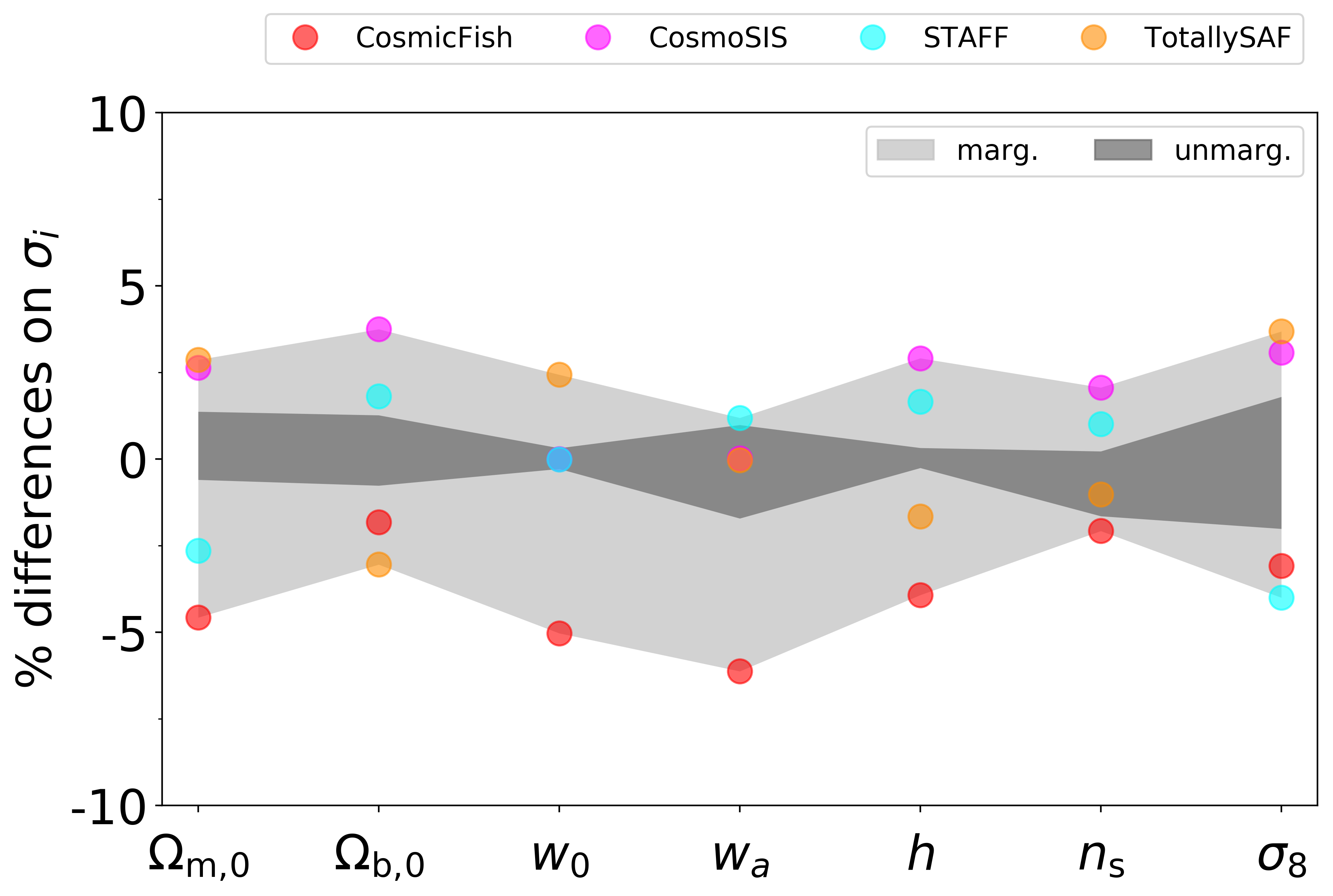} \\
\includegraphics[width=0.48\textwidth]{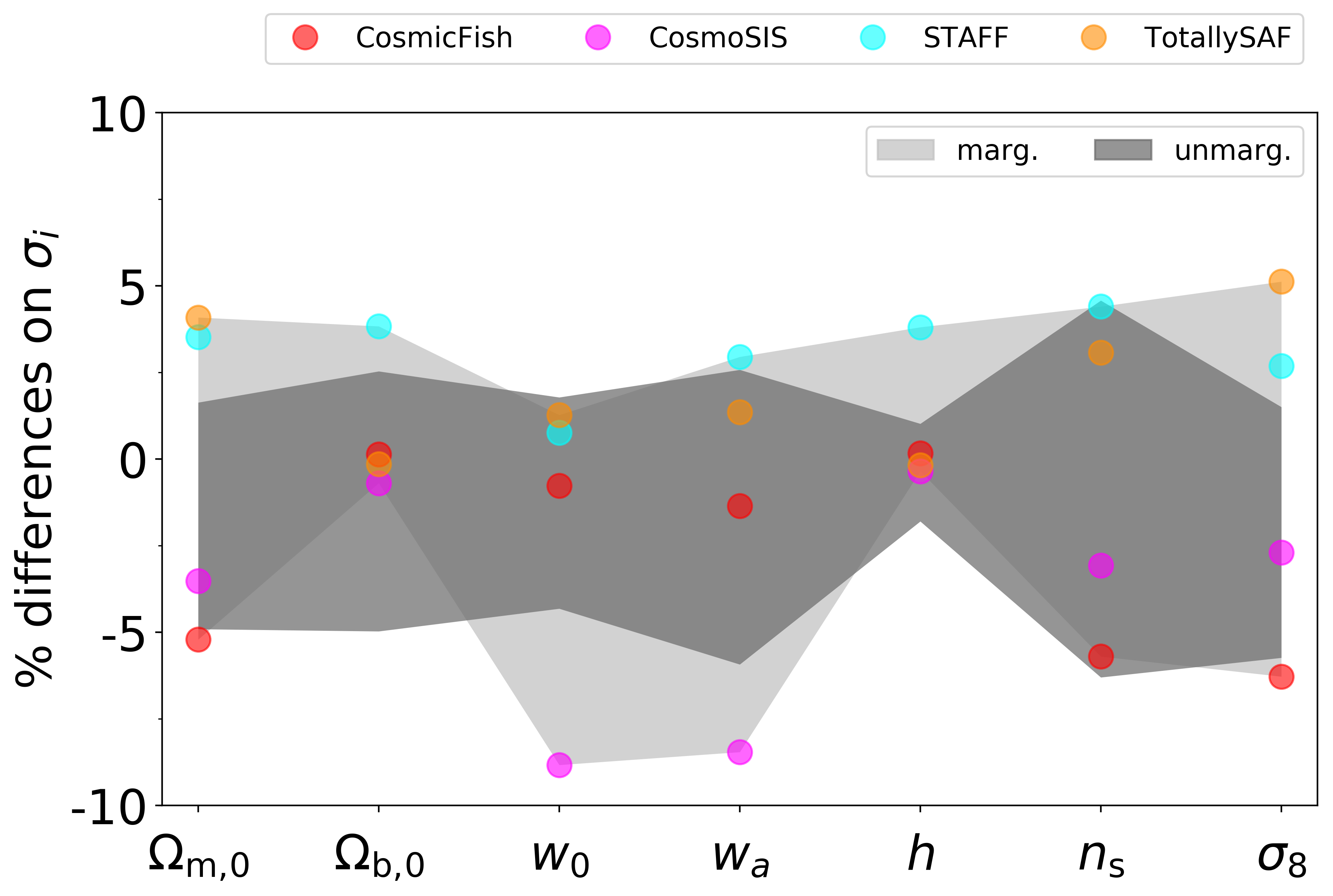} &
\includegraphics[width=0.48\textwidth]{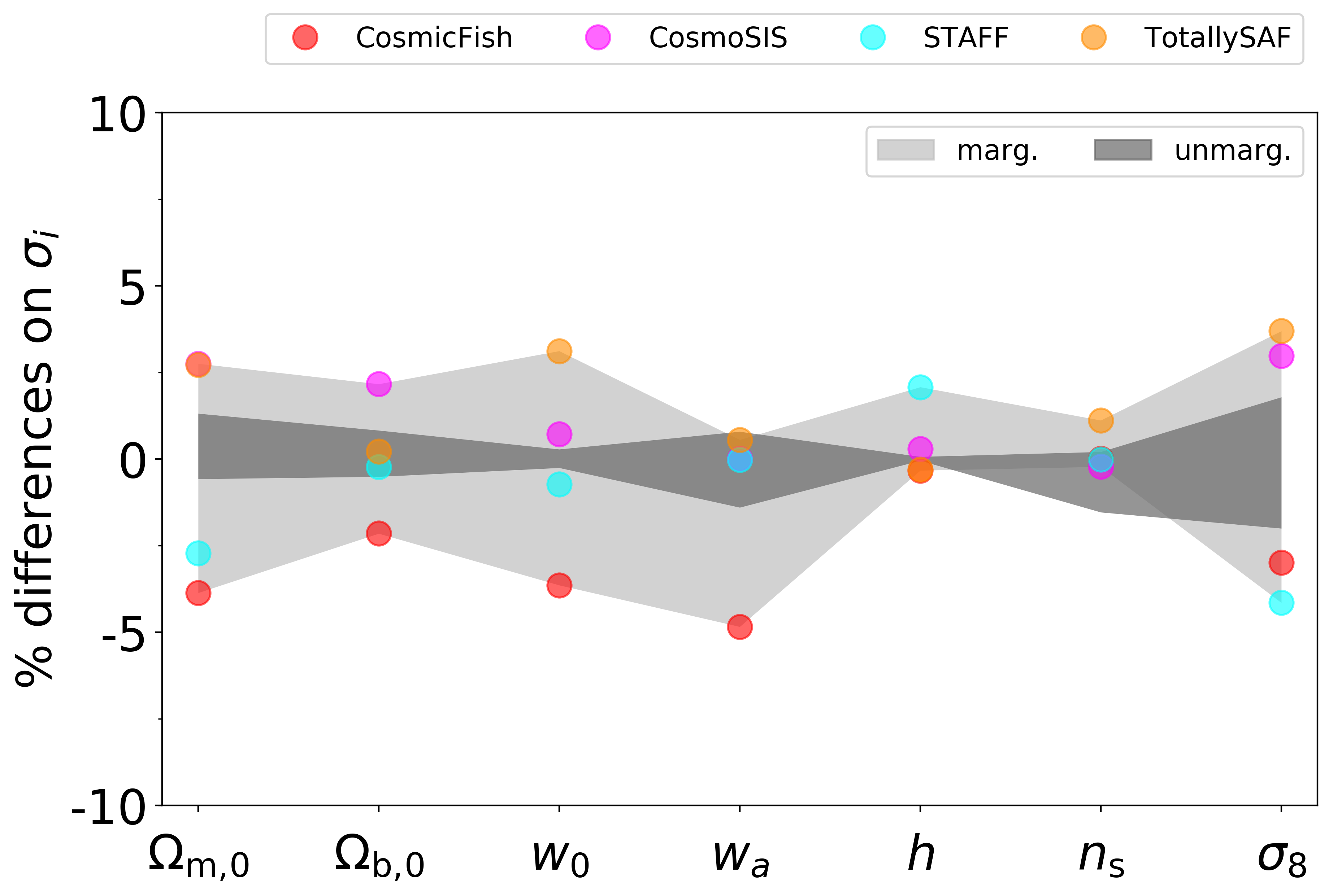} \\
\end{tabular}
\caption{Percentage difference in the errors on the final parameters among \texttt{CosmicFish}, \texttt{CosmoSIS}, \texttt{STAFF}, {and \texttt{TotallySAF}} for
the GC$_{\rm ph}$+WL+XC$^{(\rm GC_{\rm ph},WL)}$ combination (top panels) and the GC$_{\rm s}$+GC$_{\rm ph}$+WL+XC$^{(\rm GC_{\rm ph},WL)}$ combination (bottom panels). We assume here a flat $(w_0, w_a)$ cosmology. This comparison is done both in the pessimistic and optimistic configurations (left and right panels, respectively); we note, however, that for the top panels, since no spectroscopic galaxy clustering is included, there is no need to perform a cut in $z$ (as the pessimistic setting would require). The light grey band correspond to the marginal errors on the parameters, while the dark grey band indicates the errors when we do not marginalise over the other parameters.} \label{fig:xccompplot1}
\end{figure}

In \autoref{fig:xccompplot1} we show the comparison of the {four} codes for the pessimistic and optimistic configurations (left and right panels respectively) for the {combination of observables GC$_{\rm ph}$+WL+XC$^{(\rm GC_{\rm ph},WL)}$ and the full combination GC$_{\rm s}$+GC$_{\rm ph}$+WL+XC$^{(\rm GC_{\rm ph},WL)}$ described in \autoref{sec:xccases}.} In all cases investigated here we use the non-linear corrections given by the `TakaBird' formalism and marginalise over intrinsic alignment and galaxy bias nuisance parameters, included in the analysis as described in \autoref{sec:recipexc}. We find a good agreement between the {four} codes, with the relative difference on the errors on cosmological parameters on each pair of codes always within $10\%$. 

In addition to the results shown here, the comparison was performed using also a different neutrino mass ($\sum{m_\nu}=0.15\ {\rm eV}$), in a $\Lambda$CDM cosmology, combining subgroups of observables, {for nonflat models,} and without intrinsic alignment nuisance parameters, obtaining similar results in all cases. We also compared the combination of GC$_{\rm ph}$ and WL without the inclusion of the cross correlation, again finding compatible results between the {four} codes. 

\subsection{Lessons learned}\label{sec:lessons}

The process of comparison and validation among different codes involved a series of tests, choices, and lessons learned through the process. Here we describe some of these in detail, as a list of suggestions meant to facilitate comparison with any other Fisher matrix code. As usual, we highlight the main points, separating them among different probes.

\subsubsection{Galaxy clustering}
\label{sec:fisher-caveat-GC}
We list here some critical points that can be responsible for the different results one can find when comparing different galaxy clustering Fisher matrix codes, together with general tips to keep in mind when writing a new forecasting code.

\begin{enumerate}
\item{\textit{Input}}\\
Input quantities are needed in order to evaluate the observed galaxy power spectrum and these are the matter power spectra produced with \texttt{CAMB} or \texttt{CLASS}. 
If input files are produced independently we suggest to switch off the default spline in \texttt{CAMB}. This is to avoid, where possible, unnecessary interpolations that may introduce instabilities. In the case of \texttt{CLASS}, one would have to modify the source code in order to do the same.
\item{\textit{Interpolations}}\\
In order to compute reliable derivatives of $P_{\rm obs}(k,\mu;z)$, needed in \autoref{eq:GC:pk}, an interpolation scheme is necessary: we can choose to interpolate in the normal basis $\left\{k, P_{\delta\delta}(k,z) \right\}$, or in the logarithmic basis $\left\{\log_{10} k, \log_{10} P_{\delta\delta}(k,z) \right\}$. We found that the latter method produces more reliable results, as it is the best way to follow quantities that vary over orders of magnitude. Moreover, we found that it is best to interpolate such logarithmic tables with cubic splines, and to avoid any other unnecessary interpolation.
\item{\textit{Derivatives}}\\
The observed galaxy power spectrum in \autoref{eq:GC:pk} is governed by different parameters with respect to which we need to take the derivatives. For some of the parameters an analytical derivative can be constructed. However, we suggest considering fully numerical derivatives for all the parameters. To explain the reason for this let us take the particular example of the derivatives with respect to the Hubble parameter and the angular diameter distance. In the literature it is found that it is possible to take the derivative of the matter power spectrum with respect to $\ln H(z)$ and $\ln D_{\rm A}(z)$ by using the chain rule, i.e. 
\be
\frac{{\rm d} P_{\delta\delta}(k,z)}{{\rm d} \ln H(z) } = \frac{{\rm d} P_{\delta\delta}(k,z)}{{\rm d}k }\frac{{\rm d}k}{{\rm d} \ln H(z)}; 
\ee
we note that $k$ depends on $H(z)$ via \autoref{eq:kref-muref}.
This is a semi-analytical derivative because the derivative of $P_{\delta\delta}(k,z)$ with respect to $k$ has to be computed numerically, whereas the ${\rm d} k/{\rm d} \ln H(z)$ can be calculated analytically. 
During the code comparison we found that the semi-analytical derivatives of $P_\text{obs}(k,\mu; z)$ with respect to $\ln H(z)$ and $\ln D_A(z)$ do not necessarily agree with the numerical derivatives for $k> 0.1\,{\rm h\,Mpc}^{-1}$.
Moreover, we found that the semi-analytical derivative is not stable when varying the $k$-binning, while the numerical derivative is quite stable.
For the latter reason it is safer to produce numerical derivatives with respect to $\ln H(z)$ and $\ln D_A(z)$. This also has the advantage of facilitating the procedure of passing from linear modelling to non-linear corrections to $P_\text{obs}(k,\mu; z)$ in \autoref{eq:GC:pk}.

For a generic function $f(x)$,\footnote{In the present case, this is for instance the matter power spectrum or the shear tomographic angular power spectrum.} the three-point stencil derivative of $f$ with respect to $x$ at $x_0$ is given by
\begin{equation}
\left.\frac{{\rm d} f}{{\rm d}x}\right|_{x=x_0}\simeq \frac{f(x_0 + \varepsilon x_0) - f(x_0 - \varepsilon x_0)}{2 \varepsilon x_0},
\end{equation}
with $\varepsilon \ll 1$ an increment, usually an arbitrary choice. The three-point stencil approximates the derivative up to order $\varepsilon^2$ errors. To be more precise, another choice is represented by the five-point stencil
\begin{equation}
\left.\frac{{\rm d} f}{{\rm d}x}\right|_{x=x_0}\simeq \frac{-f(x_0 + 2\varepsilon x_0)+8f(x_0 + \varepsilon x_0) -8 f(x_0 - \varepsilon x_0)+f(x_0 -2 \varepsilon x_0)}{12 \varepsilon x_0}.
\end{equation}
When dealing with numerical derivatives, the choice of the step $\varepsilon$ is crucial. On the one hand, if it is chosen too small, the subtraction yields a large rounding error. Actually, all the finite difference formulae are ill-conditioned, and due to cancellations they produce a value of zero if $\varepsilon$ is small enough. On the other hand, if $\varepsilon$ is too large, the calculation of the slope of the secant line is more accurately calculated, but the estimate of the slope of the tangent by using the secant could be worse. Hence, we recommend checking the convergence of the derivatives when choosing the steps, in order to obtain realistic errors and FoM values, and reach agreement among different implementations of the same fisher recipe. Moreover, in this respect, we have to warn the reader that single-probe results can be step-dependent, but they become step-independent when \Euclid probes are combined together (see \autoref{sec:fisher-caveat-XC}). This is physically due to the difficulty for the single probe to  simultaneously constrain many parameters, for the GC case, in particular, a total of 24 different parameters, i.e. 4 shape parameters and 20 redshift-dependent ones.

\item{\textit{Integrals}}\\
During the code comparison we did not find any issue concerning the integration method. However, in order to avoid unnecessary interpolations, we suggest to consider the same $k$ binning as in the input files. 
\item{\textit{k-binning in input files}}\\
In order to validate the code we suggest to check stability of the code when using input files having different step sizes in $\log_{10} k$ space. We have tested the stability against several possible step sizes, namely 20 steps per decade in $k$, 50 steps, and 200 steps; as our default we have chosen the case of 50 steps per decade in $k$. 

\item{\textit{Stability of the results}}\\
As a final test, we performed several stability checks in order to ensure that the codes used in the code comparison give reliable results. One of the main challenges in the Fisher matrix approach is to consider an appropriate step for the derivatives of the observed matter power spectrum with respect to the parameters. In our analysis, we performed a systematic analysis  considering, firstly the step in the derivatives with respect to the shape parameters ${\omega_{\rm m,0}},\,h, {\omega_{\rm b,0}}, n_{\rm s}$, and, separately, we let the step for the derivatives vary with respect to $\ln D_{\rm A}(z)$ and $\ln H(z)$ only.\footnote{Concerning the other redshift-dependent parameters, i.e. $\ln[f\sigma_8(z)]$ and $\ln[b\sigma_8(z)]$, we did not perform any stability check because the parameters enter quadratically into the observed power spectrum, and the derivatives are well-behaved.}

In \autoref{fig:foms-shape-da-step-C3-PRn} left panel, we show the FoMs when varying the step for the derivatives of the observed galaxy power spectrum with respect to the shape parameters for the $w_0w_a$flat projection case; we neglected to show all the other projection cases as they manifest a similar behaviour. The analysis has been performed by varying one parameter at a time. In \autoref{fig:foms-shape-da-step-C3-PRn} (right panel), we show the FoMs when varying the step for the derivatives of the observed galaxy power spectrum with respect to the $\ln D_{\rm A}$ for all the projection cases.

\begin{figure}[h!]
\center 
\begin{tabular}{c}
\includegraphics[width=0.5\textwidth]{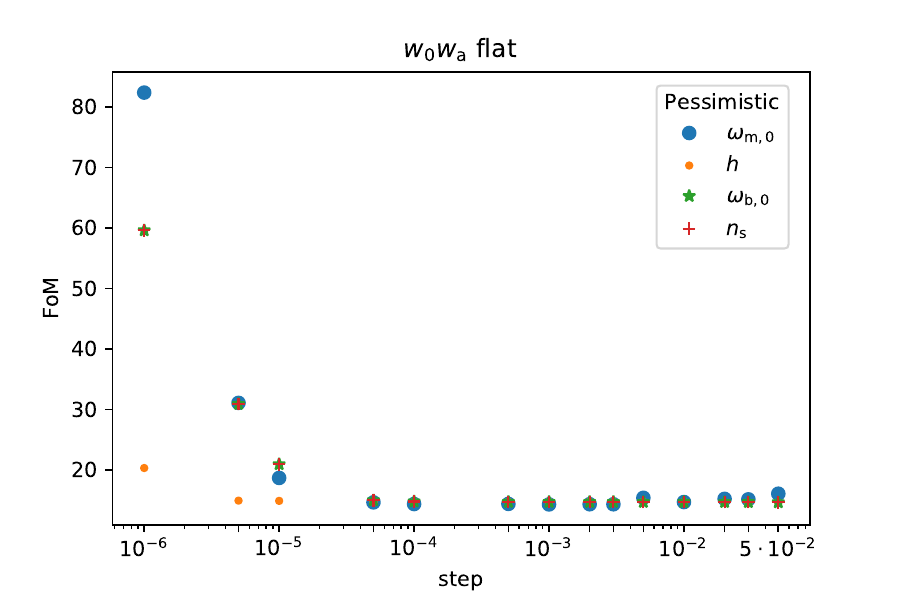} 
\includegraphics[width=0.5\textwidth]{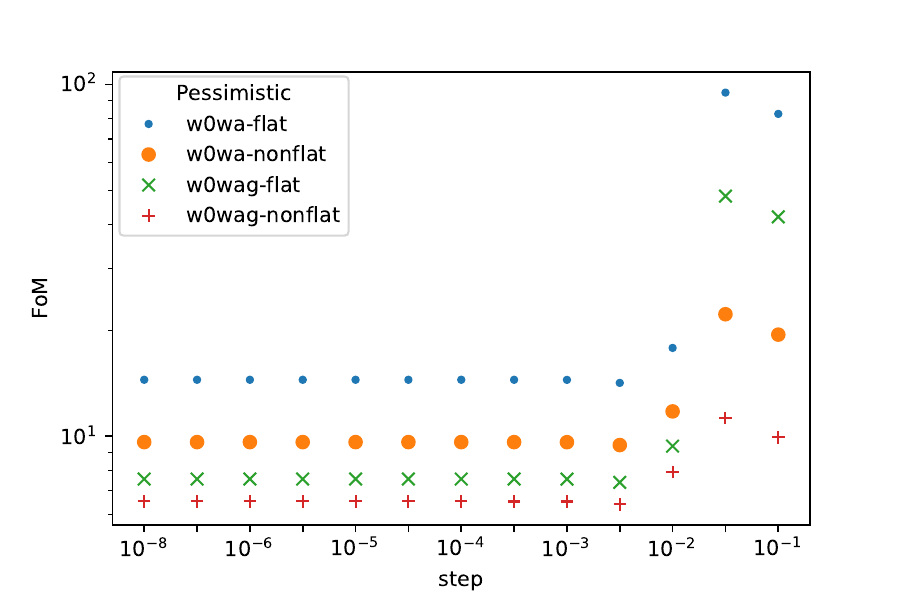}
\end{tabular}
\caption{The left panel shows the FoM as a function of the steps for the shape parameters, whereas  the right panel  shows the FoM as a function of the steps for the $\ln D_{\rm A}(z)$ parameter. Both results are for spectroscopic GC$_{\rm s}$ only. }
\label{fig:foms-shape-da-step-C3-PRn}
\end{figure}

\end{enumerate}

\subsubsection{Weak lensing}
\label{sec:fisher-caveat-WL}
Computing the weak lensing Fisher matrix involves several different steps, which must be dealt with using great care to avoid biasing the final result. The code comparison challenge has allowed us to identify several critical issues, and we briefly comment upon them in the following.
\begin{enumerate}
\item \textit{ Fiducial shear power spectra} 
The tomographic angular power spectrum of cosmic shear, $C^{\epsilon\epsilon}_{ij}(\ell)$, is given in \autoref{eq: defcij}. Neither of the two integrands, the various kernels or the power spectrum, are analytic functions. Hence, each code in the code comparison has to compute them numerically, and then interpolate using an intermediate look-up table. We found that the lensing kernel of \autoref{eq: kijggend}, $K_{ij}^{\gamma \gamma}(z)$, is a well-behaved function of redshift, and the results (in terms of spectra and Fisher matrices) are stable provided that the number of points in the $\{z,K_{ij}^{\gamma \gamma}\}$ table $N_z\geq 300$.

The matter power spectrum, on the contrary, is a much more involved function to compute. This is because it depends on two variables, namely the scale $k$ and the redshift $z$, and it presents an oscillating behaviour because of the baryon acoustic oscillation wiggles. Moreover, we evaluate the various $P_A(k, z)$ ($A = \{\delta \delta, \delta {\rm I}, {\rm II}\}$) at $k_{\ell} = (\ell + 1/2)/r(z)$, which means spanning a large range in $k$. On the other hand, for the various $a = \{\gamma\gamma, {\rm I}\gamma, {\rm II}\}$, the product $K^a_{ij}(z)P_A(k_{\ell}, z)$ is suppressed at both small and high $z$ because the lensing kernel is only significantly non-zero over a limited redshift range. To equally well sample the support of the integrand, and to model accurately the baryon acoustic oscillations, we find that a good strategy is to create a table with $\{\log_{10}{k}, z, \log_{10}{P_A(k, z)}\}$ over the range $0.0005 \le k/{\rm Mpc}^{-1} \le 35$ and $0 \le z \le 2.5$ with $N_k \times N_z$ entries. The results are stable with the same $N_z$ as before and $N_k = 800$. We stress that we do not use $h$ units, i.e.\ we take $k$ in ${\rm Mpc}^{-1}$ and $P_A(k, z)$ in ${\rm Mpc}^3$ in order to avoid unnecessary complications when taking derivatives with respect to the dimensionless Hubble constant $h$.\\

\item \textit{  Numerical versus semi-analytic derivatives}
  The derivatives of the shear power spectra are a critical ingredient of the WL Fisher matrix. Two different approaches are possible, as shown by the identity
\begin{align}
\frac{\partial C^{\epsilon\epsilon}_{ij}(\ell)}{\partial\theta_{\mu}} &= \frac{\partial}{\partial \theta_{\mu}}\left\{\sum_{A,a}
\int_{z_{\rm min}}^{z_{\rm max}}\!\!{\rm d}z\,K^{a}_{ij}(z)P_{A}[k_{\ell}(z), z]\right\},\\
&=\sum_{A,a}
\int_{z_{\rm min}}^{z_{\rm max}}\!\!{\rm d}z\,\frac{\partial}{\partial\theta_{\mu}}\left\{K^{a}_{ij}(z)P_{A}[k_{\ell}(z), z]\right\}.
\end{align}
According to this relation, one can either~: $(i)$ first compute the integral and then the derivatives, for instance through finite-difference methods; or $(ii)$ first compute the derivatives of the integrand and then integrate the result. The former choice fully relies on numerical techniques, while the latter is partly analytic, since the derivatives of the kernel may be transformed into integrals of analytic quantities (at least for the models we have considered in this paper). Both approaches have their pros and cons. The fully numerical approach is quite generic and does not require any intervention from the user if a different cosmological model is adopted (apart from changes in the Hubble rate expression), but it is prone to the instability of numerical derivatives. On the contrary, the `semi-analytic' approach transforms almost all the derivatives into integrals (the only exception being those of the matter power spectrum). However, the user has to adjust the code to update the expressions for the derivatives of the Hubble parameter and the radial comoving distance each time a new cosmological model is considered. Of the {seven} codes participating to the challenge, {three} (\texttt{CCCP}, {\texttt{CCCPy}}, and \texttt{STAFF}) adopt the semi-analytic approach, three (\texttt{CosmicFish}, \texttt{CosmoSIS}, {and \texttt{TotallySAF}}) employ a fully numerical method, and a single code (\texttt{FisherMathica}) uses a mixed technique. Provided care is taken when computing the numerical derivatives, we find all the approaches provide consistent results. Choosing among the two schemes is therefore the users choice and does not affect results.\\

\item \textit{  Numerical differentiation methods} 
Regardless of the choice one decides to adopt for derivative computation, the method implemented to compute the numerical derivatives of the quantities of interest plays a critical role in the estimate of the Fisher matrix. Finite differences methods are the typical choice. 
\\
As already mentioned in point 3. of~\autoref{sec:fisher-caveat-GC}, particular care should be taken in the choice of the step $\varepsilon$ of the derivatives. 
In fact, when dealing for example with derivatives of the tomographic cosmic-shear angular power spectrum $C^{\epsilon\epsilon}_{ij}(\ell)$, the situation is much more complicated. This is due to the fact that even the most careful choice of $\varepsilon$ is not, in general, hold true over the whole range of $\ell$ modes, and for any $ij$ bin pair combination. As an example, consider the baryon fraction $\Omega_{{\rm b},0}$: its main impact on the matter power spectrum, and, consequently, on the shear signal, is at BAO scales, corresponding today to {$k_{{\rm BAO},0} \simeq 2\pi/150\,\mathrm{Mpc}^{-1}$}. However, this physical scale shows up at different angular multipoles, depending on the redshift bin considered. Therefore, a step $\varepsilon$ optimised to capture at best the response of $C^{\epsilon\epsilon}_{ij}(\ell)$ to $\Omega_{{\rm b},0}$ at $i=j=1$ and $\ell=k_{{\rm BAO},0}r(z=0)$, may itself (and indeed typically will) lead to potentially significant numerical errors for a different combination of $ij$ and $\ell$ mode.

As a possible solution to this issue, \cite{Camera:2016owj} proposed an alternative method, first implemented in the \texttt{CCCP} code (see \autoref{sec:allcodes}). Let us consider a parameter $x$: for each combination of $ij$ and $\ell$ -- in our example, the generic function $f(x)$ -- we sample the $x$-line in parameter space in, say, 15 $x_i$ points around the fiducial value $x_0$ (including $x_0$). In the hypothesis that the neighbourhood is small enough, we interpolate the $x_i$ values with a straight line. Then, by the definition of a derivative, the slope of the linear interpolation is the derivative of the function at the fiducial value, i.e.\ ${\rm d}f/{\rm d}x|_{x_0}$. We can test our ansatz by checking whether the spread between the linearly fitted $f_{\rm fit}(x_i)$ and the true values $f(x_i)$ is less than some required accuracy, e.g.\ $1\%$. If this requirement is not met, we zoom in on the sampled $x$-range by cutting out a few values on the edges, until we reach the requested accuracy.

Of the {seven} codes participating in the comparison challenge, four of them implemented the aforementioned technique to compute derivatives at different levels. In particular, \texttt{CCCP}, {\texttt{CCCPy}}, and \texttt{STAFF} use the technique to compute the derivatives of the matter power spectrum, while the rest of derivatives are computed semi-analytically. \texttt{FisherMathica} uses this method for the matter power spectrum derivative, while the $C^{\epsilon\epsilon}_{ij}(\ell)$ derivatives are computed with standard three- and five-point stencil methods. On the other hand, \texttt{CosmicFish}, \texttt{CosmoSIS}, {and \texttt{TotallySAF}} rely on three-point stencil. 

The choice of derivative method impacts the ability to make statements on the stability of the results. 
For example, \texttt{CCCP}, {\texttt{CCCPy}}, and \texttt{STAFF} do not rely on the classical finite difference method, so there is no need to check their stability against the step $\varepsilon$ used in the numerical differentiation; while some concern might be related to the range used in the fitting procedure,  we have tested this and found to have no impact on the results. However, $\varepsilon$ plays a key role in \texttt{CosmicFish}, \texttt{CosmoSIS}, {and \texttt{TotallySAF}} so that a detailed analysis of how the results change with $\varepsilon$ has been carried out. We find that for $4\%\leq \varepsilon \leq 10\%$ both the marginalised errors on the cosmological parameters and the FoM are stable (i.e. unchanged over this range). A similar test has also been done for \texttt{FisherMathica} where stability is achieved for $\varepsilon \simeq 3\%$ no matter whether three- or five-point stencil derivatives are used; we note that the weaker requirement on $\varepsilon$ is also related to the fact that \texttt{CosmicFish}, \texttt{CosmoSIS}, {and \texttt{TotallySAF}} are also used for the probe combination, which is more demanding in terms of stability than cosmic shear.

We find that the FoM and marginalised errors on $\{\Omega_{{\rm m},0}, w_0, w_{\rm a}, \sigma_8\}$ are stable with respect to the methodology and the step size, whereas this is not the case for $\{\Omega_{{\rm b},0}, h, n_{\rm s}\}$. For the unstable parameters this is not surprising, since weak lensing is less sensitive to these parameters. However, what is really difficult to control is the relative orientation of the contours in the 2D parameter projections. Indeed, we find that these orientations -- governed by the off-diagonal elements of the Fisher matrix -- turn out to be quite sensitive to the choice of $\varepsilon$ and to whether a three- or five-point stencil is used. However, provided the above quoted requirements on the step are met, stability of the results is achieved. \\

\item \textit{  Number of multipoles and spacing}
The WL Fisher matrix is obtained by summing up contributions from different multipoles $\ell$ spanning the wide range $10 \le \ell \le \ell_{\rm max}$. In determining the number and values of the $\ell$ modes to use in the calculation two options are possible: one can use $N_\ell$ linearly equispaced bins in $\ell$; or one can use $N_{\lambda}$ logarithmically equispaced bins in $\lambda = \log{\ell}$. We have explored both possibilities when evaluating the quantities of interest (i.e.\ $C^{\epsilon\epsilon}_{ij}(\ell)$ and its derivatives).\footnote{If the bins are linearly spaced, the centre of the $i$\,-\,th bin is $\ell_i = (\ell_{i\rm L} + \ell_{i\rm U})/2$ with $(\ell_{i\rm L}, \ell_{i\rm U})$ the lower and upper limit of the bin. For logarithmically equally spaced bins, we set $\ell_i = [{\rm dex}(\lambda_{i\rm L}) + {\rm dex}(\lambda_{i\rm U})]/2$ where $(\lambda_{i\rm L}, \lambda_{i\rm U})$ are the bin boundaries.} We found that $N_{\ell} = N_{\lambda} = 100$ gives consistent results among the two approaches for both $\ell_{\rm max} = 1500$ and $\ell_{\rm max} = 5000$. Furthermore, we found that reducing the number of logarithmic bins up to $N_{\lambda} = 60$ changes the FoM and the marginalised constraints by less than $1\%$, so that speeding up the computation is possible with a small degradation in accuracy. We nevertheless recommend using $N_{\lambda} = 100$ logarithmic bins, since they more finely sample the shear power spectra.
\end{enumerate}

\subsubsection{Probe combination}
\label{sec:fisher-caveat-XC}

In this section we list some of the critical points that can lead to different results when comparing outputs from the probe combination codes. These involve both the construction of the Fisher matrix and the operations needed to obtain cosmological constraints from it, as well as possible issues arising when obtaining the relevant quantities from the Boltzmann solvers.

\begin{enumerate}
\item \textit{  Matrix inversion}
The final sensitivity expressed through  the correlation matrix obtained by inverting the Fisher matrix is at the heart of the methodology. This has a drawback: matrix inversion can be tricky when the condition number is large i.e.\ the matrix is then ill-conditionned.\footnote{{The condition number is defined to measure how much the output value of the function can change for a small change in the input argument.}} In the case of weak lensing and galaxy clustering the condition number is of the order of $2 \times10^6$. In such a case, in principle, a precision better than $0.5\times10^{-7}$ is needed for a secure inversion (i.e. no numerical instability) of the matrix with an accuracy better than 10\%. However, in practice the required accuracy depends on the matrix. 
This means that each Fisher matrix needs a specific treatment to quantify the accuracy needed to avoid numerical instability. The ability to achieve this accuracy was tested by perturbing the Fisher matrix with a Gaussian-distributed random percentage of all the elements of the WL Fisher matrix (using the TakaBird non-linear prescription) according to the approach proposed by \cite{Yahia-Cherif}. The 10$\%$ agreement level on the Fisher matrix is reached at $>99\%$ degree of confidence when using a (relative) perturbation of $4\times10^{-5}$. The 65$\%$ confidence level is reached at $10^{-4}$. More investigations were done in the spectroscopic GC linear case: all elements of the Fisher were perturbed one by one. Some of the elements can be perturbed by 100$\%$ without changing the FoM by more than 10$\%$, whereas the most sensitive elements to the perturbations can give differences in the FoM higher than 10$\%$ for a perturbation of the order of $5\times10^{-4}$. 

The above results have implications on the accuracy required for the derivatives entering the Fisher matrix. Standard methods of derivation are the N-point stencils with a fixed step. A single step size for all parameters may not be appropriate. Two sources of errors exist: the truncation errors that increase with the step size; and the numerical noise that increases when the step size diminishes. The outputs of  the Boltzmann code may be subject to a significant numerical noise for small steps: in the calculation of the power spectrum $P_{\delta\delta}(k)$ this problem was noted at a step-size  below $10^{-5}$ using \texttt{CAMB} and $10^{-4}$ with \texttt{CLASS}.  The fact that the outputs have a different sensitivity depending on the parameter does not ease the determination of an optimal step. An adaptive scheme can then be used but at the price of a significant increase of the computational time.

This issue is extremely relevant for parameters that are not well constrained by the probe in use; this is because
poorly constrained parameters lead to elements of the Fisher matrix that have small numerical values, and therefore lead to complications when inverting the matrix. We  therefore find that this problem is most significant  when using the observational probes separately, while the combination of these probes has a weaker dependence on the choice of the step size.\\

\item \textit{  Scale sampling}
The computation of $C(\ell)$ for the considered observables is performed by obtaining the matter power spectrum from the Boltzmann solver, sampled in the redshift and scale ranges considered. The comparison performed highlighted how 
results can depend on this sampling. In particular, we noticed that non-linear corrections are particularly sensitive to the number of scales at which the matter power spectrum is computed; this is due to the fact that we compute the power spectrum up to $k_{\rm max}=70 \, {\rm Mpc}^{-1}$ and the highest scales are sampled more sparsely due to the logarithmic step in $k$ used in the Boltzmann solver. Since the codes interpolate between these $k$ values when performing 
the integral needed to compute the $C(\ell)$, a step that is too large  in the sampling leads to artificial oscillations in the matter power spectrum that  can significantly impact the results. 

We recommend therefore to check carefully that the chosen step is small enough so that no artificial oscillations are introduced.\\

\item \textit{  Change of parameter basis}
The codes investigated for probe combination (\texttt{CosmicFish}, \texttt{CosmoSIS}, \texttt{STAFF}, {and \texttt{TotallySAF}}) deal with the choice of the primary parameters in different ways; while the first uses \texttt{CAMB} parameters $\vartheta=\{\omega_{\rm{b,0}},\ \omega_{\rm{c,0}}, \ h, \ n_{\rm{s}},\ A_{\rm{s}}\}$ as the primary set and can then project on the chosen set of derived parameters, {\texttt{CosmoSIS}} is able to work both with this set and the target set of this paper, i.e. $\theta_{\rm final}=\{\Omega_{\rm{m,0}},\ \Omega_{\rm{b,0}},\ h,\ n_{\rm{s}},\ \sigma_{\rm{8}}\}$. During our comparison effort we found at first that while the agreement of the two codes was optimal when both codes used the first basis, the same agreement was not reached when using the latter for \texttt{CosmoSIS} and using the projection of \texttt{CosmicFish}. This comes from the fact that varying one parameter in the $\theta_{\rm final}$ basis, while keeping the others fixed, implies a variation of more than one parameter in the $\vartheta$ set; as an example, a variation of $\Omega_{\rm{m,0}}$ would obviously imply a variation of $\omega_{\rm{c,0}}$ in the $\vartheta$ set, with $\omega_{\rm{b,0}}$ fixed in order  not to vary $\Omega_{\rm{b,0}}$, but it would also require a variation of $A_{\rm{s}}$, needed to keep $\sigma_8$ at its fiducial value. In order to fix this issue, \texttt{CosmicFish} was modified in order to be able to effectively work in the $\theta_{\rm final}$ basis, i.e. using as input the full set of $\vartheta$ parameters corresponding to each variation of the $\theta_{\rm final}$ set. We therefore recommend to pay attention to this issue when the parameters varied in the Fisher code do not coincide with those varied in the Boltzmann code that is used.  
\end{enumerate}

\section{Results}\label{sec:results}
In this section we show the results of the validated Fisher-matrix codes for the cosmological parameters of interest (baseline and extensions), whose fiducial parameters have been illustrated in \autoref{tab:cosmofid}:
\be
\theta_\textrm{final} = \left\{\Omega_{\rm b,0},\,\Omega_{\rm m,0},\,h,\,n_{\rm s},\,\sigma_8, \, \, \tau, \sum{m_\nu} , \,\Omega_{\rm DE,\,0},\,w_0,\,w_{a},\,\gamma\right\} \,\,\, .
\label{eq:9parametersRecall}
\ee
We do this for the following scenarios:
\begin{itemize}
\item $\Lambda\text{CDM}$ within a spatially flat (and non-flat) geometry;
\item $w_0$, $w_a$ within a spatially flat (and non-flat) geometry;
\item $w_0$, $w_a, \gamma$ within a spatially flat (and  non-flat) geometry.
\end{itemize}
In the $\Lambda\text{CDM}$ case $\Omega_{\rm DE,\,0} \equiv \Omega_\Lambda$, the equation-of-state parameter and its derivative are fixed to $w_0=-1$ and $w_a=0$, and  the modified gravity parameter $\gamma$ is fixed to $\gamma = 6/11$.\footnote{We round this value to 0.55 in code implementations.}  For the flat case  $\Omega_{\rm DE,\,0} \equiv 1 -\Omega_{\rm m,0}$. All results refer to a fixed value of the neutrino mass of $\sum m_\nu = 0.06$ eV and to the fixed value of $\tau$ reported below \autoref{tab:cosmofid}. The latter does not have any influence on the cosmological observables considered, and is simply mentioned here so that anyone interested in validating their own code can perform a consistent comparison.

As illustrated in \autoref{sec:recipexc}, we compute cross-correlation terms only between the photometric galaxy clustering and the weak-lensing survey. 
We recall here for convenience that in \autoref{sec:xccases} we have identified two configurations, a pessimistic (optimistic) setting, corresponding to: (a) a stronger (weaker) cut of the maximum angular mode $\ell_{\rm max}$, for the $C(\ell)$ of WL, GC$_{\rm ph}$, and the cross-correlation of the 
two probes; and (b) a stronger (weaker) cut when adding GC$_{\rm s}$. For the pessimistic case, the WL non-linear regime is cut at $\ell_{\rm max}=1500$.  We further limit the maximum multipole used for the analysis of GC$_{\rm ph}$ to $\ell_{\rm max}=750$, with a further cut in redshift of $z < 0.9$ when this probe is combined with ${\rm GC_{\rm s}}$ to safely neglect cross terms between photometric and spectroscopic data. Consequently, in the pessimistic case the cross-correlation analysis is limited to the lowest of these multipoles, with $\ell^{\rm XC}_{\rm max}=750$. For GC$_{\rm s}$ we reach a value of $k_{\rm max}=0.25 h\, {\rm Mpc^{-1}}$. The optimistic case extends the regime of GC$_{\rm s}$ to $k_{\rm max}=0.30 h \,{\rm Mpc^{-1}}$, fixing the parameters $\sigma_{\rm p}$ and $\sigma_{\rm v}$; the range used for GC$_{\rm ph}$ is also extended up to $\ell^{\rm XC}_{\rm max}=3000$ with no cut in redshift. As discussed in \autoref{sub:covariancexc}, this choice of ${\rm GC_{\rm ph}}$ is quite ideal, since it entirely neglects non-Gaussian terms, which are nevertheless important to consider for this probe and become important earlier than for WL; it does, however, show the potential of such a probe, if one were able to also include such high multipoles. The optimistic WL setting extends the non-linear regime up to $\ell_{\rm max}=5000$. We can summarise this as follows~:
\begin{align}
 {\textit{\textit{pessimistic settings:}}} &  k_{\rm max} ({\rm GC_{\rm s}}) = 0.25\ h\ {\rm Mpc}^{-1} \,\,\, , \nonumber \\
 & \ell_{\rm max}({\rm WL})=1500 \,\,\,,  \nonumber \\ 
 & \ell_{\rm max}({\rm GC_{\rm ph}})=\ell_{\rm max}({\rm XC^{(\rm GC_{\rm ph},WL)}})=750 \,\,\, ,  \nonumber \\
&  {\rm GC_{\rm ph}\,\,\,} \text{for  } z < 0.9  \,\,\, {\rm{when}} \,\, {\rm{combined }} \,\, {\rm{with}} \,\, {{\rm GC_{\rm s}}} \,\,; \\
 {\textit{\textit{optimistic settings:}}}  &  k_{\rm max} ({\rm GC_{\rm s}}) = 0.3\ h\ {\rm Mpc}^{-1}, {\rm{with}} \,\, {\rm{fixed }} \,\, \sigma_{\rm p} \,\, {\rm{and }} \,\, \sigma_{\rm v} \,\,\, , \nonumber \\
 & \ell_{\rm max}({\rm WL})=5000 \,\,\, , \nonumber \\ 
 & \ell_{\rm max}({\rm GC_{\rm ph}})=\ell_{\rm max}({\rm XC^{(\rm GC_{\rm ph},WL)}})=3000 \,\,\, .
\label{def:optimistic}
\end{align}
For GC$_{\rm s}$, we have further tested two more settings, as they can provide a valuable reference for comparison: 
\begin{align}
 {\textit{\textit{linear setting:}}} & \,\,\, {\rm{linear \,\, recipe \,\, with}} \,\,\, k_{\rm max} ({\rm GC_{\rm s}}) = 0.25\ h\ {\rm Mpc}^{-1} \,\,\,; \\ 
 {\textit{\textit{intermediate setting:}}} & \,\,\, {\rm{non-linear \,\, recipe \,\, with}} \,\,\, k_{\rm max} ({\rm GC_{\rm s}}) = 0.3\ h\ {\rm Mpc}^{-1} {\rm{varying}} \,\, \sigma_{\rm p} \,\, {\rm{and}} \,\, \sigma_{\rm v} . \label{def:intermediate}
\end{align}
We have also verified that when considering a higher value in neutrino mass of $\sum m_\nu = 0.15$ eV, the results stay substantially the same, with variations of less than $1\%$ on the relative error for each parameter.
We do not consider here a linear or intermediate setting for weak lensing, whose power is very much related to the inclusion of non-linear scales.


\subsection{$\Lambda$CDM}
\begin{figure}[h]
\center 
\begin{tabular}{c}
\includegraphics[width=0.5\textwidth]{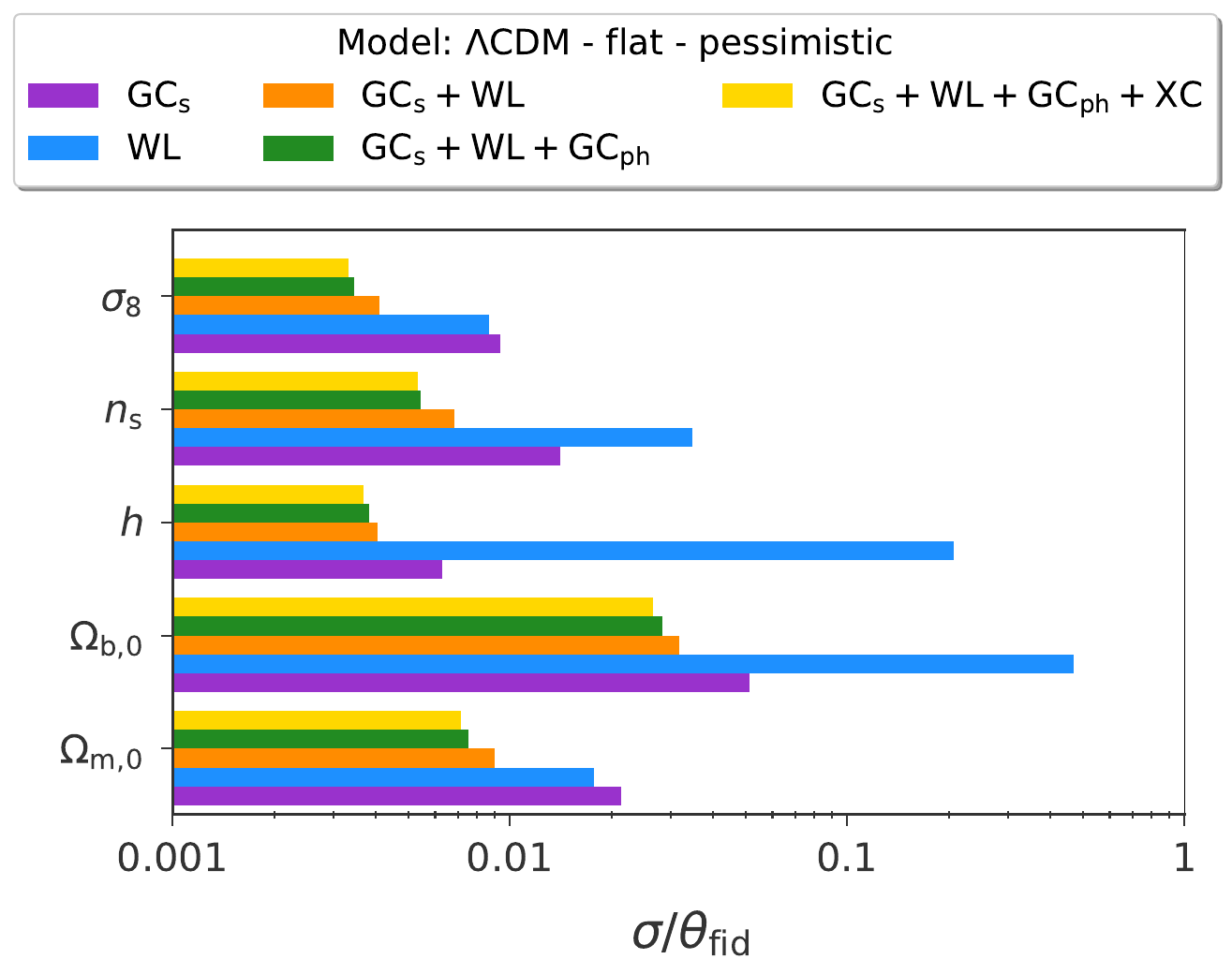}\includegraphics[width=0.5\textwidth]{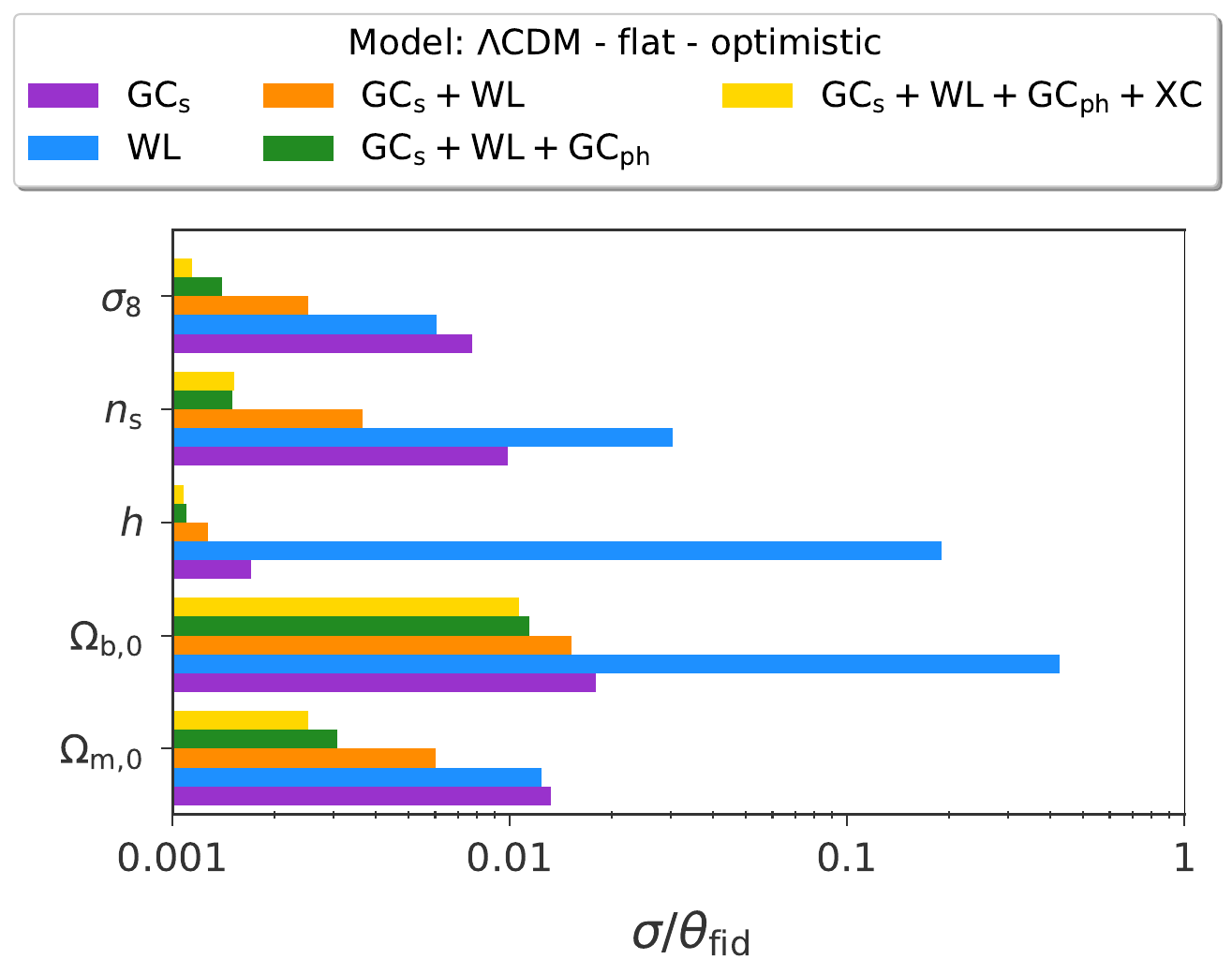} \\
\includegraphics[width=0.5\textwidth]{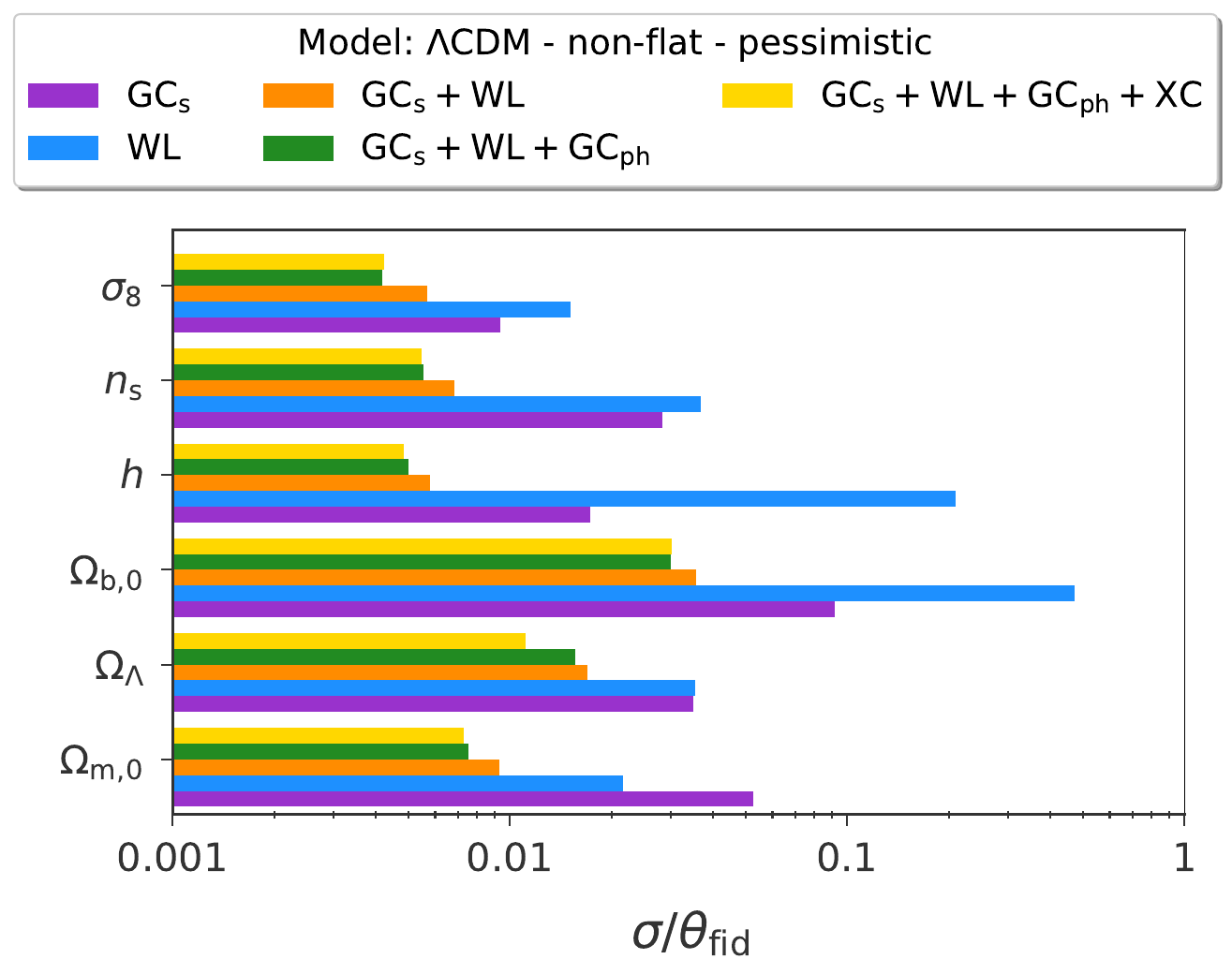}\includegraphics[width=0.5\textwidth]{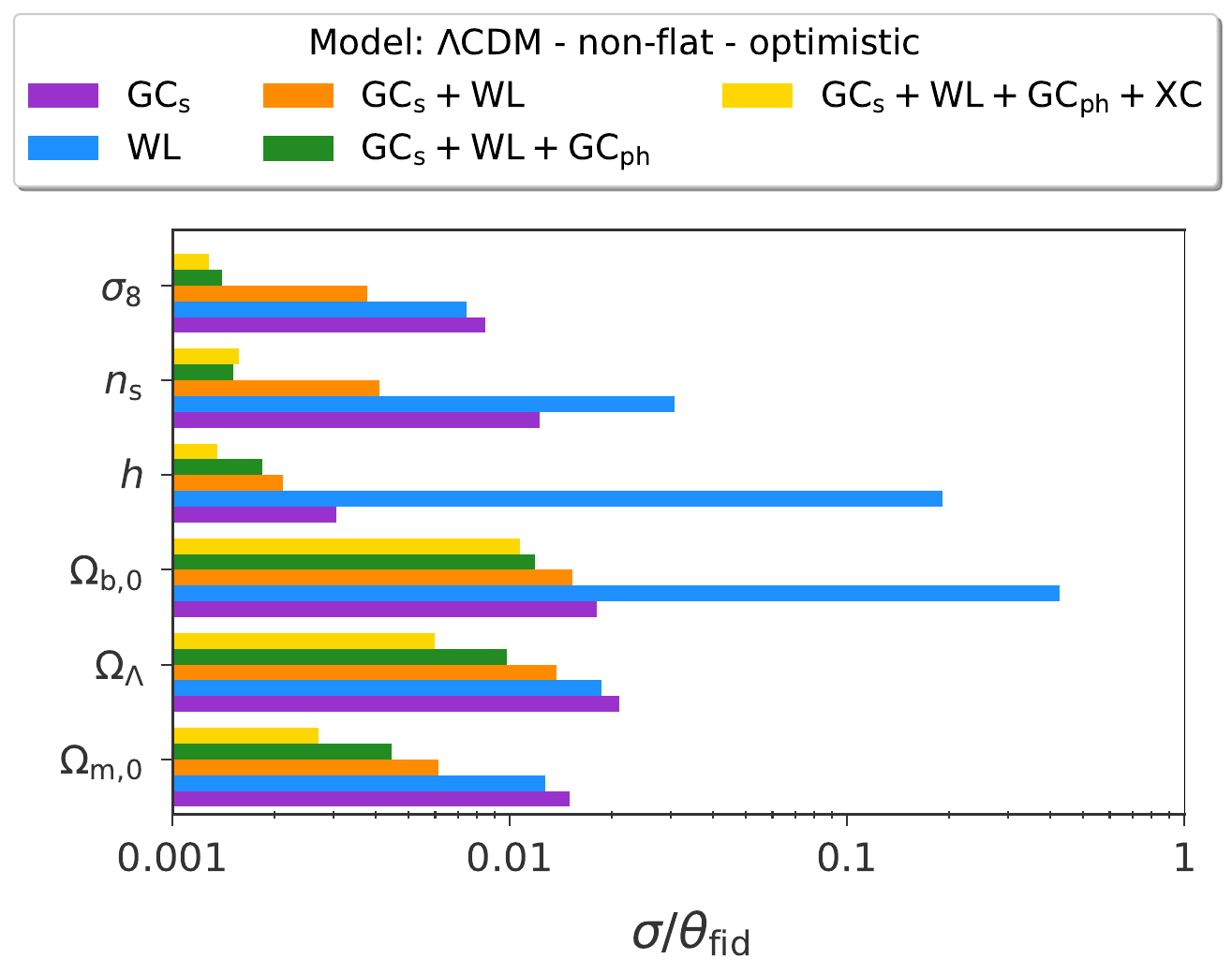}
\end{tabular}
\caption{Marginalised $1 \sigma$ errors on cosmological parameters (from the Fisher matrix for an indicative code), relative to their corresponding fiducial value for a flat (upper panels) and non-flat (lower panels) spatial geometry, in $\Lambda$CDM. Left (right) panels correspond to pessimistic (optimistic) settings, as described in the text. Each histogram refers to different observational probes. {For both a spatially flat and a spatially non-flat cosmology, we show results for GC$_{\rm s}$, WL, GC$_{\rm s}$+WL, GC$_{\rm s}$+WL+GC$_{\rm ph}$, and GC$_{\rm s}$+WL+GC$_{\rm ph}$+XC.}}
\label{fig:err-GC-WL-lcdm}
\end{figure}

We start by presenting results for a cosmological constant $\Lambda$CDM scenario. The marginalised $1\sigma$ errors on cosmological parameters, relative to the corresponding fiducial values, are shown in \autoref{fig:err-GC-WL-lcdm} for the pessimistic (left panels) and optimistic (right panels) settings. Top (bottom) panels refer to a spatially flat (non-flat) cosmology. Relative marginalised errors are also detailed in \autoref{tab:rel-errors-lcdm}.  

\begin{table}[h]
\renewcommand{\arraystretch}{1.5}
\caption{Marginalised $1 \sigma$ errors on cosmological parameters (from the Fisher matrix for an indicative code), relative to their corresponding fiducial value for a flat spatial geometry, in $\Lambda$CDM. We show results for the linear, the pessimistic, and the optimistic settings, as described in the text.} 
  \centering
\begin{tabularx}{\textwidth}{Xlllll}
\hline
\rowcolor{cosmiclatte} \multicolumn{6} {c}{{\bf{$\boldsymbol\Lambda$CDM flat}}}\\ 
 & \multicolumn{1}{c}{$\Omega_{\rm m,0}$} & \multicolumn{1}{c}{$\Omega_{\rm b,0}$} & \multicolumn{1}{c}{$h$} & \multicolumn{1}{c}{$n_{\rm s}$} & \multicolumn{1}{c}{$\sigma_{8}$} \\
\hline
\rowcolor{lavender(web)} \multicolumn{6} {l}{{{Linear setting}}} \\ 
  GC$_{\rm s}$ & 0.016 & 0.026 & 0.0031 & 0.0093 & 0.0071  \\
\hline
\rowcolor{lavender(web)} \multicolumn{6} {l}{{{Pessimistic setting}}} \\ 
  GC$_{\rm s}$ & 0.021 & 0.051 &  0.0063&   0.014 &  0.0094\\
  WL & 0.018 & 0.47 & 0.21 & 0.035 & 0.0087  \\
  GC$_{\rm s}$+WL & 0.009 & 0.032 & 0.0041 & 0.0068 & 0.0041 \\
  GC$_{\rm ph}$+WL & 0.0095 & 0.052 & 0.029 & 0.0098 & 0.0044\\
  GC$_{\rm s}$+WL+GC$_{\rm ph}$ & 0.0075 & 0.028& 0.0038 &0.0055 &0.0035 \\
  WL+GC$_{\rm ph}$+XC$^{^{\rm (GC_{\rm ph},WL)}}$ & 0.0081& 0.052 &0.027 &0.0085 &0.0038\\
  GC$_{\rm s}$+WL+GC$_{\rm ph}$+XC$^{^{\rm (GC_{\rm ph},WL)}}$ & 0.0071 &0.026 & 0.0037 &0.0053 &0.0033 \\
  \hline
\rowcolor{lavender(web)}    \multicolumn{6} {l}{{{Optimistic setting}}} \\ 
 GC$_{\rm s}$ & 0.013 &0.018 &0.0017 &0.0099& 0.0077 \\
WL & 0.012 &0.42& 0.20 & 0.030 &0.0061 \\
 GC$_{\rm s}$+WL & 0.0060  & 0.015 &0.0013 &0.0036 &0.0026 \\
 GC$_{\rm ph}$+WL & 0.0038 & 0.046 & 0.020 & 0.0037 &0.0017\\
GC$_{\rm s}$+WL+GC$_{\rm ph}$ & 0.0031 & 0.011 & 0.0011& 0.0015&  0.0014\\
WL+GC$_{\rm ph}$+XC$^{^{\rm (GC_{\rm ph},WL)}}$ &0.0028 & 0.046 & 0.020 & 0.0036 &0.0013\\
GC$_{\rm s}$+WL+GC$_{\rm ph}$+XC$^{^{\rm (GC_{\rm ph},WL)}}$ & 0.0025 & 0.011 & 0.0011 & 0.0015 & 0.0011\\
   \hline   
\end{tabularx}
\label{tab:rel-errors-lcdm}
\end{table}

\begin{table}[h]
\renewcommand{\arraystretch}{1.5}
\caption{Same as \autoref{tab:rel-errors-lcdm} for a spatially non-flat $\Lambda$CDM cosmology.}
\centering
\begin{tabularx}{\textwidth}{Xllllll}
\hline
\rowcolor{cosmiclatte} \multicolumn{7} {c}{{\bf{$\boldsymbol\Lambda$CDM non-flat }}}\\ 
 & \multicolumn{1}{c}{$\Omega_{\rm m,0}$} & \multicolumn{1}{c}{$\Omega_{\rm DE,0}$} & \multicolumn{1}{c}{$\Omega_{\rm b,0}$} & \multicolumn{1}{c}{$h$} & \multicolumn{1}{c}{$n_{\rm s}$} & \multicolumn{1}{c}{$\sigma_{8}$} \\
\hline
\rowcolor{lavender(web)}\multicolumn{7} {l}{{{Linear setting}}} \\ 
GC$_{\rm s}$ & 0.022 & 0.021& 0.033   & 0.0067 & 0.013 & 0.0073 \\
\hline
\rowcolor{lavender(web)}\multicolumn{7} {l}{{{Pessimistic setting}}} \\ 
GC$_{\rm s}$ & 0.053 & 0.035 &0.092 &  0.017 & 0.028 & 0.0094  \\
WL & 0.021 & 0.035 & 0.47 & 0.21& 0.035& 0.015\\
GC$_{\rm s}$+WL & 0.0093 & 0.017 & 0.036& 0.0058& 0.0068 & 0.0057\\
GC$_{\rm ph}$+WL & 0.011 & 0.021& 0.054& 0.030 & 0.010 & 0.0044\\
GC$_{\rm s}$+WL+GC$_{\rm ph}$ & 0.0075 & 0.016 & 0.030 & 0.0050 &0.0055 &0.0042\\
WL+GC$_{\rm ph}$+XC$^{^{\rm (GC_{\rm ph},WL)}}$ & 0.0082 & 0.011& 0.054 &0.029& 0.0098& 0.0044 \\
GC$_{\rm s}$+WL+GC$_{\rm ph}$+XC$^{^{\rm (GC_{\rm ph},WL)}}$ &0.0073 & 0.011 &0.030 & 0.0048& 0.0055& 0.0042\\
  \hline
\rowcolor{lavender(web)}   \multicolumn{7} {l}{{{Optimistic setting}}} \\ 

 GC$_{\rm s}$ & 0.015 & 0.018 & 0.021 & 0.0031 & 0.012 & 0.0085  \\
WL &0.013 & 0.019 & 0.43 & 0.20 & 0.031 & 0.0075\\
GC$_{\rm s}$+WL& 0.0061 & 0.014 &  0.015 &  0.0021 & 0.0041 & 0.0038\\
GC$_{\rm ph}$+WL& 0.0075 & 0.015 & 0.047 & 0.021 & 0.0037 & 0.0019 \\
GC$_{\rm s}$+WL+GC$_{\rm ph}$ & 0.0045 & 0.0098& 0.012& 0.0018& 0.0015& 0.0014\\
WL+GC$_{\rm ph}$+XC$^{^{\rm (GC_{\rm ph},WL)}}$ &  0.0030 & 0.0064 &0.046 & 0.020& 0.0037& 0.0014\\
GC$_{\rm s}$+WL+GC$_{\rm ph}$+XC$^{^{\rm (GC_{\rm ph},WL)}}$ & 0.0027 & 0.0060 & 0.011& 0.0014 &0.0016 &0.0013\\
  \hline
\end{tabularx}
\label{tab:rel-errors-lcdmnonflat}
\end{table}

The combination of GC$_{\rm s}$ and WL is very powerful in breaking degeneracies among cosmological parameters and reduces the error on all parameters down to at least 0.3$\%$ in the flat case and at least the percent level in the non-flat case, both in the pessimistic and in the optimistic settings.  
When the information of GC$_{\rm ph}$ is added (green bars in \autoref{fig:err-GC-WL-lcdm}) the errors are further reduced for all cases and parameters; on the other hand, the gain in adding cross-correlation terms (XC) among GC$_{\rm ph}$ and WL (yellow bars) is very small in the $\Lambda$CDM scenario, and could even degrade the results relative to an optimistic case when added in a pessimistic setting. However where results get worse when adding more information, we attribute this to numerical instability of the Fisher matrix formalism itself (even despite the high accuracy we achieve in this paper).

This situation  changes for more complex models, where we see that XC has a much larger (positive) impact. 
When comparing the pessimistic and optimistic errors (\autoref{tab:rel-errors-lcdm}), GC$_{\rm s}$ is the probe that gains most from the inclusion of higher modes.  Finally, relative errors for a non-flat $\Lambda$CDM cosmology, shown in \autoref{fig:err-GC-WL-lcdm} (lower panels) and in \autoref{tab:rel-errors-lcdmnonflat} are, as expected, larger, since we are adding an extra parameter.


\subsection{Changing the background: $w_0$, $w_a$}
\begin{figure}[h!]
\center 
\begin{tabular}{c}
\includegraphics[width=0.5\textwidth]{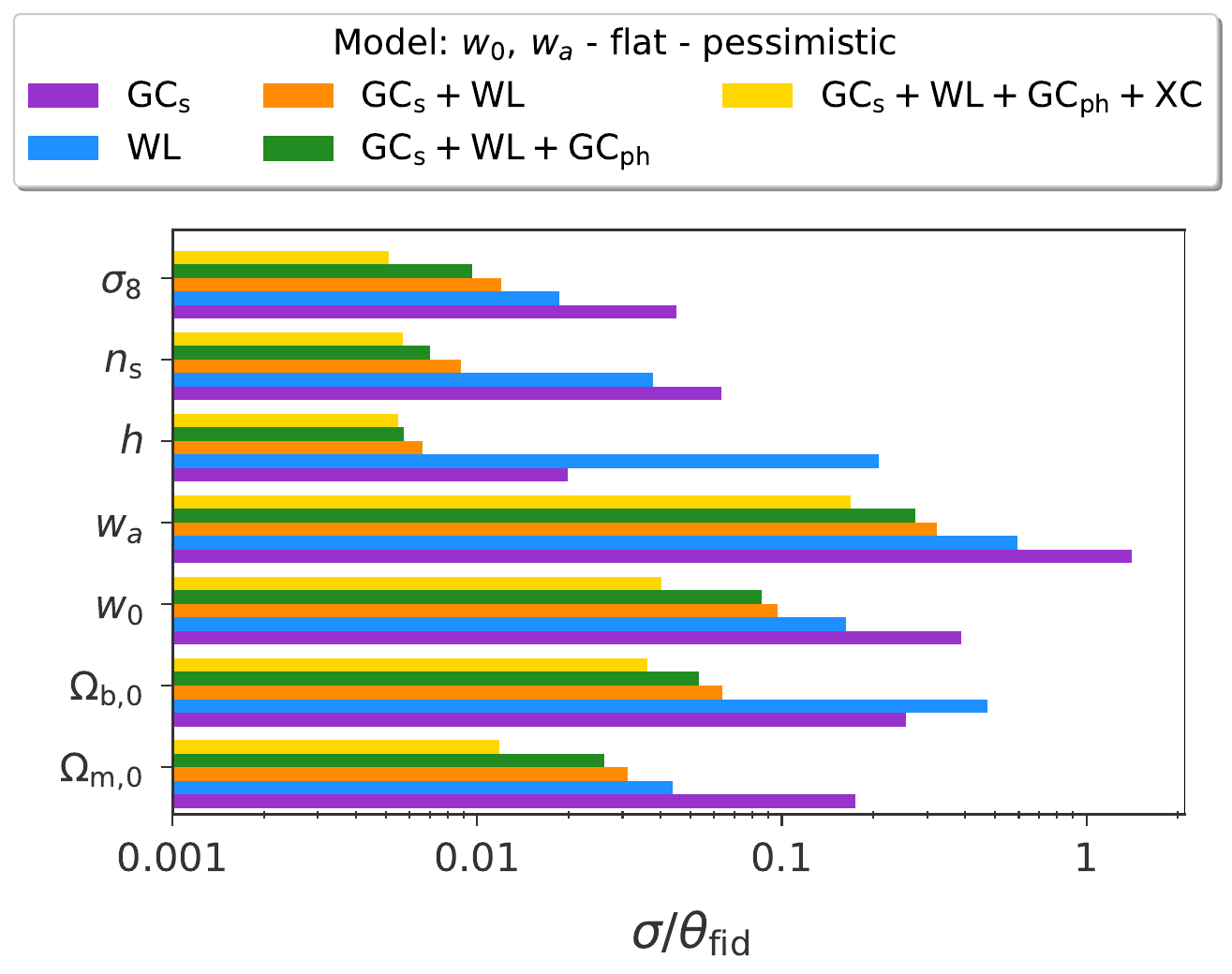} \includegraphics[width=0.5\textwidth]{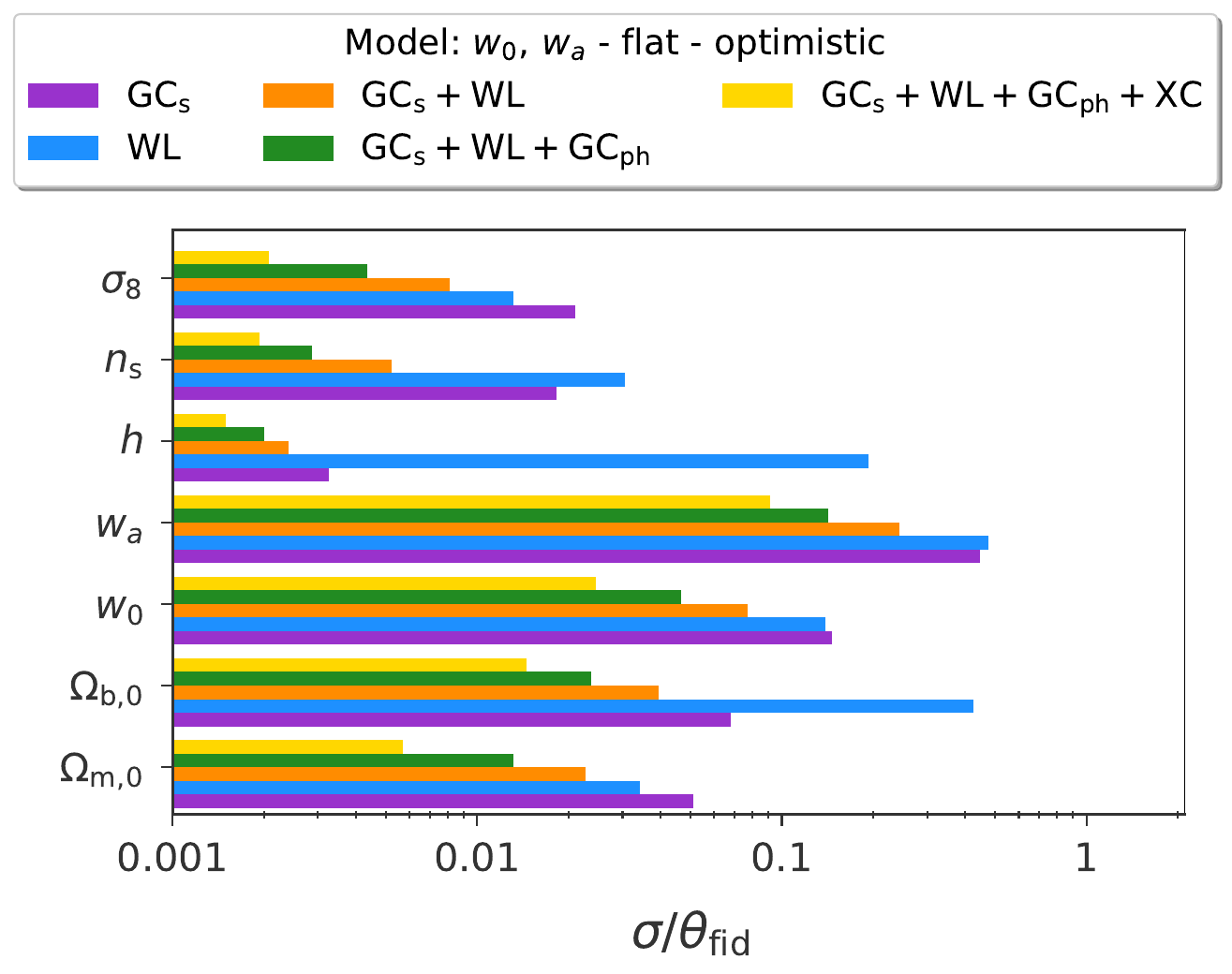} \\
\includegraphics[width=0.5\textwidth]{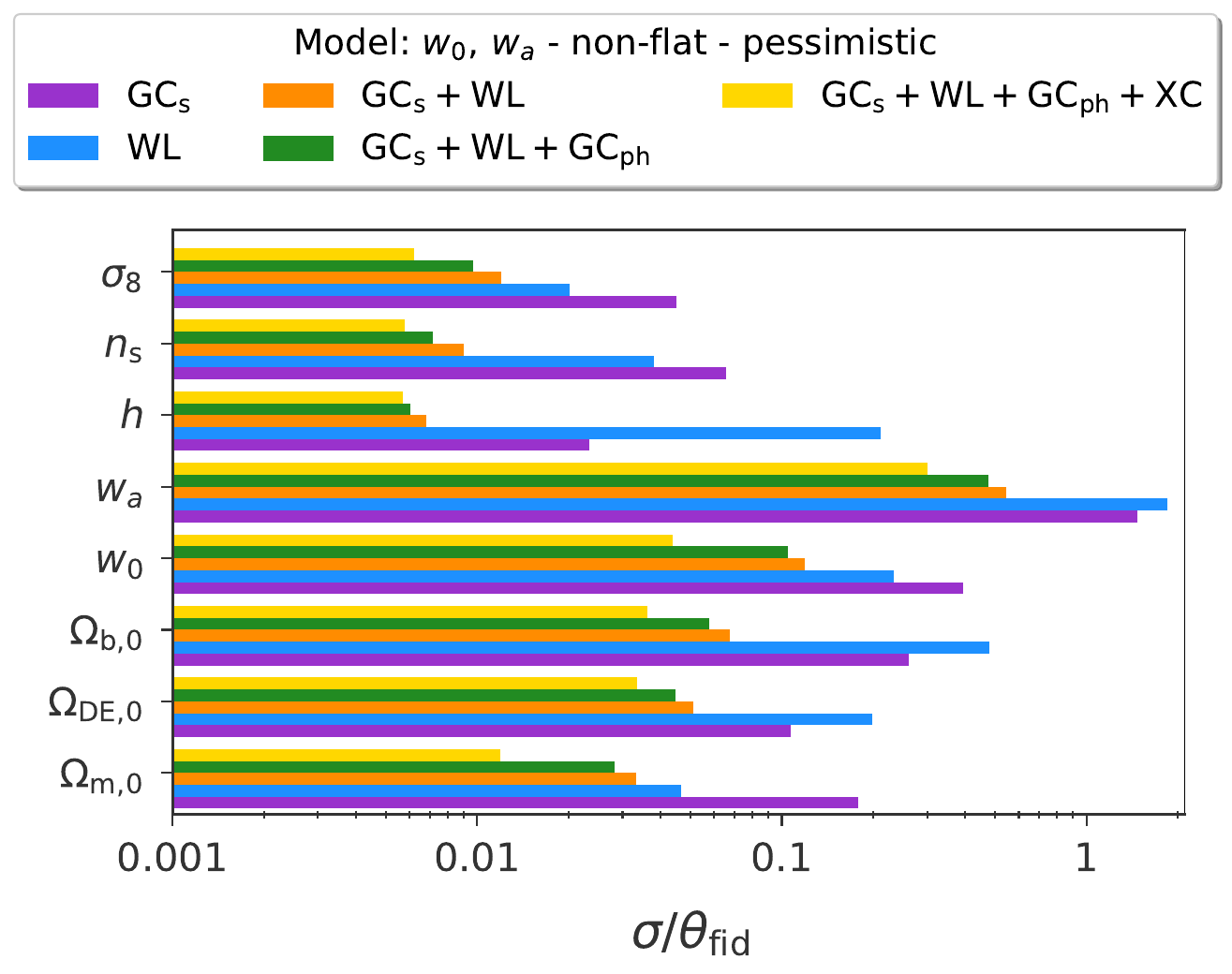}\includegraphics[width=0.5\textwidth]{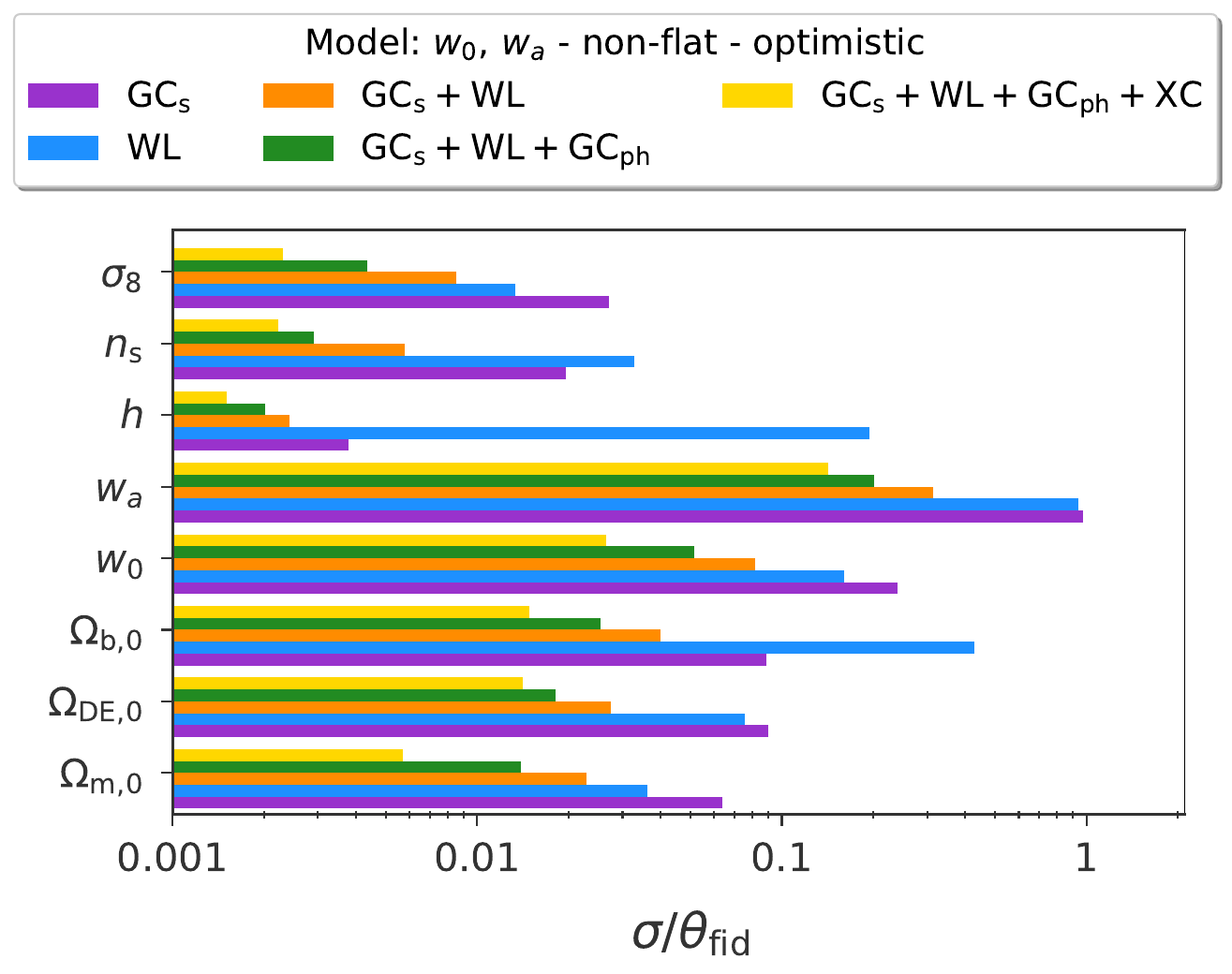}
\end{tabular}
\caption{Marginalised $1 \sigma$ errors on cosmological parameters, relative to their corresponding fiducial values for a flat (upper panels) and non-flat (lower panels) spatial geometry, in the ($w_{0}, w_a$) cosmology. Left (right) panels correspond to pessimistic (optimistic) settings, as described in the text. The histogram refers to different observational probes. {For both a spatially flat and a spatially non-flat cosmology, we show results for GC$_{\rm s}$, WL, GC$_{\rm s}$+WL, GC$_{\rm s}$+WL+GC$_{\rm ph}$, and GC$_{\rm s}$+WL+GC$_{\rm ph}$+XC}. For $w_a$ we show the absolute error, since a relative error is not possible for a fiducial value of $0$.} 
\label{fig:err-GC-WL-w0wa}
\end{figure}

We now move to the case in which the equation of state of a dark-energy component is not $w=-1$ at all times (as in a cosmological constant case), but can vary in time. We consider the case in which such variation in time is parameterised by $w_0$, $w_a$, as defined in \autoref{eq:weos}.
Results on the marginalised errors forecasted for the \Euclid survey are shown in \autoref{fig:err-GC-WL-w0wa}. As before, left (right) panels in \autoref{fig:err-GC-WL-w0wa}  refer to the pessimistic (optimistic) settings, while top (bottom) panels refer to a flat (non-flat) cosmology. Fisher forecasts assume a $\Lambda$CDM fiducial model, around which contours are drawn. Marginalised errors for different combinations of the observables and different settings are shown in more detail in \Cref{tab:rel-errors-w0waflat,tab:rel-errors-w0wanonflat}.
The uncertainty on all cosmological parameters is reduced when GC$_{\rm s}$ and WL are combined. Furthermore, unlike what happens in the $\Lambda$CDM case, cross-correlations have a significant impact on the estimation of the errors, and help to further tighten uncertainties, in particular on $w_0$. 
\begin{table}[h!]
\renewcommand{\arraystretch}{1.5}
\caption{Same as \autoref{tab:rel-errors-lcdm} for flat $w_0, w_a$ cosmology.}
\centering
\begin{tabularx}{\textwidth}{Xlllllll}
\hline
\rowcolor{jonquil} \multicolumn{8} {c}{{\boldsymbol{${w_0, w_a} $ {\bf{ flat}}}}}\\ 
 & \multicolumn{1}{c}{$\Omega_{\rm m,0}$} & \multicolumn{1}{c}{$\Omega_{\rm b,0}$} & \multicolumn{1}{c}{$w_0$} & \multicolumn{1}{c}{$w_a$} & \multicolumn{1}{c}{$h$} & \multicolumn{1}{c}{$n_{\rm s}$} & \multicolumn{1}{c}{$\sigma_{8}$} \\
\hline
\rowcolor{lavender(web)}\multicolumn{8} {l}{{{Linear setting}}} \\ 
GC$_{\rm s}$ & 0.080 & 0.10 & 0.20 & 0.66 & 0.0063 & 0.030 & 0.024 \\
\hline
\rowcolor{lavender(web)}\multicolumn{8} {l}{{{Pessimistic setting}}} \\ 
GC$_{\rm s}$ & 0.17  & 0.26 & 0.39 & 1.4 & 0.020 & 0.063 & 0.045 \\
WL & 0.044 & 0.47&  0.16& 0.59& 0.21& 0.038& 0.019\\
GC$_{\rm s}$+WL & 0.031& 0.064& 0.097& 0.32 &0.0066 &0.0089& 0.012\\
GC$_{\rm ph}$+WL &0.036 &0.068& 0.14& 0.47& 0.031& 0.011& 0.013\\
GC$_{\rm s}$+WL+GC$_{\rm ph}$ & 0.026& 0.053& 0.086& 0.27& 0.0057 &0.0070 &0.0096 \\
WL+GC$_{\rm ph}$+XC$^{^{\rm (GC_{ph},WL)}}$ & 0.011& 0.054& 0.042& 0.17& 0.029& 0.010&0.0048\\
GC$_{\rm s}$+WL+GC$_{\rm ph}$+XC$^{^{\rm (GC_{ph},WL)}}$ & 0.012& 0.036& 0.040& 0.17& 0.0055& 0.0057& 0.0051\\
\hline
\rowcolor{lavender(web)}\multicolumn{8} {l}{{{Optimistic setting}}} \\ 
GC$_{\rm s}$ & 0.051 & 0.068&  0.15 &  0.45 &  0.0032 & 0.018 & 0.021 \\
WL & 0.034 & 0.42  & 0.14 & 0.48 & 0.20 & 0.030 & 0.013 \\
GC$_{\rm s}$+WL & 0.022 &  0.039 & 0.077 & 0.24 & 0.0024 & 0.0052 & 0.0081 \\
GC$_{\rm ph}$+WL & 0.015 & 0.048& 0.063& 0.21& 0.021& 0.0043 &0.0049\\
GC$_{\rm s}$+WL+GC$_{\rm ph}$ & 0.013& 0.024& 0.047& 0.14& 0.0020& 0.0029& 0.0043 \\
WL+GC$_{\rm ph}$+XC$^{^{\rm (GC_{ph},WL)}}$ &0.0059 & 0.046& 0.027& 0.10& 0.020&0.0039 &0.0022\\
GC$_{\rm s}$+WL+GC$_{\rm ph}$+XC$^{^{\rm (GC_{ph},WL)}}$ & 0.0057& 0.015& 0.025& 0.092&0.0015 &0.0019 &0.0021\\
\hline
\end{tabularx}
\label{tab:rel-errors-w0waflat}
\end{table}
\begin{table}[h!]
\renewcommand{\arraystretch}{1.5}
\caption{Same as \autoref{tab:rel-errors-lcdm} for a non-flat $w_0, w_a$ cosmology.}
\centering
\begin{tabularx}{\textwidth}{Xllllllll}
\hline
\rowcolor{jonquil} \multicolumn{9} {c}{$\boldsymbol{w_0, w_a}$ {\bf{non-flat}} }\\ 
 & \multicolumn{1}{c}{$\Omega_{\rm m,0}$} &\multicolumn{1}{c}{$\Omega_{\rm DE,0}$} & \multicolumn{1}{c}{$\Omega_{\rm b,0}$} &  \multicolumn{1}{c}{$w_0$} & \multicolumn{1}{c}{$w_a$} & \multicolumn{1}{c}{$h$} & \multicolumn{1}{c}{$n_{\rm s}$} & \multicolumn{1}{c}{$\sigma_{8}$} \\
\hline
\rowcolor{lavender(web)}\multicolumn{9} {l}{Linear setting} \\
GC$_{\rm s}$ & 0.083 & 0.075 &0.10 &  0.23 & 0.89 & 0.0068 & 0.031 & 0.027 \\
\hline
\rowcolor{lavender(web)}\multicolumn{9} {l}{Pessimistic setting} \\ 
GC$_{\rm s}$ & 0.18 & 0.11 & 0.26 &  0.39 & 1.5 & 0.023 & 0.066 & 0.045 \\
WL & 0.047 & 0.20 &  0.48 & 0.23& 1.85 & 0.21& 0.038& 0.020\\
GC$_{\rm s}$+WL & 0.033 & 0.051 & 0.067 & 0.12 & 0.55 & 0.0068 & 0.0090 & 0.012 \\
GC$_{\rm ph}$+WL & 0.036 &0.083& 0.068& 0.15 &0.75& 0.032& 0.011& 0.014\\
GC$_{\rm s}$+WL+GC$_{\rm ph}$ & 0.028& 0.045& 0.058& 0.11& 0.47& 0.0060 &0.0071& 0.0097\\
WL+GC$_{\rm ph}$+XC$^{^{\rm (GC_{ph},WL)}}$ &0.011 &  0.064& 0.054& 0.050& 0.53& 0.029& 0.011& 0.0073\\
GC$_{\rm s}$+WL+GC$_{\rm ph}$+XC$^{^{\rm (GC_{ph},WL)}}$ & 0.012& 0.033& 0.036& 0.044& 0.30 &0.0057 &0.0058& 0.0062\\
\hline
\rowcolor{lavender(web)}\multicolumn{9} {l}{Optimistic setting} \\ 
GC$_{\rm s}$ & 0.064 & 0.090 & 0.089 &  0.24 & 0.97 & 0.0038 & 0.020 & 0.027 \\
WL & 0.036 & 0.076 & 0.43 & 0.16 & 0.94 & 0.19 & 0.033 & 0.013\\
GC$_{\rm s}$+WL & 0.023 & 0.027 & 0.040 & 0.082 & 0.31 & 0.0024 & 0.0058 & 0.0085\\
GC$_{\rm ph}$+WL & 0.015 &  0.024 & 0.049 & 0.063 & 0.24 & 0.021 &  0.0047 & 0.0050\\
GC$_{\rm s}$+WL+GC$_{\rm ph}$ & 0.014 &0.018& 0.025& 0.052 &0.20 & 0.0020& 0.0029& 0.0044\\
WL+GC$_{\rm ph}$+XC$^{^{\rm (GC_{ph},WL)}}$ & 0.0059& 0.018& 0.047& 0.029& 0.17& 0.020& 0.0042 &0.0025\\
GC$_{\rm s}$+WL+GC$_{\rm ph}$+XC$^{^{\rm (GC_{ph},WL)}}$ & 0.0057 & 0.014& 0.015& 0.026& 0.14& 0.0015 &0.0022 &0.0023\\
\hline
\end{tabularx}
\label{tab:rel-errors-w0wanonflat}
\end{table}

For this set of cosmologies, we can also estimate the FoM for $w_0, w_a$, as defined in \autoref{FoM}, which is an estimate of the performance of the experiment in constraining a specific set of parameters. Results for the FoM for $w_0, w_a$ are shown in \autoref{w0wa_table_FoM} for both the single probes (GC$_{\rm s}$, GC$_{\rm ph}$, WL separately) and their different combinations, within the pessimistic and optimistic settings described in \autoref{sec:results}. 

The combination of all three probes (without cross-correlations) reaches a FoM of 122 (376) in the pessimistic (optimistic) settings, for a spatially flat cosmology. Including cross-correlations has a substantial (positive) impact on the FoM, which reaches 441 in the pessimistic setting (i.e. enhancing it by a factor of 3.5) and improves up to 1245 in an optimistic setting (i.e. a factor of 3.3 larger than in absence of cross-correlations). This demonstrates the importance, for a $w_0, w_a$ scenario, of including cross-correlations in order to fully exploit future \Euclid data. This is confirmed when looking at the correlation matrix, plotted for this model (in the optimistic setting) in \autoref{fig:correlationmatrix} (this plot is obtained following figure 4 in \citealt{2017PDU....18...73C}). The left panel shows the photometric survey combination, without cross-correlation. The matrix is not diagonal, indicating the presence of correlations among cosmological parameters. When XC terms are taken into account (right panel), the correlation matrix non-diagonal terms are reduced, i.e. correlations among the corresponding parameters are removed; this in turn allows for better constraints on those parameters.  
\begin{figure}[h!]
\center 
\begin{tabular}{c}
\includegraphics[width=0.5\textwidth]{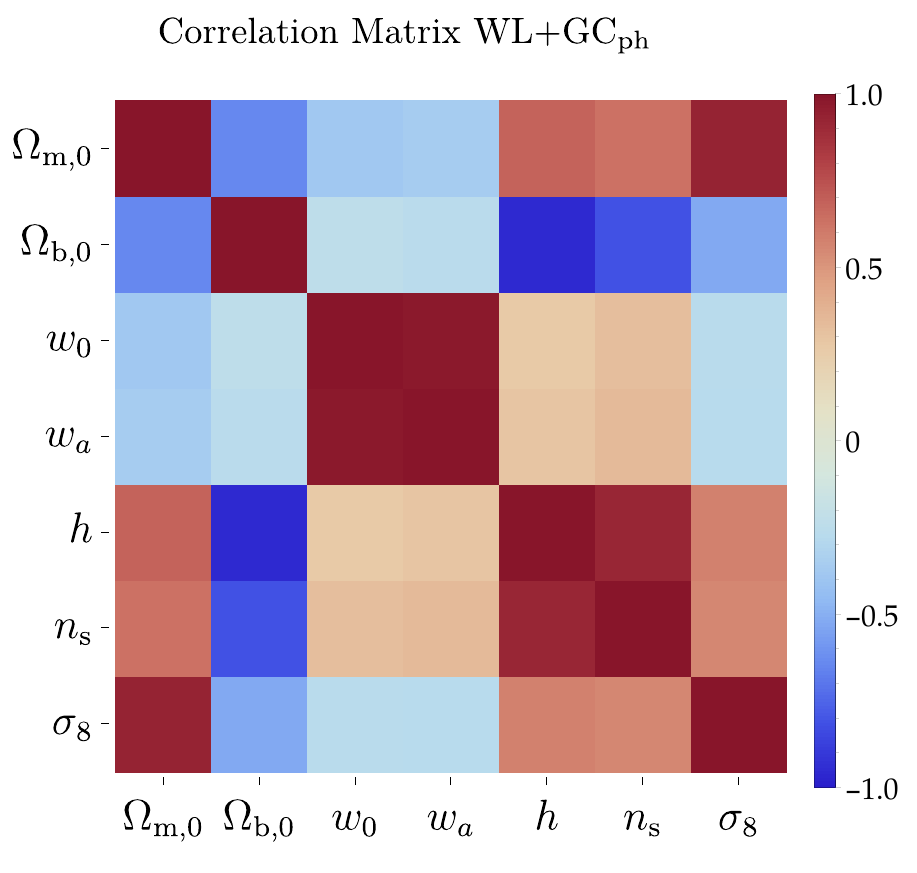} \includegraphics[width=0.5\textwidth]{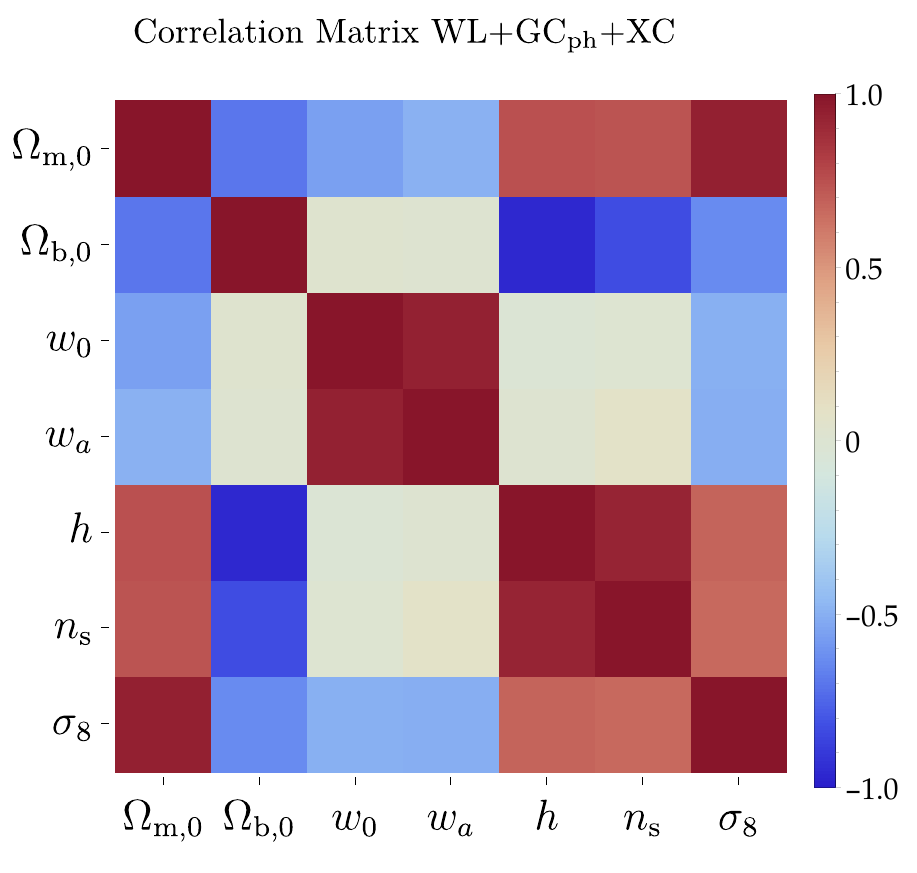}
\end{tabular}
\caption{
Correlation matrix $P$ defined in \autoref{eq:correlation_def}, obtained from the covariance matrix in a ($w_{0}, w_a$) flat cosmology, for an optimistic setting. 
The left panel refers to the photometric survey combination, without cross-correlation. The right panel adds the cross-correlation term. 
The inclusion of cross-correlation between GC$_{\rm ph}$ and WL significantly reduces the correlation among the dark-energy parameters $w_0, w_a$ and the standard parameters $n_{\rm s}$, $h$, and $\Omega_{\rm{b,0}}$. The $\mathrm{FoC}$ reduces from $1907$ in the case with no cross-correlations to $824$ in the $\rm{GC_{\rm ph}}$+WL+$\rm{XC^{(GC_{\rm ph},WL)} }$ case. This, in turn, leads to reduced uncertainties on all parameters.}
\label{fig:correlationmatrix}
\end{figure}

The impact of GC$_{\rm ph}$ is substantially more significant in the optimistic setting, i.e. ideally, there is much gain in being able to include as many multipoles as possible in the analysis, in order to retain information from the GC$_{\rm ph}$. It is important to recall, however, as discussed in \autoref{sub:covariancexc}, that the optimistic setting for this probe entirely neglects non-Gaussian terms in the analysis and still includes all multipoles up to $\ell_{\rm max} = 3000$ for this specific probe; this might be regarded as too optimistic, as non-Gaussian terms are estimated to become important earlier than that, and earlier than for cosmic shear. We still include results for this ideal setting here to show the potential power that GC$_{\rm ph}$ could have, provided we could extend the analysis to such high multipoles. 

A final remark on WL: not including IA would give  FoM of (27, 6) (flat and non flat), slightly higher than the one shown in the table,  in the pessimistic setting for the flat and non-flat cosmologies, respectively, and values (54, 15) in the optimistic setting. However the absence of IA makes the result more unreliable and unrealistic and therefore we do not include these values in the table, but only report them here in the text for comparison with other analyses. 

\begin{figure}[h]
\center \includegraphics[width=0.90\textwidth]{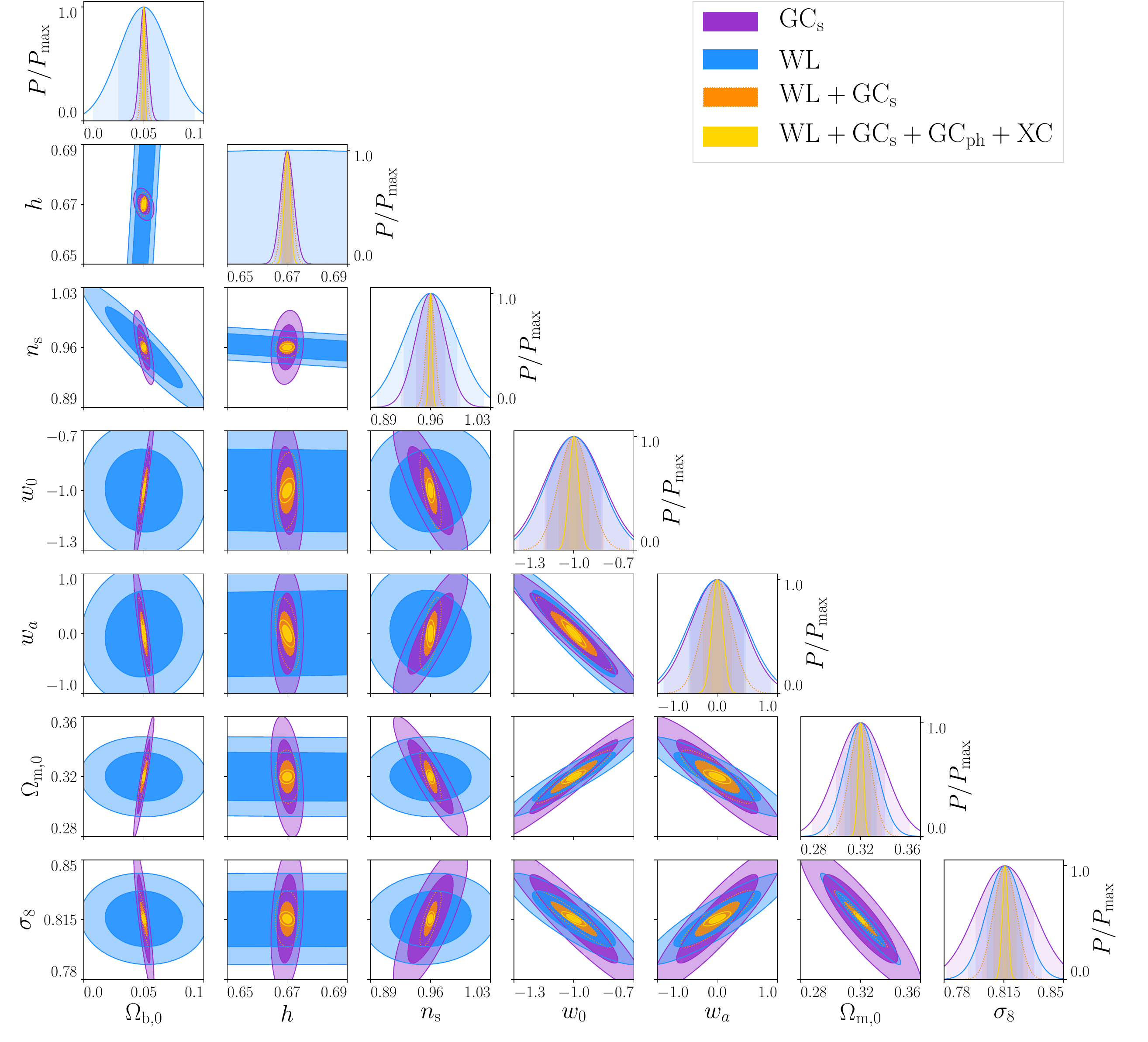}
\caption{Fisher-matrix-marginalised contours for the \Euclid space mission for a $w_0, w_a$ flat cosmology, for GC$_{\rm s}$ (purple), WL (blue), their combination (orange), and with the addition of $\rm{GC_{\rm ph}}$ and its cross-correlation with WL (yellow). All combinations correspond to an optimistic setting, as defined in \autoref{def:optimistic}.
} \label{fig:ellipses-w0wa}
\end{figure}

\begin{figure}[h]
\center 
\includegraphics[width=0.48\textwidth]{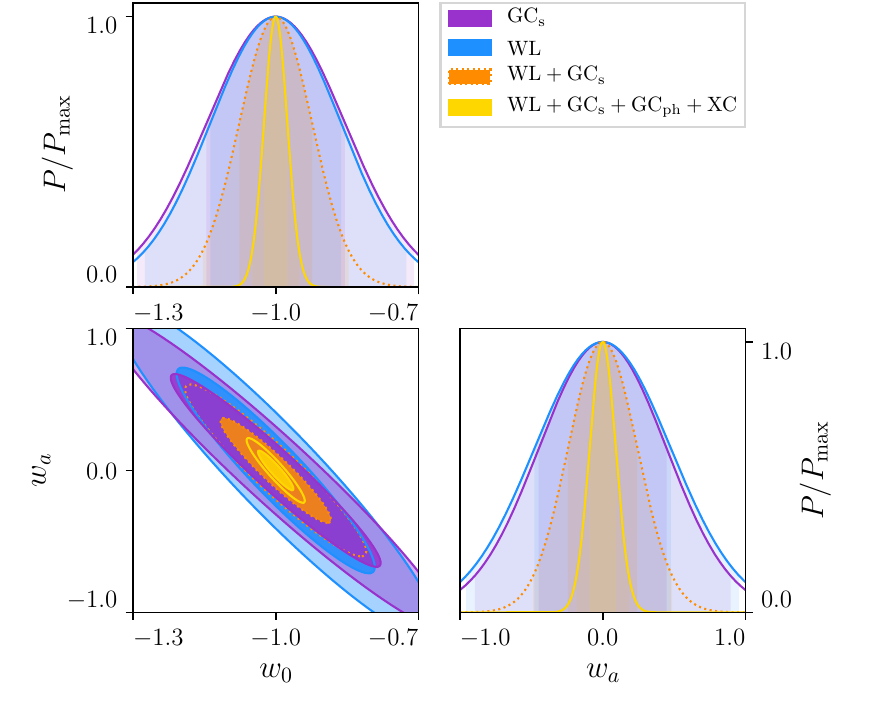}
\includegraphics[width=0.48\textwidth]{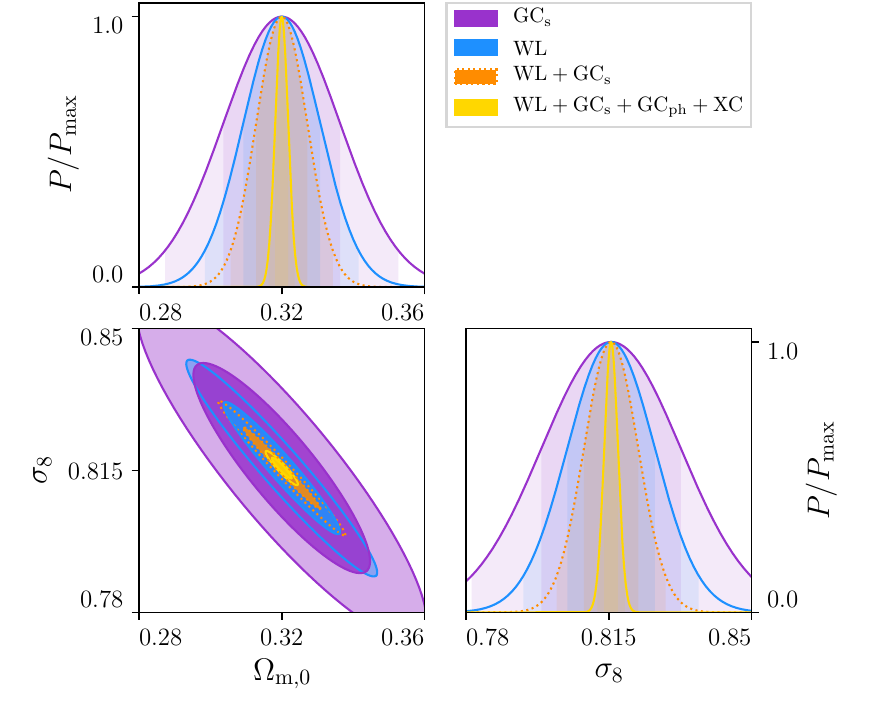}
\caption{Same as in \autoref{fig:ellipses-w0wa}, focusing on the $w_0, w_a$ plane (left panel) and the $\Omega_{\rm m, 0},  \sigma_{\rm 8}$ plane (right panel).} \label{fig:ellipses-w0wa_zoom}
\end{figure}

\begin{table}[h!]
\renewcommand{\arraystretch}{1.5}
\caption{ Figure of merit (FoM) values for the $w_0$ and $w_a$ parameters in flat and non-flat cosmologies for GC$_{\rm s}$, GC$_{\rm ph}$, WL, and different combinations of these observables, including cross-correlations between photometric galaxy clustering and weak lensing. The top (bottom) half of the table refers to the pessimistic (optimistic) settings, as defined in \autoref{sec:results}. For WL alone, not including IA would give a higher (but more unreliable) FoM of (27, 6) in the pessimistic setting, (54, 15) in the optimistic setting, for the flat and non-flat cosmologies respectively. We further note that for the photometric survey only (the combination $\rm{GC_{\rm ph}}$ + WL + $\rm{XC^{(GC_{\rm ph},WL)} }$ in the table), no cut in $z$ is needed in either pessimistic or optimistic settings. When, however both $\rm{GC_{\rm ph}}$ and GC$_{\rm s}$ probes are combined, the pessimistic setting only includes $\rm{GC_{\rm ph}}$ for $z < 0.9$, since we lack an estimate of their cross-correlation, as discussed in \autoref{sec:results}.} 
\centering
\begin{tabularx}{\textwidth}{Xlll}
\hline
\rowcolor{jonquil} \multicolumn{1}{c}{$\boldsymbol{w_0,  w_a}$ {\bf FoM}} &  & {\bf{Flat}}  & {\bf{Non-flat}} \\                                                                                    
\hline
\rowcolor{lavender(web)}\multicolumn{4} {l}{Linear setting} \\ 
$\rm{GC_{\rm s}}$    &   &  	  40 &  19   \\
\hline
\rowcolor{lavender(web)}\multicolumn{4} {l}{Pessimistic setting} \\ 
GC$_{\rm s}$   &     &  14 &  10  \\
WL   &    & 23 &  5 \\
\hline
GC$_{\rm s}$+WL  &      &99 & 40 \\
GC$_{\rm ph}$+WL  &      & 64 & 14\\
GC$_{\rm s}$+WL+GC$_{\rm ph}$ &   & 123 & 49 \\
WL+GC$_{\rm ph}$+XC$^{^{\rm (GC_{ph},WL)}}$ &   & 367 & 59 \\
GC$_{\rm s}$+WL+GC$_{\rm ph}$+XC$^{^{\rm (GC_{ph},WL)}}$ & & 377 & 128 \\
\hline
\rowcolor{lavender(web)}\multicolumn{4} {l}{Optimistic setting} \\ 
GC$_{\rm s}$   &    &   55  &  {19}  \\
WL   &    & 44 &  {12} \\
\hline
GC$_{\rm s}$+WL  &      & 157 &  87\\
GC$_{\rm ph}$+WL   &     & 235 & {129} \\
GC$_{\rm s}$+WL+GC$_{\rm ph}$ &   & 398 & 218 \\
WL+GC$_{\rm ph}$+XC$^{^{\rm (GC_{ph},WL)}}$ &  & 1033 & 326 \\
GC$_{\rm s}$+WL+GC$_{\rm ph}$+XC$^{^{\rm (GC_{ph},WL)}}$  &  & 1257 & 500 \\
\hline
\end{tabularx}
  \label{w0wa_table_FoM}
\end{table}
For the $w_0$, $w_a$ scenario we also show the contour plots in \autoref{fig:ellipses-w0wa} (and an expanded  in \autoref{fig:ellipses-w0wa_zoom}), for the optimistic setting and flat cosmology. 
As expected, parameters characterising dark energy, such as $w_0$, $w_a$, are degenerate with the spectral index $n_{\rm{s}}$, since they both tilt the matter power spectrum at a certain scale. The combination of different probes, as in GC$_{\rm s}$ + WL, helps to break this degeneracy and to tighten constraints on both parameters.

\begin{figure}[h]
\center \includegraphics[width=0.90\textwidth]{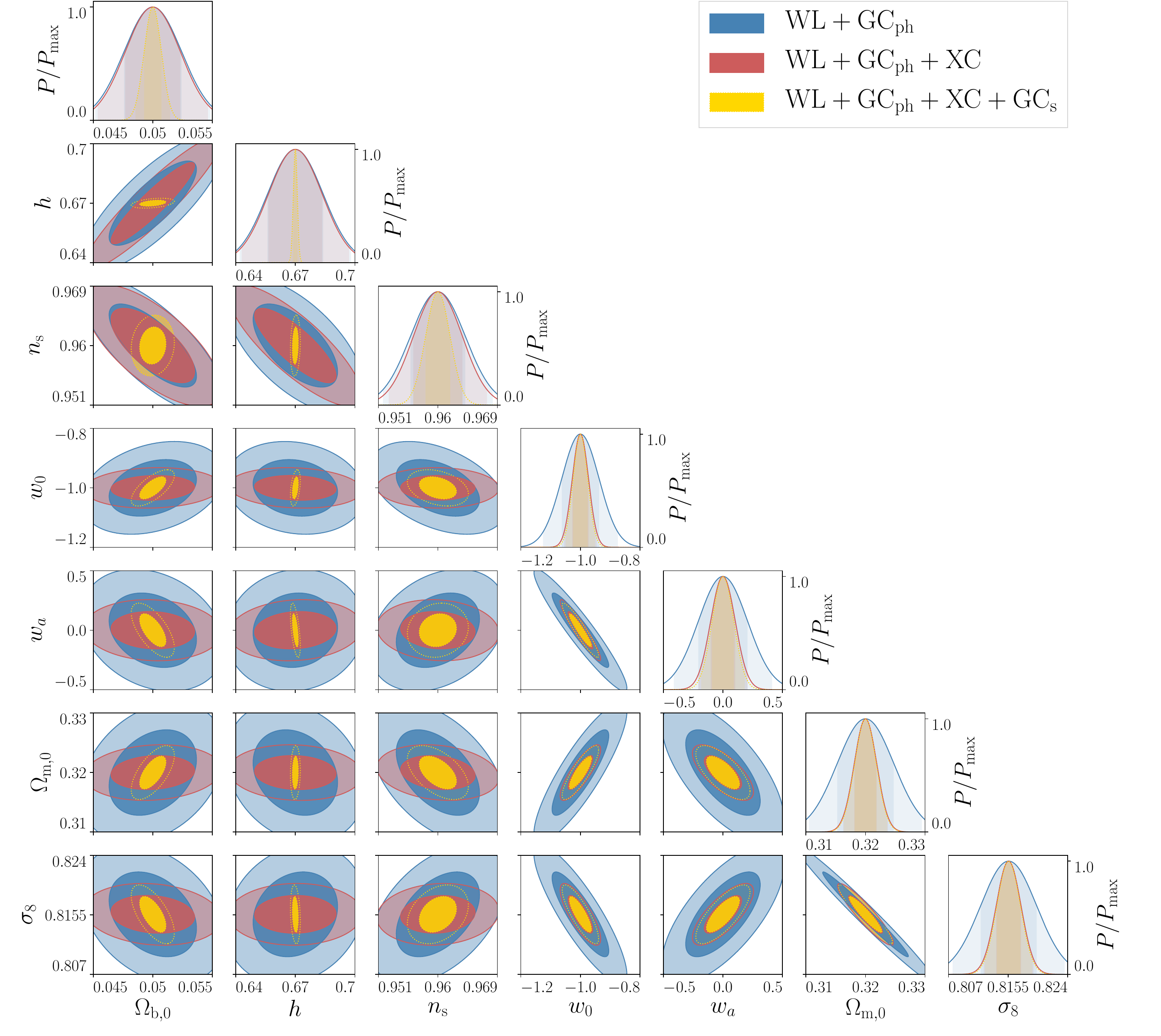}
\caption{Fisher-matrix-marginalised contours for the \Euclid space
  mission for a $w_0, w_a$ cosmology, for a flat cosmology and with an optimistic setting,
of the photometric survey GC$_{\rm ph}$ + WL with (dark red) and without (blue)
their cross-correlation term XC. The combination in yellow is the same as in \autoref{fig:ellipses-w0wa} and includes all probes.}
  \label{fig:ellipses-XC}
\end{figure}

Furthermore, in \autoref{fig:ellipses-XC} we show the photometric survey alone (blue) including $\rm{GC_{\rm ph}}$ and WL, and the impact of their cross-correlations (red), in the absence of any spectroscopic information; the ellipses overlap with the full (yellow) combination, highlighting the impact of adding GC$_{\rm s}$ to the photometric survey.

A comparison of yellow and red contours shows that the inclusion of cross-correlations in the photometric survey causes the impact of GC$_{\rm s}$ to remain quite significant on  $\Omega_{{\rm b},0}$, $h$, $n_{\rm s}$; however GC$_{\rm s}$ has less impact on $w_0$, $w_a$, $\Omega_{{\rm m},0}$, and $\sigma_8$, where the yellow contours almost overlap with the red ones. As discussed above, this is an ideal setting, but again shows the potential of the photometric probe if one could extend the analysis to high multipoles.


\subsection{Changing the growth: $w_0$, $w_a, \gamma$}

\begin{figure}[h]
\center 
\begin{tabular}{c}
\includegraphics[width=0.5\textwidth]{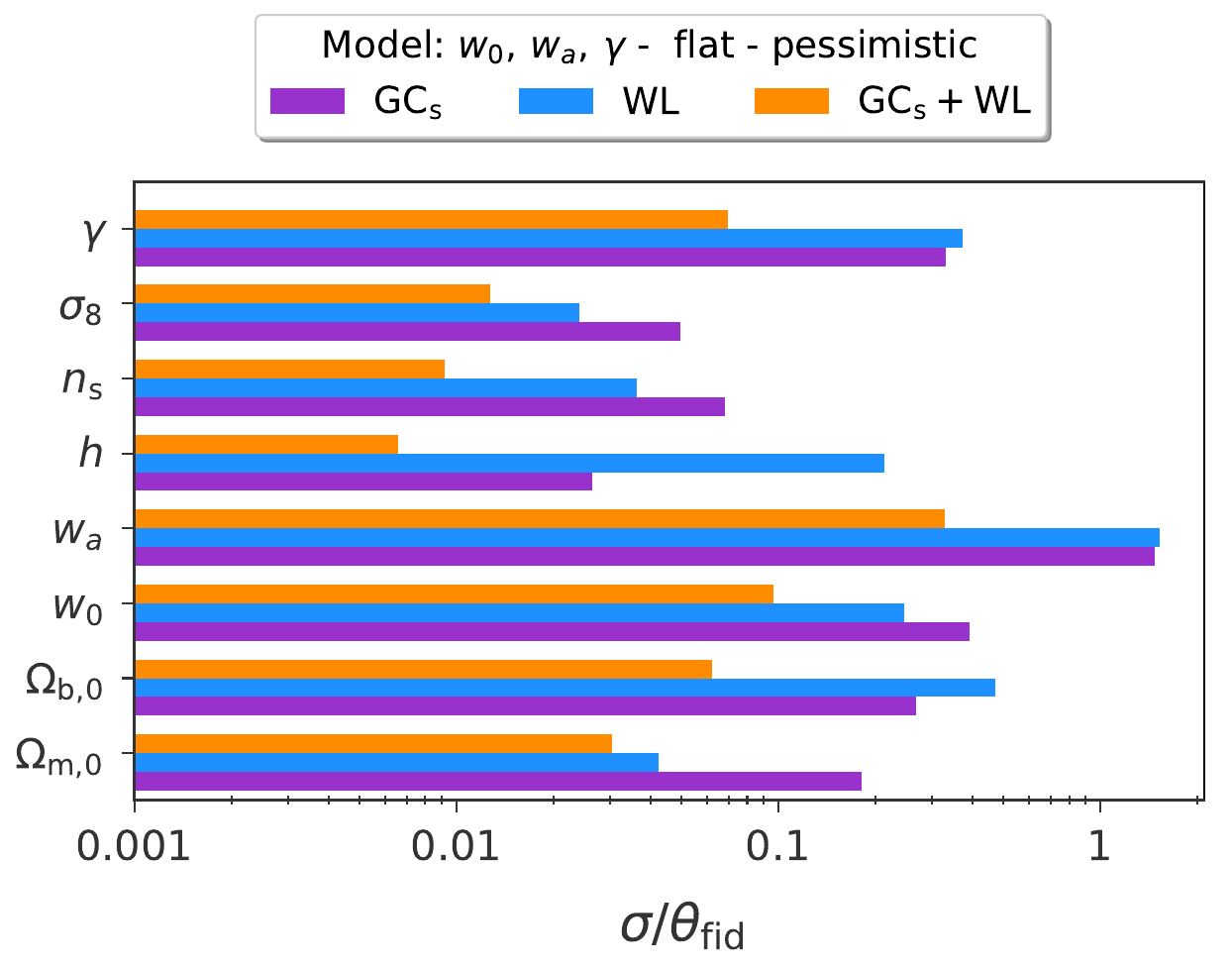}\includegraphics[width=0.5\textwidth]{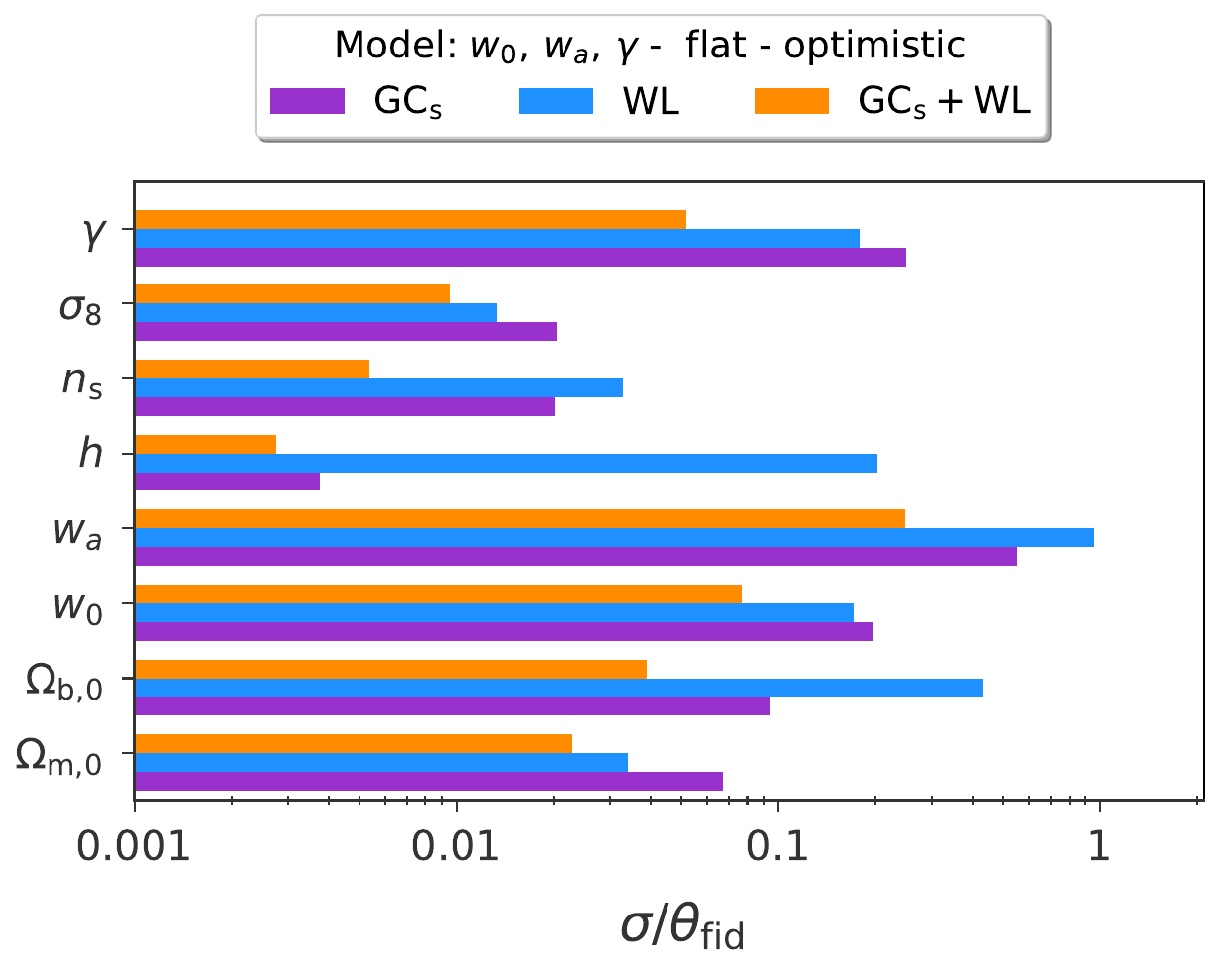} \\
\includegraphics[width=0.5\textwidth]{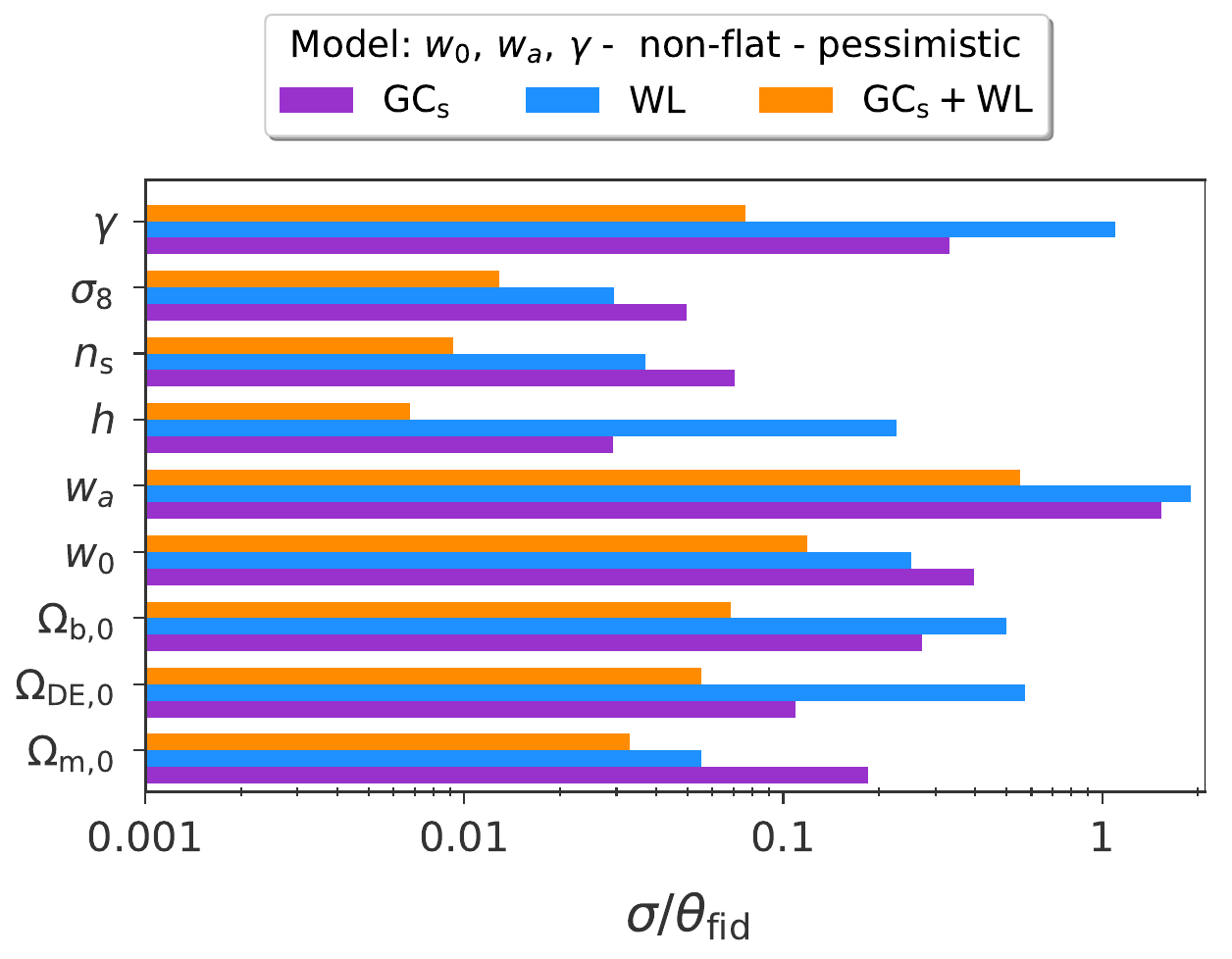}\includegraphics[width=0.5\textwidth]{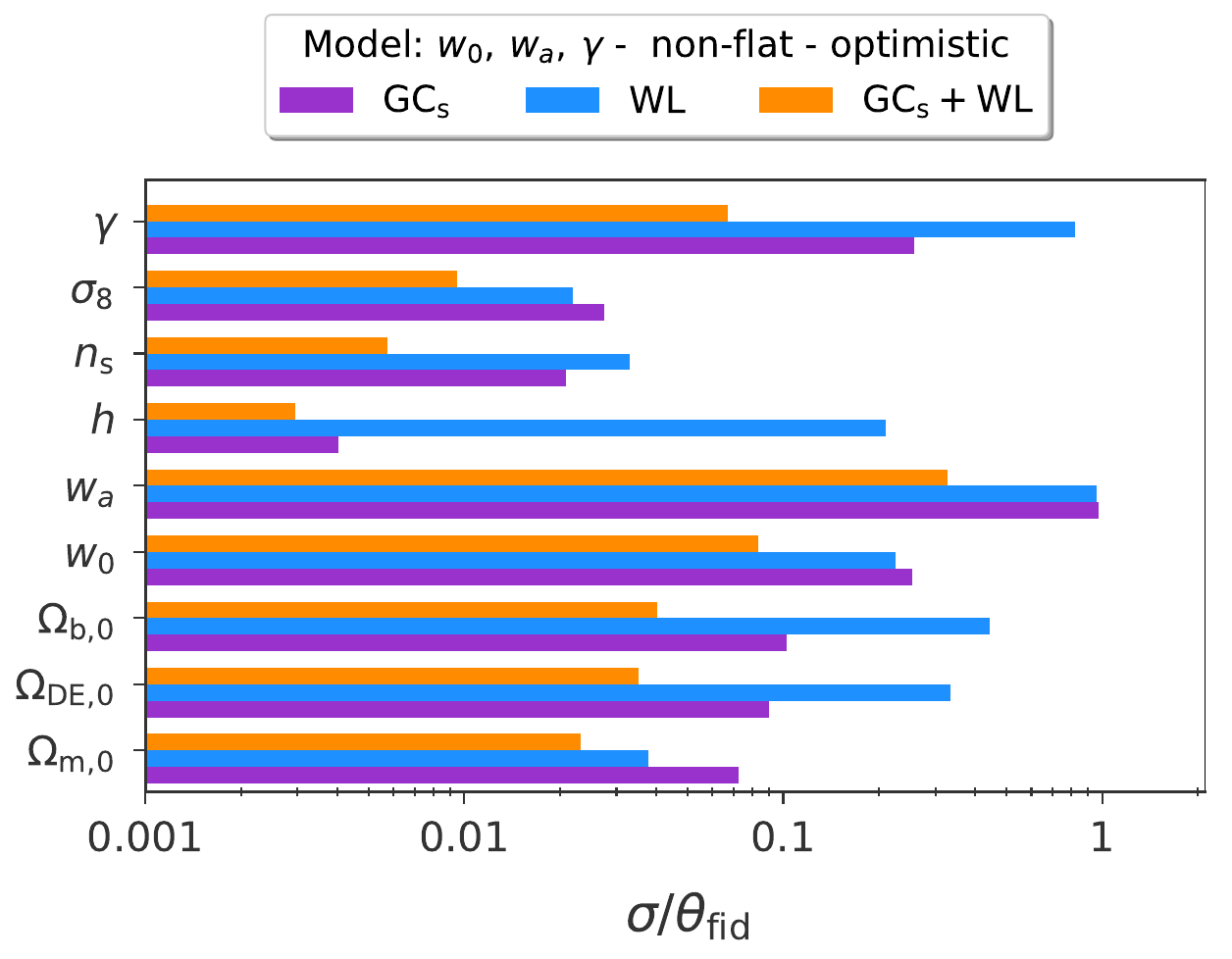}
\end{tabular}
\caption{
Marginalised $1 \sigma$ errors on cosmological parameters, relative to their corresponding fiducial values for a flat (upper panels) and non-flat (lower panels) spatial geometry, in ($w_{0}, w_a$, $\gamma$) cosmology. Left (right) panels correspond to pessimistic (optimistic) settings, as described in the text. The histogram refers to different observational probes: GC$_{\rm s}$ (purple), WL (blue), GC$_{\rm s}$+WL (orange); none of the available codes allow us to include $\rm{GC_{\rm ph}}$ or cross-correlation, while also allowing $\gamma$ to vary. For $w_a$ we show the absolute error, since a relative error is not possible for a fiducial value of $0$.} 
\label{fig:err-GC-WL_w0wagamma}
\end{figure}

We now consider the case in which both the background and the growth of structure are changing, as described in \autoref{sec:cosmo-growth}. 
In the $w_0, w_a, \gamma$ scenario the relative marginal errors for GC$_{\rm s}$, WL, and their combination, are shown in \autoref{fig:err-GC-WL_w0wagamma} for  flat and non-flat spatial cosmologies, for pessimistic (left panels) and optimistic (right panels) settings. The corresponding errors are also shown in \Cref{tab:rel-errors-w0wagflat,tab:rel-errors-w0wagnonflat}.
Including the extra parameter $\gamma$ into the analysis degrades the errors, as expected, since we are adding extra degeneracies. For a pessimistic setting, GC$_{\rm s}$ and WL are comparable, with the exception of $\Omega_{\rm m,0}$ (better determined by WL) and $h$ (better determined by GC$_{\rm s}$). The combination of the two probes is able to push the accuracy down to the few percent level even in the pessimistic setting, allowing us to also determine  $w_a$, which is nearly unconstrained by the single probes. 
For the optimistic case we have an improvement of  approximately a factor of $3$ for GC$_{\rm s}$  and of approximately $2$  for WL, with respect to the pessimistic case.

\begin{table}[h!]
\renewcommand{\arraystretch}{1.5}
\caption{Same as \autoref{tab:rel-errors-lcdm} for a flat $w_0, w_a, \gamma$ cosmology.}
\centering
\begin{tabularx}{\textwidth}{Xllllllll}
\hline
\rowcolor{thistle}\multicolumn{9} {c}{$\boldsymbol{w_0, w_a, \gamma}$ \, {\bf  flat}}\\ 
 & \multicolumn{1}{c}{$\Omega_{\rm m,0}$} & \multicolumn{1}{c}{$\Omega_{\rm b,0}$} & \multicolumn{1}{c}{$w_0$} & \multicolumn{1}{c}{$w_a$} & \multicolumn{1}{c}{$h$} & \multicolumn{1}{c}{$n_{\rm s}$} & \multicolumn{1}{c}{$\sigma_{8}$} & \multicolumn{1}{c}{$\gamma$} \\
\hline
\rowcolor{lavender(web)}\multicolumn{9} {l}{Linear setting} \\
GC$_{\rm s}$ & 0.10 & 0.13 & 0.28  & 0.83 & 0.0073 & 0.036 & 0.028 & 0.17 \\
\hline
\rowcolor{lavender(web)}\multicolumn{9} {l}{Pessimistic setting} \\ 
GC$_{\rm s}$ & 0.18 & 0.27 & 0.39 & 1.5 & 0.026 & 0.068 & 0.050 &  0.33 \\
WL & 0.042 & 0.47 & 0.25 & 1.5 & 0.21 &  0.036 &  0.024 &  0.37\\
GC$_{\rm s}$+WL & 0.030 & 0.062 & 0.096 & 0.33 & 0.0066 & 0.0092 & 0.013 &  0.070 \\
\hline
\rowcolor{lavender(web)}\multicolumn{9} {l}{Optimistic setting} \\ 
GC$_{\rm s}$ & 0.067  & 0.094 & 0.20 & 0.55 & 0.0037 & 0.020 & 0.021 & 0.25 \\
WL &0.034 & 0.43 & 0.17 & 0.96 & 0.20 & 0.033 & 0.013 & 0.18 \\
GC$_{\rm s}$+WL & 0.023 & 0.039 & 0.077 & 0.25 & 0.0028 & 0.0054 & 0.0095  & 0.052\\
\hline
\end{tabularx}
\label{tab:rel-errors-w0wagflat}
\end{table}
\begin{table}[h!]
\renewcommand{\arraystretch}{1.5}
\caption{Same as \autoref{tab:rel-errors-lcdm} for a non-flat $w_0, w_a, \gamma$ \, cosmology.}
\centering
\begin{tabularx}{\textwidth}{Xlllllllll}
\hline
\rowcolor{thistle} \multicolumn{10} {c}{$\boldsymbol{w_0, w_a, \gamma}$ \,{\bf non-flat}}\\ 
 & \multicolumn{1}{c}{$\Omega_{\rm m,0}$} & \multicolumn{1}{c}{$\Omega_{\rm DE,0}$} &\multicolumn{1}{c}{$\Omega_{\rm b,0}$} &  \multicolumn{1}{c}{$w_0$} & \multicolumn{1}{c}{$w_a$} & \multicolumn{1}{c}{$h$} & \multicolumn{1}{c}{$n_{\rm s}$} & \multicolumn{1}{c}{$\sigma_{8}$} & \multicolumn{1}{c}{$\gamma$} \\
\hline                                                          
\rowcolor{lavender(web)}\multicolumn{10} {l}{Linear setting}\\
GC$_{\rm s}$ & 0.11 &  0.075  & 0.13 & 0.28 & 0.92 & 0.0073 & 0.036 &  0.028 & 0.20\\
\hline
\rowcolor{lavender(web)}\multicolumn{10} {l}{Pessimistic setting} \\ 
GC$_{\rm s}$ & 0.18 & 0.11 &0.27 & 0.40 & 1.5 & 0.029 &  0.070 &  0.050 & 0.33\\
WL & 0.055 & 0.57 &  0.50 & 0.25& 1.9 & 0.23 & 0.037 & 0.029 & 1.1\\
GC$_{\rm s}$+WL & 0.033 & 0.055 & 0.068 & 0.12 & 0.55  & 0.0068  & 0.0092  &0.013  &0.076\\
\hline
\rowcolor{lavender(web)}\multicolumn{10} {l}{Optimistic setting} \\ 
GC$_{\rm s}$ & 0.072 & 0.090 & 0.10 &  0.25 & 0.98 & 0.0040 & 0.021 & 0.027 & 0.26\\
WL & 0.038 & 0.33 & 0.44 & 0.22 &  0.96 & 0.21 & 0.033 & 0.022 & 0.82 \\
GC$_{\rm s}$+WL & 0.023 & 0.035 &  0.040 &  0.084 &  0.33 &  0.0030 &  0.0057 & 0.0095 &  0.067\\
\hline
\end{tabularx}
\label{tab:rel-errors-w0wagnonflat}
\end{table}


\subsection{Overview results for galaxy clustering and weak lensing}
We have so far shown results for different probe combinations, for a given model. We finally collect here results for a given probe and different models, to have at hand how they perform when comparing different models and settings.
In \autoref{tab:rel-errors-allcases-gconly} we show the four settings tested for GC$_{\rm s}$ only, for all models and parameters.
\begin{table}
\renewcommand{\arraystretch}{1.5}
\caption{Marginalised $1 \sigma$ errors on cosmological parameters, relative to their corresponding fiducial value for a flat spatial geometry, in a $w_0, w_a, \gamma$ cosmology. We show results for the linear, pessimistic, and optimistic settings as well as the intermediate setting described in \autoref{def:intermediate}, for GC$_{\rm s}$ only.} 
\centering
\begin{tabularx}{\textwidth}{Xlllllllll}
\hline
\multicolumn{10}{c}{\bf Spectroscopic galaxy clustering, GC$_{\rm s}$} \\
\hline
Setting & \multicolumn{1}{c}{$\Omega_{\rm m,0}$} & \multicolumn{1}{c}{$\Omega_{\rm b,0}$} & \multicolumn{1}{c}{$\Omega_{\rm DE,0}$} & \multicolumn{1}{c}{$w_0$} & \multicolumn{1}{c}{$w_a$} & \multicolumn{1}{c}{$h$} & \multicolumn{1}{c}{$n_{\rm s}$} & \multicolumn{1}{c}{$\sigma_{8}$} & \multicolumn{1}{c}{$\gamma$} \\   
\hline
\rowcolor{cosmiclatte}\multicolumn{10} {c}{$\boldsymbol{\Lambda}${\bf CDM flat}}  \\
  Linear & 0.016 & 0.026 & \multicolumn{1}{c}{--} & \multicolumn{1}{c}{--} & \multicolumn{1}{c}{--} & 0.0031 & 0.0093 & 0.0071& \multicolumn{1}{c}{--} \\
  Pessimistic & 0.021 & 0.051 & \multicolumn{1}{c}{--} & \multicolumn{1}{c}{--} & \multicolumn{1}{c}{--} &  0.0063 & 0.014 & 0.0094 & \multicolumn{1}{c}{--} \\
  Intermediate & 0.019 & 0.047 & \multicolumn{1}{c}{--} & \multicolumn{1}{c}{--} & \multicolumn{1}{c}{--} &  0.0058 & 0.012 & 0.0085 & \multicolumn{1}{c}{--} \\
  Optimistic & 0.013 & 0.018 & \multicolumn{1}{c}{--} & \multicolumn{1}{c}{--} & \multicolumn{1}{c}{--} &  0.0017 & 0.0099 & 0.0077 & \multicolumn{1}{c}{--} \\
     \hline 
\rowcolor{cosmiclatte}\multicolumn{10} {c}{$\boldsymbol{\Lambda}${\bf CDM non-flat}} \\
Linear & 0.022 & 0.033 &  0.021 & \multicolumn{1}{c}{--} & \multicolumn{1}{c}{--} & 0.0067 & 0.013 &  0.0073 & \multicolumn{1}{c}{--} \\
Pessimistic & 0.053 & 0.092 & 0.035 & \multicolumn{1}{c}{--} & \multicolumn{1}{c}{--} & 0.017 & 0.028 & 0.0094 & \multicolumn{1}{c}{--} \\
Intermediate & 0.048 & 0.084 & 0.033 & \multicolumn{1}{c}{--} & \multicolumn{1}{c}{--} & 0.016 & 0.026 & 0.0086 & \multicolumn{1}{c}{--} \\
Optimistic & 0.015 & 0.018 & 0.021 & \multicolumn{1}{c}{--} & \multicolumn{1}{c}{--} & 0.0031 & 0.012 & 0.0085 & \multicolumn{1}{c}{--} \\
  \hline
\rowcolor{jonquil}  \multicolumn{10} {c}{$\boldsymbol{w_0, w_a} \, ${\bf flat}} \\
Linear & 0.080 & 0.10 & \multicolumn{1}{c}{--} & 0.20 & 0.66 & 0.0063 & 0.030 & 0.024  & \multicolumn{1}{c}{--} \\
Pessimistic & 0.17 & 0.26 & \multicolumn{1}{c}{--} &  0.39 & 1.4 & 0.020 & 0.063 & 0.045  & \multicolumn{1}{c}{--} \\
Intermediate & 0.14 & 0.21 & \multicolumn{1}{c}{--} &  0.31 & 1.1 & 0.017 & 0.048 & 0.036  & \multicolumn{1}{c}{--} \\
Optimistic & 0.051 & 0.068 & \multicolumn{1}{c}{--} &  0.15  & 0.45  & 0.0032 & 0.018 & 0.021  & \multicolumn{1}{c}{--} \\
 \hline
\rowcolor{jonquil}\multicolumn{10} {c}{$\boldsymbol{w_0, w_a} \, ${\bf non-flat}} \\
Linear & 0.083 & 0.10 & 0.075 & 0.23 & 0.89 & 0.0068 & 0.031 & 0.027 & \multicolumn{1}{c}{--} \\
Pessimistic & 0.18 & 0.26 & 0.11 & 0.39 & 1.5 & 0.023 & 0.066 & 0.045 & \multicolumn{1}{c}{--} \\
Intermediate &0.14 & 0.21 & 0.093 & 0.32 & 1.2 & 0.020 & 0.050 & 0.036 & \multicolumn{1}{c}{--} \\
Optimistic & 0.064 & 0.089 & 0.090 & 0.24 & 0.97 & 0.0038 & 0.020 & 0.027 & \multicolumn{1}{c}{--} \\
\hline
\rowcolor{thistle}\multicolumn{10} {c}{$\boldsymbol{w_0, w_a, \gamma} \, ${\bf flat}} \\
Linear & 0.10 & 0.13 & \multicolumn{1}{c}{--} & 0.28  & 0.83 & 0.0073 & 0.036 & 0.028 & 0.17 \\
Pessimistic & 0.181 & 0.27 & \multicolumn{1}{c}{--} &  0.39 & 1.5 & 0.026 & 0.068 & 0.050  & 0.33 \\
Intermediate & 0.14 & 0.21 & \multicolumn{1}{c}{--} &  0.30 & 1.2  & 0.021 & 0.051 & 0.040 & 0.29\\
Optimistic & 0.067  & 0.094 & \multicolumn{1}{c}{--} &  0.20 & 0.55 & 0.0037 & 0.020 & 0.021 & 0.25 \\
\hline
\rowcolor{thistle}\multicolumn{10} {c}{$\boldsymbol{w_0, w_a, \gamma} \, ${\bf non-flat}} \\
Linear & 0.11 & 0.13 & 0.075  & 0.28 & 0.92 & 0.0073 & 0.036  & 0.028 & 0.20\\
Pessimistic & 0.18  & 0.27 & 0.11 & 0.40 & 1.5 & 0.029 &  0.070  & 0.050 & 0.33\\
Intermediate & 0.14 & 0.22 & 0.094  & 0.32 & 1.2  & 0.024 & 0.053 & 0.040 & 0.29\\
Optimistic & 0.072 & 0.10 & 0.090 & 0.25 & 0.98 & 0.0040 & 0.021 & 0.027 & 0.26\\
\hline
\end{tabularx}
\label{tab:rel-errors-allcases-gconly}
\end{table}
In \autoref{tab:rel-errors-allcases-wlonly} we show the overview for WL only, for which we recall that we only tested the non-linear settings, since this is essential to grasp any information from this probe. 

\begin{table}
\renewcommand{\arraystretch}{1.5}
\caption{Marginalised $1 \sigma$ errors on cosmological parameters, relative to their corresponding fiducial value for a flat spatial geometry, in a $w_0, w_a, \gamma$ cosmology. We show results for the pessimistic and optimistic settings for WL only.} 
\centering
\begin{tabularx}{\textwidth}{Xlllllllll}
\hline
\multicolumn{10}{c}{\bf Weak lensing cosmic shear, WL} \\
\hline
Setting & \multicolumn{1}{c}{$\Omega_{\rm m,0}$} & \multicolumn{1}{c}{$\Omega_{\rm b,0}$} & \multicolumn{1}{c}{$\Omega_{\rm DE,0}$} & \multicolumn{1}{c}{$w_0$} & \multicolumn{1}{c}{$w_a$} & \multicolumn{1}{c}{$h$} & \multicolumn{1}{c}{$n_{\rm s}$} & \multicolumn{1}{c}{$\sigma_{8}$} & \multicolumn{1}{c}{$\gamma$} \\  
\hline
\rowcolor{cosmiclatte}\multicolumn{10} {c}{$\boldsymbol{\Lambda}${\bf CDM flat}}  \\
  Pessimistic & 0.017 & 0.47 & \multicolumn{1}{c}{--} & \multicolumn{1}{c}{--} & \multicolumn{1}{c}{--} &  0.21 & 0.034 & 0.0087 & \multicolumn{1}{c}{--} \\
  Optimistic & 0.012 & 0.42 & \multicolumn{1}{c}{--} & \multicolumn{1}{c}{--} & \multicolumn{1}{c}{--} &  0.20 & 0.031 & 0.0061 & \multicolumn{1}{c}{--} \\
     \hline 
\rowcolor{cosmiclatte}\multicolumn{10} {c}{$\boldsymbol{\Lambda}${\bf CDM non-flat}} \\
Pessimistic & 0.021 & 0.47 & 0.035 & \multicolumn{1}{c}{--} & \multicolumn{1}{c}{--} & 0.21 & 0.035 & 0.016 & \multicolumn{1}{c}{--} \\
Optimistic & 0.012 & 0.26 & 0.032 & \multicolumn{1}{c}{--} & \multicolumn{1}{c}{--} & 0.20 & 0.032 & 0.0074 & \multicolumn{1}{c}{--} \\
  \hline
\rowcolor{jonquil}  \multicolumn{10} {c}{$\boldsymbol{w_0, w_a} \, ${\bf flat}} \\
Pessimistic & 0.043 & 0.47 & \multicolumn{1}{c}{--} &  0.16 & 0.59 & 0.21 & 0.036 & 0.019  & \multicolumn{1}{c}{--} \\
Optimistic & 0.034 & 0.43 & \multicolumn{1}{c}{--} &  0.14  & 0.49  & 0.20 & 0.033 & 0.013  & \multicolumn{1}{c}{--} \\
 \hline
\rowcolor{jonquil}\multicolumn{10} {c}{$\boldsymbol{w_0, w_a} \, ${\bf non-flat}} \\
Pessimistic & 0.044 & 2.6 & 0.035 & 0.23 & 1.8 & 0.21 & 0.037 & 0.021 & \multicolumn{1}{c}{--} \\
Optimistic & 0.035 & 1.0 & 0.032 & 0.15 & 0.92 & 0.21 & 0.033 & 0.013 & \multicolumn{1}{c}{--} \\
\hline
\rowcolor{thistle}\multicolumn{10} {c}{$\boldsymbol{w_0, w_a, \gamma} \, ${\bf flat}} \\
Pessimistic & 0.042 & 0.47 & \multicolumn{1}{c}{--} &  0.25 & 1.5 & 0.21 & 0.036 & 0.024  & 0.33 \\
Optimistic & 0.034  & 0.43 & \multicolumn{1}{c}{--} &  0.17 & 0.96 & 0.20 & 0.033 & 0.013 & 0.18 \\
\hline
\rowcolor{thistle}\multicolumn{10} {c}{$\boldsymbol{w_0, w_a, \gamma} \, ${\bf non-flat}} \\
Pessimistic & 0.055  & 7.8 & 0.037 & 0.25 & 1.9 & 0.23 &  0.034  & 0.029 & 1.1\\
Optimistic & 0.038 & 4.6 & 0.033 & 0.23 & 0.96 & 0.21 & 0.033 & 0.022 & 0.82\\
\hline
\end{tabularx}
\label{tab:rel-errors-allcases-wlonly}
\end{table}
Finally, the combination of all probes, including cross-correlations, is summarised in \autoref{tab:rel-errors-allcases-xc} for convenience, for the available models.
\begin{table}
\renewcommand{\arraystretch}{1.5}
\caption{Marginalised $1 \sigma$ errors on cosmological parameters, relative to their corresponding fiducial value for a flat spatial geometry, in a $w_0, w_a$ cosmology. We show results for the pessimistic and optimistic settings for the full probe combination GC$_{\rm s}$+WL+GC$_{\rm ph}$+XC$^{^{\rm (GC_{\rm ph},WL)}}$.} 
\centering
\begin{tabularx}{\textwidth}{Xlllllllll}
\hline
\multicolumn{10}{c}{\bf All probe combination GC$_{\rm s}$+WL+GC$_{\rm ph}$+XC$^{^{\rm (GC_{\rm ph},WL)}}$} \\
\hline
Setting & \multicolumn{1}{c}{$\Omega_{\rm m,0}$} & \multicolumn{1}{c}{$\Omega_{\rm b,0}$} & \multicolumn{1}{c}{$\Omega_{\rm DE,0}$} & \multicolumn{1}{c}{$w_0$} & \multicolumn{1}{c}{$w_a$} & \multicolumn{1}{c}{$h$} & \multicolumn{1}{c}{$n_{\rm s}$} & \multicolumn{1}{c}{$\sigma_{8}$} & \multicolumn{1}{c}{$\gamma$} \\   
\hline
\rowcolor{cosmiclatte}\multicolumn{10} {c}{$\boldsymbol{\Lambda}${\bf CDM flat}}  \\
  Pessimistic & 0.0071 & 0.027 & \multicolumn{1}{c}{--} & \multicolumn{1}{c}{--} & \multicolumn{1}{c}{--} &  0.0037 & 0.0053 & 0.0033 & \multicolumn{1}{c}{--} \\
  Optimistic & 0.0025 & 0.011 & \multicolumn{1}{c}{--} & \multicolumn{1}{c}{--} & \multicolumn{1}{c}{--} &  0.0011 & 0.0015 & 0.0011 & \multicolumn{1}{c}{--} \\
     \hline 
\rowcolor{cosmiclatte}\multicolumn{10} {c}{$\boldsymbol{\Lambda}${\bf CDM non-flat}}  \\
  Pessimistic & 0.0073 & 0.030 & 0.011 & \multicolumn{1}{c}{--} & \multicolumn{1}{c}{--} &  0.0048 & 0.0055 & 0.0042 & \multicolumn{1}{c}{--} \\
  Optimistic & 0.0027 & 0.011 & 0.0060 & \multicolumn{1}{c}{--} & \multicolumn{1}{c}{--} &  0.0014 & 0.0016 & 0.0013 & \multicolumn{1}{c}{--} \\
     \hline 
\rowcolor{jonquil}  \multicolumn{10} {c}{$\boldsymbol{w_0, w_a} \, ${\bf flat}} \\
Pessimistic & 0.012 & 0.036 & \multicolumn{1}{c}{--} &  0.040 & 0.17 & 0.0055 & 0.0057 & 0.0051  & \multicolumn{1}{c}{--} \\
Optimistic & 0.0057 & 0.015 & \multicolumn{1}{c}{--} &  0.025  & 0.092  & 0.0015 & 0.0019 & 0.0021  & \multicolumn{1}{c}{--} \\
 \hline
\rowcolor{jonquil}  \multicolumn{10} {c}{$\boldsymbol{w_0, w_a} \, ${\bf non-flat}} \\
Pessimistic & 0.012 & 0.036 & 0.033 &  0.044 & 0.30 & 0.0057 & 0.0058 & 0.0062  & \multicolumn{1}{c}{--} \\
Optimistic & 0.0057 & 0.015 & 0.014 &  0.027  & 0.14  & 0.0015 & 0.0022 & 0.0023  & \multicolumn{1}{c}{--} \\
 \hline
\end{tabularx}
\label{tab:rel-errors-allcases-xc}
\end{table}
If we increase the maximum wavenumber $k$, i.e. comparing the pessimistic with the intermediate settings, we clearly gain more information because more scales are taken into account. We observe a decrease on the errors of about $10\%$ for the set of parameters in both $\Lambda$CDM flat and non-flat cases. For the other set of parameters, i.e. $w_0w_{a}$ and $w_0w_{a}\gamma$, for both flat and non-flat cases, the errors decrease by about $25-30\%$, with the exception of $\gamma$ for which we gain  $15\%$ information only. 

Finally, comparing the intermediate with the optimistic setting, i.e. keeping  $k_{\rm max} =0.3h\,{\rm Mpc}^{-1}$ but assuming perfect knowledge on the non-linear parameters $\sigma_{\rm p}$ and $\sigma_{\rm v}$, we gain information over all the cosmological parameters. For example, considering the projection case $w_0, w_{a}$ flat, the errors on the dark energy parameters decrease by about 100$\%$ and 150$\%$, for $w_0$ and $w_{a}$, respectively. If curvature is also considered in the analysis, the errors on $w_0$ and $w_{a}$ decrease by  $28\%$ only.

\section{Conclusions}
\label{sec:conclusions}
In this paper we have presented the validation of Fisher-matrix forecasts for the \Euclid satellite, as a result of activity carried out by the inter-science task force for forecasting, joining expertise from different science working groups (SWGs) of the \Euclid Collaboration. In particular, we have included in this analysis consideration of \Euclid primary probes, spectroscopic galaxy clustering and weak lensing cosmic shear, as well as the cross-correlation between WL and photometric galaxy clustering. Our target was to provide reliable numerical codes for each probe, which can be then used for a variety of forecasts and specifications -- including applications beyond the scope of this paper. We also provide full documentation and tools that will allow any new code owner to perform their own validation

At the start of this work, some numerical codes were already available, although they were providing different results, both in amplitude and in orientation. A few of them were producing similar results due to a similar validation carried out, separately, within the theory working group (TWG). However the recipe used for the different probes had not yet been developed or validated. Our task was therefore to join expertise from different SWGs and validate a common recipe for each probe, including ${\rm GC_{\rm s}}$, WL, ${\rm GC_{\rm ph}}$ and the cross correlation across each redshift bin for the two photometric probes. The first result of this analysis is therefore the documentation of each of these recipes in \autoref{Sec:Fisher}, which are now defined for the \Euclid Collaboration and for anyone else who would like to implement and/or validate their own numerical code. The development of the recipe allowed an improvement in all pre-existing codes, making sure that they are all aiming at describing the same features. In \autoref{sec:required-accuracy} we have also elaborated on the accuracy that  is required in Fisher matrix forecasts, i.e. on the parameter uncertainties, on the accuracy of the Fisher matrix approximation itself, and on how it is computed. We have concluded that to reach $\leq 10\%$ precision on parameter errors and FoM and $\leq 2\%$ precision on the orientation of the contours, the corresponding required fractional Fisher matrix precision (both in calculation of elements, and its inversion) is approximately $10^{-4}$, which is a reasonable aim for numerical implementations.

As a second step, we validated the way in which the implementation was performed, for each recipe and numerical code. In \autoref{sec:codecomparison} we provided a detailed documentation of each step, describing what works best, and how different codes compare to each other. If readers wish to compare and validate their own Fisher matrix code(s) for \Euclid, they should follow these instructions to do so.
As part of the code-comparison process, all codes were adapted to use common inputs. We make 
these inputs, alongside our outputs, available with this paper, describing them and providing instructions for their use in \autoref{sec:appendix}. 
One main product of this work is then the validation of several codes, some of which will be made publicly available so that they can be reliably used, in their present version, for forecasts. A description of all validated codes is provided in \autoref{sec:codecomparison} where we highlight their differences (in terms of which probe they consider, which programming language they use, which features they can deal with -- or example external input), as summarised in \autoref{tab:forecastcodes}. All validated codes match with an error on the parameters uncertainty of less than $10 \%$ which, as initially stated, was our requirements. We have further developed an appropriate visualisation that allows us to check how close the outputs of the codes are with respect to each other, for each probe: these are shown in \Cref{fig:resu-GC-zdep,fig:resu-GC-proj3} for ${\rm GC_{\rm s}}$, \autoref{fig:resu-WL-1} for WL, and \autoref{fig:xccompplot1} for probe combination. In \autoref{sec:appendix} we also provide the plotting scripts, to allow readers to compare  their own results or in order to overlap \Euclid Fisher matrices with those of other experiments.

Many tests have been carried out along the course of this work for all probes, involving different numerical strategies (for example how to calculate the derivatives), as well as different recipes (for example how to model the non-linear contribution, and how to take different cuts depending on the emerging presence of non-Gaussian contributions or baryons). A summary of the lessons we learned while performing such validation has been presented in \autoref{sec:lessons} for each probe. Further investigation will certainly be necessary to optimise the modelling at non-linear scales and will be an object of study for a new taskforce. In this paper we have identified  pessimistic and optimistic settings, summarised in \autoref{sec:results} and \autoref{def:optimistic}, for which we have shown final results. 

We have focused on the cosmological context described in \autoref{Sec:Cosmology} which includes: $\Lambda$CDM; a $(w_0, w_a)$ scenario in which the equation of state of dark energy is changing with time; the case $(w_0, w_a, \gamma)$ in which the growth is parameterised by $\gamma$ and allowed to change
compare to its fiducial value in general relativity. All cosmologies were tested within a spatially flat and non-flat geometry. For all models we present a series of histogram bar plots (\autoref{fig:err-GC-WL-lcdm} for $\Lambda$CDM and \autoref{fig:err-GC-WL-w0wa} for $w_0, w_a$; \autoref{fig:err-GC-WL_w0wagamma} for $w_0, w_a, \gamma$) that allow marginalised 1$\sigma$ errors on cosmological parameters to be visualised for different probe combinations. 

In the  $(w_0, w_a)$ case we can define a FoM \autoref{FoM} for which we show our final results in \autoref{w0wa_table_FoM} for different probe combinations. This table represents one of the main results of this paper. The FoM for the full combination of probes reaches 377 (1257) for a flat cosmology within a pessimistic (optimistic) setting. Values decrease as expected if a non-flat cosmology is allowed. 

We have demonstrated that the impact of cross-correlations in redshift among ${\rm GC_{\rm ph}}$ and WL is particularly relevant in models beyond $\Lambda$CDM: XC improves the FoM for a $(w_0, w_a)$ scenario by a factor of $\sim 3.6$ ($\sim 3.3$) in a pessimistic (optimistic) setting with respect to the case in which primary probes and ${\rm GC_{\rm ph}}$ are combined as if they were independent probes. We have visually shown how including XC removes degeneracies among parameters  (\autoref{fig:correlationmatrix}). Within the $\Lambda$CDM scenario, instead, XC can have a negligible impact.  

All \Euclid specifications used in this paper are detailed in \autoref{tab:GC-specifications} for the spectroscopic survey and in \autoref{tab:WL-specifications} for the photometric survey. In particular, the expected number density of observed H$\alpha$ emitters for the spectroscopic survey shown in \autoref{tab:gc_nofz} has been updated since the Red Book \citep{RB} to match new observations of number densities and new instrument and survey specifications. Further modifications, such as the different areas and depth specifications motivated from the \Euclid flagship simulations used within the Science Performance Verification (SPV), may still be possible in the future and will now be straightforward to implement, given a validated forecast pipeline. Since these simulation-derived quantities will change before publication of any SPV results, we chose to remain as close as possible to the Red Book specifications with regard to the photometric survey; in future work photometric values will also be updated.

Most importantly, we have provided a validated framework and reliable numerical codes to perform forecasts. Although we restricted the analysis to primary probes and cross correlations, reliable codes are now available as a baseline for the analysis of other probes (such as the CMB) and for further extensions beyond the cosmologies investigated in this paper. Our results and numerical codes can be implemented for all future updates in the specific settings chosen for \Euclid, and are currently used, for example, for the process of SPV.

\section*{Acknowledgements}
\addcontentsline{toc}{section}{Acknowledgements}

{\sl Authors' contributions}: All
authors have significantly contributed to this publication and/or
made multi-year essential contributions to the {\it Euclid}
project which made this work possible. The paper-specific main
contributions of the lead authors are as follows:
\emph{Coordinators and corresponding contacts}: Thomas Kitching, Valeria Pettorino, Ariel S\'anchez. \emph{Galaxy Clustering Task}: Ariel G. S\'anchez, Domenico Sapone, Carmelita Carbone, Santiago Casas, Katarina Markovic, Alkistis Pourtsidou, Safir Yahia-Cherif, Elisabetta Majerotto, Victoria Yankelevich. \emph{Weak Lensing Task}: Vincenzo F. Cardone, Santiago Casas, Stefano Camera, Matteo Martinelli, Isaac Tutusaus, Thomas Kitching, Martin Kilbinger. \emph{Cross-correlations Task}: Matteo Martinelli, Isaac Tutusaus, Alain Blanchard,  Martin Kunz, Fabien Lacasa, Vincenzo F. Cardone, Sebastien Clesse, St\'ephane Ili\'{c}, Martin Kilbinger, Marco Raveri, Ziad Sakr. \emph{Cosmological context}: Valeria Pettorino, Eric Linder, Martin Kunz, Matteo Martinelli, Martin Kilbinger, Alkistis Pourtsidou, Ariel G. S\'anchez, Thomas Kitching, Carmelita Carbone, Ziad Sakr. \emph{Paper editing}: Valeria Pettorino, Ariel G. S\'anchez, Thomas Kitching, Domenico Sapone, Matteo Martinelli, Santiago Casas, Katarina Markovic, Isaac Tutusaus, Stefano Camera, Alkistis Pourtsidou, Vincenzo Cardone, Ziad Sakr. 

We acknowledge \AckInstitutions
Stefano Camera is supported by the Italian Ministry of Education, University and Research (MIUR) through Rita Levi Montalcini project `\textsc{prometheus} -- Probing and Relating Observables with Multi-wavelength Experiments To Help Enlightening the Universe's Structure', and by the `Departments of Excellence 2018-2022' Grant awarded by MIUR (L.\ 232/2016). Carmelita Carbone acknowledges financial support from the European Research Council through the Darklight Advanced Research Grant (n. 291521), and from the MIUR PRIN 2015 `Cosmology and Fundamental Physics: illuminating the Dark Universe with Euclid'. Santiago Casas and Safir Yahia-Cherif acknowledge support from french space agency CNES. Sebastien Clesse aknowledges support from the Belgian Fund for Research F.R.S-FNRS.  Martin Kunz and Fabien Lacasa acknowledge financial support from the Swiss National Science Foundation. Eric Linder acknowledges support by NASA ROSES grant 12-EUCLID12-0004. Katarina Markovic carried out some of the work at the Jet Propulsion Laboratory, California Institute of Technology, under a contract with the National Aeronautics and Space Administration (80NM0018D0004), and acknowledges support from the UK Science \& Technology Facilities Council through grant ST/N000668/1, and from the UK Space Agency through grant ST/N00180X/1. Matteo Martinelli acknowledges support from the D-ITP consortium, a program of the NWO that is funded by the OCW. Alkistis Pourtsidou is a UK Research and Innovation Future Leaders Fellow and also acknowledges support from the UK Science \& Technology Facilities Council through grant ST/S000437/1. Domenico Sapone acknowledges financial support from the Fondecyt Regular project number 1200171. Isaac Tutusaus acknowledges support from the Spanish Ministry of Science, Innovation and Universities through grant ESP2017-89838-C3-1-R, and the H2020 programme of the European Commission through grant 776247. Victoria Yankelevich acknowledges financial support by the Deutsche Forschungsgemeinschaft through the Transregio 33 `The Dark Universe', the International Max Planck Research School for Astronomy and Astrophysics at the Universities of Bonn and Cologne and the Bonn-Cologne Graduate School for Physics and Astronomy. Thomas D. Kitching acknowledges support from a Royal Society University Research Fellowship. St\'ephane Ili\'c acknowledges financial support from the European Research Council under the European Union's Seventh Framework Programme (FP7/2007--2013)/ERC Grant Agreement No. 617656 ``Theories and Models of the Dark Sector: Dark Matter, Dark Energy and Gravity.'' Elisabetta Majerotto acknowledges financial support from the Swiss National Science Foundation.

\begin{appendix}
\section{How you can perform your own code comparison}  \label{sec:appendix}
In this Appendix we address the following two questions that the reader may have:
\begin{description}
\item[- ] I have another Fisher matrix code, how can I validate my own code and its output?
\item[- ] How can I overplot \Euclid marginalised posterior contours with other contours (for example from another experiment)?
\end{description}

In \autoref{Sec:Fisher} we described the recipes required to calculate forecasts for \Euclid's main cosmological probes, and their combination; in \autoref{sec:codecomparison} we showed how to validate different codes against each other and the presented plots that allow to visualise the comparison. 
We now provide the tools needed to build a Fisher matrix and compare it with representative results of this paper.
To address the second question, we also provide plotting routines and reference Fisher matrices to reproduce a plot like the one in \autoref{fig:ellipses-w0wa}. These tools can be found in the public repository \url{https://github.com/euclidist-forecasting/fisher_for_public}. We describe its content  in detail below.

\subsection{Content of the repository}

The public \Euclid IST:F repository contains representative Fisher matrices obtained following the recipes of \autoref{Sec:Fisher}. One \Euclid IST:F reference Fisher matrix is provided for each probe and cosmology.

These matrices are in the \texttt{All\_results} folder, divided in the \texttt{optimistic} and \texttt{pessimistic} subfolders, corresponding
to the two sets of specifications used in this paper.
These subfolders, further contain a \texttt{flat} and \texttt{non-flat} 
folder each.
Within these, the files containing the Fisher matrices follow the naming convention
\begin{eqnarray}
{\rm \texttt{EuclidISTF\_Observables\_CosmoModel\_SpecificationsCase.txt}}\nonumber
\end{eqnarray}
For example, a Fisher matrix whose file name is
\begin{eqnarray}
{\rm \texttt{EuclidISTF\_GCph\_WL\_XC\_w0wa\_flat\_optimistic.txt}}\nonumber
\end{eqnarray}
corresponds to the combined Fisher matrix for photometric galaxy clustering (GCph), weak lensing (WL) and their cross correlation (XC), obtained in the flat case (flat) for the flat $w_0,w_a$ cosmology (w0wa), using the optimistic set of specifications (optimistic). Each Fisher matrix contains a header listing the cosmological parameter of the matrix and their order.

In addition, the repository contains a set of \texttt{Python} scripts that can be used to plot the contour ellipses and compare them with user-provided matrices.

\subsection{How to produce a Fisher matrix using IST:F input} \label{ISTF:input}

The \Euclid Fisher matrices contained in the public repository are obtained following the recipes described in \autoref{Sec:Fisher} for the cases analysed in \autoref{sec:codecomparison} and \autoref{sec:results}. Users who wish to validate their own code can build their Fisher matrices following these recipes, using the input files for cosmological quantities that are also included within the repository. These input files have been generated using \texttt{CAMB}, version {August 2018}. The input files containing the cosmological quantities used to create the Fisher matrices can
be found at \url{http://pc.cd/0wJotalK}.
The link contains two folders: \texttt{GC} and \texttt{WL} which are described below.

\paragraph{{\bf GC folder}} 
The GC folder contains the zip file \texttt{ISTF\_GC.zip}. Inside you will find: 
\begin{itemize}
\item the file {\texttt{fiducial\_rate\_growth.dat}}, containing three columns with $z$, growth rate $f(z)$ -- as defined in \autoref{eq:fgrowth} -- and growth function $D(z)$ -- as defined in \autoref{eq:growthfactor-gc-val}; 
\item the subfolder \texttt{Pk}, containing  the $P_{\delta\delta}(k,z)$, as defined in \autoref{eq: pddlinMath};
\item the subfolder \texttt{Pk-nw} containing the $P_{\rm nw}(k,z)$ that appears in \autoref{eq:pk_dw}.
\end{itemize}
Each of the two subfolders contains:
\begin{itemize}
\item the fiducial folder, including the $P_{\delta\delta}(k,z)$ for the fiducial cosmology, with one file for each redshift at which this spectrum is computed. Each file has a suffix \texttt{z\_ii}, where \texttt{ii} runs from \texttt{00}
to \texttt{03} corresponding to the four redshift bins used for the \Euclid ${\rm GC_{\rm s}}$ probe $z=\{1.0,\,1.2,\,1.4,\,1.65\}$,
\item the power spectra used to calculate the derivatives, grouped into folders with names \texttt{par\_step\_eps\_1p0E-2}, with \texttt{step} being the name of the parameter, \texttt{par} being either \texttt{pl} indicating a positive step or \texttt{mn} indicating a negative step. These folders contain, in particular, the power spectra for each step in each cosmological parameter. The steps amplitudes are relative; for each parameter the absolute step will be ${\rm \texttt{fidpar}} (1+{\rm \texttt{eps}})$ with \texttt{fidpar} the fiducial value of the parameter and \texttt{eps} the value of the relative step.
\end{itemize}
All the power spectrum files have three columns, being: 
\begin{equation*}
k \, [h/{\rm Mpc}] \,\,\,\, ; \,\,\,\,   {\rm linear} \,\,P(k,z)/\sigma_8^2(z)\, [{\rm Mpc}^3/h^3] \,\,\,\, ;  \,\,\,\, \sigma_8(z)\, . \nonumber 
\end{equation*}
\paragraph{{\bf WL folder}}
The WL folder contains:
\begin{itemize}
\item the file \texttt{scaledmeanlum-E2Sa.dat}, which contains $z$ (first column) and the ratio $\langle L \rangle(z)/L_{\star} (z)$ (second column) described in \autoref{eq: effeia};
\item \texttt{ISTF\_WL\_Flat.zip}, the matter power spectrum $P_{\delta\delta}(k,z)$ for the flat case;
\item \texttt{ISTF\_WL\_NonFlat.zip}, the matter power spectrum $P_{\delta\delta}(k,z)$ for the non-flat case;
\end{itemize}
Each of these two zip files contains: 
\begin{itemize}
\item the file \texttt{pkz-Fiducial.txt}, with the fiducial power spectrum;
\item the spectra relative to the positive (negative) steps in cosmological parameters, e.g. \texttt{ns\_pl\_eps\_5p0E-2} indicates a step on $n_s$  in the positive direction of amplitude $5.0\times10^{-2}$ (the steps amplitudes are relative; for each parameter the absolute step will be ${\rm \texttt{fidpar}} (1+{\rm \texttt{eps}})$ with \texttt{fidpar} the fiducial value of the parameter and \texttt{eps} the value of the relative step).
\end{itemize}
Notice that with respect to the GC input files, here many more steps for numerical derivatives are computed; this is because most of the IST:F codes for WL use the derivative scheme described in \autoref{sec:ccwl}. All the $P_{\delta\delta}(k,z)$ files have four columns: 
\begin{equation*}
z \,\,\,\, ; \,\,\,\,   k \, [h/{\rm Mpc}] \,\,\,\, ;  \,\,\,\, P_{\delta\delta}(k,z)\, {\rm linear}\, [{\rm Mpc}^3/h^3] \,\,\,\, ; \,\,\,\,  P_{\delta\delta}(k,z)\, {\rm non}-{\rm linear} \, [{\rm Mpc}^3/h^3]\, . \nonumber 
\end{equation*}

\subsubsection{How to include an external matrix}
In order to compare their own Fisher matrix with the IST:F results, users can use the scripts provided in the following. For that to work, users should add their matrix in the folder corresponding to the case of interest, following the same organisation of the \texttt{EuclidISTF} matrices. The filenames of the new matrices, must follow the convention of the repository, i.e. with a prefix stating the
name of the code used, and a suffix containing the observables included in the matrix, the cosmological model used
and the specification settings. As an example, a Fisher matrix provided by the user for WL in the optimistic case, for a flat $w_0,w_a$ cosmology should have the path

\begin{verbatim}
All_results/optimistic/flat/USERFISHER_WL_w0wa_flat_optimistic.txt
\end{verbatim}

The full description of the settings, with all the corresponding specifications used to obtain the matrices, can be
found in \autoref{sec:codecomparison}.
Furthermore, each matrix file provided by users must contain the header listing the cosmological parameters and
their order (see any of the \texttt{EuclidISTF} matrices for an example).

\subsection{How to compare an external Fisher matrix}
Within the main repository \url{https://github.com/euclidist-forecasting/fisher_for_public}, we also provide a comparison package  which consists of:
\begin{itemize}
\item a common trunk, \texttt{comparison.py} found in the parent folder.
\item three applications (\texttt{appGC.py}, \texttt{appWL.py}, \texttt{appXC.py}) found in \texttt{PlottingScripts/} handling the different comparisons
\item a set of plotting and analysis libraries found in \texttt{PlottingScripts/}
\end{itemize}
The python script performing the comparison between an external Fisher matrix and the reference IST:F ones can be launched from the parent folder using

\begin{verbatim}
python comparison.py [-o OBSERVABLES ] [-f USERFISHER ]
\end{verbatim}
The \texttt{-o} and \texttt{-f} arguments are optional.

\texttt{USERFISHER} are the labels of the user provided Fisher matrices. Notice that the user's matrices must be placed in the correct folders of the repository and must follow the filename notation described above, with \texttt{USERFISHER} the prefix of the filename, i.e. the code name.
The script will produce plots, similar to the comparison plots of \autoref{sec:codecomparison}, only for those cases for which an external matrix to compare with is found.
In case no \texttt{USERFISHER} argument is provided, a test matrix will be used. This test matrix is contained within the repository only for the cross-correlation comparison cases (see \autoref{sec:ccxc}).

The \texttt{OBSERVABLES} can be one or more of the cases considered in \autoref{sec:codecomparison}, i.e.
\begin{itemize}
\item GC for spectroscopic galaxy clustering
\item WL
\item XC 
\end{itemize}

The script can perform one, two or all cases at once. The latter is the default selection if no observable is specified.

Results will be placed in results folders in each of the subcases locations, e.g. in the parent folder one will find.
\begin{verbatim}
All_results/optimistic/flat/results_XC
\end{verbatim}
These results folders are automatically generated by the comparison script.

As an example, let's assume that a user wants to compare a WL matrix computed in the optimistic case, for a flat $w_0,w_a$ cosmology. The user will add the matrix as

\begin{verbatim}
All_results/optimistic/flat/MyName_WL_w0wa_flat_optimistic.txt
\end{verbatim}

Running the script with the command

\begin{verbatim}
python comparison.py -o WL -f MyName
\end{verbatim}

\noindent will compare the new matrix with the \Euclid IST:F for this specific case and cosmology, which will be found in 

\begin{verbatim}
All_results/optimistic/flat/results_WL .
\end{verbatim}

Notice that, as said above, the script does not produce comparison results for those cases in which no external matrix is provided. In the example above, where only the WL, optimistic matrix for $w_0,w_a$ is provided, the script will produce the results only for this specific case also running 
\begin{verbatim}
python comparison.py -f MyName
\end{verbatim}

\subsection{How to obtain results for the IST:F and external matrices}

The python script performing the IST:F ellipses plot can be launched from the parent folder using

\begin{verbatim}
 python ellipses.py [-o OBSERVABLES ] [-f USERFISHER ] [-n USERNAME ] [-c USERCOLOR]
\end{verbatim}
The \texttt{-o}, \texttt{-f}, \texttt{-n} and \texttt{-c} arguments are optional.

In case no arguments are provided,  the script will reproduce \autoref{fig:ellipses-w0wa} from the paper.

With the option \texttt{-o} a list of \texttt{OBSERVABLES} can be specified, separated by spaces. Available options are: 
\begin{itemize}
	\item \texttt{All} produces a plot like \autoref{fig:ellipses-w0wa} for different observables;
	\item \texttt{XC} produces a plot focusing on \texttt{XC} correlations, like \autoref{fig:ellipses-w0wa};
	\item \texttt{GC} produces a plot containing only $\textrm{GC}_{\rm s}$ contours;
	\item \texttt{WL} produces a plot containing only $\textrm{WL}$ contours.
\end{itemize}

With the option \texttt{-f}, a \texttt{USERFISHER} can be specified, which is the \textit{relative path} to the Fisher matrix provided by the user. Notice that the user's matrices must be placed in the correct folders of the repository and must follow the filename notation described above.

If an external user Fisher matrix is provided, the script will overplot it on top of the other ellipses using  a colour and a label name provided by the user. 
Default colour: \texttt{black}, default name: \texttt{USER}.

The user can change the colour and the label name of its Fisher matrix, with the option \texttt{-c} and \texttt{-n}, respectively.

Ellipses plots will be placed in the following folder, e.g.
\begin{verbatim}
All_results/optimistic/flat/results_ellipses
\end{verbatim}

Additionally to the elliptical contours, the routine also produces a file called:
\texttt{w0waCDM-flat-optimistic\_bounds.txt}
containing the relative errors on all the parameters of interest and a file
\texttt{w0waCDM-flat-optimistic\_FoMs.txt}
containing the FoM for $w_0$, $w_{\rm a}$ corresponding to each of the Fisher matrices which were plotted.

The \texttt{.ini} files inside the directory \texttt{PlottingScripts/} store all the parameters of the plot, and provide a way for the user to gain finer control of the plotting routine, if needed.

\end{appendix}

\bibliographystyle{aa}
\bibliography{\filepath{references}}

\end{document}